\documentclass[a4paper,fleqn,usenatbib]{mnras}

\usepackage{newtxtext,newtxmath}

\usepackage[T1]{fontenc}
\usepackage{ae,aecompl}
\usepackage{epsfig}
\usepackage{epstopdf}

\usepackage{graphicx}	
\usepackage{amsmath}	
\usepackage{amssymb}	
\usepackage{wasysym}
\usepackage{latekpackages/booktabs/booktabs}
\usepackage{csvsimple}
\usepackage{subfig}
\usepackage{tablefootnote}
\usepackage{csquotes}
\usepackage{textcomp}
\usepackage[english]{babel}
\usepackage{lscape}
\usepackage{longtable}
\usepackage{xcolor}
\usepackage{soul}
\usepackage[version=4]{mhchem}

\newcommand{\Rp}{R_{\textnormal{p}}}
\newcommand{\Rs}{R_{\star}}
\newcommand{\RpRs}{\Rp/\Rs}
\newcommand{\water}{H$_2$O\,\,}

\newcommand{\mic}{\,\muup \mathrm{m}}

\newcommand{\corr}{}

\title[Library of Exoplanet Atmospheres]{A Library of Self-Consistent Simulated Exoplanet Atmospheres}

\author[Goyal et al.]{Jayesh M. Goyal$^{1,2}$\thanks{E-mail: jgoyal@astro.cornell.edu},
Nathan Mayne$^{2}$, Benjamin Drummond$^{2,3}$, David K. Sing$^{4,5,2}$,
\newauthor Eric H\'ebrard$^{2}$, Nikole Lewis$^{1}$, Pascal Tremblin$^{6}$, Mark W. Phillips$^{2}$, Thomas Mikal-Evans$^{7}$,
\newauthor Hannah R. Wakeford$^{8}$
\\
$^{1}$Department of Astronomy and Carl Sagan Institute, Cornell University, 122 Sciences Drive, Ithaca, NY, 14853, USA \\
$^{2}$Astrophysics Group, School of Physics and Astronomy, University of Exeter, Exeter EX4 4QL, UK \\
$^{3}$Met Office, Fitzroy Road, Exeter, EX1 3PB, UK\\
$^{4}$Department of Earth and Planetary Sciences, Johns Hopkins University, Baltimore, MD, USA \\
$^{5}$Department of Physics \& Astronomy, Johns Hopkins University, Baltimore, MD, USA\\
$^{6}$Malson de la Simulation, CEA-CNRS-INRIA-UPS-UVSQ, USR 3441, Centre detude de Saclay, France \\
$^{7}$Kavli Institute for Astrophysics and Space Research, Massachusetts Institute of Technology, 77 Massachusetts Avenue, Cambridge, MA 02139, USA \\
$^{8}$School of Physics, University of Bristol, HH Wills Physics Laboratory, Tyndall Avenue, Bristol BS8 1TL, UK\\
}

\date{Accepted 2020 July 29. Received 2020 July 29; in original form 2020 April 7}

\pubyear{2019}

\begin{document}
\label{firstpage}
\pagerange{\pageref{firstpage}--\pageref{lastpage}}
\maketitle

\begin{abstract}

We present a publicly available library of model atmospheres with radiative-convective equilibrium Pressure-Temperature ($P$-$T$) profiles fully consistent with equilibrium chemical abundances, and the corresponding emission and transmission spectrum with R$\sim$5000 at 0.2 \textmu m decreasing to R$\sim$35 at 30 \textmu m, for 89 hot Jupiter exoplanets, for four re-circulation factors, six metallicities and six C/O ratios. We find the choice of condensation process (local/rainout) alters the $P$-$T$ profile and thereby the spectrum substantially, potentially detectable by JWST.  We find H$^-$ opacity can contribute to form a strong temperature inversion in ultra-hot Jupiters for C/O ratios $\geq$ 1 and can make transmission spectra features flat in the optical, alongside altering the entire emission spectra. We highlight how adopting different model choices such as thermal ionisation, opacities, line-wing profiles and the methodology of varying the C/O ratio, effects the $P$-$T$ structure and the spectrum. We show the role of Fe opacity to form primary/secondary inversion in the atmosphere. We use WASP-17b and WASP-121b as test cases to demonstrate the effect of grid parameters across their full range, while highlighting some important findings, concerning the overall atmospheric structure, chemical transition regimes and their observables. Finally, we apply this library to the current transmission and emission spectra observations of WASP-121b, which shows H$_2$O and tentative evidence for VO at the limb, and H$_2$O emission feature indicative of inversion on the dayside, with very low energy redistribution, thereby demonstrating the applicability of library for planning and interpreting observations of transmission and emission spectrum.

\end{abstract}

\begin{keywords}
planets and satellites: atmospheres -- planets and satellites: composition -- planets and satellites: gaseous planets -- techniques: spectroscopic
\end{keywords}



\section{Introduction}
\label{section:Introduction}

\begin{figure}
\centering
\includegraphics[scale=0.5]{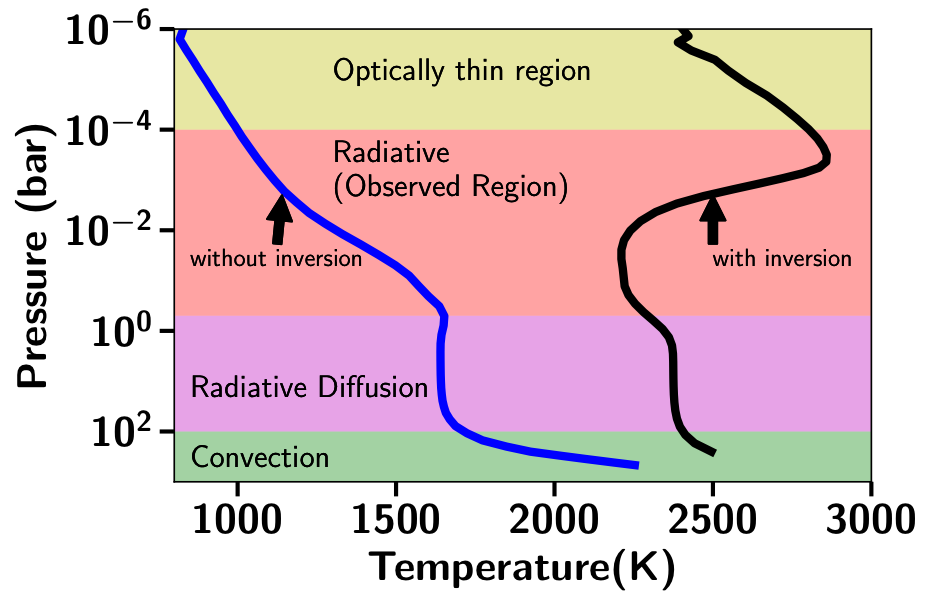}
 \caption{Illustrative figure to show dominant processes in different parts of a hot Jupiter atmosphere with radiative-convective equilibrium $P$-$T$ profiles, with (black) and without inversion (blue).}
    \label{fig:ch5_pt}
\end{figure}

The thermal or the pressure-temperature ($P$-$T$) structure of a planetary atmosphere is a result of constant feedback between radiative, advective, and chemical processes. Determining the $P$-$T$ structure of a planet is important for understanding underlying thermochemical and dynamical processes. The $P$-$T$ structure of the planetary atmosphere also governs its spectral signatures, when remotely observed using telescopes, spacecrafts or satellites. Therefore, it is necessary to constrain the $P$-$T$  structure of a planet's atmosphere, to understand the various physical processes occurring within them as shown in Figure \ref{fig:ch5_pt}. Constraining the three dimensional $P$-$T$ structure of the planetary atmosphere is ideally required, but the complexity and the computational resources required for such a model, especially when constructing a library of forward model simulations, motivates one dimensional (1D) $P$-$T$ profiles. 

In \citet{Goyal2018},  \citet{Errgoyal2019} and \citet{Goyal2019} we used isothermal $P$-$T$ profiles and corresponding equilibrium chemical abundances for computing transmission spectra for a wide range of exoplanet atmospheres. This extensive library of more than a million model simulations is publicly available and has proved to be very useful in interpreting observations of various exoplanets using the Hubble Space Telescope (HST), Spitzer and Very Large Telescope (VLT) \citep[see for e.g][]{Wakeford2018,Alam2018,Zhang2019,Carter2020} and for planning future observations\footnote{\url{https://exoctk.stsci.edu/generic}} using various telescope facilities, including the James Webb Space Telescope (JWST). However, the assumption of an isothermal $P$-$T$ profiles is only accurate for a small region of the atmosphere, specifically the high-altitude, low-pressure regions probed by transmission spectra. In fact, the atmosphere even in this region may rarely be exactly isothermal, but current transmission spectra observations cannot differentiate between such small temperature changes in $P$-$T$ profiles \citep[][]{Fortney2005, Heng2017, Goyal2018}, which may be revealed by higher precision, higher resolution, and broader wavelength coverage observations.  Furthermore, truly isothermal $P$-$T$ profiles would give rise to a simple black body emission spectrum, devoid of spectral features. The emission spectrum is much more strongly and directly dependent on the temperature than the transmission spectrum (Emission Flux $\propto$ T$^4$). Moreover, emission spectrum reveals the dayside/nightside of the exoplanet atmosphere, providing altogether different information than the transmission spectrum, which measures the limb. Emission spectrum also probes deeper (higher pressure) regions of the planetary atmosphere in comparison to transmission spectrum. Therefore, to identify features in the emission spectrum as well as to constrain the $P$-$T$ profile using these features, we require computation of a more accurate non-isothermal $P$-$T$ profile.

1D $P$-$T$ profiles of irradiated H$_2$/He dominated planetary atmospheres with sufficiently high equilibrium temperatures are expected to reach a radiative-convective equilibrium condition \citep[see for e.g][]{Marley1996, Burrows1997, Seager1998, Sudarsky2003, Marley2014, Molliere2015, Gandhi2017, Malik2018}. This is basically a result of two factors, first the exposure to strong irradiation from the host star, rapidly forcing perturbations in the $P$-$T$ structure back to a radiative equilibrium in the lower pressure regions (< $\sim$100 bar), and second the dominance of convection in the deep atmosphere forcing perturbations in the $P$-$T$ structure to convective equilibrium in the higher pressure regions (> $\sim$100 bar). Therefore, combining the lower pressure regions, and higher pressure regions, these atmospheres are likely to exist close to a radiative-convective equilibrium, termed RCE hereafter. Additionally, assuming chemical equilibrium to model hot Jupiter atmospheres is likely a reasonable assumption, due to the high temperature of these planets, especially where chemical timescales are likely to be short, particularly for temperatures above $\sim$ 2000K. The assumption of chemical equilibrium is also a good starting point to constrain the main atmospheric constituents for lower temperature planets \citep[see for e.g][]{Burrows1999, Lodders2002, Visscher2006, Marley2014, Madhureview2016}. 

RCE $P$-$T$ profiles for different planets can vary depending on their gravity, host star distance and spectral type, circulation in the planet's atmosphere and the chemical composition of the atmosphere. This can then lead to a wide range of possible spectra for a given planet, governed by its $P$-$T$ profile. Therefore, a library of RCE $P$-$T$ profiles and the corresponding chemical abundances and simulated spectra is required to interpret the observations of exoplanet atmospheres and constrain the important physical processes occurring within them. In this work we compute $P$-$T$ profiles in radiative - convective equilibrium for 89 observationally significant exoplanets, along with their corresponding equilibrium chemical abundances, simulated transmission and emission spectra with resolution of R$\sim$5000 at 0.2 \textmu m decreasing to R$\sim$35 at 30 \textmu m, and contribution functions, which are publicly available here \footnote{\url{https://drive.google.com/drive/folders/1zCCe6HICuK2nLgnYJFal7W4lyunjU4JE}}$^,$\footnote{\url{https://noctis.erc-atmo.eu:5001/fsdownload/hq0z4udQJ/goyal2020}}.

We start by describing the 1D-2D atmosphere model ATMO \citep{Amundsen2014, Tremblin2015, Tremblin2016, Tremblin2017, Drummond2016, Goyal2018} and the procedure to compute RCE $P$-$T$ profiles, followed by recent additions to ATMO in Section \ref{ch5:model_des}. We detail the numerical setup of this grid with RCE $P$-$T$ profiles in Section \ref{ch5:Grid Setup}, including the chemistry and opacity setup. In Section \ref{ch5:grid_par_space} we describe the parameter space of the grid. We discuss the results in Section \ref{ch5:results}, where we show the sensitivity of model simulations to different model choices by comparing RCE $P$-$T$ profiles, equilibrium chemical abundances and the transmission/emission spectra in Section \ref{ch5:model_choice_sensitivity}. In Section \ref{ch5:high_level_irrad} we show the effects caused by high levels of irradiation by using extremely irradiated hot Jupiter WASP-121b as the test case. Sensitivity to grid parameters, namely; recirculation factor, metallicity and C/O ratio is discussed is Section \ref{ch5:grid_param_sensitivity} using WASP-17b and WASP-121b as test cases. In Section \ref{ch5:obs_int}, transmission and emission spectrum observations of WASP-121b are interpreted using the grid of models presented in this work.  Finally, we summarize and conclude in Section \ref{ch5:summary}. The implementation and validation of thermal ionization,  H$^-$ and Fe opacity used in ATMO is detailed in Appendix \ref{app:high_temp}.

\section{Model Description}
\label{ch5:model_des}
As in \citet{Goyal2018} and \citet{Goyal2019}, in this work we use \texttt{ATMO}, a 1D-2D radiative-convective equilibrium model for planetary atmospheres. A nice comparison between ATMO and other models, before some of the additions described in this paper, is shown in \citet{Baudino2017} and \citet{Malik2018}.  In this section we describe some of the details of ATMO along with recent additions to the model.   

\subsection{Computing $P$-$T$ profiles in Radiative-Convective equilibrium}
\label{subsec:compute_pt}

We here discuss the basic methodology used in ATMO to compute $P$-$T$ profiles in radiative-convective equilibrium, consistently with equilibrium chemistry. Radiative transfer and absorption cross-section calculations along with equilibrium chemistry computations performed iteratively, all contribute to the computation of consistent RCE $P$-$T$ profiles, discussed in detail in \citet{Drummond2016} and \citet{Goyalthesis2019}. We initialise the model with an estimated (initial guess) $P$-$T$ profile (the starting profile can also be isothermal although this will likely take longer computation time to reach convergence). We note that the final RCE $P$-$T$ in independent of the initial starting $P$-$T$ profile. Chemical equilibrium abundances are then calculated for this $P$-$T$ profile. Using these abundances along with the absorption cross-sections of different chemical species, in the form of \textit{k}-coefficients \citep{Amundsen2014}, the total opacity is calculated for each layer of the atmosphere.  Following this, the radiative transfer equation is solved to compute incoming and outgoing radiative fluxes for each layer of the atmosphere. The same approach is followed for convective flux. This is then checked for energy flux balance in each layer of the atmosphere as well as the atmosphere as a whole, using the energy conservation equation given by 

\begin{equation} 
     F_{\textup{rad}} + F_{\textup{conv}} - \sigma T^{4}_{\textup{int}} = 0,
     \label{eq:energy conservation1}
\end{equation}

where $F_{\textup{rad}}$ the radiative flux  and $F_{\textup{conv}}$ is the convective flux, respectively,  as derived in \citet{Drummond2016} and \citet{Goyalthesis2019}. $T_{\textup{int}}$ represents the internal temperature of planet or the temperature at which the planet is cooling. Isotropic scattering is also included in the radiative flux computation. The $P$-$T$ profile is also constrained by hydrostatic equilibrium, which defines the pressure structure as a function of altitude, implemented using

\begin{equation} 
     \frac{d}{dz}(P_{\textup{gas}} + P_{\textup{turb}}) - \rho g  = 0, 
     \label{eq:hydrostatic equilibrium}
\end{equation}

where $P_{\textup{gas}}$ is pressure due to gaseous species, $P_\textup{{turb}}$ is turbulent pressure, $z$ is altitude, $\rho$ is density and $g$ is gravity. If the conditions for energy conservation and hydrostatic equilibrium are not satisfied within the required numerical accuracy, corresponding to an error in flux balance (typical value of $\sim$10$^{-3}$), the $P$-$T$ profile is perturbed within minimum and maximum step sizes ($\sim$0.1-0.9) for the next iteration. This step size is basically the multiplying factor to the temperature perturbation while the model is iterating to obtain a converged solution. All the steps are repeated for each iteration, until a $P$-$T$ profile that satisfies hydrostatic equilibrium and conservation of energy is obtained, consistent with equilibrium chemistry and corresponding opacities, for a given set of planetary characteristics. 

Figure \ref{fig:ch5_pt} shows radiative-convective equilibrium $P$-$T$ profiles with and without a temperature inversion. Unlike a static table of chemical abundances used in various models, these $P$-$T$ profiles (and all models in this grid) are consistent with equilibrium chemical abundances, as explained above. In this schematic, we also show the dominant physical processes in each region. Convection plays an important role only in the deepest parts of the irradiated hot Jupiter atmospheres (> 100 bar), which current observations cannot probe. Radiation governs the $P$-$T$ structure across a wide range of pressures. Radiative diffusion leads to an isothermal structure in the deep atmosphere as shown in Figure \ref{fig:ch5_pt}, primarily because the high opacity in this region decreases the mean free path for the photons as in stellar atmospheres \citep{Rybicki1986}. The interplay between absorption of stellar radiation due to optical opacities and planetary emission due to infrared opacities, governs the $P$-$T$ profile in the radiative region between $\sim$1 bar to 0.1 millibar, which is the only region probed by current observations, either using transmission or emission spectra. Sufficiently high optical opacity can lead to a temperature inversion in this observed radiative region. At the top of the atmosphere where the pressure becomes less than $\sim$ 0.1 millibar the atmosphere becomes optically thin and most of the radiation penetrates this part of the atmosphere.

\subsection{Recirculation factor}
\label{ch5:rcf}
In a planetary atmosphere advection due to winds is one of the major processes responsible for transporting energy. To incorporate the 3D effect of advection in 1D models as adopted by \citet{Fortney2007}, we simply reduce the incoming flux in the 1D column of the atmosphere by a factor called the \enquote{recirculation factor}, hereafter termed f$_{\textup{c}}$. It parameterises the redistribution of input stellar energy in the planetary atmosphere, by the dynamics, where a value of $1$ equates to no redistribution, while $0.5$ represents efficient redistribution. The value of 0.5 f$_{\textup{c}}$ indicates 50\% of the total incoming stellar energy is advected to the night side (the side of the planet facing away from the star), while 0.25 f$_{\textup{c}}$ indicates 75\% of the total incoming stellar energy is advected to the night side. It must  also be noted that an additional factor, the incidence angle $\theta_o$ also contributes to the reduction in this total incoming stellar energy. $\theta_o= 60^{\circ}$ equating to the dayside (the side of the planet facing towards the star) average is the most commonly adopted value of incidence angle, contributing to 50\% reduction in the total incoming stellar energy, since $\cos{60^{\circ}} = 0.5$.  

\subsection{Contribution Function}
\label{ch5:confun}
The emission spectrum represents the top of the atmosphere (ToA) flux at different wavelengths for a given planet. However, this is a combination of flux from the different layers of the atmosphere. To identify the levels of the atmosphere  contributing the most to this ToA emission, the Contribution function (CF) \citep{Chamberlain1987, Knutson2009, Drummond2018b} is calculated given by, 

\begin{equation} 
    CF = {B(\nu,T)} \frac{d(e^{-\tau_\nu)}}{d(ln(P))},  
     \label{eq:cf}
\end{equation}

where $\nu$ is the frequency, $T$ is the temperature, $B(\nu,T)$ is the Planck emission, $\tau_\nu$ is the optical depth and $P$ is the pressure at each level of the atmosphere. The vertical $P$-$T$ profile and the wavelength dependent optical depth, are the primary quantities required to calculate the contribution function. Optical depth is a function of transmittance which decreases as we go deeper in the planet's atmosphere. Therefore, the CF is higher in the region where there is a larger change in the optical depth or transmittance for a unit change in the pressure (altitude) over the same region. In simpler terms, the CF peaks in the region where the wavelength dependent optical depth is one, when the planet is being probed from the top of the atmosphere.  Although we define contribution function here, for ease of understanding and plotting we compute the Normalised Contribution Function (NCF), by normalising using  the largest value of the contribution function along the $P$-$T$ profile. 

\subsection{High Temperature Additions}
\label{ch5:high_temp_mod}
We include thermal ionization chemistry in ATMO by including ion species while computing equilibrium chemical abundances. We also included iron (Fe) and H$^-$ opacity. We detail their implementation and validation in Appendix \ref{app:high_temp}. 

\begin{figure*}
\begin{center}
 \subfloat[]{\includegraphics[width=\columnwidth]{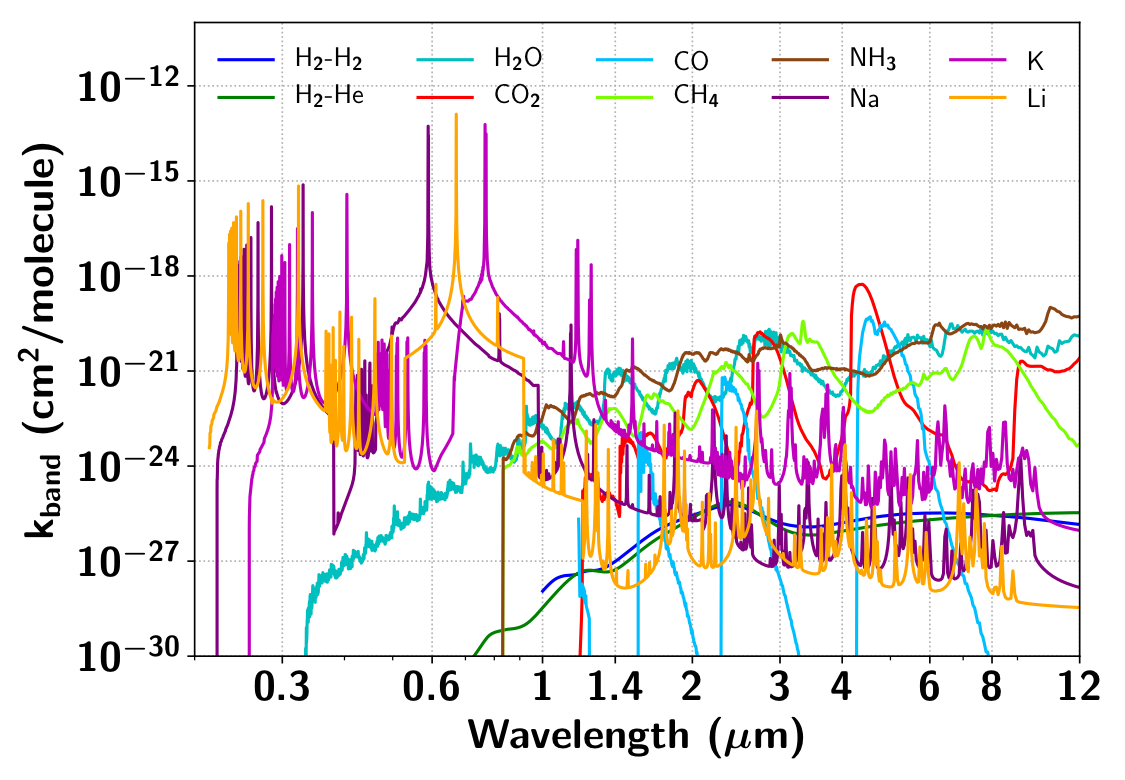}\label{ch5:ch5:fig_crossection_1}}
 \subfloat[]{\includegraphics[width=\columnwidth]{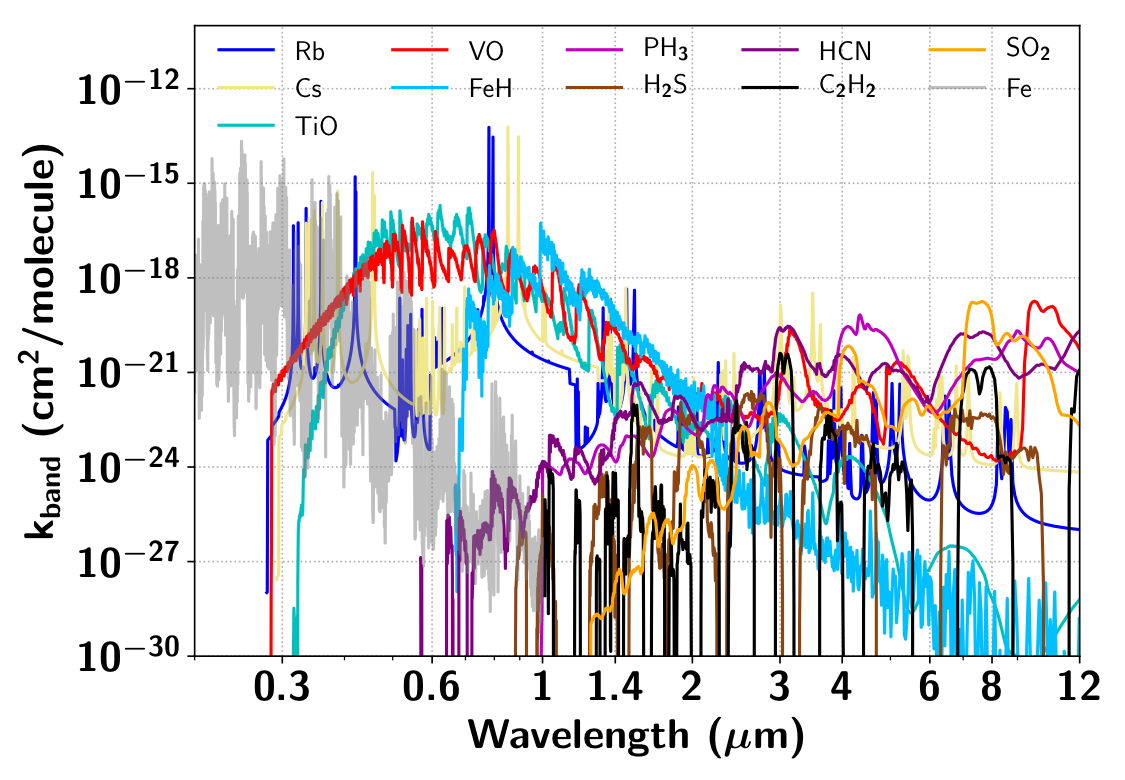}\label{ch5:ch5:fig_crossection_2}}
\end{center}
 \caption{Absorption cross-sections (k$_\textup{band}$) of all species used in ATMO in each of the 5000 correlated-\textit{k}  bands at 1 bar and 2000\,K. These are absolute cross-sections without any dependance on chemical abundances.\textbf{(a)} Cross-sections for H$_2$-H$_2$ (blue), H$_2$-He (green), H$_2$O (cyan), CO$_2$ (red), CO (deepskyblue), CH$_4$ (lawngreen), NH$_3$ (saddlebrown), Na (purple), K (magenta), Li (orange). \textbf{(b)} Cross-sections for Rb (blue), Cs (khaki), TiO (cyan), VO (red), FeH (deepskyblue), PH$_3$ (magenta), H$_2$S (saddlebrown), HCN (purple), C$_2$H$_2$ (black), SO$_2$ (orange) and Fe (grey).}
 \label{ch5:fig_crossection}
\end{figure*}

\section{Numerical setup for the Grid}
\label{ch5:Grid Setup}

Computing RCE $P$-$T$ profiles consistent with equilibrium chemistry is not a trivial task. The $P$-$T$ profile and the chemistry are intricately linked as they depend on each other. The chemical composition is largely dependent on the temperature, and the temperature is largely dependent on composition (via opacities). Moreover, the $P$-$T$ profile as well as the chemical abundance profile continuously change as the simulation progresses towards the solution. In such a scenario a large number of temperature and pressure points are encountered. Therefore, it is extremely difficult to obtain converged (satisfying all constraints) RCE $P$-$T$ profiles for values across a large parameter space. There are always the regions of the parameter space where the simulations tend not to reach a converged solution due to many factors such as the boundary conditions, numerical instabilities, non-convergence of equilibrium chemistry especially with condensation and many more. We deal with such problems by incrementally adjusting the numerical setup for some of the failed model simulations as described in the next section.

We use 50 vertical model levels with a maximum optical depth of $5\times10^5$ at 1 \textmu m. Since ATMO calculates quantities on an optical depth grid, the minimum and maximum optical depths govern the pressure domain (extent of the atmosphere). An increase in the maximum optical depth leads to an increase in the pressure domain of the $P$-$T$ profile, for a given set of parameters. The model stability when solving for radiative-convective equilibrium, consistently with equilibrium chemistry, is very sensitive to the selected top of the atmosphere (minimum) optical depth boundary condition, as the atmosphere can become very sparse (less dense) in this region. Therefore, the top of the atmosphere optical depth is varied to achieve convergence. The typical values used for top of the atmosphere optical depth are 10$^{-2}$, 10$^{-5}$, 10$^{-10}$ and 10$^{-13}$ at 1 \textmu m. Although this is a wide range for top of the atmosphere optical depth, these extremely low optical depth regions are outside the domains of the region probed by either transmission or emission spectra. Moreover, we find that a value of 10$^{-5}$ is sufficient for most of the model simulations. The top of the atmosphere pressure is restricted to $10^{-6}$ bar, which corresponds to the top of the atmosphere minimum optical depth. Even though we vary the minimum optical depth to achieve convergence, the pressure is always set at $10^{-6}$ bar for this minimum optical depth, serving as a reference for the atmospheric $P$-$T$ profile. 

We use 32 band correlated-\textit{k} cross-sections \citep{Amundsen2014, Goyal2018} for generating consistent RCE $P$-$T$ profiles and 5000 bands to generate transmission spectra, emission spectra and contribution functions. The internal temperature of the planet (T$_\textup{{int}}$) defined in Section \ref{subsec:compute_pt} is set at 100\,K following \citep{Guillot2002, Fortney2007b}. However, we note that following some recent studies \citep{Thorngren2019,Sing2019}, this assumption might be debatable. We adopt a mixing length constant $\alpha = 1.5$ for calculating convective flux \citep{Baraffe2015}, as used in previous ATMO simulations \citep[e.g][]{Drummond2016}. To standardise the comparison of transmission spectra for a range of variables, we set the pressure at which the radius of the planet is defined at 1 millibar \citep{Lecavelier2008}, only while computing the transmission spectra. We note that there exists a  degeneracy between reference transit radius and associated reference pressure as highlighted by \citet{Lecavelier2008}. 

The target planet selection technique for this library of models is the same as that adopted in \citet{Goyal2018} with the additional constraint of the planetary equilibrium temperature. In this library only the planets with equilibrium temperatures greater than 1200\,K, as computed in the TEPCat database \citep{Tepcat2011} are selected, all shown in the Table 3 of the online supplementary material. The choice of 1200\,K is based on the capability of ATMO to obtain converged solution (RCE $P$-$T$ profile consistent with equilibrium chemistry) for a range of grid parameter space values. It was found that for planets with equilibrium temperatures less than 1200\,K a large number of model simulations across the grid parameter space fail to converge, most likely due to numerical instabilities arising due to inclusion of various condensates in our equilibrium chemistry computation. 

The input stellar spectra for each planetary model grid are taken from the BT-Settl\footnote{\url{https://phoenix.ens-lyon.fr/Grids/BT-Settl/AGSS2009/SPECTRA/}} models \citep{Allard2012, Rajpurohit2013}. These stellar spectra are selected according to the closest obtained host star temperature, gravity and metallicity from the TEPCat database \citep{Tepcat2011}. All the parameters required for model initialisation like stellar radius, planetary radius, planetary equilibrium temperature, surface gravity and semi-major axis are also adopted from the TEPCat\footnote{\url{http://www.astro.keele.ac.uk/jkt/tepcat/allplanets-ascii.txt}} database, shown in the Table 1 of the online supplementary material. 

\subsection{Chemistry Setup}
\label{ch5:chemistry}
Similar to \citet{Goyal2018} and \citet{Goyal2019}, we restrict our calculations to equilibrium chemistry in this work, but with self-consistent RCE $P$-$T$ profiles. However, in this work we also include thermal ionization in the equilibrium chemistry computation, detailed and validated in Appendix \ref{ch5:thermal_ions} . Therefore, in addition to the 258 species as used in \citet{Goyal2018} and \citet{Goyal2019} which are listed here\footnote{\url{https://drive.google.com/drive/folders/1g7Bc6pbwvLUDf-QFJOCEOFKX6glKLuqF}}, the list of species for equilibrium computation now also includes H$^+$, H$^-$, Na$^+$, K$^+$, e$^-$, C$^+$, He$^+$, Ca$^+$ and Si$^+$ ions along with additional gaseous species NaF, KF, SiO, SiS, CaH, CaOH and condensate species Na$_3$AlF$_6$ in three different forms and LiF in the crystalline form. Thus the total number of species used in the equilibrium chemistry calculation adds up to 277 for the simulations presented here.

\begin{figure*}
\begin{center}
 \subfloat[]{\includegraphics[width=\columnwidth]{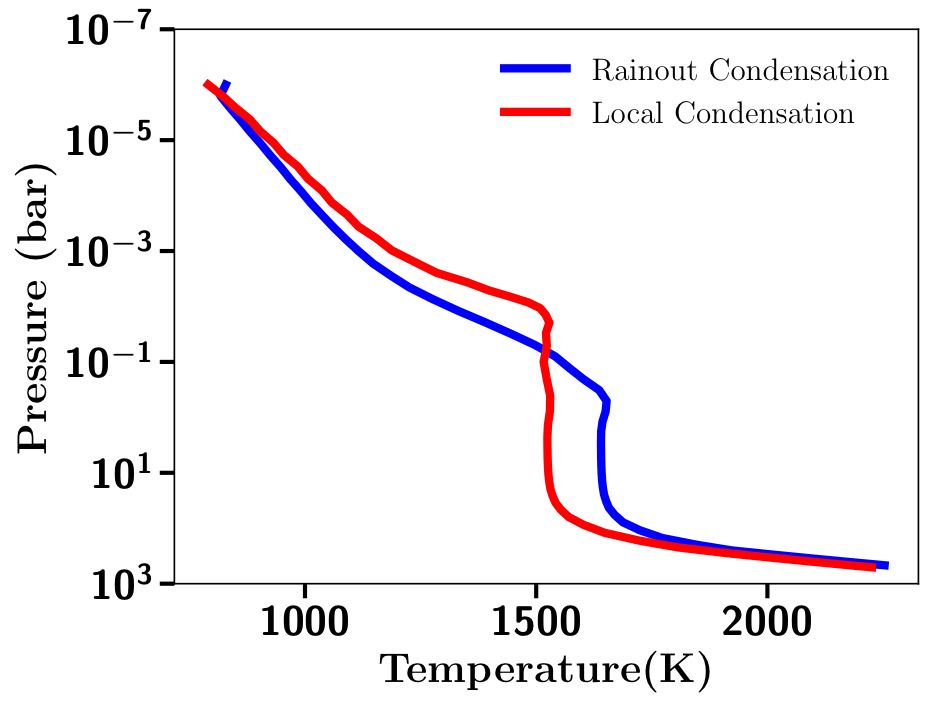}\label{ch5:fig_rainout1}}
 \subfloat[]{\includegraphics[width=\columnwidth]{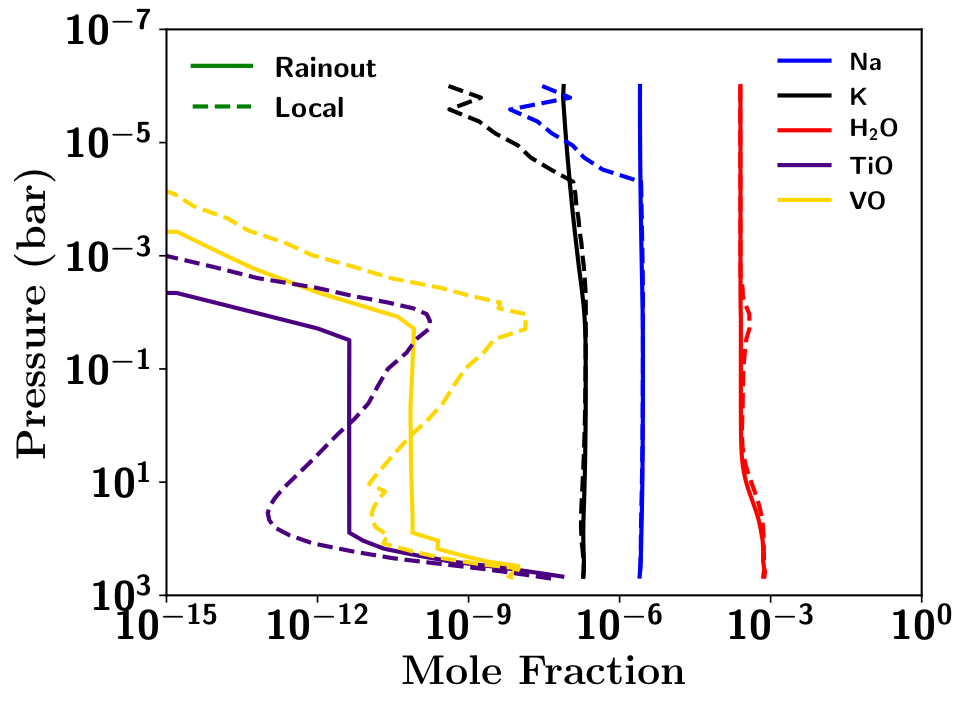}\label{ch5:fig_rainout2}}
  \newline
 \subfloat[]{\includegraphics[width=\columnwidth]{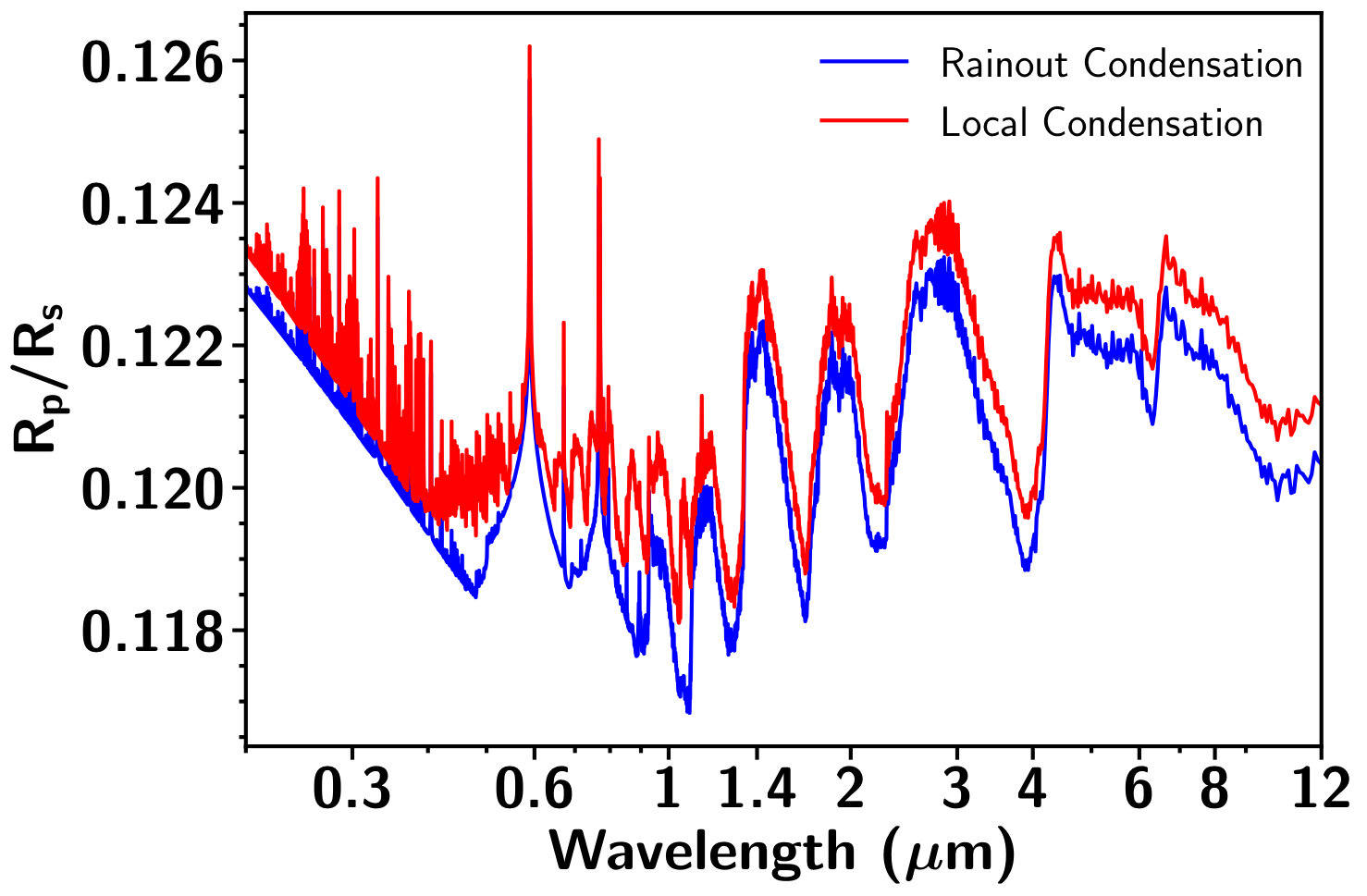}\label{ch5:fig_rainout3}}
 \subfloat[]{\includegraphics[width=\columnwidth]{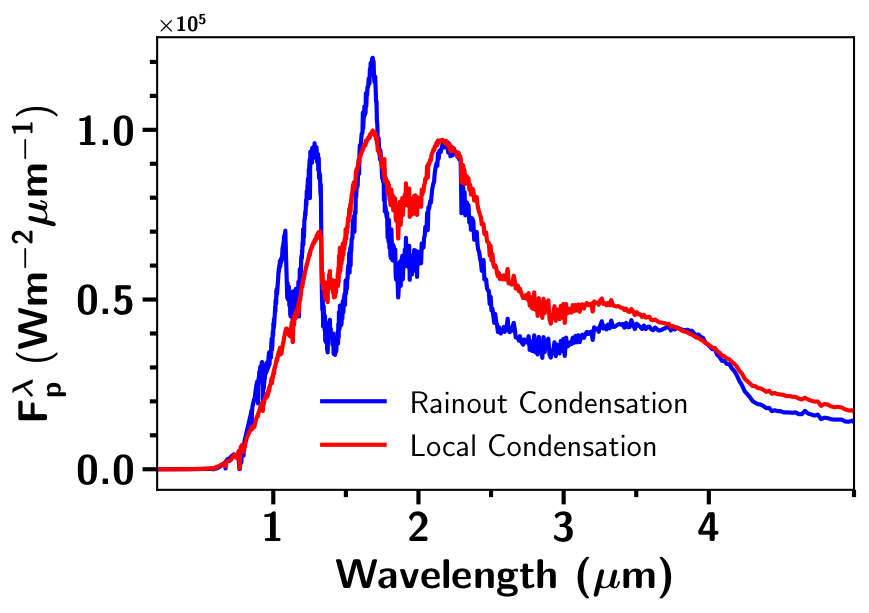}\label{ch5:fig_rainout4}}
 \newline
  \subfloat[]{\includegraphics[width=\columnwidth]{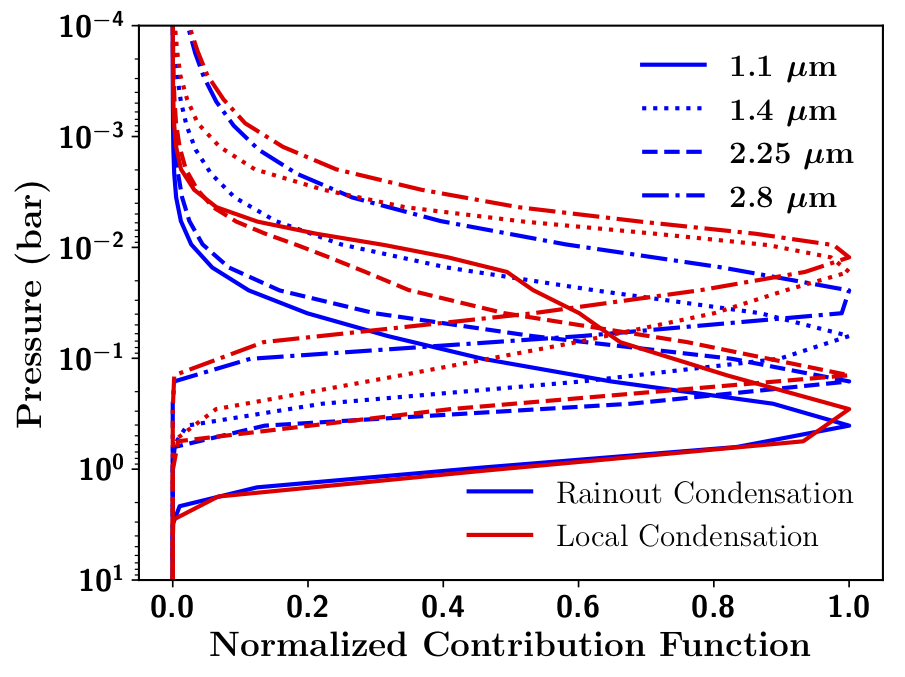}\label{ch5:fig_rainout5}}
 \end{center}
 \caption[Figure comparing RCE $P$-$T$ profiles and the corresponding equilibrium chemical abundances, transmission and emission spectra, and contribution functions, with local and rainout condensation approach]{\textbf{(a)} Figure showing RCE $P$-$T$ profiles for WASP-17b at 0.25 f$_{\textup{c}}$ and solar metallicity and solar C/O ratio (0.55), with rainout (blue) and local condensation (red). \textbf{(b)}  Figure showing equilibrium chemical abundances of important species using $P$-$T$ profiles shown in Figure \ref{ch5:fig_rainout1}, with rainout (solid) and local condensation (dashed). \textbf{(c)} Figure showing transmission spectra using $P$-$T$ profiles shown in Fig. \ref{ch5:fig_rainout1} and chemical abundances shown in Figure \ref{ch5:fig_rainout2}, with rainout (blue) and local condensation (red). \textbf{(d)} Figure showing emission spectra using $P$-$T$ profiles shown in Fig. \ref{ch5:fig_rainout1} and chemical abundances shown in Figure \ref{ch5:fig_rainout2}, with rainout (blue) and local condensation (red) .\textbf{(e)} Figure showing contribution function for emission spectra shown in Fig. \ref{ch5:fig_rainout4}, with rainout (blue) and local condensation (red) at 1.1 (solid), 1.4 (doted), 2.25 (dashed) and 2.8 (dot-dash)$\mic$.} 
\end{figure*}

\subsection{Opacity Setup}
\label{ch5:opacity}
The potential of a particular species to absorb/emit radiation at a particular spectral interval is governed by its absorption cross-sections. Compared to the simulations presented in \citet{Goyal2018} and \citet{Goyal2019}, the opacities due to H$^-$ and Fe have been newly included in this work, in addition to H$_2$-H$_2$ and H$_2$-He collision induced absorption (CIA) opacities, and opacities due to H$_{2}$O, CO$_2$, CO, CH$_4$, NH$_3$, Na, K, Li, Rb, Cs, TiO, VO, FeH, CrH, PH$_3$, HCN, C$_{2}$H$_{2}$, H$_{2}$S and SO$_{2}$. The implementation of H$^-$ opacity is described in Appendix \ref{ch5:h-_imp} and that of Fe in Appendix \ref{ch5:fe_imp}. The line list sources for all opacities used in this library and the type/sources of pressure broadening, are shown in Table 1 and Table 2 of the supplementary material, respectively. The effects of these newly added opacities, H$^-$ and Fe on the $P$-$T$ profiles and thereby the spectra are discussed in Section \ref{ch5:effect_h-_opacity} and \ref{ch5:effect_fe_opacity}, respectively. 

Absorption cross-sections for all the species included in ATMO  in each of the 5000 correlated-\textit{k} bands at 1 bar and 2000\,K are shown in Figure \ref{ch5:fig_crossection}. These plots aid identifying the major absorbing species in various parts of the spectrum. However, it must be noted that the final opacities are the product of absorption cross-sections and chemical abundances. Therefore, although species such as TiO,VO, FeH and Fe have strong absorption cross-sections in the optical their contribution to total absorption will be zero if they condense or don't form at any given temperature.

\begin{figure*}
\begin{center}
 \subfloat[]{\includegraphics[width=\columnwidth]{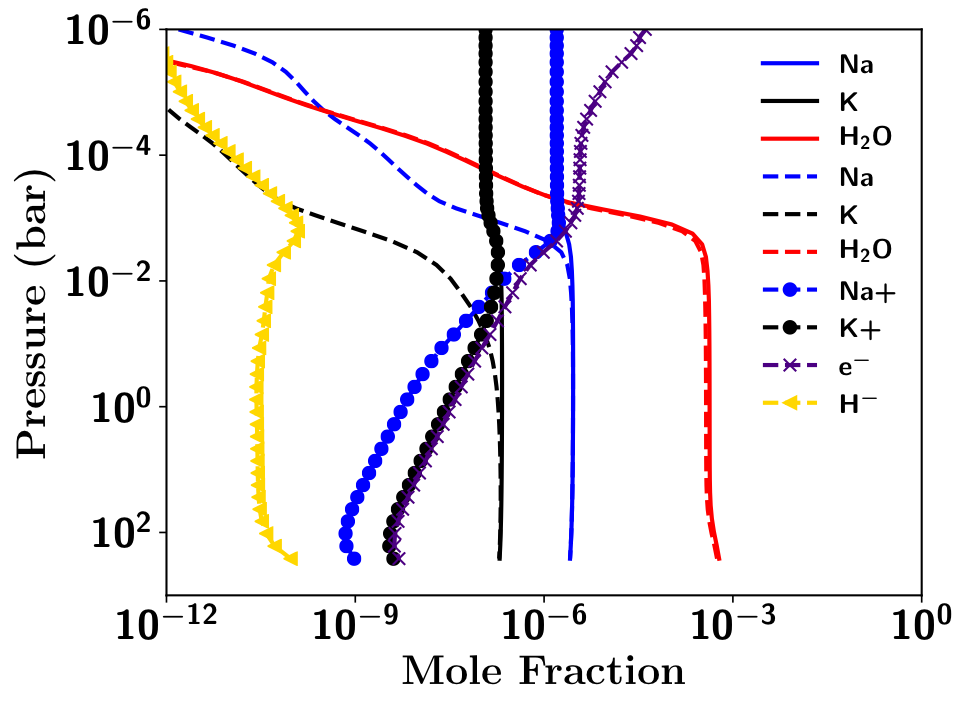}\label{ch5:fig_ion_1}}
 \subfloat[]{\includegraphics[width=\columnwidth]{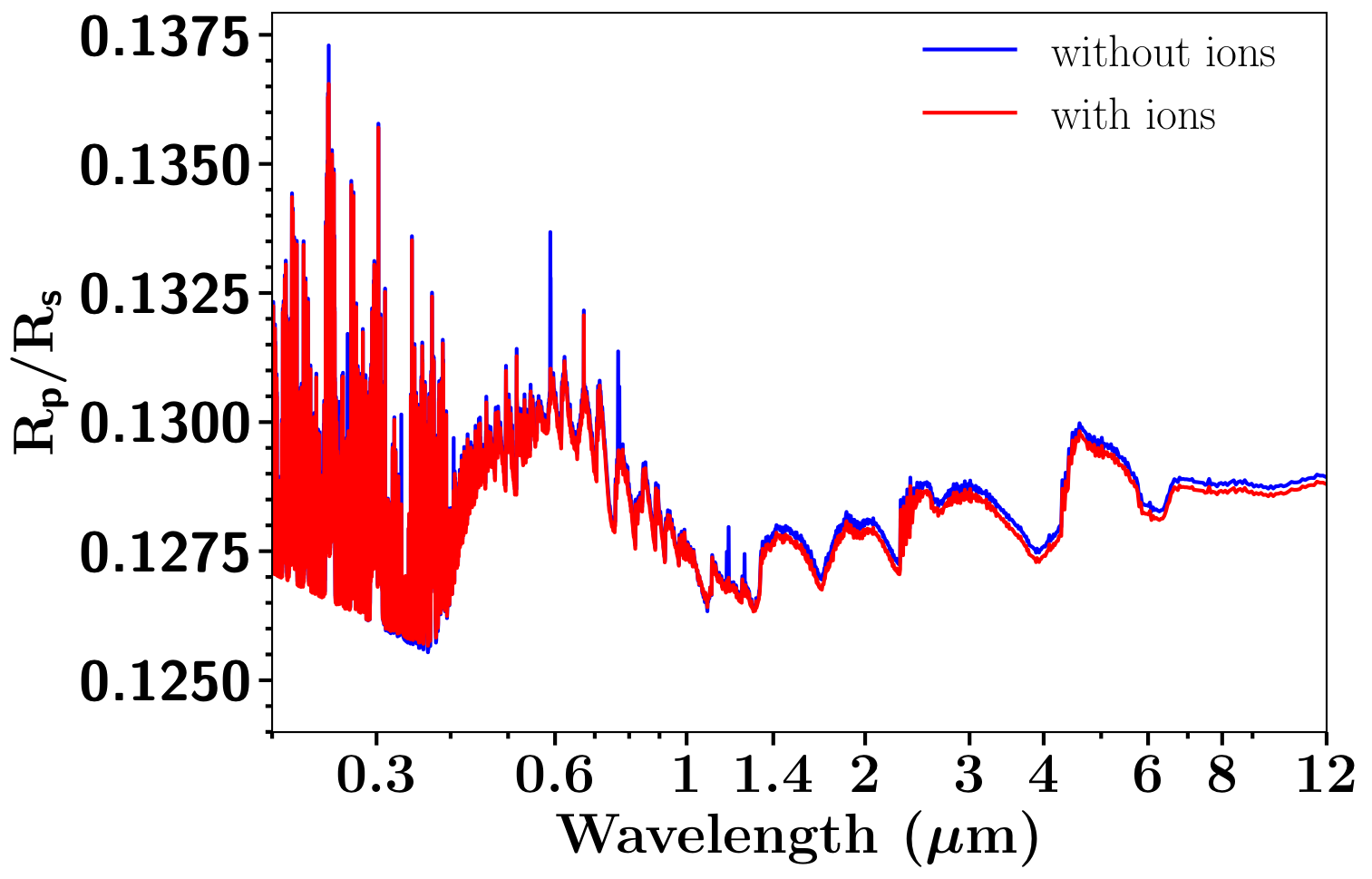}\label{ch5:fig_ion_2}}
\end{center}
 \caption[Figure showing the effect of including thermal ionic species in the model for planets with high levels of irradiation]{\textbf{(a)} Figure showing equilibrium chemical abundances of various species without (solid) and with (dashed) thermal ionic species (ionic species are shown using marked lines) included in the model for WASP-121b at solar metallicity, solar C/O ratio (0.55) and 0.5 f$_{\textup{c}}$. \textbf{(b)} Figure showing transmission spectra models, with and without thermal ionic species at solar metallicity, solar C/O ratio and 0.5 f$_{\textup{c}}$.}
\end{figure*}

\section{Grid Parameter Space}
\label{ch5:grid_par_space}
 The library of models with RCE $P$-$T$ profiles and corresponding equilibrium chemical abundances, transmission spectra, emission spectra and contribution functions for various planets are computed at four different recirculation factors (f$_{\textup{c}}$ = 0.25, 0.5, 0.75, 1.0), six metallicities (0.1, 1, 10, 50, 100, 200; all in $\times$\,solar) and six C/O ratios (0.35, 0.55, 0.7, 0.75 1.0,1.5), giving a total of 144 model RCE $P$-$T$ profiles and spectra per planet. However, as mentioned earlier, not all models in the parameter space achieve convergence, thus leading to absence of some models in the library ($\sim$5-10\%) which varies for different planets. The RCE $P$-$T$ profiles and corresponding equilibrium chemical abundances used to generate our transmission and emission spectra are the same, i.e they span a similar parameter space. Since we are presenting here a library of model atmospheres and corresponding spectra, they are intended to cover a wide range of possibilities and all the extremes.
 
The choice of recirculation factor (f$_{\textup{c}}$) covers all possible scenarios from no recirculation (1.0) to extremely fast winds (0.25). We do not extend the grid to metallicities greater than 200$\times$\,solar, because above this metallicity the atmosphere becomes abundant in species other that H$_2$ and He, such as CO$_2$, H$_2$O, CO etc. This would require the inclusion of pressure broadening effects due to these species, and no existing studies have solved this problem due to lack of lab-based observational data \citep{Fortney2016}. The choice of C/O ratios in the grid is guided by important transition regimes as found by previous studies \citep{Madhusudhan2012, Molliere2015, Goyal2018, Molaverdikhani2019}.  

As described in \citet{Goyal2018,Errgoyal2019, Goyal2019}, there are two approaches to treat condensation in our library of models, rainout and local condensation. While computing equilibrium chemical abundances to obtain RCE $P$-$T$ profiles with rainout condensation, each layer is dependent on other layers, specifically only on layers that lie at higher pressures, in contrast to the local condensation approach where each layer is independent. This makes the assumption of rainout with RCE $P$-$T$ profiles more realistic for a planetary atmosphere as compared to just local condensation. Therefore, we generate RCE $P$-$T$ profiles with rainout condensation for the library of models presented in this work. However, we show the differences in the $P$-$T$ profiles and the spectra due to the different approaches in the next section. 

The structure of the pressure broadened line wings of Na and K can have a substantial effect on the $P$-$T$ profiles and thereby the emission spectrum of brown dwarfs and hot Jupiter exoplanet atmospheres \citep{Burrows2000, Allard2003}. Even with their high resolution measurements, the shape of the pressure broadened wings are still a matter of debate for Brown dwarfs \citep{Burrows2002, Burgasser2003, Allard2003}. For hot Jupiters we have very recently started to observationally probe the line wings of Na/K \citep{Nikolov2018}. In Section 2.1 of the online supplementary material of this paper we show the differences in the $P$-$T$ profiles and the spectra due to these two line wing profiles for Na and K. The differences are negligible and unlikely to be detectable by observations. We adopt the Na and K line wing profiles from \citet{Allard2003} for the library of models presented in this work, which includes detailed quantum mechanical calculations while computing these profiles \citep{Amundsenthesis}. 

Although convection plays an important role in determining the $P$-$T$ profile of brown dwarfs, we see for hot Jupiter exoplanet atmospheres the effect of convection on the $P$-$T$ structure of the observable atmosphere is negligible. This is because of the strong irradiation on these planets from their host stars, which reduces the radiative time scale, thus making the atmospheric $P$-$T$ profile almost entirely dependent on the top of the atmosphere irradiation (along with the atmospheric pressure, temperature and heat capacity), at-least in the region where observations can probe the atmosphere. Therefore, we conclude that it is not necessary to include convection while computing RCE $P$-$T$ profiles for irradiated hot Jupiter exoplanet atmospheres. However, in this work all the $P$-$T$ profiles  include parameterized convection for completeness, as it is computationally inexpensive.

As described in \citet{Drummond2019} there are three different methodologies to vary C/O ratio relative to the solar C/O ratio. In the simulations presented in \citet{Goyal2018} and \citet{Goyal2019}, we varied C/O ratio by varying O/H. However, varying C/O by varying C/H can lead to differences in $P$-$T$ profiles and equilibrium chemical abundances as shown by \citet{Drummond2019}.  Therefore, we investigated the effect of varying C/O ratio, by varying O/H and those by varying C/H using WASP-17b as the test case, with the results presented in the online supplementary material Section 2.3. Although there are some differences in the results obtained using these two different methodologies, in the parameter space we consider, they are smaller compared to the effects of other parameters in the grid and other model choices (for e.g. local or rainout condensation). Therefore, to enable fair comparison between the model spectra generated in our previous works \citep{Goyal2018,Goyal2019} and keep the library of models consistent with them, we again in this work adopt the methodology of varying C/O ratio by varying O/H. Ideally, we could use O/H and C/H as separate parameters in the grid, however, that increases the size of the grid substantially and makes it computationally expensive for a large number of exoplanets. Therefore, we select one methodology over the other.

\section{Results}
\label{ch5:results}

In this section we initially show the effects of some of the important model choices on the RCE $P$-$T$ profiles and thereby the equilibrium chemical abundances, transmission and emission spectra, which could effect interpretation of observations. Additional tests showing the sensitivity of some of the other model choices such as the Na and K line-wing profiles, VO line-list sources and the methodology to vary C/O ratio either by varying O/H or C/H, are all detailed in the online supplementary material of this paper. Decoupled emission spectrum are also shown and described in the online supplementary material, to aid identification of different spectral features in the emission spectra. In the following sections, we show the sensitivity to the choice of condensation process (rainout/local), high levels of irradiation due to thermal ionization, addition of H$^-$ and Fe opacity, and formation of temperature inversion in the atmosphere. Finally, we show the sensitivity of the model simulations, i.e $P$-$T$ profiles, chemical abundances, spectra and the contribution functions, to all the grid parameters across their full range, using WASP-121b and WASP-17b as test cases.

\subsection{Sensitivity to Model Choices}
\label{ch5:model_choice_sensitivity}

\subsubsection{Comparing model simulations with rainout and local condensation}
\label{ch5:rainout_vs_local}
\citet{Goyal2019} have demonstrated that the adoption of either a rainout or local condensation approach results in differences in the transmission spectra when adopting an isothermal P-T profile. Here, we investigate the effects of these two condensation approaches on the RCE $P$-$T$ profiles and thereby the equilibrium chemical abundances, and transmission/emission spectra. Fig. \ref{ch5:fig_rainout1}, \ref{ch5:fig_rainout2}, \ref{ch5:fig_rainout3}, \ref{ch5:fig_rainout4} and \ref{ch5:fig_rainout5}, show the RCE $P$-$T$ profiles, equilibrium chemical abundances, transmission spectra, emission spectra and Normalised Contribution Functions (NCF), respectively, for WASP-17b at 0.25 f$_{\textup{c}}$, solar metallicity and solar C/O ratio, adopting both rainout and local condensation. A difference of $\sim$100-300\,K can be seen between the 1 and 100 mbar pressure levels in Fig. \ref{ch5:fig_rainout1}, with the $P$-$T$ profile adopting the local condensation approach having higher temperatures. However, in the deeper atmosphere ($\sim$ 0.1-1000 bar) the $P$-$T$ profile with rainout condensation is hotter by $\sim$200\,K. This difference in temperature leads to lower abundance of Na and K in the upper atmosphere and a higher abundance of TiO/VO in the local condensation case, as compared to rainout condensation case, shown in Fig. \ref{ch5:fig_rainout2}. This higher TiO/VO abundance can also be seen in the transmission spectra  shown in Figure \ref{ch5:fig_rainout3} where the spectra with local condensation show TiO/VO features in the optical, missing in the rainout condensation case. This also strengthens the findings of \citet{Goyal2019}, that TiO/VO features can reveal dominant physical process (rainout or local condensation) in the planet's atmosphere. 

The differences in the emission spectra shown in Fig. \ref{ch5:fig_rainout4} are substantial, primarily due to the differences in the $P$-$T$ structure. At 1.1$\mic$ deeper parts of the atmosphere ($\sim$1 bar) are being probed as seen in the NCF shown in Fig. \ref{ch5:fig_rainout5}, mainly because none of the species have a strong opacity to absorb in this region. Therefore, it can be seen in the emission spectra that the flux is higher for the rainout condensation as compared to local condensation, since the temperature at  $\sim$1 bar is higher for the simulation adopting rainout condensation. At 1.4$\mic$ the wavelength of one of the strongest water opacity bands, we see the NCF moves to the upper atmosphere and the difference between the pressure levels being probed using emission spectrum in the case of the rainout (0.1 bar) and local condensation (0.01bar) simulations are also substantially different. This is mainly due to the large difference in the temperatures between the rainout and local condensation simulations between 0.1 and 0.01bar. The difference in the peak pressure level of the NCF as observed between 1.1 and 1.4$\mic$, the wing and core of water band, respectively, can be observed more strongly at 2.25 and 2.8$\mic$, since this is the region of peak emission for a body with temperature similar to the equilibrium temperature of WASP-17b. Since the differences in the spectra are substantial, it might be possible to distinguish between emission spectra due to rainout and local condensation, and therefore constrain the $P$-$T$ profiles and thereby the condensation processes using JWST. 

\begin{figure*}
\begin{center}
 \subfloat[]{\includegraphics[width=\columnwidth]{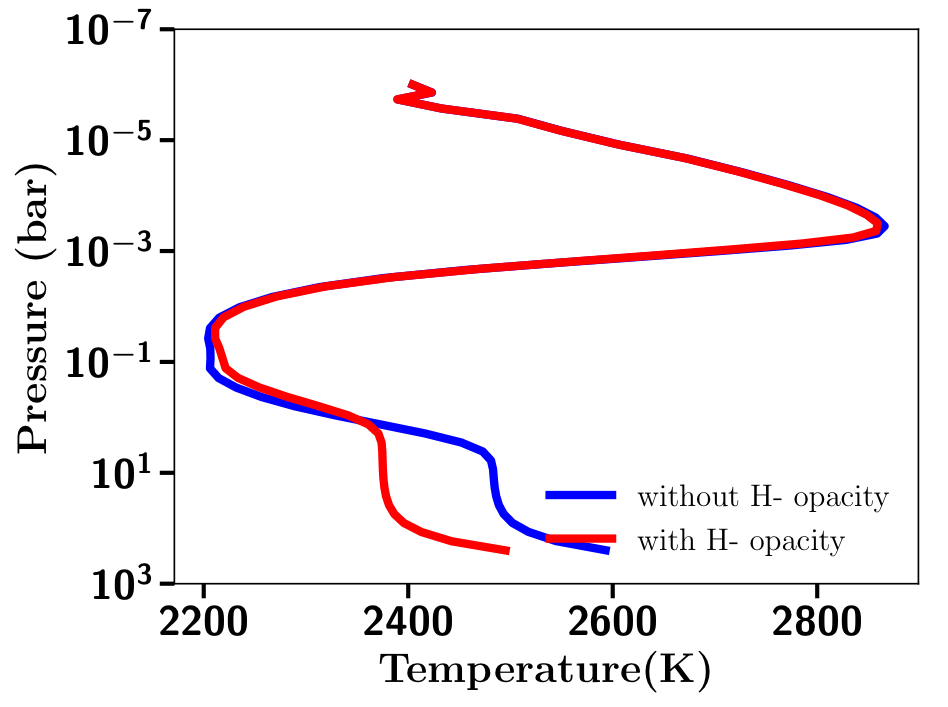}\label{ch5:fig_H-_1}}
 \subfloat[]{\includegraphics[width=\columnwidth]{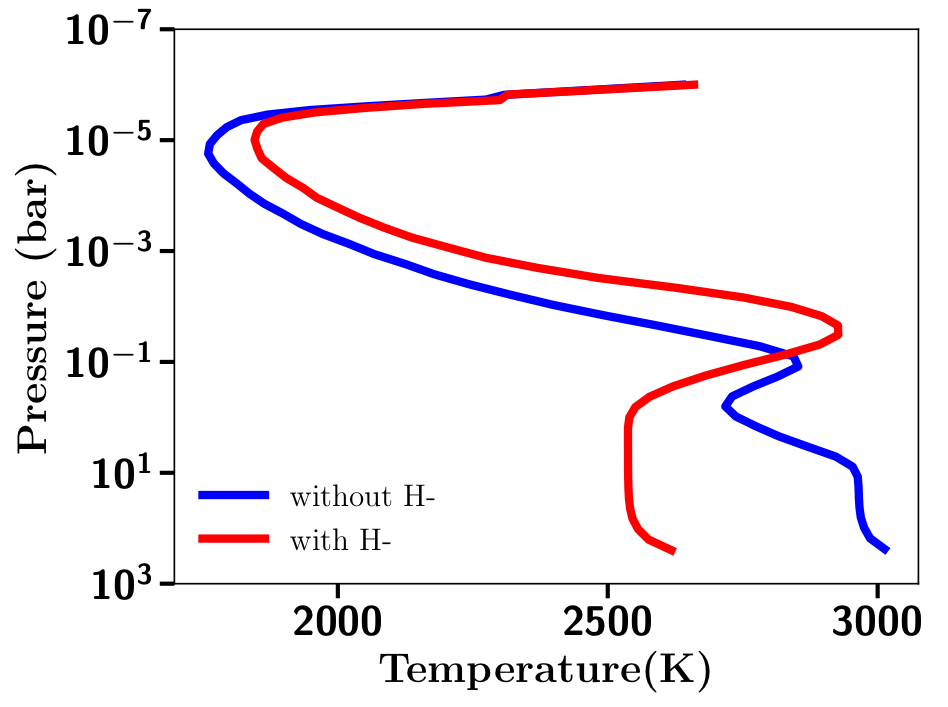}\label{ch5:fig_H-_2}}
 \newline
 \subfloat[]{\includegraphics[width=\columnwidth]{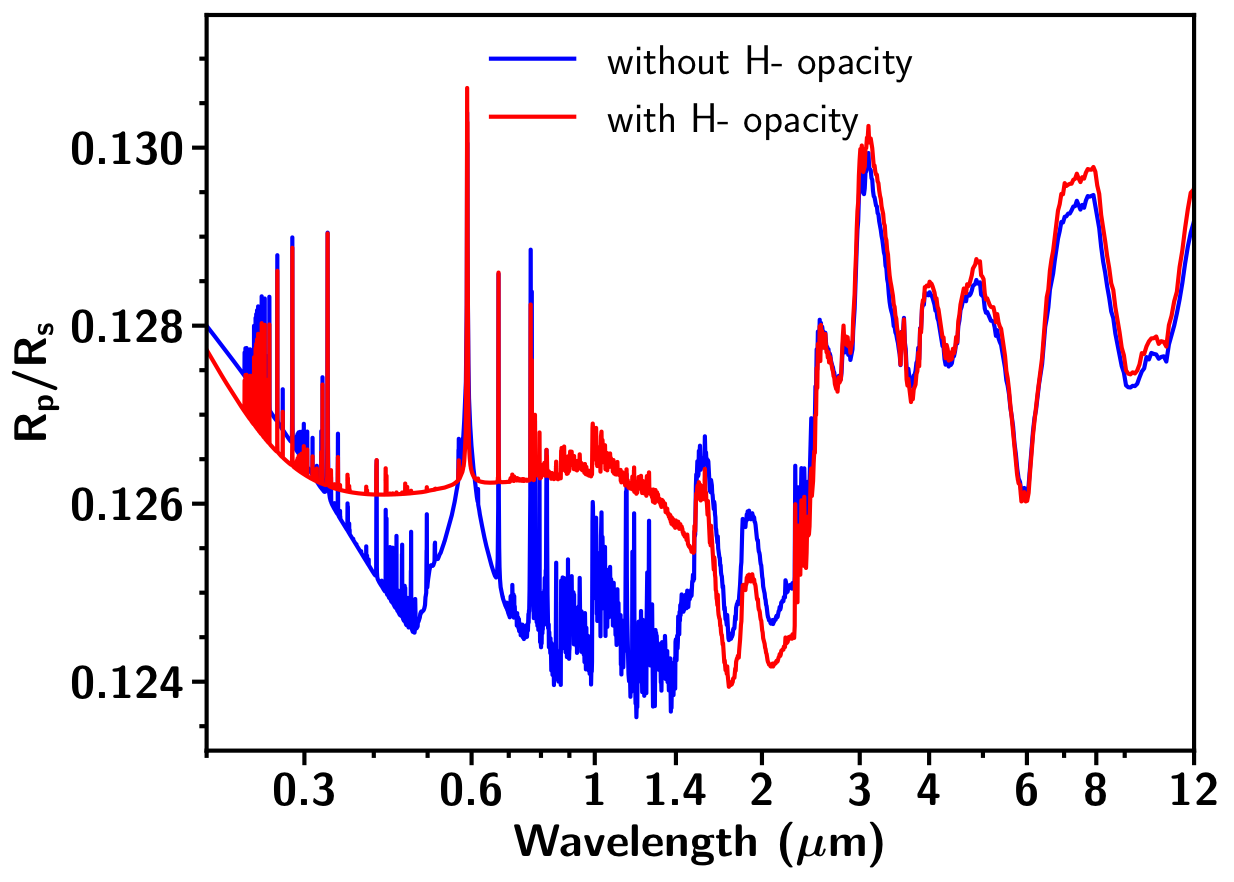}\label{ch5:fig_H-_3}}
 \subfloat[]{\includegraphics[width=\columnwidth]{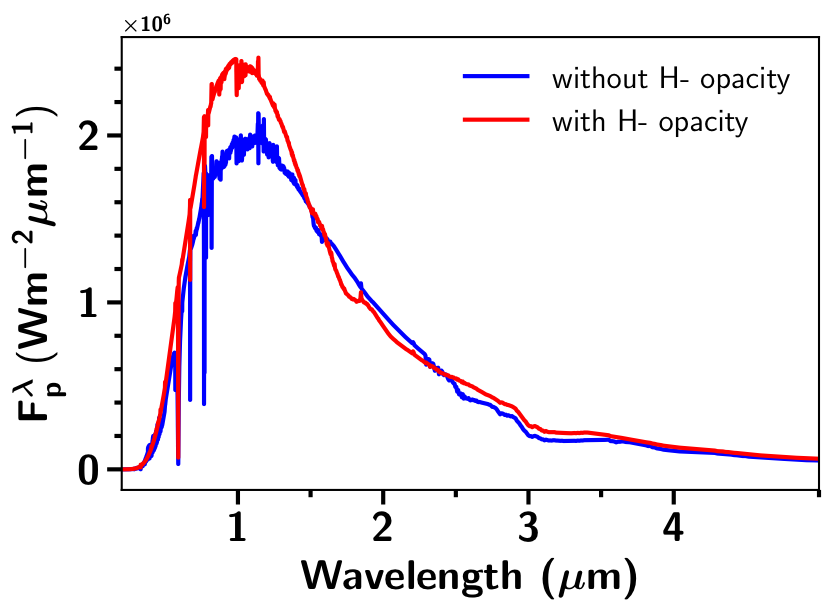}\label{ch5:fig_H-_4}}
\end{center}
 \caption[Figure showing the effect of including H$^-$ opacity in the model for planets with high levels of irradiation]{\textbf{(a)} Figure showing RCE $P$-$T$ profiles without (blue) and with (red) H$^-$ opacity included in the model for WASP-121b at solar metallicity, solar C/O ratio and 0.5 f$_{\textup{c}}$. Fe opacity is not included in both these models (see Section \ref{ch5:effect_h-_opacity} for details). \textbf{(b)} Same as Figure \ref{ch5:fig_H-_1} but at C/O ratio of 1.5 and 1.0 f$_{\textup{c}}$. \textbf{(c)} Figure showing transmission spectra using $P$-$T$ profiles shown in Figure \ref{ch5:fig_H-_2} and corresponding chemical abundances with and without H$^-$ opacity. \textbf{(d)} Figure showing emission spectra using $P$-$T$ profiles shown in Figure \ref{ch5:fig_H-_2} and corresponding chemical abundances with and without H$^-$ opacity.}
 \label{ch5:fig_H-}
\end{figure*}

\subsection{Effects caused by high levels of irradiation}
\label{ch5:high_level_irrad}

\subsubsection{Effect of Thermal Ionisation}
\label{ch5:eff_ionisation}
Thermal ionisation of certain species can have a substantial effect on the chemistry of the planetary atmospheres, depending on the atmospheric temperature. Figure \ref{ch5:fig_ion_1} shows the equilibrium chemical abundances of certain important species for an extremely irradiated hot Jupiter WASP-121b, with (dashed line) and without (solid line) thermal ionic species included in the equilibrium chemistry computation. H$^-$ opacity has been included in the model simulation with thermal ionic species and Fe opacity in both. The abundance of Na decreases by about 3 orders of magnitude in the upper atmosphere (transmission spectra probed region) when thermal ionic species are included, as Na is ionized to form Na$^+$. Figure \ref{ch5:fig_ion_1} shows that the abundance of Na$^+$ becomes almost equal to Na without thermal ionic species model simulation, in the upper atmosphere (P < 10$^{-3}$ bar). Similar behavior can be observed for K and K$^+$ ions. The abundance of H$_2$O also drastically decreases for P < 10$^{-3}$ bar, however, is still substantial in the deeper atmosphere probed by emission spectrum. Inclusion of  thermal ionic species also has effects on the transmission spectra, where the narrow Na and K features (cores) seen in the model spectra without thermal ionic species, disappear in the model spectra with thermal ionic species as shown in Figure \ref{ch5:fig_ion_2}. We note that the equilibrium abundance of H$_2$O at very high temperatures and low pressures as for WASP-121b is almost similar (see Figure \ref{ch5:fig_ion_1}), with and without thermal ionic species, because, thermal decomposition of H$_2$O is taken care of in the equilibrium chemistry computation (by neutral products H$_2$ and O$_2$), even if ions are not included in the computation.

\begin{figure*}
\begin{center}
 \subfloat[]{\includegraphics[width=\columnwidth]{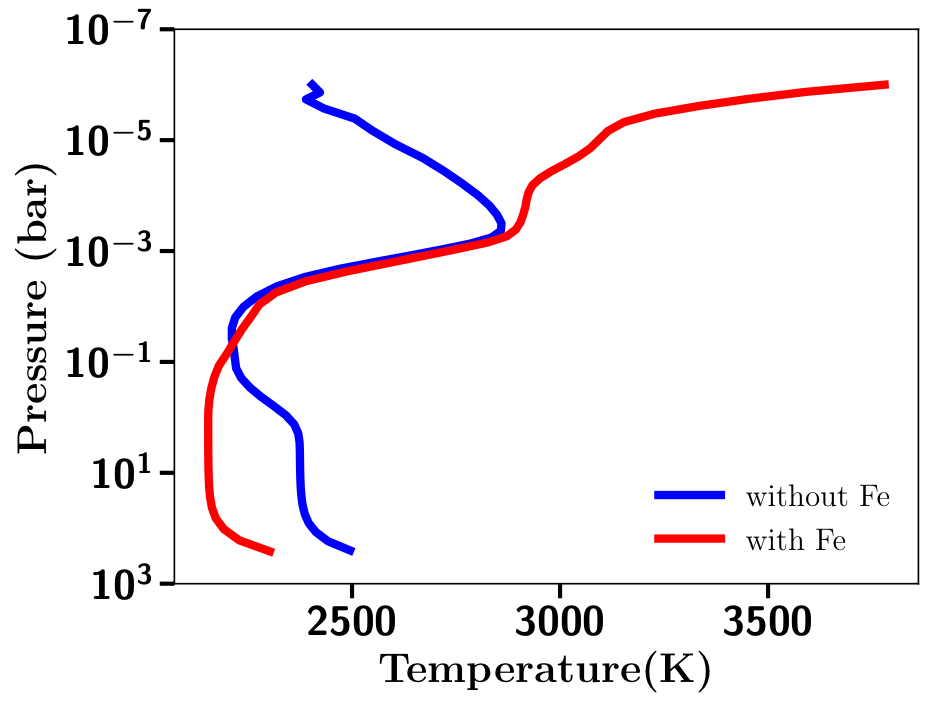}\label{ch5:fig_Fe_1}}
 \subfloat[]{\includegraphics[width=\columnwidth]{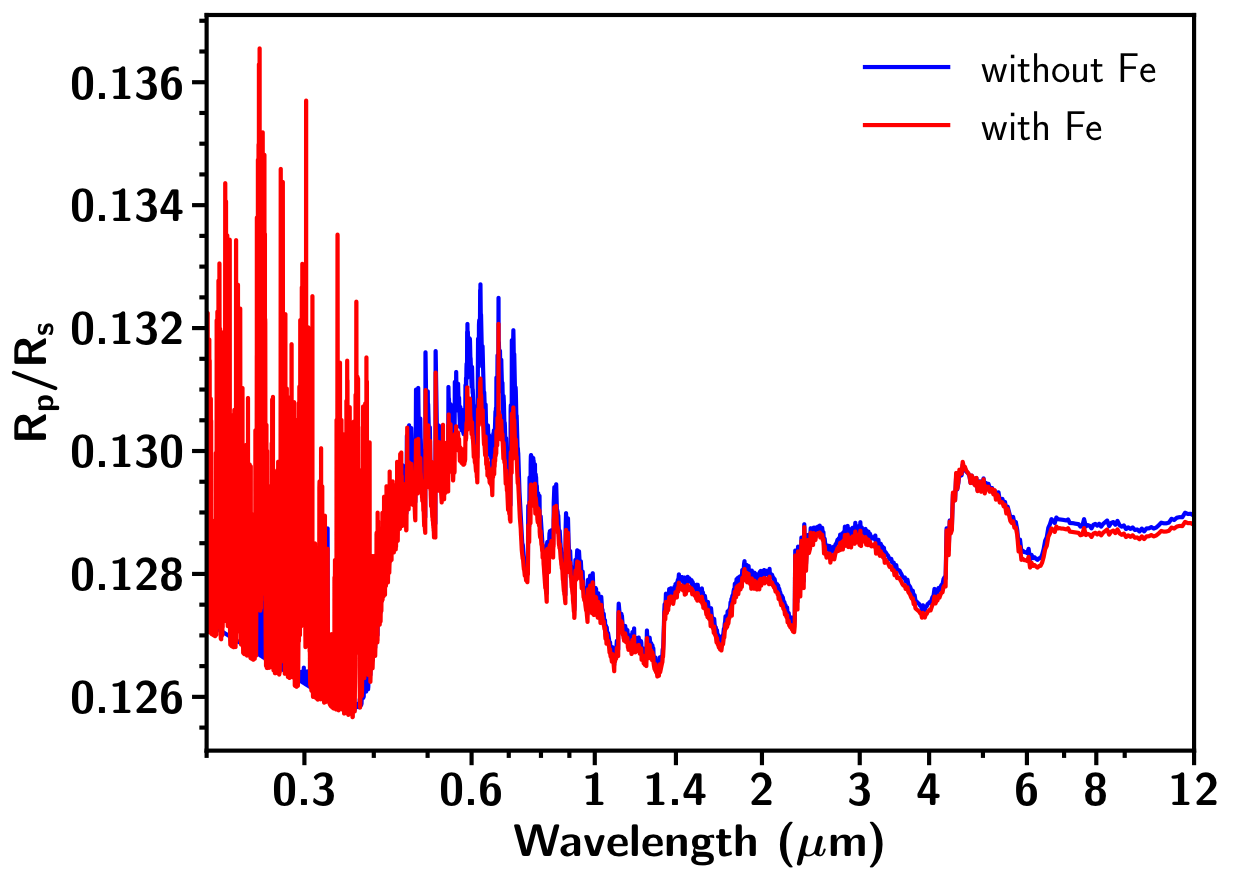}\label{ch5:fig_Fe_2}}
 \newline
  \subfloat[]{\includegraphics[width=\columnwidth]{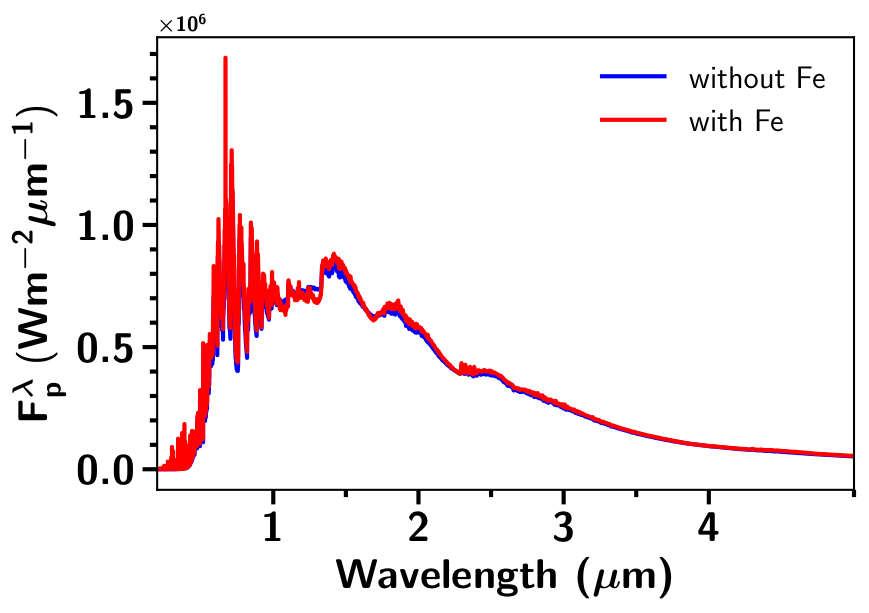}\label{ch5:fig_Fe_3}}
\end{center}
 \caption[Figure showing the effect of including Fe opacity in the model for planets with high levels of irradiation]{\textbf{(a)} Figure showing RCE $P$-$T$ profiles without (blue) and with (red) Fe opacity included in the model for WASP-121b at solar metallicity, solar C/O ratio and 0.5 f$_{\textup{c}}$.  \textbf{(b)} Figure showing transmission spectra using $P$-$T$ profiles shown in Figure \ref{ch5:fig_Fe_1} and corresponding chemical abundances with and without Fe opacity. \textbf{(c)} Figure showing emission spectra using $P$-$T$ profiles shown in Figure \ref{ch5:fig_Fe_1} and corresponding chemical abundances with and without Fe opacity.}
 \label{ch5:fig_Fe}
\end{figure*}

\subsubsection{Effect of H$^-$ opacity}
\label{ch5:effect_h-_opacity}
H$^-$ opacity contributes to the absorption of radiation in hot Jupiters via bound-free and free-free cross-sections as explained in Appendix \ref{ch5:h-_imp}. There is also a strong observational evidence for H$^-$ opacity in \citet{Evans2019}. To understand the effect of H$^-$ opacity we compute $P$-$T$ profiles with and without H$^-$ opacities included in the model. We also don't include Fe opacity specifically for models in this tests, to isolate the effects of H$^-$ opacity. Figure \ref{ch5:fig_H-_1} show these $P$-$T$ profiles at 0.5 f$_{\textup{c}}$ with solar metallicity and C/O ratio for WASP-121b.  It can be noticed from the $P$-$T$ profiles that H$^-$ opacity tends to cool the deeper atmosphere (> 1 bar) by about 100\,K  which increases to 400\,K with an f$_{\textup{c}}$ value of 1 (not shown here). This can be attributed to an increase in H$^-$ abundance in the lower atmosphere (> 1 bar), with a higher f$_{\textup{c}}$ value, as the temperature increases. In Figure \ref{ch5:fig_H-_2} the $P$-$T$ profile obtained using a f$_{\textup{c}}$ value of 1.0, solar metallicity and a C/ O ratio of 1.5 is shown. This has been particularly chosen to show the extreme effects due to H$^-$ opacity. Figure \ref{ch5:fig_H-_2} shows that there is a substantial difference in $P$-$T$ profiles with and without H$^-$ opacity. 

Without H$^-$ opacity there is a weak temperature inversion as compared to that with H$^-$ opacity. At high C/O ratios (> 1) as shown in Figure \ref{ch5:fig_co_wasp121_2}, the abundance of TiO/VO decreases dramatically, the major absorbers likely to cause an inversion in extremely irradiated hot Jupiters like WASP-121b. Therefore, other species start contributing to form a temperature inversion. Without H$^-$ opacity the strong inversion is not sustained but there is a weak inversion due to Na and K \citep[as shown in][]{Molliere2015} discussed in detail in Section \ref{ch5:wasp121_co_oxy}. With H$^-$ opacity the strength of inversion increases dramatically at  f$_{\textup{c}}$ = 1.0 and C/O ratio of 1.5 as shown in Figure \ref{ch5:fig_H-_2}, because the H$^-$ abundance between pressure levels $\sim$1 to 10$^{-2}$ bar increases (see Figure \ref{ch5:fig_co_wasp121_2} for trends even though at f$_{\textup{c}}$ = 0.5), such that its opacity can create an inversion similar to TiO/VO. However, thermal inversion due to H$^-$ opacity lies  deeper (higher pressure) in the atmosphere ($\sim$1 to 10$^{-2}$ bar), in comparison to the inversion due to TiO/VO  ($\sim$10$^{-1}$ to 10$^{-3}$ bar). To investigate the potential of H$^-$ opacity by its own to form temperature inversions at different C/O ratios, we performed some model simulations by removing dominant optical absorbers such as TiO, VO, Na, K, Li, Rb, Cs, Fe and FeH. We find H$^-$ opacity is able to create a strong thermal inversion for C/O ratios $\geq$ 1, we also see a very weak temperature inversion ($\sim$100\,K) at C/O ratio of 0.75 and almost negligible at solar C/O ratio (0.55), highlighting the C/O ratio dependency of H$^-$ opacity to create thermal inversions. 

The effect of H$^-$ opacity on the transmission spectrum is also substantial as shown in the Figure \ref{ch5:fig_H-_3} where it tends to obstruct the deeper atmosphere due to its high opacity as shown for a high C/O ratio case for WASP-121b. In this case it tends to mute the wings of Na and K mimicking the effect of cloud at optical wavelengths. Figure \ref{ch5:fig_H-_4} shows the emission spectrum for WASP-121b with and without H$^-$ opacity using the $P$-$T$ profile shown in Figure \ref{ch5:fig_H-_2}. Figure \ref{ch5:fig_H-_4} shows that there is a substantial difference in the emission spectrum with and without H$^-$ opacity, due to difference in $P$-$T$ profiles.  

\begin{figure*}
\begin{center}
 \subfloat[]{\includegraphics[width=\columnwidth]{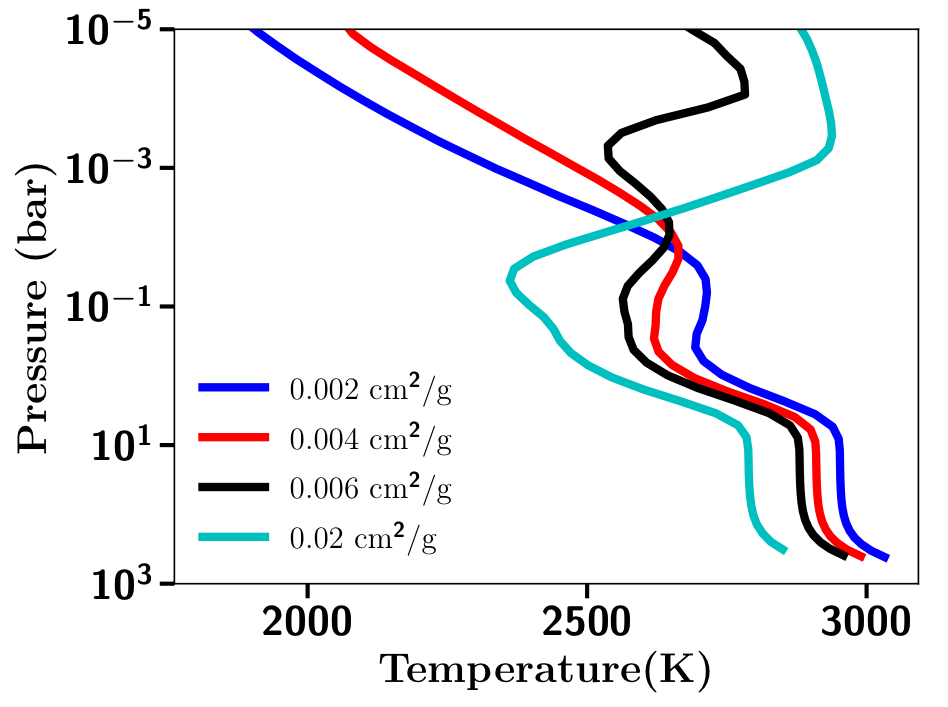}\label{ch5:fig_greyabs_1}}
 \subfloat[]{\includegraphics[width=\columnwidth]{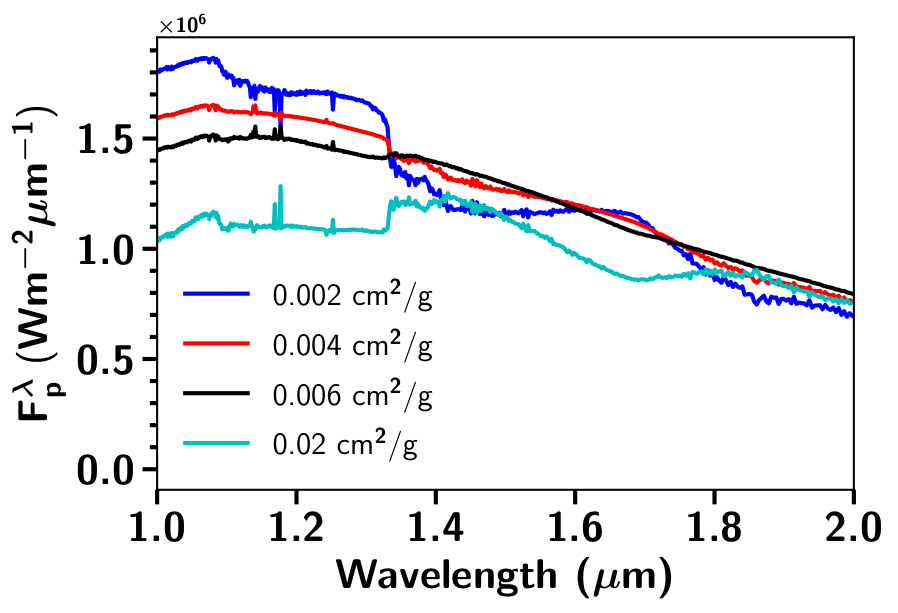}\label{ch5:fig_greyabs_2}}
\end{center}
 \caption[Figure showing RCE $P$-$T$ profiles and emission spectra with different amount of optical grey opacity added throughout the atmosphere]{\textbf{(a)} Figure showing $P$-$T$ profiles for WASP-121b with 0.8 f$_{\textup{c}}$, solar metallicity and solar C/O ratio, with different amount of optical grey opacity added throughout the atmosphere. \textbf{(b)} Figure showing emission spectra with different amount of optical grey opacity using $P$-$T$ profiles shown in Figure \ref{ch5:fig_greyabs_1}.}
\end{figure*}

\begin{figure*}
\begin{center}
 \subfloat[]{\includegraphics[width=\columnwidth]{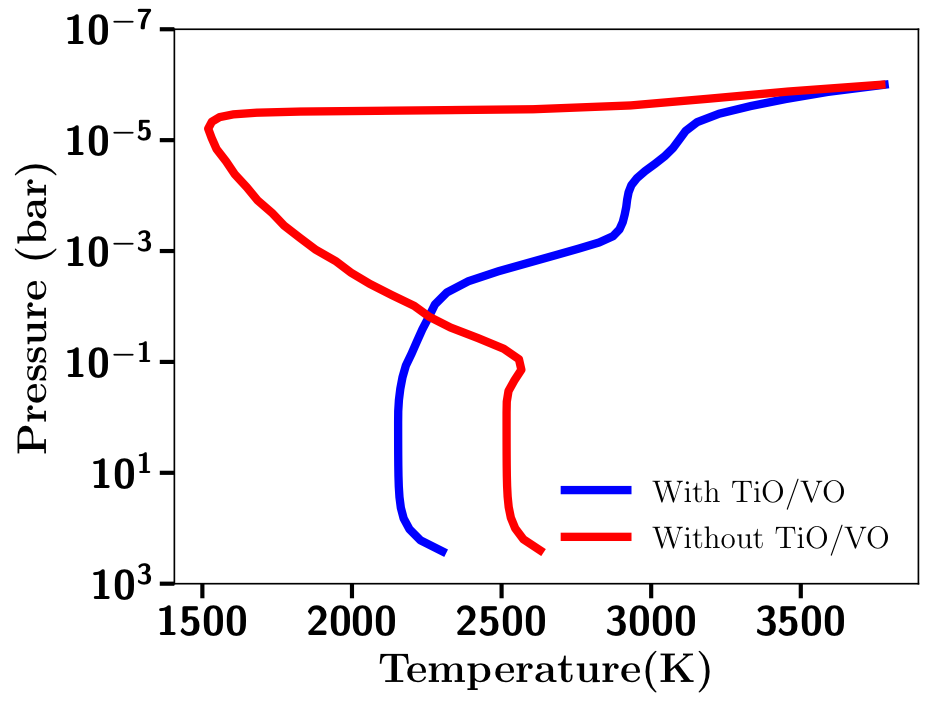}\label{ch5:fig_tiovo_1}}
 \subfloat[]{\includegraphics[width=\columnwidth]{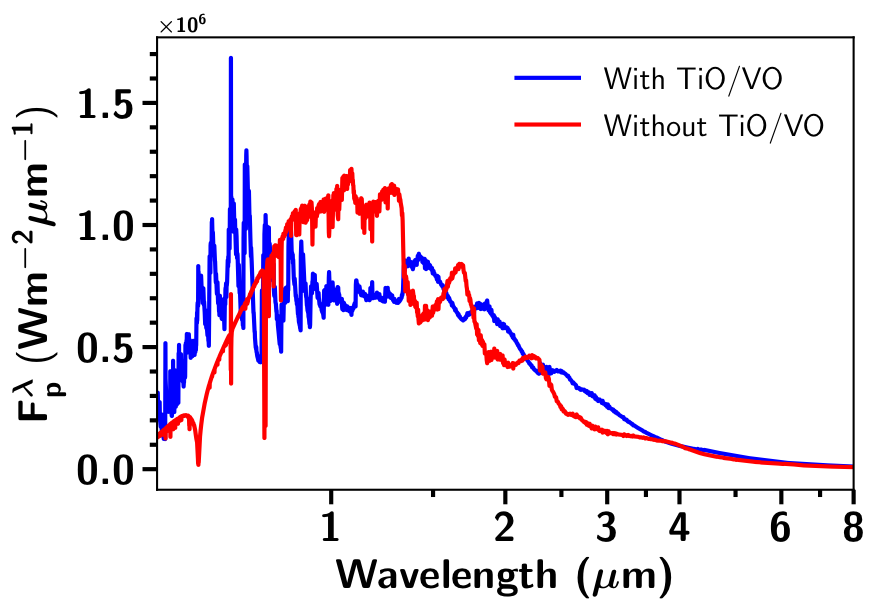}\label{ch5:fig_tiovo_2}}
\end{center}
 \caption[Figure comparing RCE $P$-$T$ profiles with and without TiO/VO opacities, but with Fe opacity]{\textbf{(a)} Figure showing RCE $P$-$T$ profiles with all opacities (blue) including Fe and without TiO/VO opacities (red), at solar metallicity and solar C/O ratio with 0.5 f$_{\textup{c}}$ for WASP-121b. \textbf{(b)} Figure showing emission spectra using $P$-$T$ profiles shown in Figure \ref{ch5:fig_tiovo_1} with all opacities (blue) including Fe and without TiO/VO opacities (red) for WASP-121b.}
 \label{ch5:fig_fe_withouttiovo}
\end{figure*} 

\subsubsection{Effect of Fe opacity}
\label{ch5:effect_fe_opacity}
Gaseous iron (Fe) opacity dramatically effects the $P$-$T$ profiles of extremely irradiated hot Jupiters \citep{Lothringer2018}. It tends to heat the upper, lower pressure, atmosphere as shown in the Fig. \ref{ch5:fig_Fe_1} and cool the lower, high pressure, atmosphere. The cross-section of Fe is quite large in the optical leading to strong absorption and therefore the heating in the upper atmosphere. This blocking of radiation higher up in the atmosphere is one of the causes of the cooling in the deeper atmosphere similar to H$^-$ . The transmission spectra in Figure \ref{ch5:fig_Fe_2}, shows that Fe opacity is important in the optical part of the spectrum, generally dominated by Rayleigh scattering slope, with very sharp spectral features. The emission spectrum in Figure \ref{ch5:fig_Fe_3} with and without Fe opacity, shows that the strength of the thermal inversion in the emission increases after adding Fe opacity. Thus Fe opacity can play an important role in creating secondary inversions, but in the lower pressure regions ($\sim$ 0.1 mbar) as compared to that created due to H$^-$ or TiO/VO opacity (see Section \ref{ch5:inv_fe_without_tio} for more details). The absence of H$_2$O due to thermal dissociation also contributes to this secondary inversion at extremely low pressures, as there is no strong infrared emitting species available to re-emit the energy absorbed by Fe. The difference between the emission spectrum shown in Figure \ref{ch5:fig_Fe_3} is mainly due to the increase in the size of emission features, due to a larger temperature difference between the lower (high pressure) and the upper (low pressure) atmosphere in simulations including Fe opacity, as compared to those without Fe opacity. 

\subsubsection{Inversions in hot Jupiters without TiO/VO}
\label{ch5:inversion}
The formation of temperature inversion in a irradiated planetary atmosphere is governed by the interplay between absorbed stellar radiation and planetary emission. The presence of temperature inversions in hot Jupiter exoplanet atmospheres have been predicted for a long time \citep{Hubeny2003, Fortney2008}, primarily due to TiO/VO opacities, as shown in Figure \ref{ch5:fig_tiovo_1}. The presence of temperature inversion due to TiO/VO also leads to substantial changes in the emission spectrum of the planet as shown in Figure \ref{ch5:fig_tiovo_2}. However, it is only very recently that a definitive evidence of an inversion layer has been seen in the atmosphere of an ultra-hot Jupiter WASP-121b \citep{Evans2017}.  Although, there is a tentative evidence of VO in the transmission spectrum of WASP-121b \citep{Evans2018}, it is still unclear what opacity source is causing this additional absorption creating a temperature inversion in WASP-121b, Therefore, in this section, the opacity required to produce an inversion, its impact on the $P$-$T$ profile and thereby the emission spectra is investigated. 

\begin{figure*}
\begin{center}
 \subfloat[]{\includegraphics[width=\columnwidth]{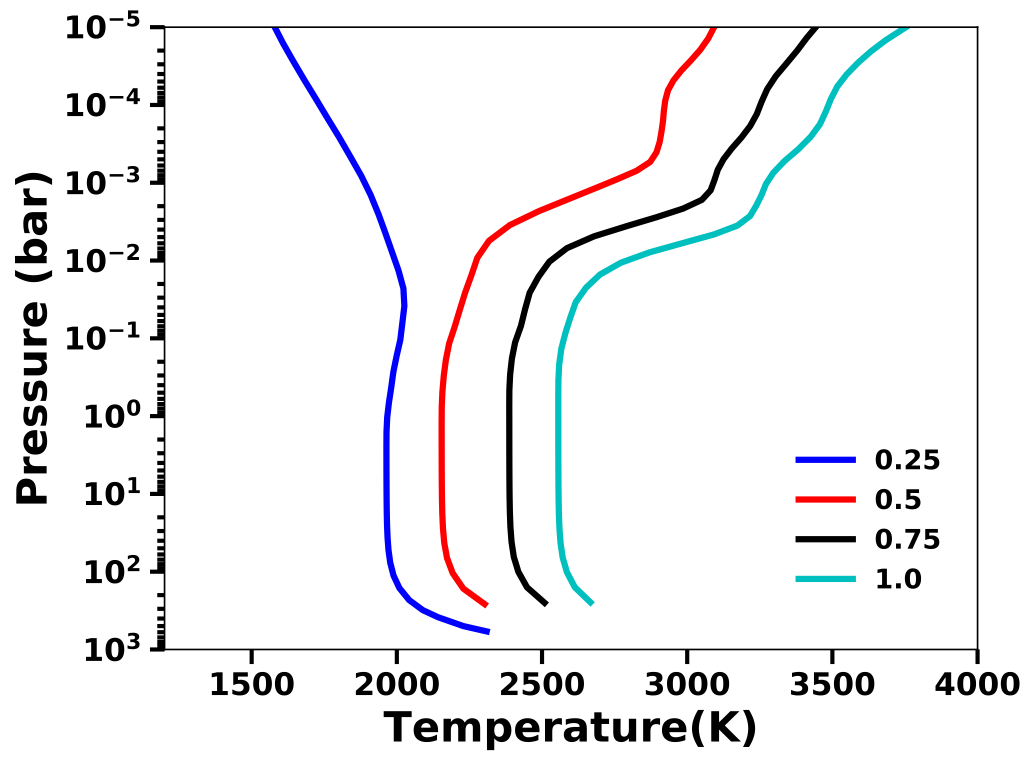}\label{ch5:fig_rc_1}}
 \subfloat[]{\includegraphics[width=\columnwidth]{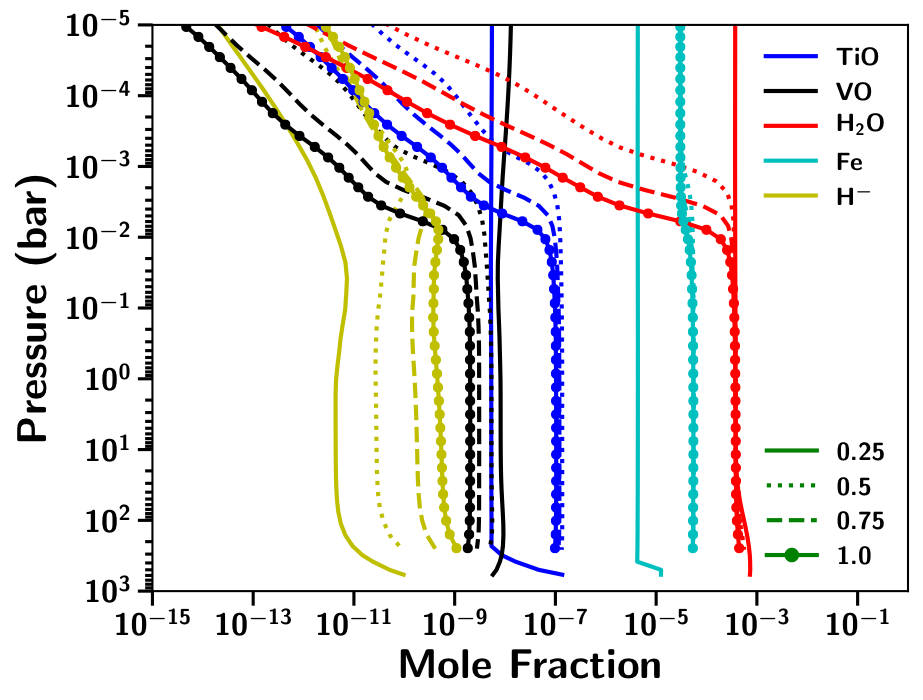}\label{ch5:fig_rc_2}}
 \newline
  \subfloat[]{\includegraphics[width=\columnwidth]{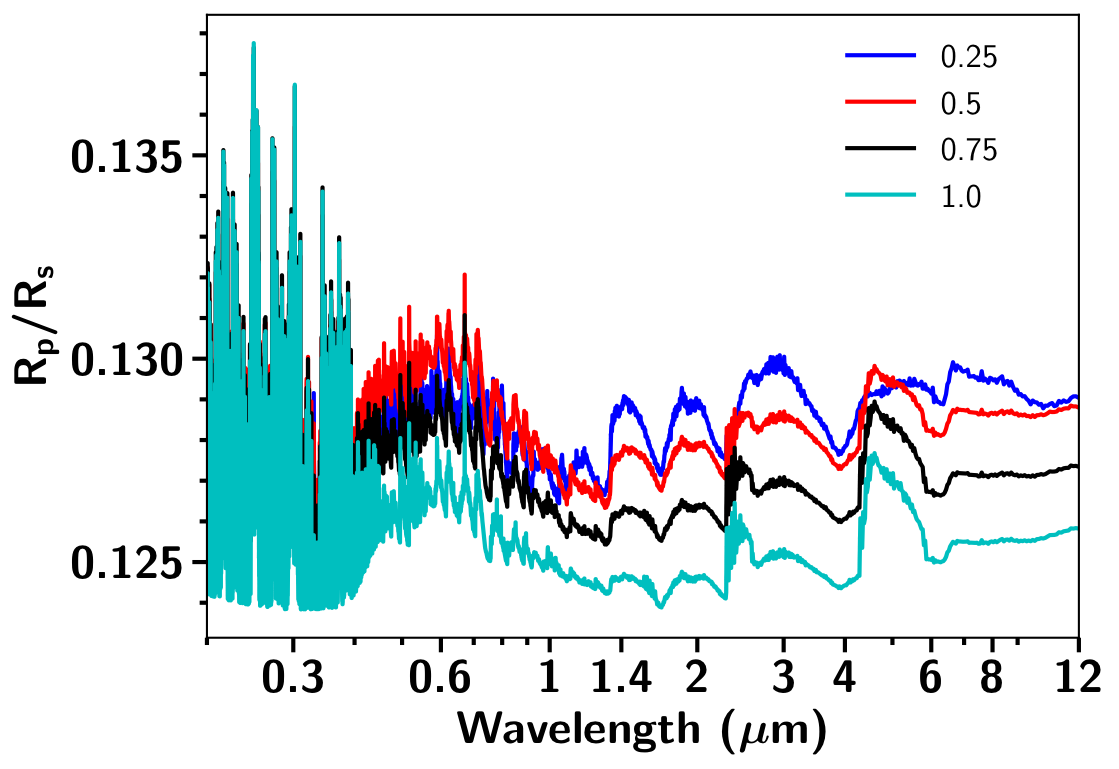}\label{ch5:fig_rc_3}}
 \subfloat[]{\includegraphics[width=\columnwidth]{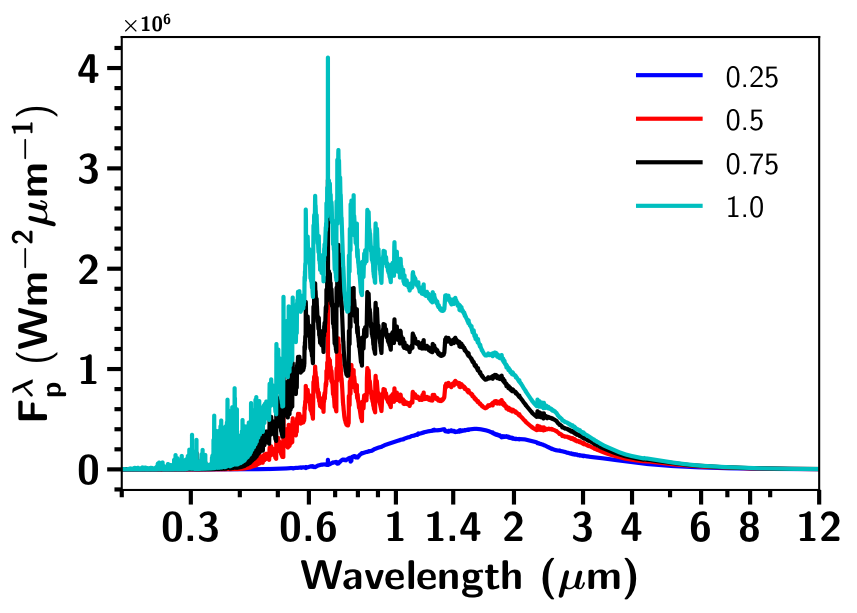}\label{ch5:fig_rc_4}}
  \newline
  \subfloat[]{\includegraphics[width=0.55\columnwidth]{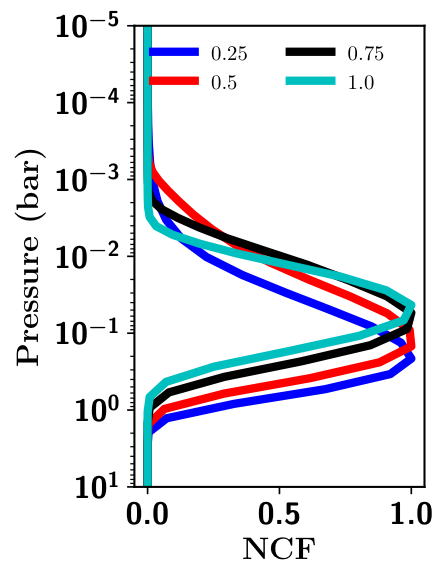}\label{ch5:fig_rc_5}}
 \subfloat[]{\includegraphics[width=0.55\columnwidth]{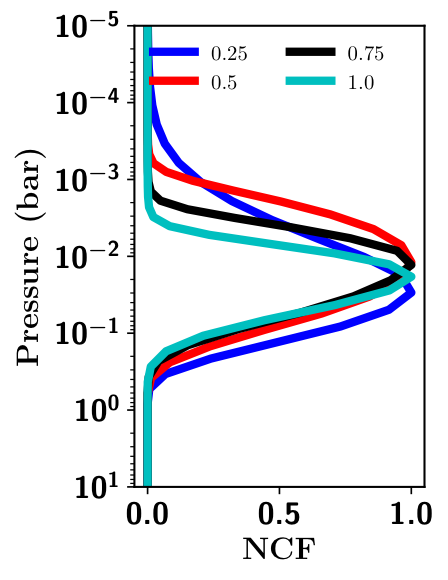}\label{ch5:fig_rc_6}}
\end{center}
 \caption[Figure showing RCE $P$-$T$ profiles and the corresponding equilibrium chemical abundances, transmission and emission spectra, for a range of recirculation factors]{\textbf{(a)} Figure showing $P$-$T$ profiles for a range of f$_{\textup{c}}$ values (0.25, 0.5, 0.75 and 1) at solar metallicity and solar C/O ratio (0.55) for WASP-121b \textbf{(b)} Figure showing equilibrium chemical abundances for some important species for various f$_{\textup{c}}$ values obtained using $P$-$T$ profiles shown in Figure \ref{ch5:fig_rc_1}. \textbf{(c)} Figure showing transmission spectra for various f$_{\textup{c}}$ values obtained using $P$-$T$ profiles shown in Figure \ref{ch5:fig_rc_1} and corresponding equilibrium chemical abundances shown in Figure \ref{ch5:fig_rc_2}. \textbf{(d)} Figure showing planetary emission spectra for various f$_{\textup{c}}$ values obtained using $P$-$T$ profiles shown in Figure \ref{ch5:fig_rc_1} and corresponding equilibrium chemical abundances shown in Figure \ref{ch5:fig_rc_2}. \textbf{(e)} Figure showing normalized contribution function at 1.7$\mic$ for a range of f$_{\textup{c}}$ values for emission spectra as shown in Figure \ref{ch5:fig_rc_4}. \textbf{(f)} Same as Figure \ref{ch5:fig_rc_5} but at 1.4$\mic$.}
 \label{ch5:fig_rc_a}
\end{figure*}

\paragraph{Temperature inversions with grey and H$^-$ opacity:}
\label{ch5:inv_grey_opacity}
Following \citet{Burrows2008} and \citet{Spiegel2009}, we add an arbitrary grey absorbing opacity across the optical  wavelengths (0.44 - 1$\mic$) throughout the atmosphere (all model layers) of WASP-121b with 0.8 f$_{\textup{c}}$, solar metallicity and C/O ratio. Varying the magnitude of this opacity then allows us to explore the evolution of the $P$-$T$ profile from being non-inverted to being inverted as a function of opacity, along with the evolution of the emission spectrum for these different atmospheric structures. The value of 0.8 f$_{\textup{c}}$ is chosen as the best fit value to observations in \citet{Evans2017}. For this particular test along with a grey absorbing opacity in the optical, only the H$_2$-H$_2$ and H$_2$-He CIA, and H$_{2}$O, CO$_2$, CO, CH$_4$, NH$_3$, Na, K, Li, Rb, Cs opacities were used for simplicity. Figure \ref{ch5:fig_greyabs_1} and \ref{ch5:fig_greyabs_2} show the $P$-$T$ profile and emission spectra, respectively, for varying levels of grey opacity for WASP-121b. We note that we have omitted TiO/VO opacity in this model simulation. It can be seen that as the grey opacity increases from 0.002 to 0.02 cm$^2$/g the $P$-$T$ profile changes from being non-inverted to inverted. We also compute the ratio 

\begin{equation} 
    \gamma = \frac{\kappa_{\textup{vis}}}{\kappa_{\textup{IR}}},  
     \label{eq:kap_ratio}
\end{equation}

where $\kappa_{\textup{vis}}$ and $\kappa_{\textup{IR}}$ are the mean opacities in the visible and IR wavelengths, as used in previous works \citep{Molliere2015, Gandhi2019}, to quantify the strength of the inversion. We compute $\kappa_{\textup{IR}}$ using the Planck mean opacity for the temperature of each atmospheric level, whereas $\kappa_{\textup{vis}}$ is computed using the Planck mean opacity at the stellar effective temperature of WASP-121, which is $\sim$6500\,K. At the photospheric level ($\sim$0.1 bar) the $\gamma$ values are 4.78, 5.5, 6.06 and 8.3 for grey opacity values of 0.002, 0.004, 0.006 and 0.02 cm$^2$/g, respectively. The strength of the inversion increases as the value of $\gamma$ increases. However, we note that this method to quantify strength of inversion can be too simplistic, especially when comparing strength of inversion caused by different species, mainly due to two reasons. Firstly, the altitude/pressure of the inversion can vary depending on the RCE $P$-$T$ profile and the corresponding chemical equilibrium abundances, therefore using the $\gamma$ value at one particular pressure level may not be an accurate way to compare strength of inversions. Secondly, for hot planets like WASP-121b there will be some overlap between the visible (stellar spectrum) and infrared (planet spectrum), which again means this simplistic approach is not likely to provide accurate measure of the strength of the inversion.

The change in the emission spectrum is more interesting where the H$_2$O absorption feature at 1.4$\mic$ gradually changes into an emission feature, as the amount of grey opacity is increased. Moreover, with 0.006 cm$^2$/g grey opacity the spectrum almost resembles a blackbody spectrum indicating an isothermal atmosphere, which can be seen in Figure \ref{ch5:fig_greyabs_1}. 

As discussed in detail in Section \ref{ch5:effect_h-_opacity} and shown in Figure \ref{ch5:fig_H-_2}, H$^-$ opacity which almost acts like a grey opacity in the optical can produce a strong inversion in the absence of TiO/VO opacities, which only happens at high C/O ratios (>1), under chemical equilibrium conditions. Therefore, H$^-$ opacity could possibly be the opacity source that leads to inversion in the atmosphere of WASP-121b and many other ultra hot Jupiter planets. 

\paragraph{Can Fe without TiO/VO form an inversion?}
\label{ch5:inv_fe_without_tio}
Iron (Fe) has very strong opacity in the UV and the optical part of the spectrum. Therefore, we investigated whether Fe can lead to inversions without TiO/VO in extremely irradiated hot Jupiter such as WASP-121b. For this we removed TiO/VO opacities from the model while computing RCE $P$-$T$ profiles. We found that the Fe opacity we include is unable to produce a strong inversion like TiO/VO at solar metallicity as shown in Figure \ref{ch5:fig_fe_withouttiovo}. Even if we increase the metallicity to 200 times solar, leading to an increase in Fe abundance to about $\sim 30\times$ than that at solar metallicity, it does not lead to an inversion. Fe opacity however, leads to a sharp increase in the temperature at lower pressures ($\sim$ 10$^{-5}$ bar like a thermospheric secondary inversion). The primary reason behind this secondary inversion at low pressures is the low abundance of species that emit strongly in the infrared (for e.g thermal decomposition of H$_2$O), which reduces atmospheric cooling (see Figure \ref{ch5:fig_co_wasp121_2}). We also observe this secondary inversion due to various other strong optical absorbers such as Na and K, when the abundance of strong infrared emitters such as H$_2$O or CH$_4$ decreases. 

This overall result due to Fe opacity might seem to contradict that of \citet{Lothringer2018}, where they see an inversion due to Fe opacity at around 10$^{-2}$ bar (10 mbar). However, \citet{Lothringer2018} used a fiducial model with a planet at 0.025 au from a star with effective temperature of 7200\,K, about $\sim$750\,K hotter than WASP-121. Moreover, they include bound-free Fe opacity in their model simulations which becomes important for wavelengths shorter than 0.2$\mic$ \citep{Sharp2007}. The combination of all these factors may have led to differences between our results, alongside other factors such as the equilibrium chemical abundances of Fe as well as the treatment of condensation. However, in KELT-9b the planet with highest equilibrium temperature in our library of models, Fe opacity leads to formation of a strong temperature inversion starting at $\sim$0.1 bar, even at solar metallicity and without requiring TiO/VO/H$^-$. Thus, the capability of Fe opacity to form a temperature inversion is system specific showing a strong dependance on the host star effective temperature. 

We note that there are many other potential species such as AlO, CaO, SiO, NaH, AlH, SiH, MgH etc. shown in recent studies \citep{Lothringer2018, Malik2018, Gandhi2019} with strong optical opacities that could lead to the formation of an inversion in ultra-hot Jupiters. However, these opacities have not been included in the current version of our library. Among these species there have been indications of AlO in some of the planets \citep{vonEssen2019, Chubb2020}. Therefore, some of these opacities will be considered in the future version of the library.

\subsection{Sensitivity to Grid Parameters}
\label{ch5:grid_param_sensitivity}
In this section we show the sensitivity of the model simulation, i.e $P$-$T$ profiles and the spectra, to all the grid parameters, namely recirculation factor, metallicity and C/O ratio across their full range used in this library, using WASP-17b and WASP-121b as test cases. 

\subsubsection{Effect of recirculation factor}
\label{ch5:eff_rcf}
The recirculation factor (f$_{\textup{c}}$) described in Section \ref{ch5:rcf} governs the efficiency of re-distribution of energy (by winds) received  from the host star in a column. The value of 1 corresponds to no-redistribution, with an increase in redistribution as this factor decreases. Here, we show the effect of varying the recirculation factor on the $P$-$T$ profiles and thereby the chemical abundances and spectra, using the extremely irradiated hot Jupiter WASP-121b as an example. As can in seen in the $P$-$T$ profiles in Figure \ref{ch5:fig_rc_1} for WASP-121b, the strength of the atmospheric temperature inversion increases with f$_{\textup{c}}$ as expected, since more energy is available to create an inversion at higher values of f$_{\textup{c}}$. At 0.25 f$_{\textup{c}}$ the inversion is absent in the $P$-$T$ structure due to reduced irradiation (energy). Absorption due to TiO/VO is the primary reason for the inversion, but surprisingly the abundance of TiO/VO starts decreasing, as the inversion is formed and increases in strength, as can be seen in Figure \ref{ch5:fig_rc_2} for increasing f$_{\textup{c}}$. However, the abundance of H$^-$ and Fe increases with increasing f$_{\textup{c}}$. This increase in H$^-$ maintains the temperature inversion even though the abundance of TiO/VO decreases, as described in in Section \ref{ch5:effect_h-_opacity} and also Section \ref{ch5:wasp121_co_oxy}.

The transmission spectrum of WASP-121b in Figure \ref{ch5:fig_rc_3}, shows that the strength of the \water features decreases with increasing f$_{\textup{c}}$ as \water abundance decreases in the region where transmission spectra probes ($\sim$ 1mbar), since \water starts becoming thermally unstable with increasing temperatures. However, CO features start appearing near $\sim$2.5$\mic$ and broadband CO features between 4 to 6$\mic$. The strength of these CO features increases with increasing f$_{\textup{c}}$.  As expected, the flux in the planetary emission spectrum shown in Figure \ref{ch5:fig_rc_4} increases with increasing f$_{\textup{c}}$ as the temperature of emission increases. The $P$-$T$ profile at 0.25 f$_{\textup{c}}$ is very close to isothermal, therefore its emission spectrum also resembles a blackbody curve, with small dips in the strong water vapour absorption bands at 1.4, 2 and 3$\mic$. In contrast, $P$-$T$ profiles at other f$_{\textup{c}}$ values have temperature inversions, thus leading to a bump instead of dip in the strong \water and CO absorption bands.  

The normalised contribution functions (NCF) at 1.7 and 1.4$\mic$ are shown in Figure \ref{ch5:fig_rc_5} and \ref{ch5:fig_rc_6}, respectively, for WASP-121b. At 1.7$\mic$, the NCF peaks deeper in the atmosphere as compared to that at 1.4$\mic$, indicating emission at 1.7$\mic$ is from comparatively deeper parts of the atmosphere, since 1.7$\mic$ is at the edge of strong water absorption band centred at 1.4$\mic$. For the profiles with a temperature inversion (f$_{\textup{c}}$ = 0.5, 0.75, 1.0), the deeper and cooler isothermal part (not the inversion) is primarily probed at 1.7$\mic$ as compared to 1.4$\mic$ which probes the inversion layer, thus leading to a emission feature (bump) instead of a absorption feature (dip) in the emission spectrum from 1.2 to 1.7$\mic$. This also happens for other strong water absorption bands as shown in the emission spectra. The 1.4$\mic$ emission feature has been detected in the atmosphere of WASP-121b \citep{Evans2017}, but other such potential emission features indicative of inversion for wavelengths > 2$\mic$ are still to be detected for WASP-121b and can only be possible with JWST.  

At 0.25 f$_{\textup{c}}$ the inversion is absent as we reduce the irradiation received from the host star to 25\% of its original value, mimicking the transport of energy by advection (strong winds). This also motivates accurate 3D modelling of extremely irradiated hot Jupiter exoplanets to predict inversions as well as to infer wind velocities based on the presence or absence of inversions. 

\subsubsection{Effect of Metallicity}
\label{ch5:eff_metal}
Metallicity fundamentally effects the chemistry of the atmosphere and thereby its $P$-$T$ structure and observed spectrum. The effect of metallicity on the $P$-$T$ structure, chemistry, transmission and emission spectra for two different planets is discussed in this section. 

\begin{figure*}
\begin{center}
 \subfloat[]{\includegraphics[width=\columnwidth]{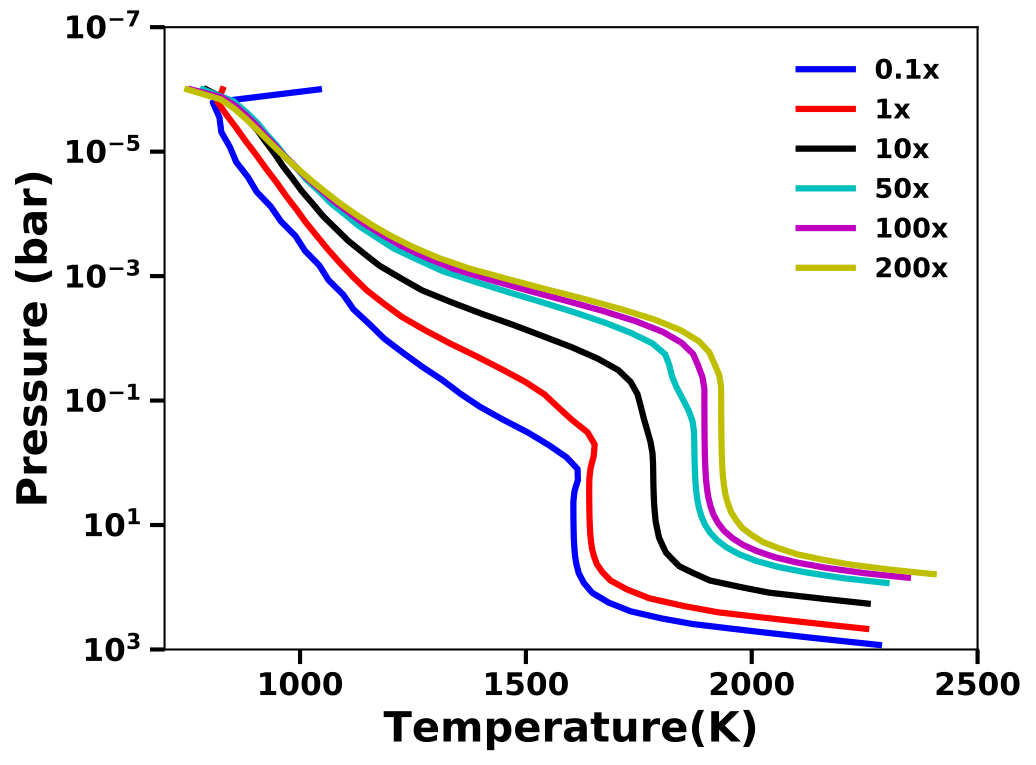}\label{ch5:fig_met_WASP-017_1}}
 \subfloat[]{\includegraphics[width=\columnwidth]{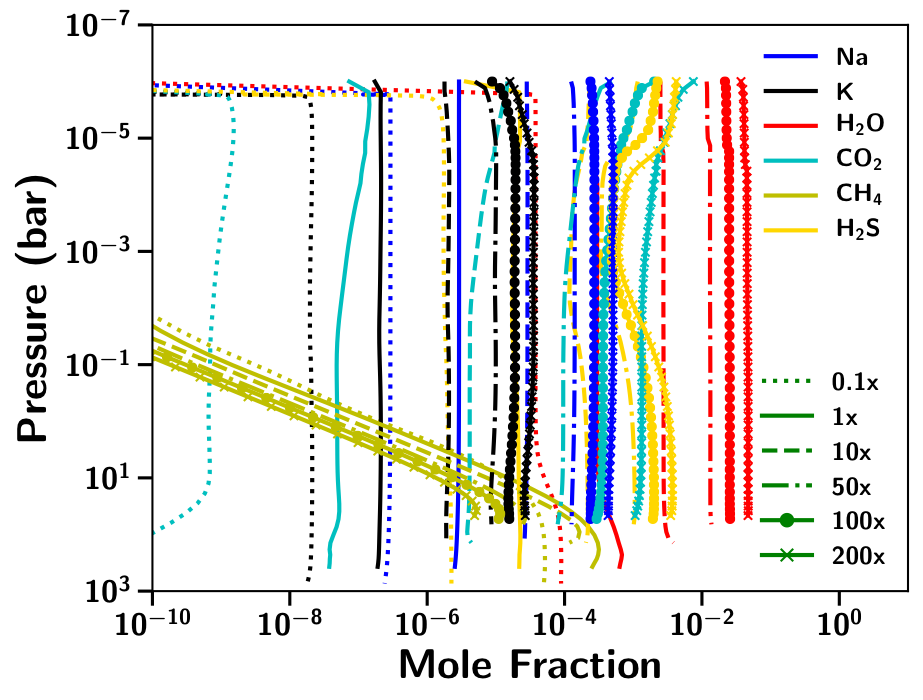}\label{ch5:fig_met_WASP-017_2}}
 \newline
  \subfloat[]{\includegraphics[width=\columnwidth]{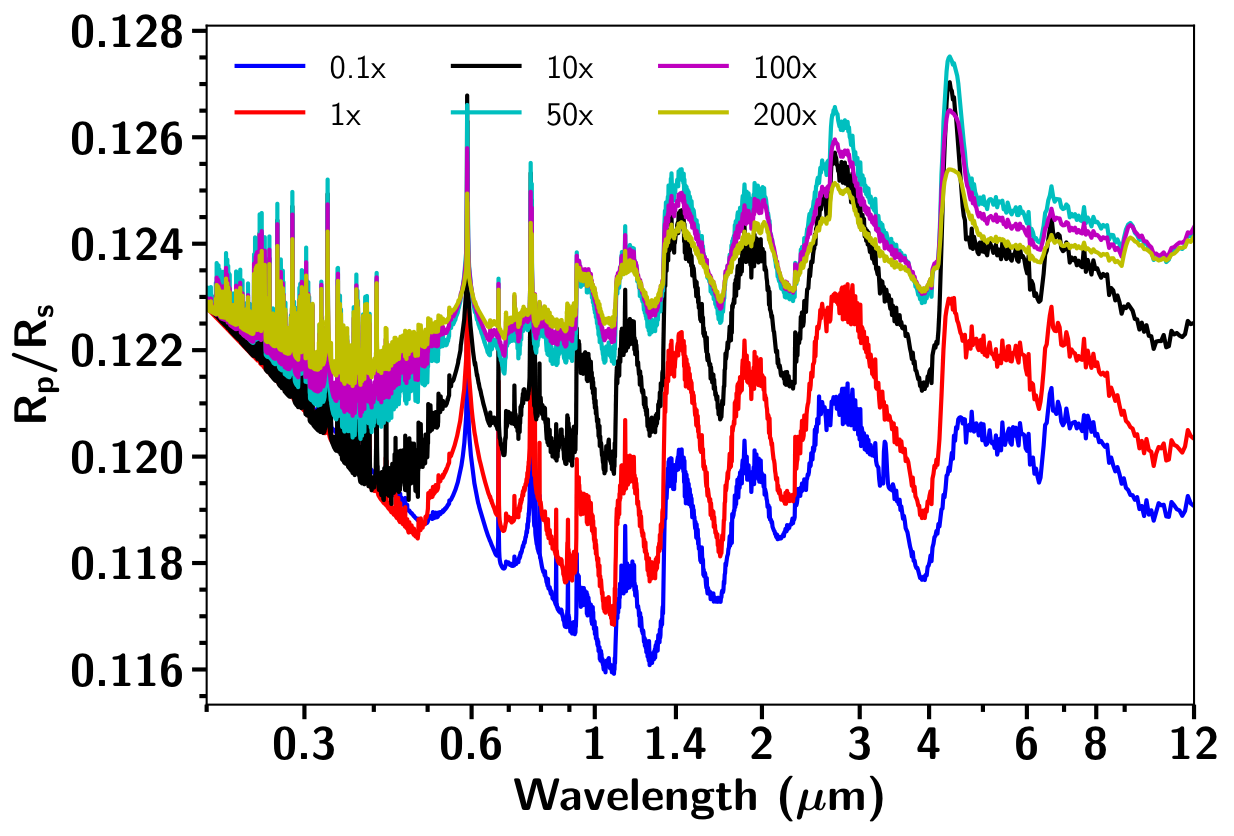}\label{ch5:fig_met_WASP-017_3}}
 \subfloat[]{\includegraphics[width=\columnwidth]{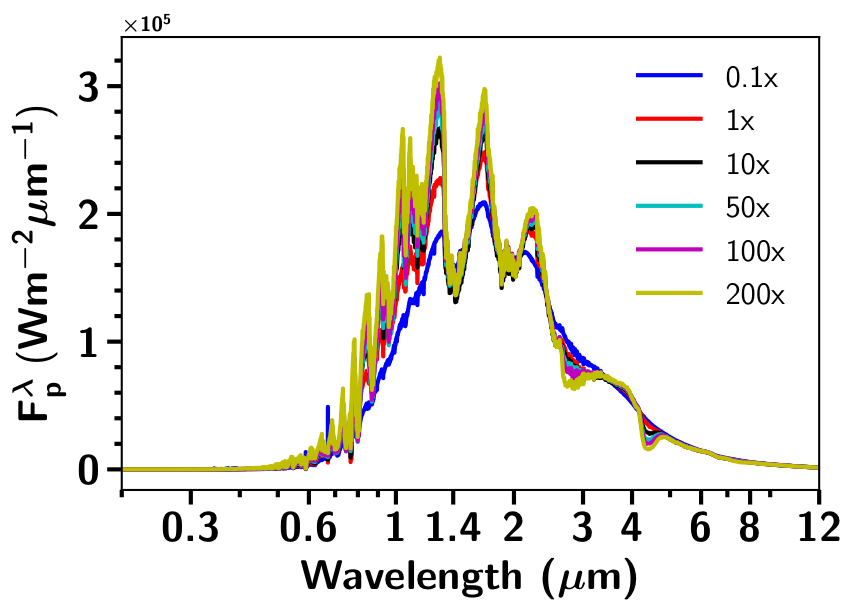}\label{ch5:fig_met_WASP-017_4}}
  \newline
  \subfloat[]{\includegraphics[width=0.5\columnwidth]{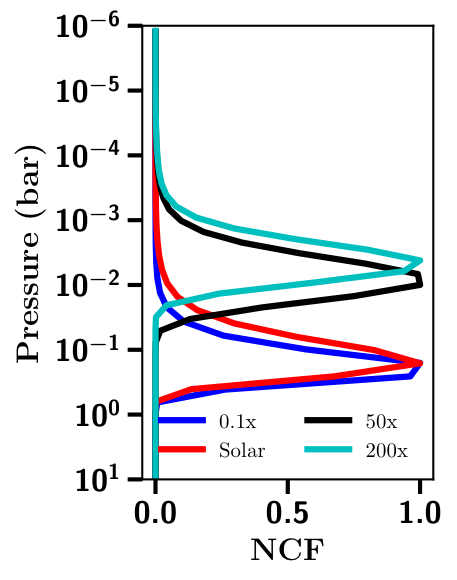}\label{ch5:fig_met_WASP-017_5}}
 \subfloat[]{\includegraphics[width=0.5\columnwidth]{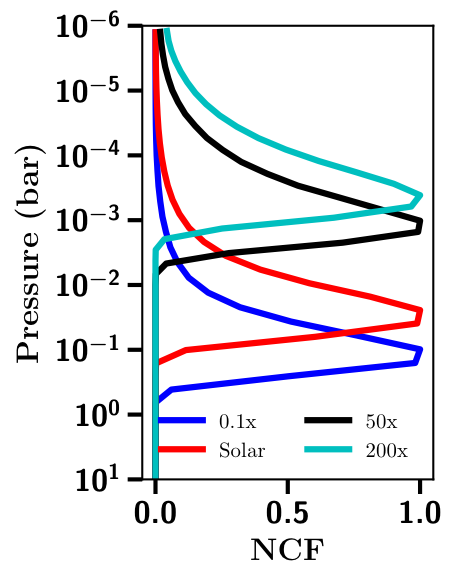}\label{ch5:fig_met_WASP-017_6}}
   \subfloat[]{\includegraphics[width=0.5\columnwidth]{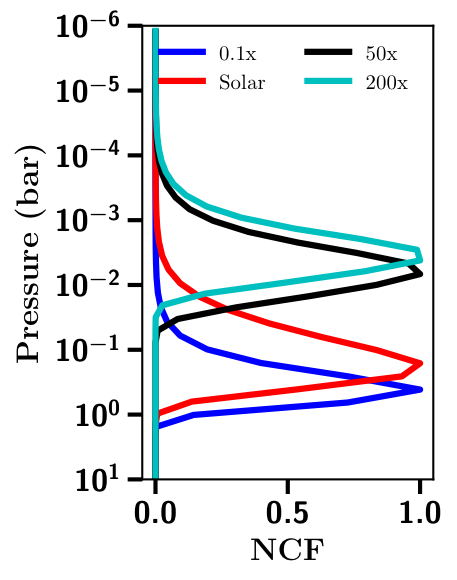}\label{ch5:fig_met_WASP-017_7}}
 \subfloat[]{\includegraphics[width=0.5\columnwidth]{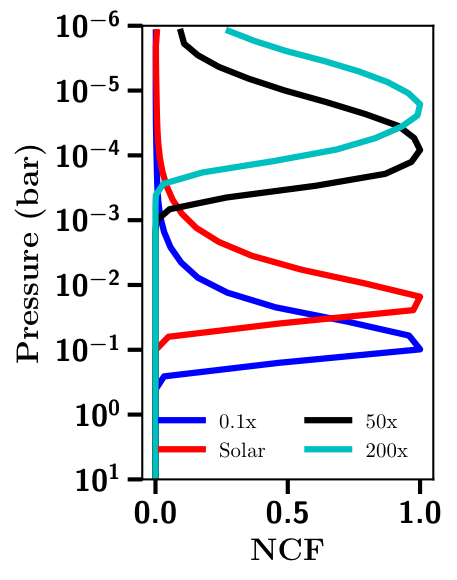}\label{ch5:fig_met_WASP-017_8}}
\end{center}
 \caption[Figure showing RCE $P$-$T$ profiles and the corresponding equilibrium chemical abundances, transmission and emission spectra, and contribution functions for a range of metallicities adopted in the consistent library, using WASP-17b as the test case]{\textbf{(a)} Figure showing $P$-$T$ profiles for a range of metallicities at 0.25 f$_{\textup{c}}$ and a solar C/O ratio (0.55) for WASP-17b. \textbf{(b)} Figure showing equilibrium chemical abundances for some important species for various metallicity values obtained using $P$-$T$ profiles shown in Figure \ref{ch5:fig_met_WASP-017_2}. \textbf{(c)} Figure showing transmission spectra for WASP-17b for different values of metallicity obtained using $P$-$T$ profiles shown in Figure \ref{ch5:fig_met_WASP-017_1} and corresponding equilibrium chemical abundances shown in Fig. \ref{ch5:fig_met_WASP-017_2}.  \textbf{(d)} Figure showing emission spectra for WASP-17b for different values of metallicity obtained using $P$-$T$ profiles shown in Figure \ref{ch5:fig_met_WASP-017_1} and corresponding equilibrium chemical abundances shown in Fig. \ref{ch5:fig_met_WASP-017_2}. \textbf{(e, f, g, h)} Figures e, f, g, h showing normalised contribution function at 2.25, 2.8, 3.8 and 4.5 $\mic$ for a range of metallicity values for emission spectra shown in Figure \ref{ch5:fig_met_WASP-017_4}.}
 \label{ch5:fig_metal_wasp17}
\end{figure*}

\paragraph{WASP-17b:}
WASP-17b is a hot Jupiter planet with an equilibrium temperature of 1755\,K. Adopting a value of  0.5 for the f$_{\textup{c}}$ the transmission spectra for this planet shows TiO/VO features due to hotter $P$-$T$ structure as compared to that using 0.25 f$_{\textup{c}}$. However, observations from \citet{Sing2016} shows the absence of TiO/VO features, therefore we restrict the following analysis to simulations adopting an f$_{\textup{c}}$ of 0.25 (although all values are available in the model grid). The C/O ratio is fixed to the solar value and the metallicity varied from across the grid range to investigate the effect of metallicity. For this planet, an increasing metallicity leads to an increase in the temperature throughout the atmosphere as shown in Figure \ref{ch5:fig_met_WASP-017_1}. This is a result of increased absorption of radiation at lower pressures, due to increased opacity, as the mean molecular weight of the atmosphere increases, driven by an increase in the abundances of species heavier than H$_2$ and He, namely H$_2$O, CO$_2$ and Na as shown in Figure \ref{ch5:fig_met_WASP-017_2}. A sharp increase in temperature for sub-solar metallicity can be seen at pressures less than $\sim10^{-5}$ bar, but it is observationally insignificant either in transmission or emission spectra, due to the low atmospheric density at these pressures. 

The effect of varying the metallicity on the transmission of WASP-17b is shown in Figure \ref{ch5:fig_met_WASP-017_3}. The transmission spectra are primarily dominated by Na, K and narrow Li features in the optical for all metallicities, with weaker TiO/VO features as metallicity increases. The infrared part of the spectrum is primarily dominated by H$_2$O features for all metallicities. The size of the H$_2$O features initially increases with an increase in metallicity as H$_2$O abundance increases shown in Figure \ref{ch5:fig_met_WASP-017_2}, but then again decreases with increasing metallicity. This is caused by the increase in the mean molecular weight of the atmosphere, leading to a decrease of the atmospheric scale height, which, in turn, shrinks the spectral features in transmission. Pressure broadened wings of Na and K are also effected by change in metallicity. Due to decreasing scale height associated with increasing metallicity, transmission spectra probes high pressure levels of the atmosphere, resulting in enhanced broadening of Na and K line wings with increasing metallicity. 
The CO$_2$ feature near 4.5$\mic$ and 2.5 - 3$\mic$ increases in amplitude, which can also be seen in the emission spectrum, primarily due to a rapid increase in the CO$_2$ abundances. This shows that even under chemical equilibrium conditions the atmosphere rapidly tends to migrate towards a CO$_2$ abundant atmosphere with increasing metallicity \citep{Moses2013}, offering potential reasons for the CO$_2$ dominated compositions of Mars and Venus in our Solar system. Even Earth in the past may have had a CO$_2$ dominated atmosphere, currently captured in the oceans and rocks by various geological processes \citep{Zahnle2010}. 
The transmission spectra can also be compared with Figure 12a in \citet{Errgoyal2019} which uses isothermal $P$-$T$ profiles. The comparison might not be completely accurate as the RCE $P$-$T$ profiles shown here are overall cooler than the equilibrium temperature of  WASP-17b (1755\,K) in the transmission spectra probed region ($\sim$1mbar). The \enquote{best-fit} model transmission spectra of WASP-17b using the grid of models presented in this paper and observations from \citet{Sing2016} is shown in Figure 6 of the online supplementary material. 

The emission spectra of WASP-17b with varying metallicity is shown in Figure \ref{ch5:fig_met_WASP-017_4}. The blackbody emission flux increases with increasing metallicity, since the overall temperature of $P$-$T$ profiles shown in Fig. \ref{ch5:fig_met_WASP-017_1} increases with increasing metallicity. This increase in temperature combined with the increase in abundances of species such as H$_2$O and CO$_2$, with increasing metallicity, leads to deeper absorption features. Similar to the transmission spectrum, the emission spectrum is also dominated by water absorption features for WASP-17b in the infrared with CO$_2$ features around 4.5 and 2.5 - 3$\mic$ region for metallicities greater than 10 times solar value. The atmospheric level which contributes most to the emission at different wavelengths can be found using the NCF. The increase in the overall temperature of the $P$-$T$ profile causes the NCF to consistently shifts towards lower pressure with increase in metallicity as shown in Figures \ref{ch5:fig_met_WASP-017_5}, \ref{ch5:fig_met_WASP-017_6}, \ref{ch5:fig_met_WASP-017_7} and \ref{ch5:fig_met_WASP-017_8}. The NCF at the core of the CO$_2$ absorption band  at 4.5$\mic$ shown in the Fig. \ref{ch5:fig_met_WASP-017_8}, peaks at lower pressures (<$\sim$ 1 mbar) in the atmosphere, for metallicities greater than 50 times solar, than the wing of the band at 3.8$\mic$ as shown in the Figure \ref{ch5:fig_met_WASP-017_7}. The dramatic drop in the emission flux between these two wavelengths is also shown in the emission spectra for metallicities greater than 50 times solar. The NCF for the 2.8 $\mic$ H$_2$O absorption band also shows similar effect as 4.5 $\mic$ CO$_2$ absorption band, shown in Figures \ref{ch5:fig_met_WASP-017_5} and \ref{ch5:fig_met_WASP-017_6}, for 2.25 and 2.8$\mic$, respectively. The atmospheric level probed by emission spectrum shifts to lower pressure levels with increasing metallicity, as shown by contribution functions in the H$_2$O  and CO$_2$ bands. The wings and cores of absorption/emission bands also probe different atmospheric layers for a given metallicity. Therefore, the absorption/emission features of various species can be used to constrain metallicity, as well as the $P$-$T$ temperature structure of the atmosphere. For some cases (for e.g Fig. \ref{ch5:fig_met_WASP-017_8}) particularly at high metallicity, NCF shows that the emission spectrum is probing quite high in the atmosphere ($\sim$10$^{-5}$ bar). In such cases, photochemistry can alter the emission spectrum via the formation of photochemical products and change in chemical abundances (and thereby the $P$-$T$ profile) at high altitudes (P<$\sim$10$^{-4}$ bar) \citep{Madhureview2016, Hobbs2019}. However, since photochemistry is a dis-equilibrium effect, it is not considered in this work and is part of our future work as discussed in the conclusions.

\begin{figure*}
\begin{center}
 \subfloat[]{\includegraphics[width=\columnwidth]{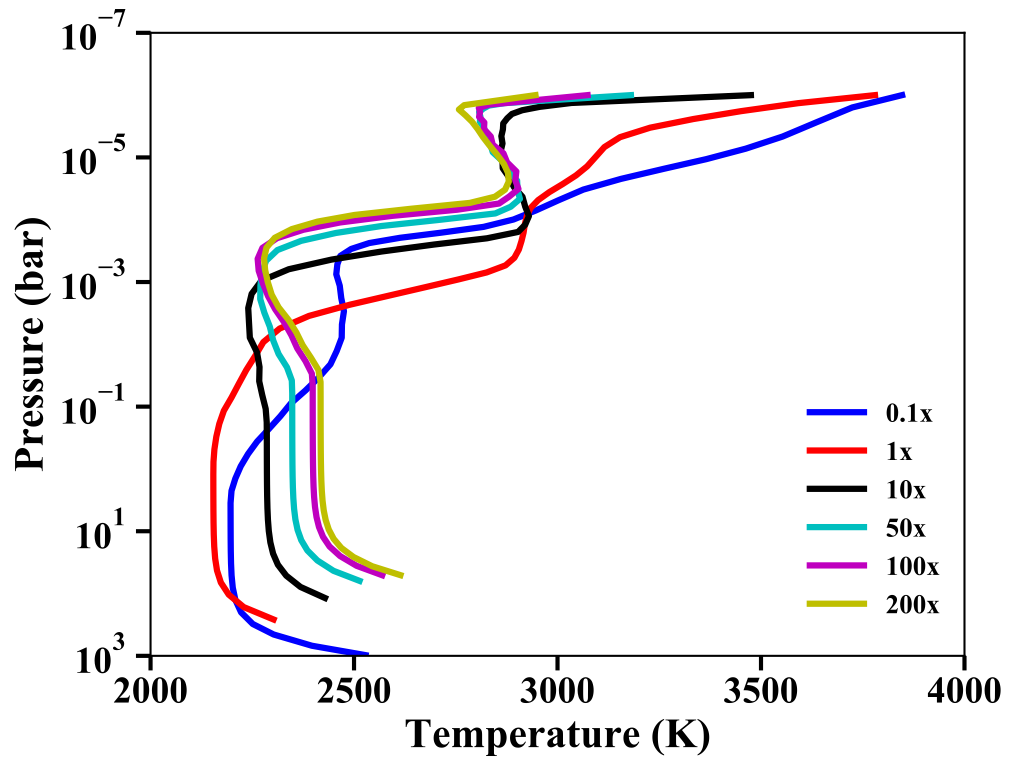}\label{ch5:fig_met_WASP-121_1}}
 \subfloat[]{\includegraphics[width=\columnwidth]{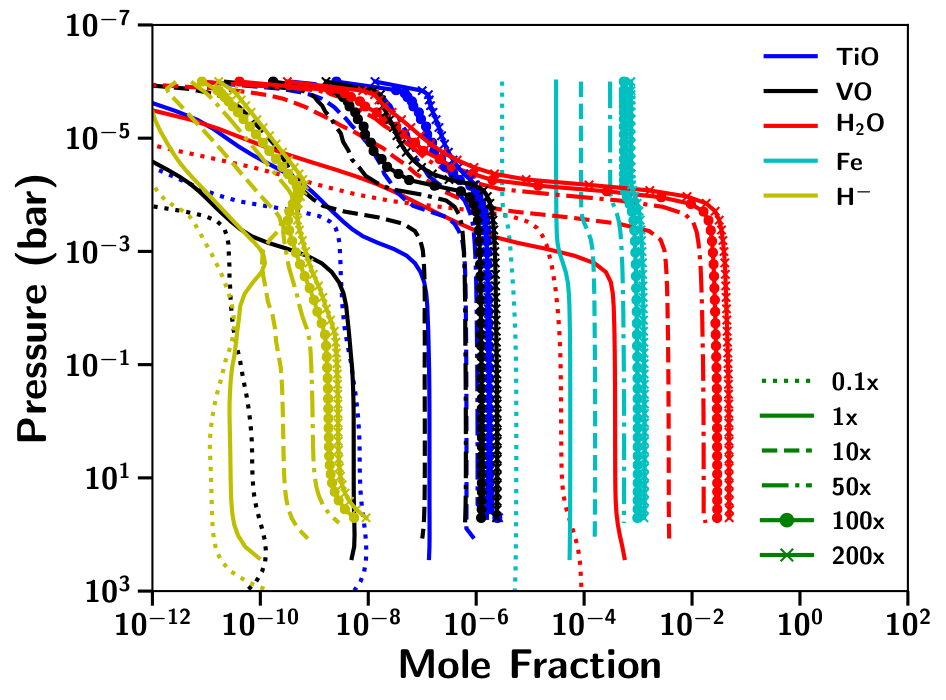}\label{ch5:fig_met_WASP-121_2}}
 \newline
  \subfloat[]{\includegraphics[width=\columnwidth]{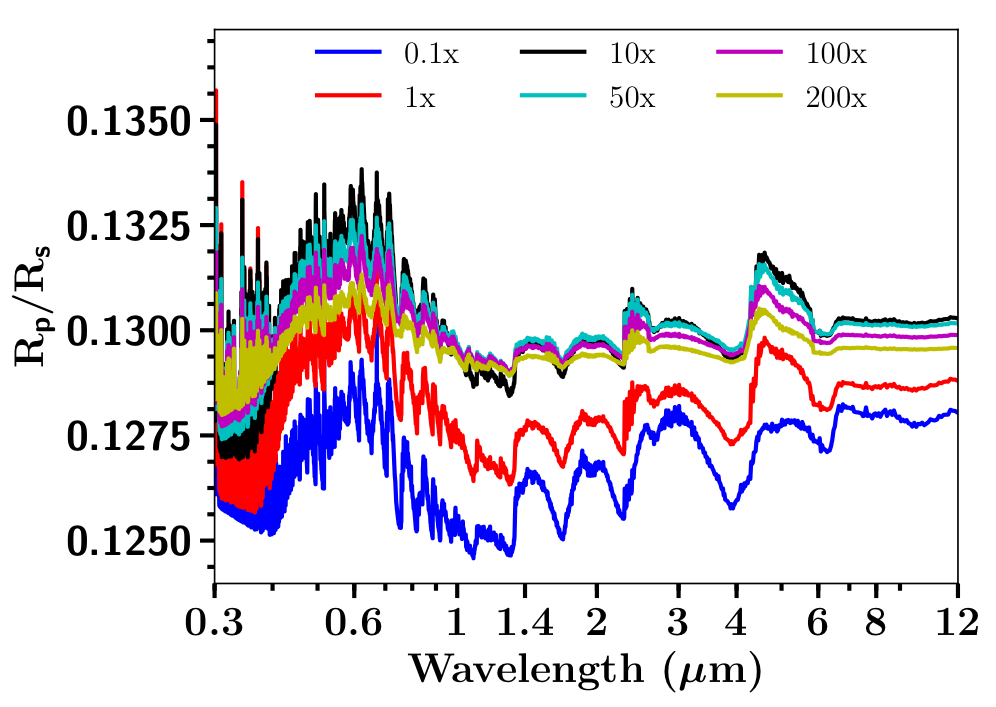}\label{ch5:fig_met_WASP-121_3}}
 \subfloat[]{\includegraphics[width=\columnwidth]{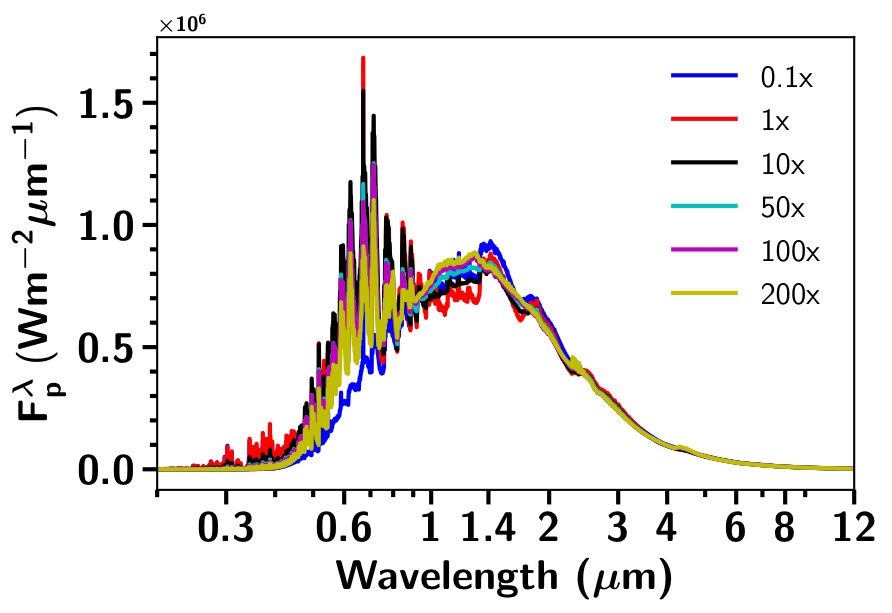}\label{ch5:fig_met_WASP-121_4}}
   \newline
  \subfloat[]{\includegraphics[width=0.5\columnwidth]{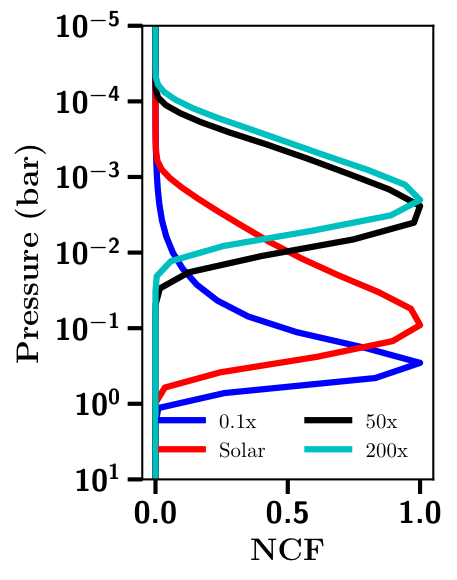}\label{ch5:fig_met_WASP-121_5}}
 \subfloat[]{\includegraphics[width=0.5\columnwidth]{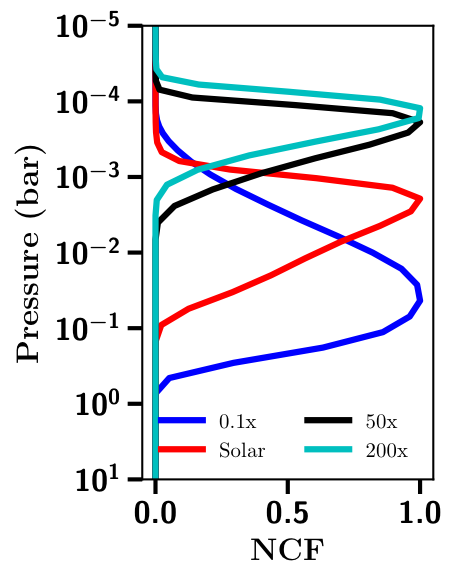}\label{ch5:fig_met_WASP-121_6}}
   \subfloat[]{\includegraphics[width=0.5\columnwidth]{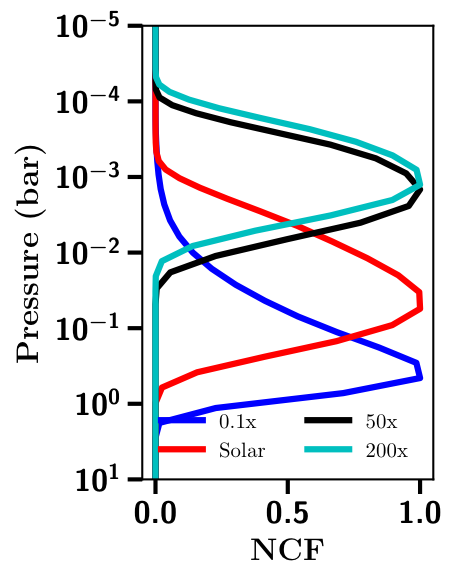}\label{ch5:fig_met_WASP-121_7}}
 \subfloat[]{\includegraphics[width=0.5\columnwidth]{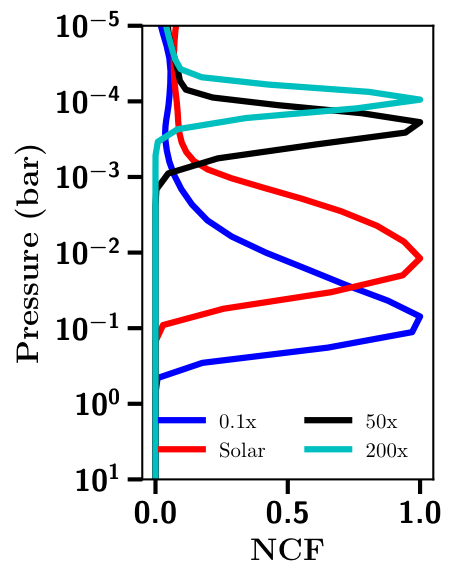}\label{ch5:fig_met_WASP-121_8}}
\end{center}
 \caption[Same as Fig. \ref{ch5:fig_metal_wasp17}, but using WASP-121b as the test case]{\textbf{(a)} Figure showing $P$-$T$ profiles for a range of metallicity at 0.5 f$_{\textup{c}}$ and a solar C/O ratio (0.55) for WASP-121b. \textbf{(b)} Figure showing equilibrium chemical abundances for some important species for various metallicity values obtained using $P$-$T$ profiles shown in Figure \ref{ch5:fig_met_WASP-121_1}. \textbf{(c)} Figure showing transmission spectra for WASP-121b for different values of metallicity obtained using $P$-$T$ profiles shown in Figure \ref{ch5:fig_met_WASP-121_1} and corresponding equilibrium chemical abundances shown in Fig. \ref{ch5:fig_met_WASP-121_2}.  \textbf{(d)} Figure showing emission spectra for WASP-121b for different values of metallicity obtained using $P$-$T$ profiles shown in Figure \ref{ch5:fig_met_WASP-121_1} and corresponding equilibrium chemical abundances shown in Fig. \ref{ch5:fig_met_WASP-121_2}. \textbf{(e, f, g, h)} Figures e, f, g, h showing normalised contribution function at 2.25, 2.8, 3.8 and 4.5 $\mic$ for a range of metallicity values for emission spectra shown in Figure \ref{ch5:fig_met_WASP-121_4}.}
 \label{ch5:fig_metal_wasp121}
\end{figure*}

\paragraph{WASP-121b:}
WASP-121b is an extremely irradiated hot Jupiter planet with an equilibrium temperature (T$_\textup{eq}$) of $\sim$2400\,K \citep{Tepcat2011}. Figures \ref{ch5:fig_met_WASP-121_1}, \ref{ch5:fig_met_WASP-121_2}, \ref{ch5:fig_met_WASP-121_3} and \ref{ch5:fig_met_WASP-121_4} show the $P$-$T$ structure, chemical abundances, transmission and emission spectra, respectively for WASP-121b at 0.5 f$_{\textup{c}}$ and solar C/O ratio. Figures \ref{ch5:fig_met_WASP-121_5}, \ref{ch5:fig_met_WASP-121_6}, \ref{ch5:fig_met_WASP-121_7} and \ref{ch5:fig_met_WASP-121_8} show the contribution function at 2.25, 2.8, 3.8 and 4.5$\mic$, respectively. 

For WASP-121b we see a temperature inversion in the $P$-$T$ profile, which moves towards higher pressure levels with increasing metallicity, primarily driven by TiO/VO absorption. At sub-solar metallicity the inversion is very weak, due to the low abundance of TiO/VO, seen in Figure \ref{ch5:fig_met_WASP-121_2}. The Fe abundance is also substantial, mainly contributing to upper atmosphere heating, as the Fe absorption cross-sections are largest in the UV-Optical spectrum. This upper atmosphere heating leads to the formation of a second inversion layer but with weak observational signatures due to the very low density of the atmosphere in this region.

The transmission spectra shown in Figure \ref{ch5:fig_met_WASP-121_3} is dominated by TiO/VO features in the optical, the size of which decreases  with increasing metallicity due to a reduction in the atmospheric scale height.  Sharp Fe features dominate the optical spectrum short-ward of $\sim$0.4$\mic$ with H$_2$O features in the infrared. The CO features are also seen in the transmission spectrum particularly around 2.5 and 4-6$\mic$, which can also be seen in the emission spectrum in Fig. \ref{ch5:fig_met_WASP-121_4}.

In the emission spectra shown in Figure \ref{ch5:fig_met_WASP-121_4} for WASP-121b, due to an inversion layer in the $P$-$T$ profile, most of the molecular features are seen as emission features as opposed to the absorption features seen for WASP-17b. The amplitude of the CO features increases with increasing metallicity. The TiO/VO features can also be seen as emission features in the emission spectra in the optical. The H$_2$O emission features dominate the infrared, where the 1.4$\mic$ feature has led to the detection of an inversion layer for the first time in an exoplanet atmosphere \citep{Evans2017}. It can also be noticed from the NCF that the wings of strong absorption bands at 2.25 (H$_2$O) and 3.8$\mic$ (CO) shown in Fig. \ref{ch5:fig_met_WASP-121_5} and \ref{ch5:fig_met_WASP-121_7}, respectively, mainly probe the region below the inversion layer, while the cores of absorption bands at 2.8 and 4.5$\mic$ shown in Fig. \ref{ch5:fig_met_WASP-121_6} and \ref{ch5:fig_met_WASP-121_8} probe the inversion layer. 

\begin{figure*}
\begin{center}
 \subfloat[]{\includegraphics[width=\columnwidth]{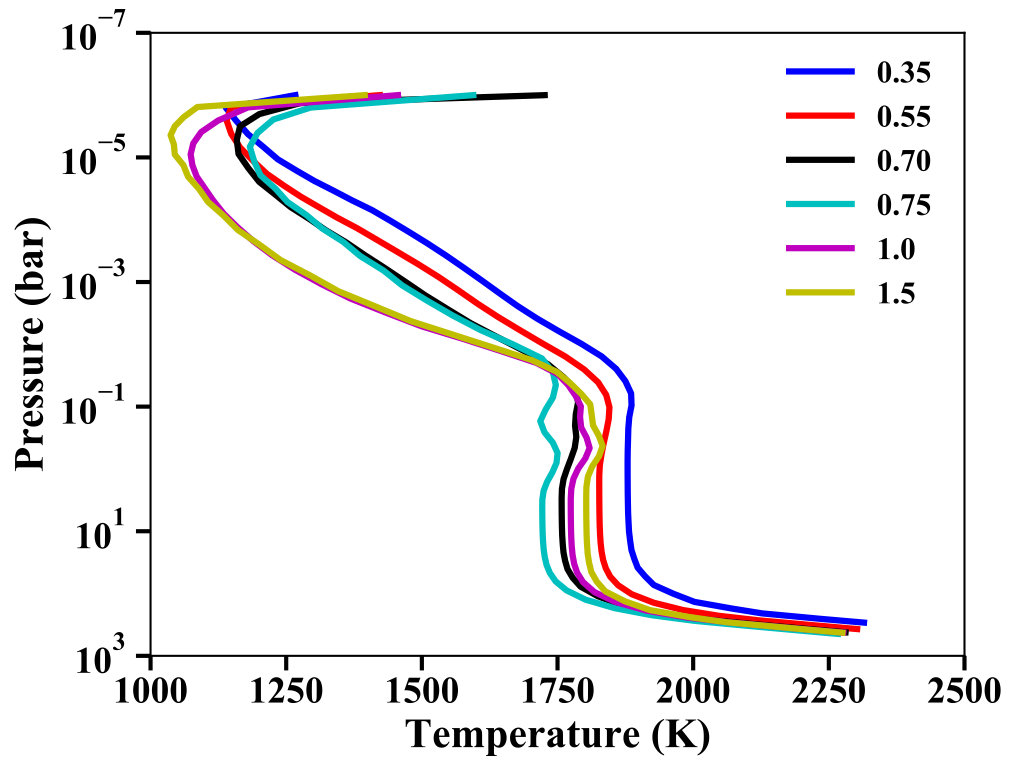}\label{ch5:fig_co_wasp17_1}}
 \subfloat[]{\includegraphics[width=\columnwidth]{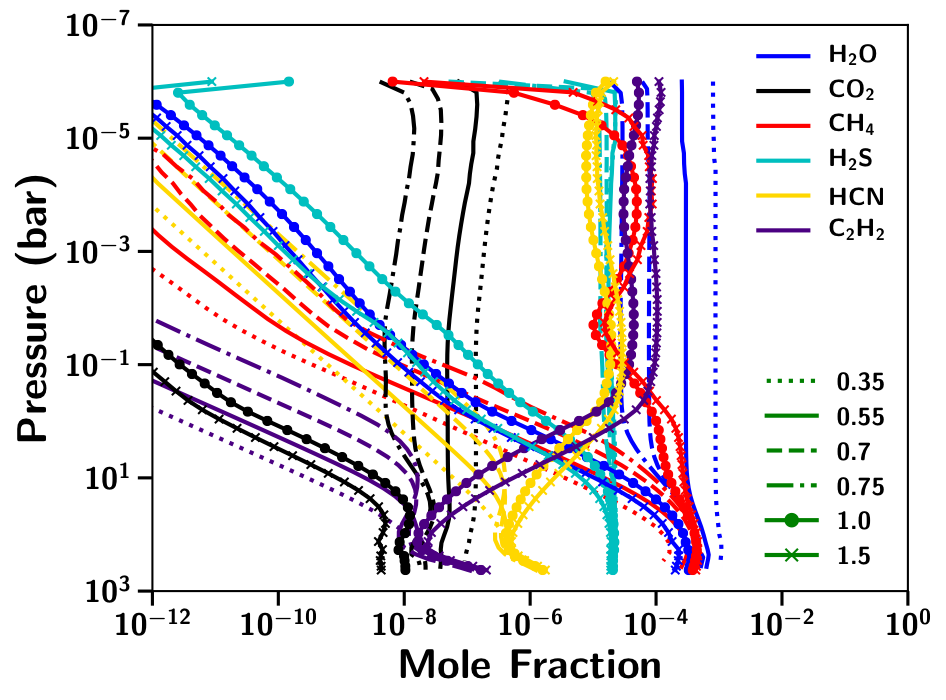}\label{ch5:fig_co_wasp17_2}}
 \newline
  \subfloat[]{\includegraphics[width=\columnwidth]{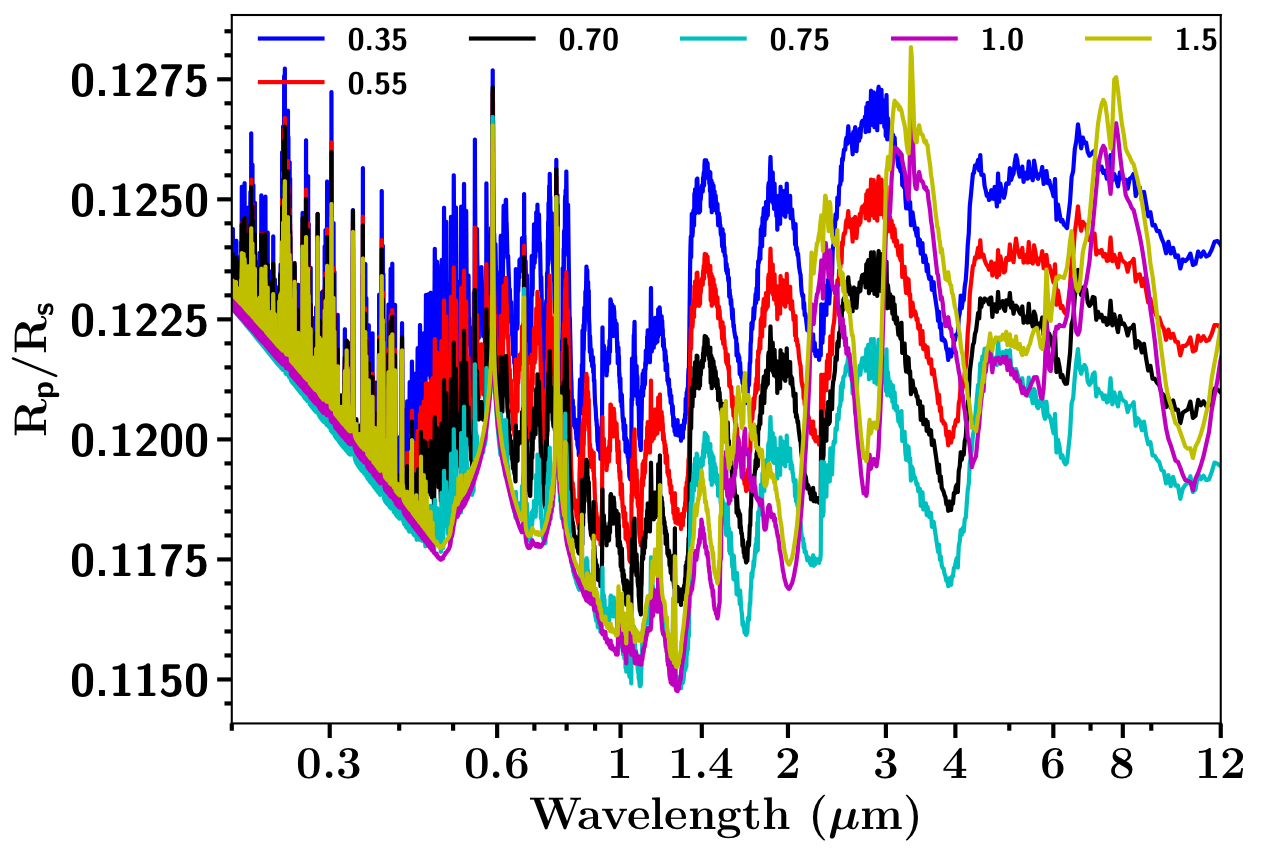}\label{ch5:fig_co_wasp17_3}}
 \subfloat[]{\includegraphics[width=\columnwidth]{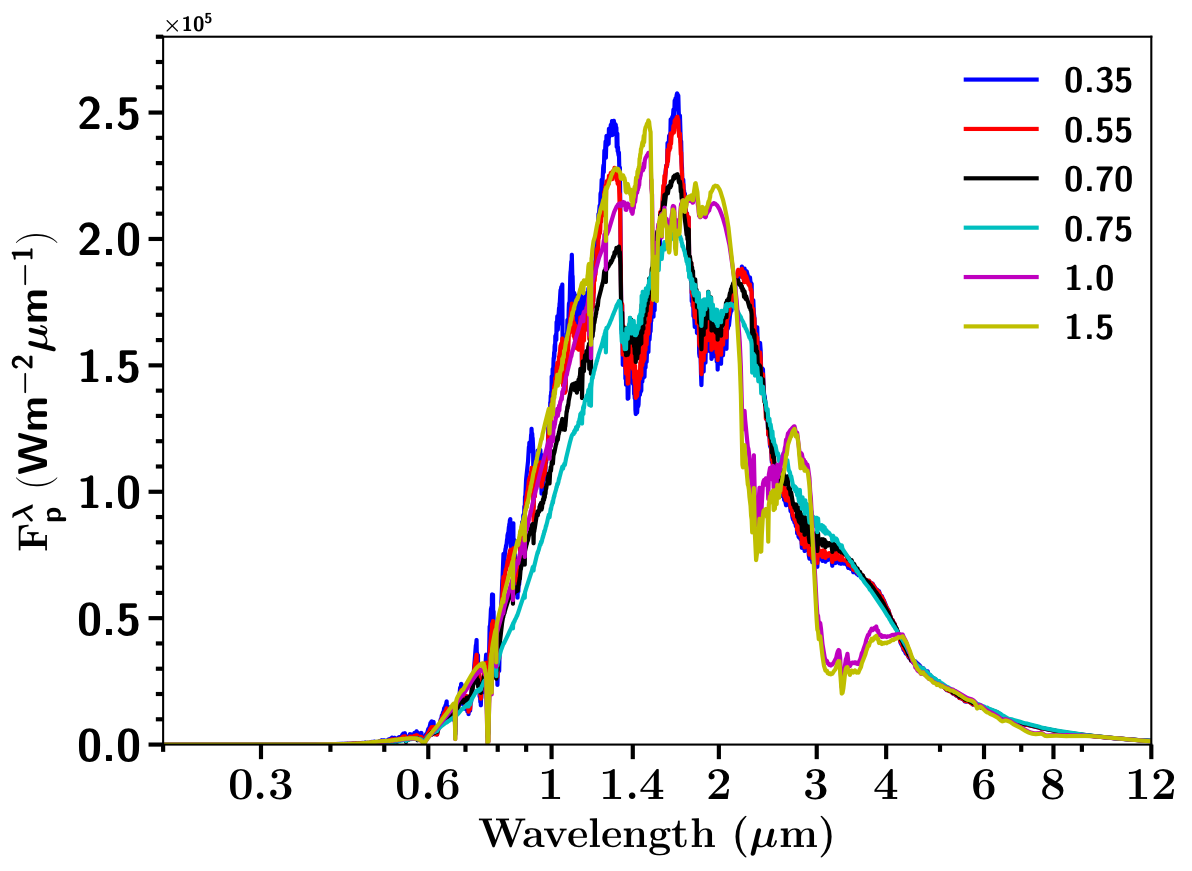}\label{ch5:fig_co_wasp17_4}}
\end{center}
 \caption[Figure showing RCE $P$-$T$ profiles and the corresponding equilibrium chemical abundances, transmission and emission spectra, for a range of C/O ratios adopted in the consistent library, using WASP-17b as the test case]{\textbf{(a)} Figure showing $P$-$T$ profiles for a range of C/O ratios at 0.5 f$_{\textup{c}}$ and solar metallicity for WASP-17b. \textbf{(b)} Figure showing equilibrium chemical abundances for some important species for various C/O values obtained using $P$-$T$ profiles shown in Figure \ref{ch5:fig_co_wasp17_1}. \textbf{(c)} Figure showing transmission spectra for WASP-17b for different values of C/O ratios obtained using $P$-$T$ profiles shown in Figure \ref{ch5:fig_co_wasp17_1} and corresponding equilibrium chemical abundances shown in Fig. \ref{ch5:fig_co_wasp17_2}.  \textbf{(d)} Figure showing emission spectra for WASP-17b for different values of C/O ratios obtained using $P$-$T$ profiles shown in Figure \ref{ch5:fig_co_wasp17_1} and corresponding equilibrium chemical abundances shown in Fig. \ref{ch5:fig_co_wasp17_2}.}
 \label{ch5:fig_co_oxy_wasp-17_a}
\end{figure*}

\begin{figure*}
\begin{center}
  \subfloat[]{\includegraphics[width=\columnwidth]{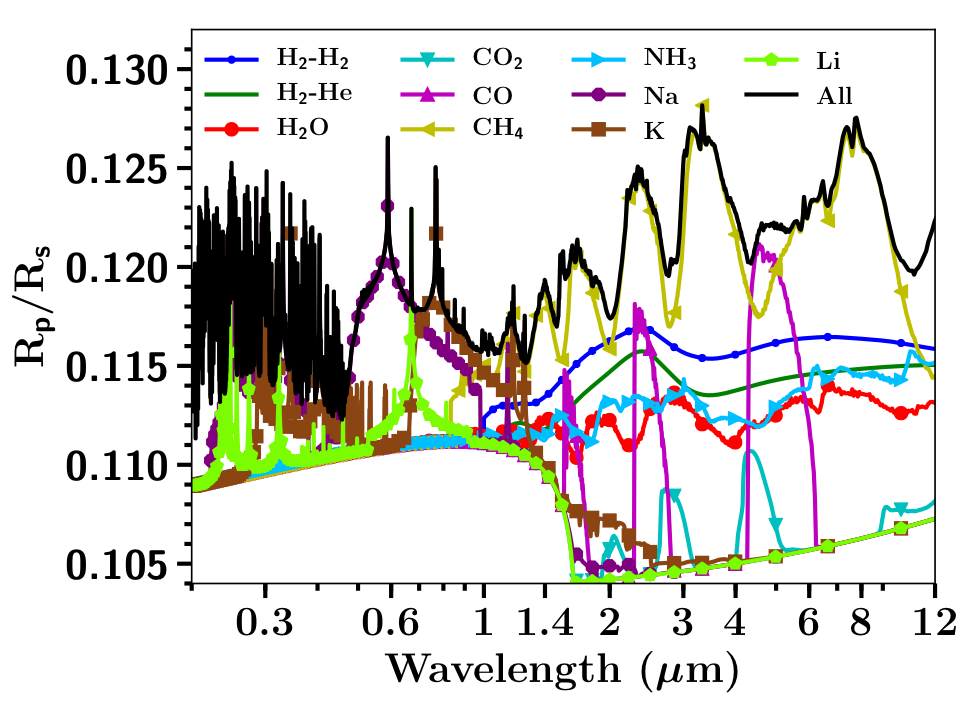}\label{ch5:fig_co_wasp17_5}}
 \subfloat[]{\includegraphics[width=\columnwidth]{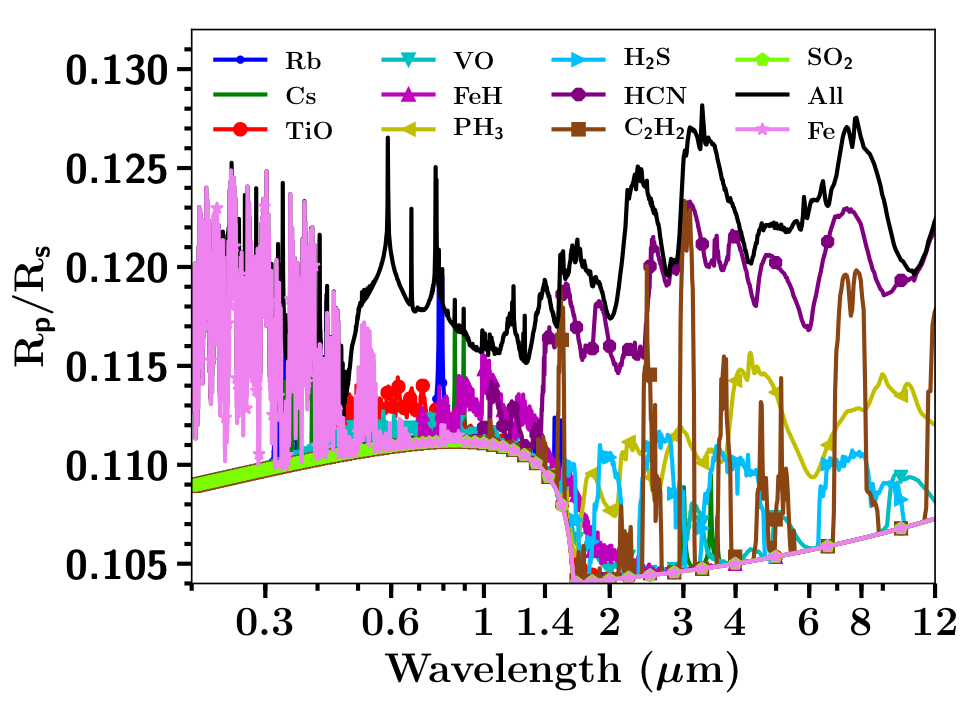}\label{ch5:fig_co_wasp17_6}}
\end{center}
 \caption[Decoupled transmission spectra, for the transmission spectra shown in Fig. \ref{ch5:fig_co_wasp17_3}]{\textbf{(a)} Figure showing transmission spectra features of each individual molecule used in \texttt{ATMO} (1 to 10) for WASP-17b transmission spectra shown in Fig. \ref{ch5:fig_co_wasp17_3} at C/O ratio of 1.5. H$_2$-H$_2$ (blue), H$_2$-He (green), H$_2$O (red), CO$_2$ (cyan), CO (magenta), CH$_4$ (yellow), NH$_3$ (lightblue), Na (purple), K (brown), Li (lightgreen) and all 21 opacities (black). The lower continuum boundary in the spectrum sharply dropping around 1.6$\mic$ is due to H$^-$ opacity. No $\RpRs$ offset was applied while plotting. Individual simulations are divided into blocks of 10 while plotting for clarity.\textbf{(b)} Same as Figure \ref{ch5:fig_co_wasp17_5} but for Rb (blue), Cs (green), TiO (red), VO (cyan), FeH (magenta), PH$_3$ (yellow), H$_2$S (lightblue), HCN (purple), C$_2$H$_2$ (brown), SO$_2$ (lightgreen), Fe (violet) and all 21 opacities (black).}
 \label{ch5:fig_co_oxy_wasp-17_b}
\end{figure*}

\subsubsection{Effect of C/O ratio}
\label{ch5:eff_co_oxy}

\paragraph{WASP-17b:}
\label{ch5:wasp017_co_oxy}
The $P$-$T$ profiles for WASP-17b for a range of C/O ratios are shown in Fig. \ref{ch5:fig_co_wasp17_1} at 0.5 f$_{\textup{c}}$ and solar metallicity. We choose to show simulations with 0.5 f$_{\textup{c}}$ instead of 0.25 f$_{\textup{c}}$ as one of the models with 0.25 f$_{\textup{c}}$ and a C/O ratio of 1.0 failed to converge in the grid. It can be seen that with increasing C/O ratio the $P$-$T$ structure consistently cools for P $\leq$ 10$^{-1}$ bar. However, for P $\geq$ 10$^{-1}$ bar the $P$-$T$ structure first cools up-to C/O ratio of 0.75 and then the temperature increases for a C/O ratio of 1 and 1.5. The sharp heating at around 10$^{-6}$ bar is due to Fe opacity as explained earlier in Section \ref{ch5:inv_fe_without_tio}. 

The change in the equilibrium chemical abundances due to the change in the C/O ratio is drastic, as it effects all the major carbon and oxygen bearing molecules. As expected the abundances of H$_2$O drop with increasing C/O ratio. Although CO$_2$ bears a carbon atom it needs two oxygen atoms per carbon atom, therefore the equilibrium abundance of CO$_2$ also drops with increasing C/O ratio, but in smaller increments as compared to H$_2$O. The abundance of carbon bearing species such as CH$_4$, HCN, C$_2$H$_2$ increases with increasing C/O ratio, while the abundance of CO is almost constant, since it has one atom of carbon and oxygen each. This transition from H$_2$O dominated spectra, to spectra dominated by various carbon bearing species occurs between C/O ratios of 0.75 and 1, slightly higher than that found with isothermal $P$-$T$ profiles where it was between 0.7-0.75 as shown in \citet{Goyal2018}. However, this value might change with change in the f$_{\textup{c}}$ value as the C/O transition point is a strong function of temperature \citep{Molliere2016, Goyal2018,Molaverdikhani2019}.

This C/O transition is also seen in the transmission spectrum shown in Fig. \ref{ch5:fig_co_wasp17_3}, where the spectrum transitions from being H$_2$O dominated to being dominated  by CH$_4$, HCN and C$_2$H$_2$ between a C/O ratio of 0.75 and 1. Fig. \ref{ch5:fig_co_wasp17_5} and \ref{ch5:fig_co_wasp17_6} show this transmission spectrum at a C/O ratio of 1.5 decoupled into various molecules. It can be seen that at a C/O ratio of 1.5, the transmission spectrum is dominated by CH$_4$ features in the infrared, with contributions from CO, HCN and C$_2$H$_2$. There is a strong HCN and C$_2$H$_2$ feature at $\sim$1.6$\mic$ and the most common CO feature at 4.5$\mic$. The emission spectrum shown in Fig. \ref{ch5:fig_co_wasp17_4} also shows this C/O transition between 0.75-1, from deep H$_2$O absorption features to more deeper CH$_4$ absorption features, in the peak region of emission around $\sim$3$\mic$

\begin{figure*}
\begin{center}
 \subfloat[]{\includegraphics[width=\columnwidth]{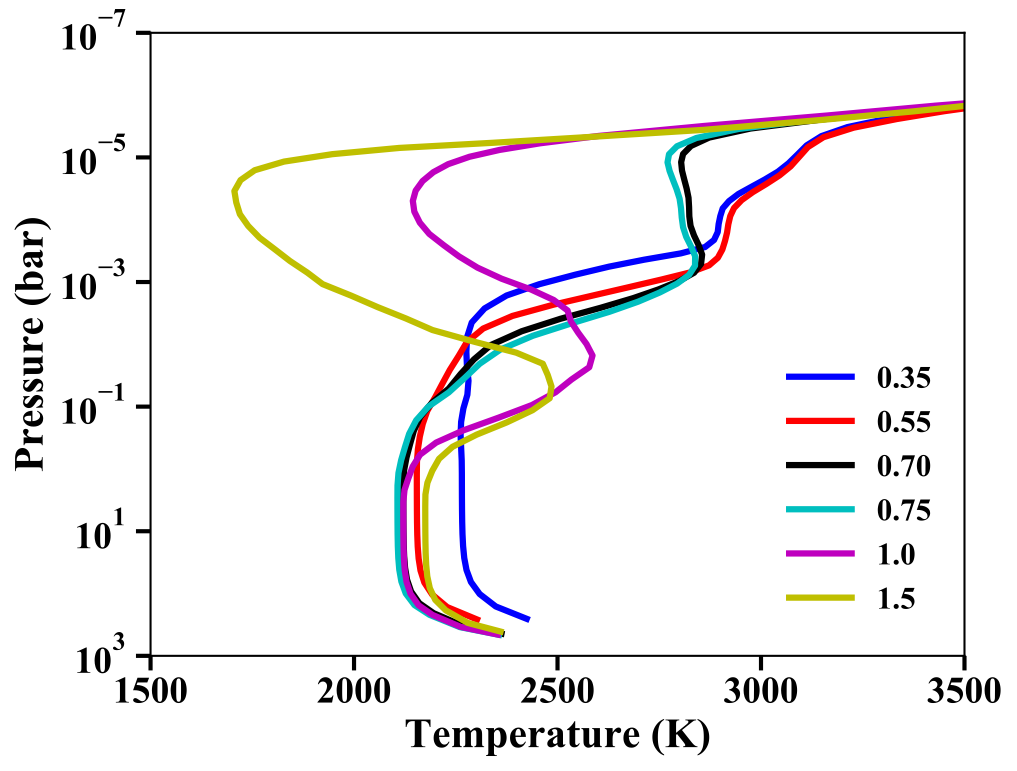}\label{ch5:fig_co_wasp121_1}}
 \subfloat[]{\includegraphics[width=\columnwidth]{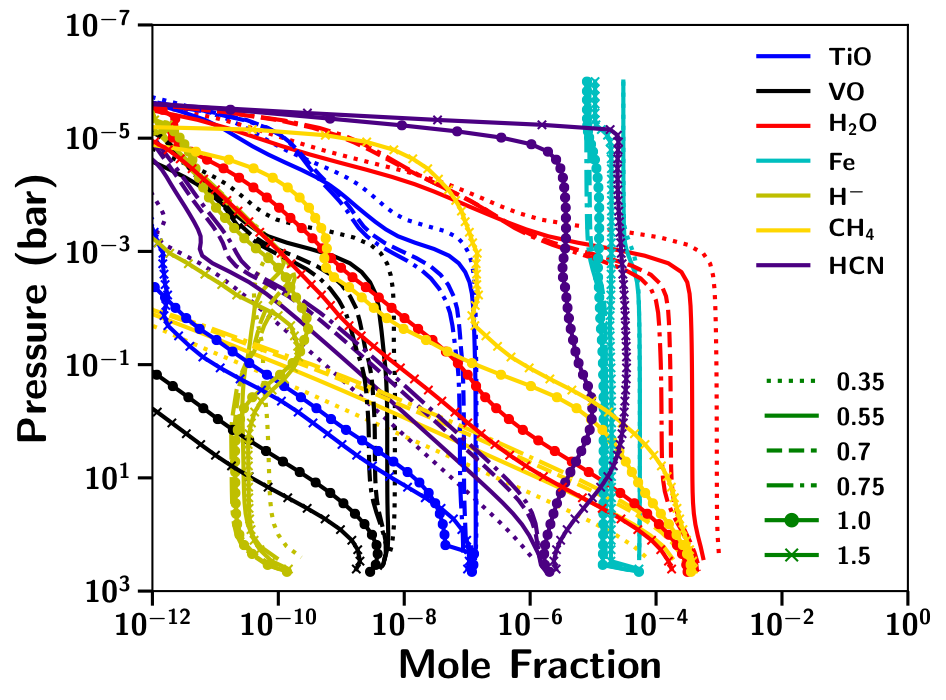}\label{ch5:fig_co_wasp121_2}}
 \newline
  \subfloat[]{\includegraphics[width=\columnwidth]{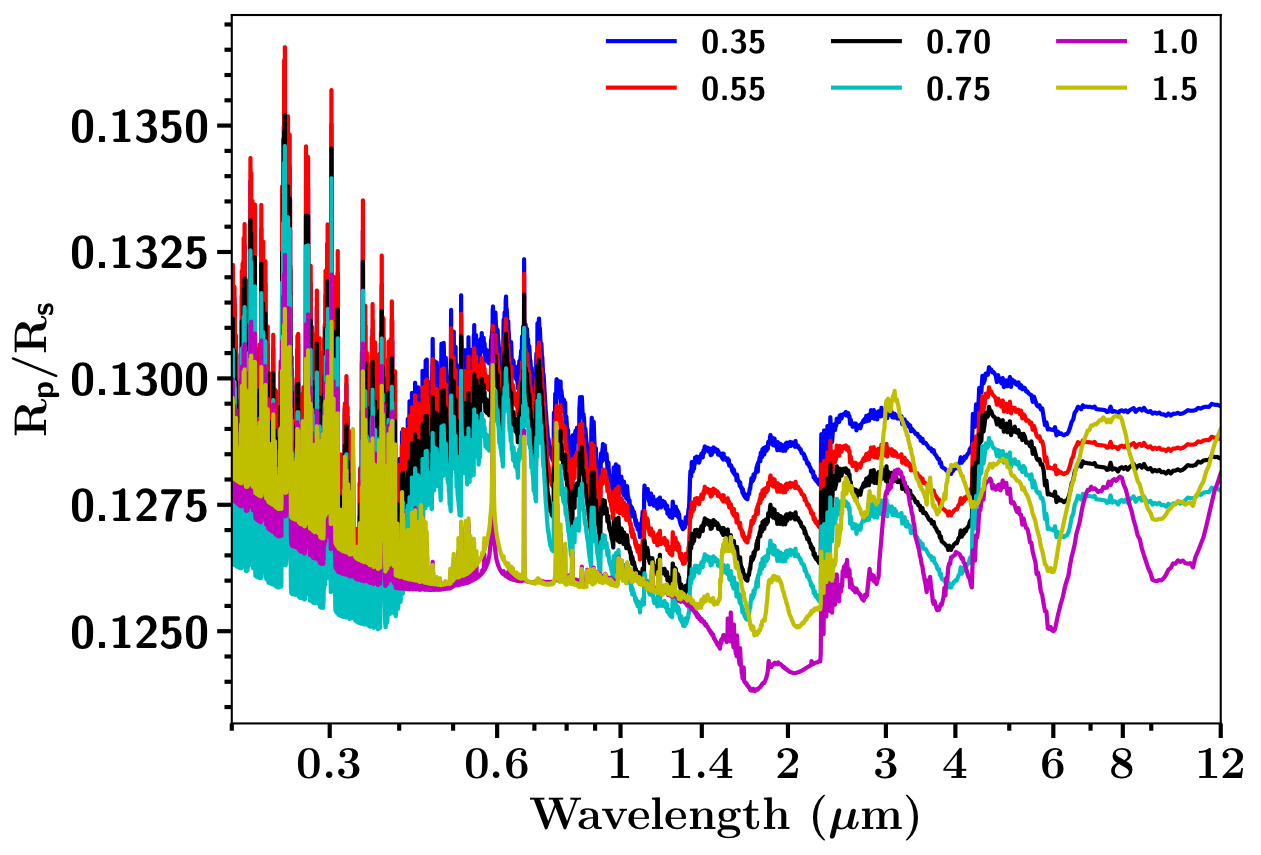}\label{ch5:fig_co_wasp121_3}}
 \subfloat[]{\includegraphics[width=\columnwidth]{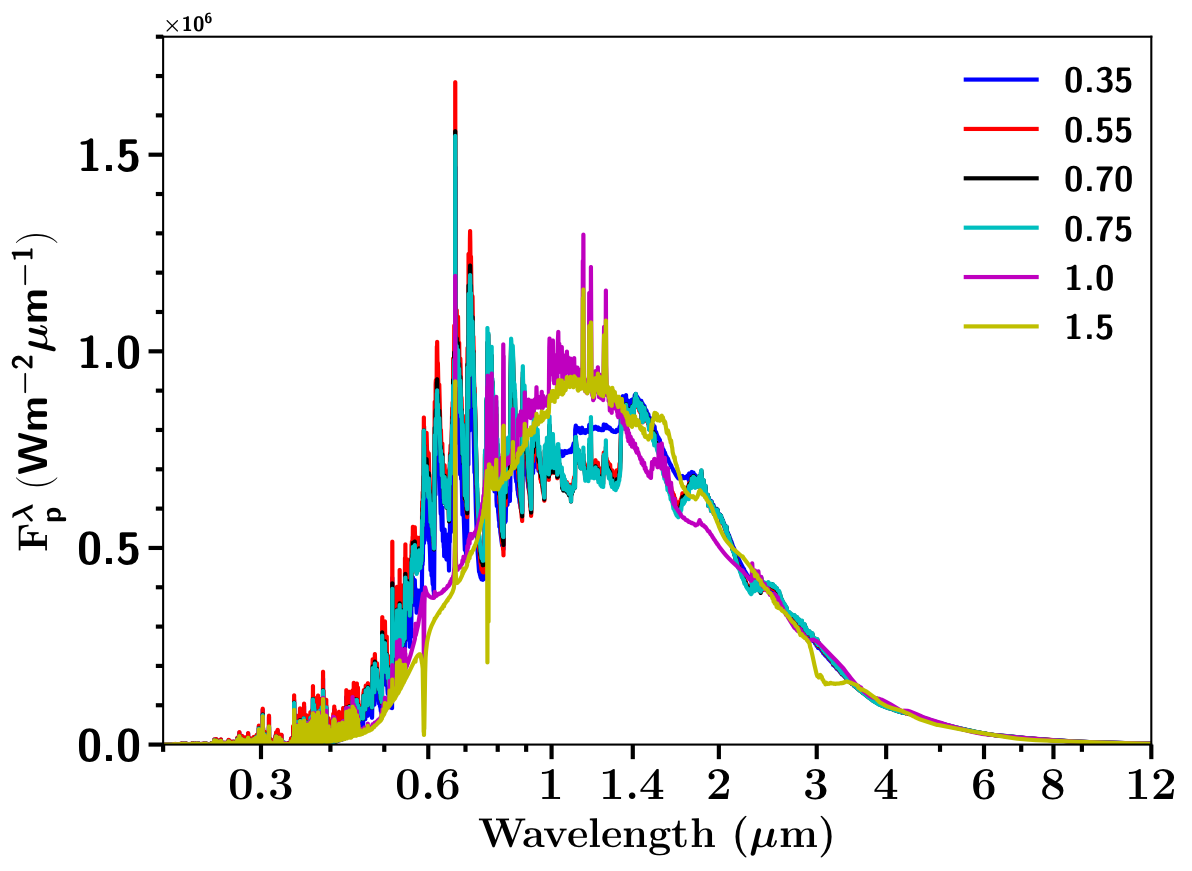}\label{ch5:fig_co_wasp121_4}}
\end{center}
 \caption[Same as Fig. \ref{ch5:fig_co_oxy_wasp-17_a}, but using WASP-121b as the test case]{\textbf{(a)} Figure showing $P$-$T$ profiles for a range of C/O ratios at 0.5 f$_{\textup{c}}$ and solar metallicity for WASP-121b. \textbf{(b)} Figure showing equilibrium chemical abundances for some important species for various C/O values obtained using $P$-$T$ profiles shown in Figure \ref{ch5:fig_co_wasp121_1}. \textbf{(c)} Figure showing transmission spectra for WASP-121b for different values of C/O ratios obtained using $P$-$T$ profiles shown in Figure \ref{ch5:fig_co_wasp121_1} and corresponding equilibrium chemical abundances shown in Fig. \ref{ch5:fig_co_wasp121_2}.  \textbf{(d)} Figure showing emission spectra for WASP-121b for different values of C/O ratios obtained using $P$-$T$ profiles shown in Figure \ref{ch5:fig_co_wasp121_1} and corresponding equilibrium chemical abundances shown in Fig. \ref{ch5:fig_co_wasp121_2}.}
 \label{ch5:fig_rc}
\end{figure*}

\begin{figure*}
\begin{center}
  \subfloat[]{\includegraphics[width=\columnwidth]{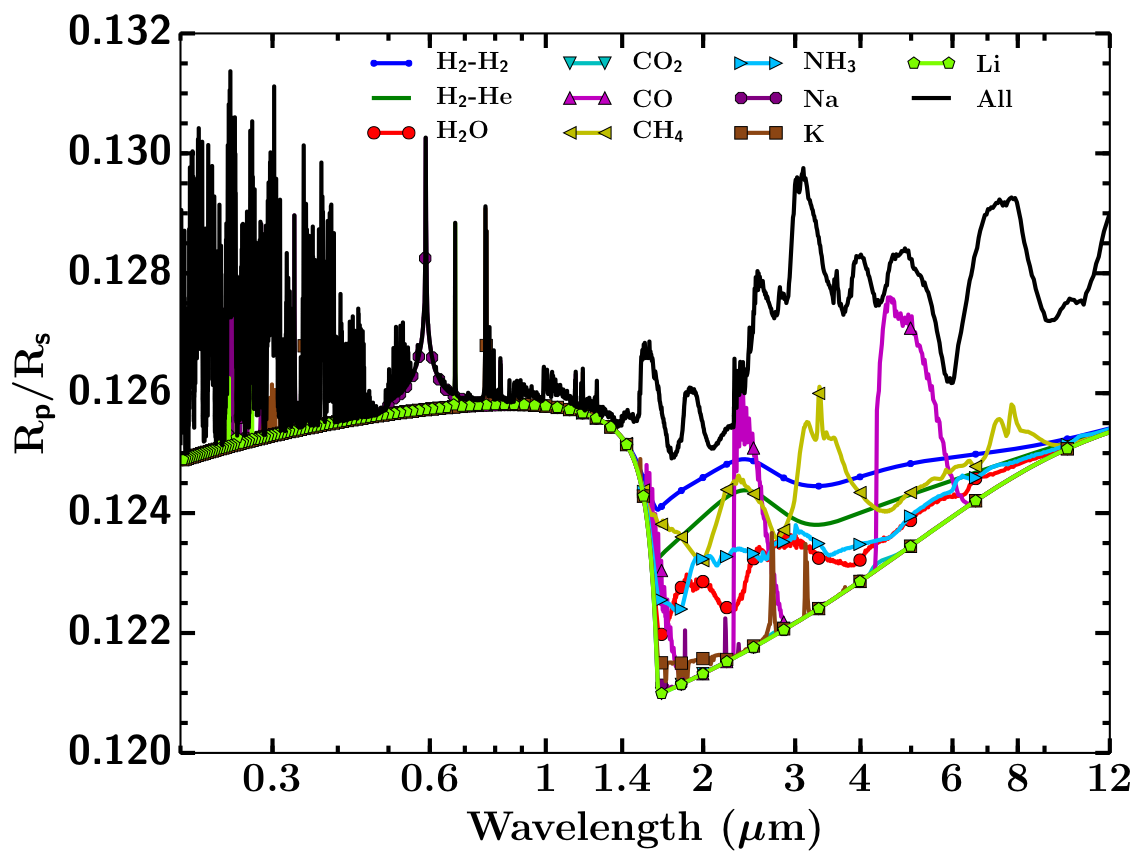}\label{ch5:fig_co_wasp121_5}}
 \subfloat[]{\includegraphics[width=\columnwidth]{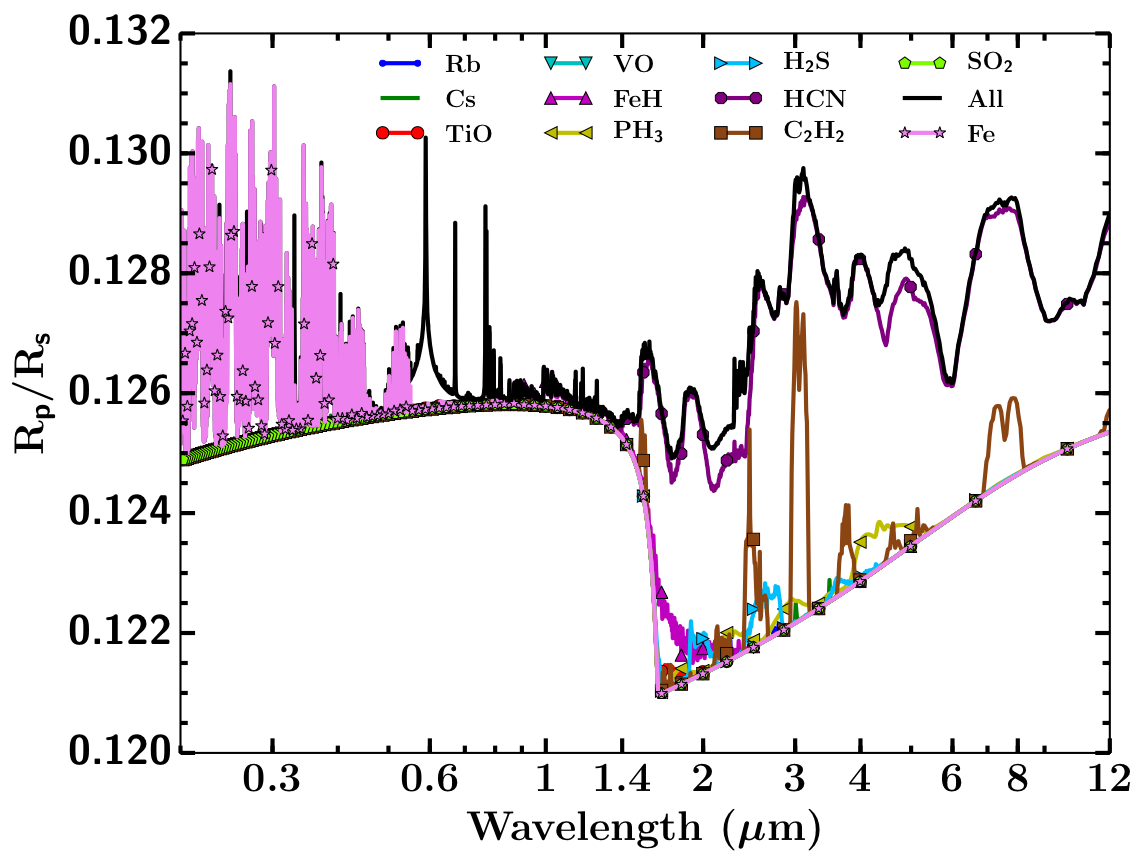}\label{ch5:fig_co_wasp121_6}}
 \newline
    \subfloat[]{\includegraphics[width=0.6\columnwidth]{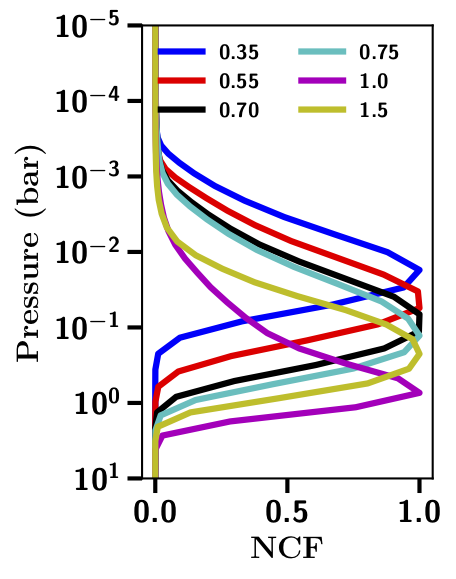}\label{ch5:fig_co_wasp121_7}}
 \subfloat[]{\includegraphics[width=0.6\columnwidth]{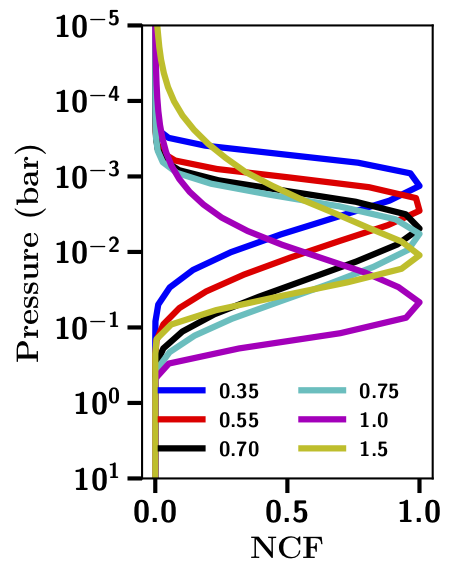}\label{ch5:fig_co_wasp121_8}}
\end{center}
 \caption[Decoupled transmission spectra, for the transmission spectra shown in Fig. \ref{ch5:fig_co_wasp121_3} and contribution functions corresponding to $P$-$T$ profiles shown in Fig. \ref{ch5:fig_co_wasp121_1} and emission spectra shown in Fig. \ref{ch5:fig_co_wasp121_4}]{\textbf{(a)} Same as Fig. \ref{ch5:fig_co_wasp17_5} but for WASP-121b at 0.5 f$_{\textup{c}}$, solar metallicity and C/O ratio of 1.5. H$^-$ opacity is quite dominant especially between 1 and 1.4$\mic$. \textbf{(b)} Same as Fig. \ref{ch5:fig_co_wasp17_6} but for WASP-121b at 0.5 f$_{\textup{c}}$, solar metallicity and C/O ratio of 1.5. \textbf{(c,d)} Figures c, d showing normalised contribution function at 1.6 and 3.1$\mic$ for a range of C/O ratios for emission spectra as shown in Figure \ref{ch5:fig_co_wasp121_4}.}
 \label{ch5:fig_rc}
\end{figure*}

\paragraph{WASP-121b:}
\label{ch5:wasp121_co_oxy}
The $P$-$T$ profiles for WASP-121b for a range of C/O ratios are shown in Fig. \ref{ch5:fig_co_wasp121_1}. With increasing C/O ratio the major temperature inversion shifts to higher pressures. The major temperature inversion refers to the inversion that has a potential observational signature unlike the inversion due to Fe opacity at extremely low pressures ($\sim$10$^{-5}$ bar) as described in Section \ref{ch5:effect_fe_opacity} and \ref{ch5:inv_fe_without_tio} and also seen in Fig. \ref{ch5:fig_co_wasp121_1}. 

The abundance of TiO/VO species which are the primary absorbers for forming the temperature inversion layer, decrease with increasing C/O ratio. Their abundance at a C/O ratio of 1.0 and 1.5 is low, still the inversion layer is maintained, albeit not as hot as at other low C/O ratios. We investigated this phenomenon by using two tests, first removing TiO/VO opacities  and second removing Na, K, TiO and VO opacities. We found that this inversion at high C/O ratio can be maintained due to Na and K opacities in the absence of TiO/VO opacities or their low abundance, as found by \citet{Molliere2015}. Furthermore, H$^-$ opacity also contributes to this inversion at high C/O ratio and high value of f$_{\textup{c}}$ (1.0) more than Na and K opacities, as discussed in detail in Section \ref{ch5:effect_h-_opacity}. The increase in the H$^-$ abundance at C/O ratio of 1.5 around the photosphere region ($\sim$0.1 to 1 bar) is the primary reason for this inversion due to H$^-$ opacity. H$^-$  features also can be clearly seen in the transmission spectrum for C/O ratio of 1.5. 

At high C/O ratios it is interesting to see that the HCN abundance is substantial even in the low pressure regions (P > $\sim$1 mbar). This can also be noticed in the transmission spectrum shown in Fig. \ref{ch5:fig_co_wasp121_3} and decoupled transmission spectrum shown in Fig. \ref{ch5:fig_co_wasp121_5} and \ref{ch5:fig_co_wasp121_6}, which is dominated by HCN features in the infrared at a C/O ratio of 1 and 1.5, as opposed to CH$_4$ for WASP-17b. This shows an important result that at high temperatures HCN dominates over CH$_4$ in the atmosphere at high C/O ratios. Therefore, HCN features provide a very strong signature to constrain high C/O ratios in exoplanet atmospheres. The transition from an H$_2$O dominated spectrum to that dominated by HCN happens between C/O ratio of 0.75-1.0. Therefore, the temperature dependence of the C/O transition as seen in \citet{Molliere2015, Goyal2018} does not seem to be holding in this case, as the transition for WASP-17b also happens in this C/O regime (0.75-1.0). The transmission spectrum also shows FeH features between  0.8 and 1.2$\mic$ at a C/O ratio of 1.0 and 1.5. The absence of TiO/VO features makes it possible for FeH to appear, without being masked. 

In the emission spectrum shown in Fig. \ref{ch5:fig_co_wasp121_4} most of the molecular features are seen in emission due to the presence of an inversion layer, as explained before. However, between a C/O ratio of 1 and 1.5 surprising differences can be seen, especially at 3.1$\mic$ which is the wavelength of a strong HCN absorption band. At a C/O ratio of 1 it is an emission feature, however it transforms to an absorption feature at a C/O ratio of 1.5. This is because at a C/O ratio of 1.5 slightly cooler upper atmosphere is being probed as can be seen in the NCF at 3.1$\mic$ in Fig. \ref{ch5:fig_co_wasp121_8} as compared to a C/O ratio of 1. Therefore, at a C/O ratio of 1.5, HCN absorbs the radiation from the lower (high pressure atmosphere) in comparison to for a C/O ratio of 1.0, which probes deeper atmosphere where it has an inversion and leads to HCN emission. Interestingly, at 1.6$\mic$ HCN leads to an emission feature both at a C/O ratio of 1 and 1.5, since at 1.6$\mic$ the inversion layer is being probed at both C/O ratios as seen in the NCF in Fig. \ref{ch5:fig_co_wasp121_7} and $P$-$T$ profile in Fig. \ref{ch5:fig_co_wasp121_1}. Therefore, HCN should be observed in emission as well as absorption at high C/O ratios (1.5) for extremely irradiated planets such as WASP-121b.

\begin{figure*}
\begin{center}
 \subfloat[]{\includegraphics[width=\columnwidth]{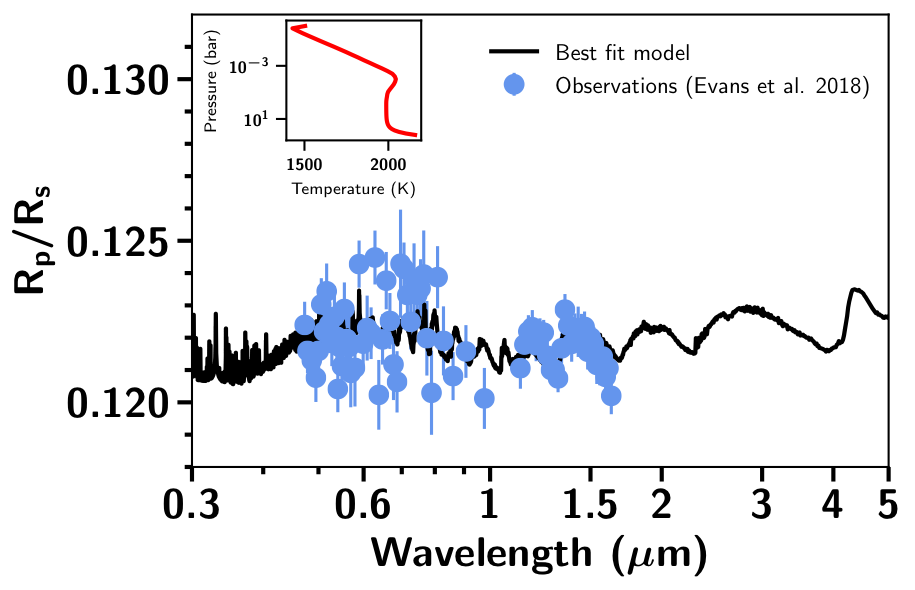}\label{ch5:best_fit1}}
 \subfloat[]{\includegraphics[width=\columnwidth]{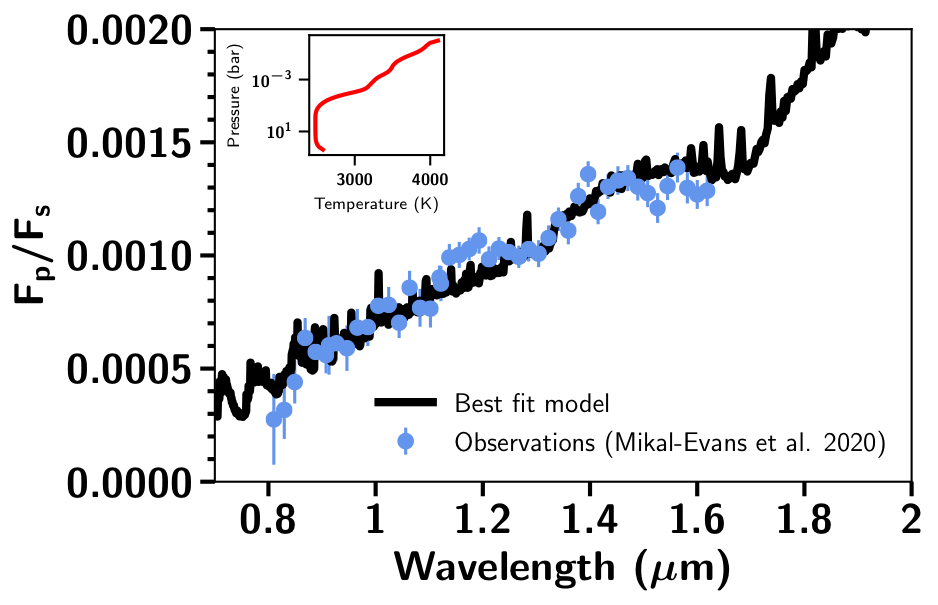}\label{ch5:best_fit2}}
  \newline
  \subfloat[]{\includegraphics[width=\columnwidth]{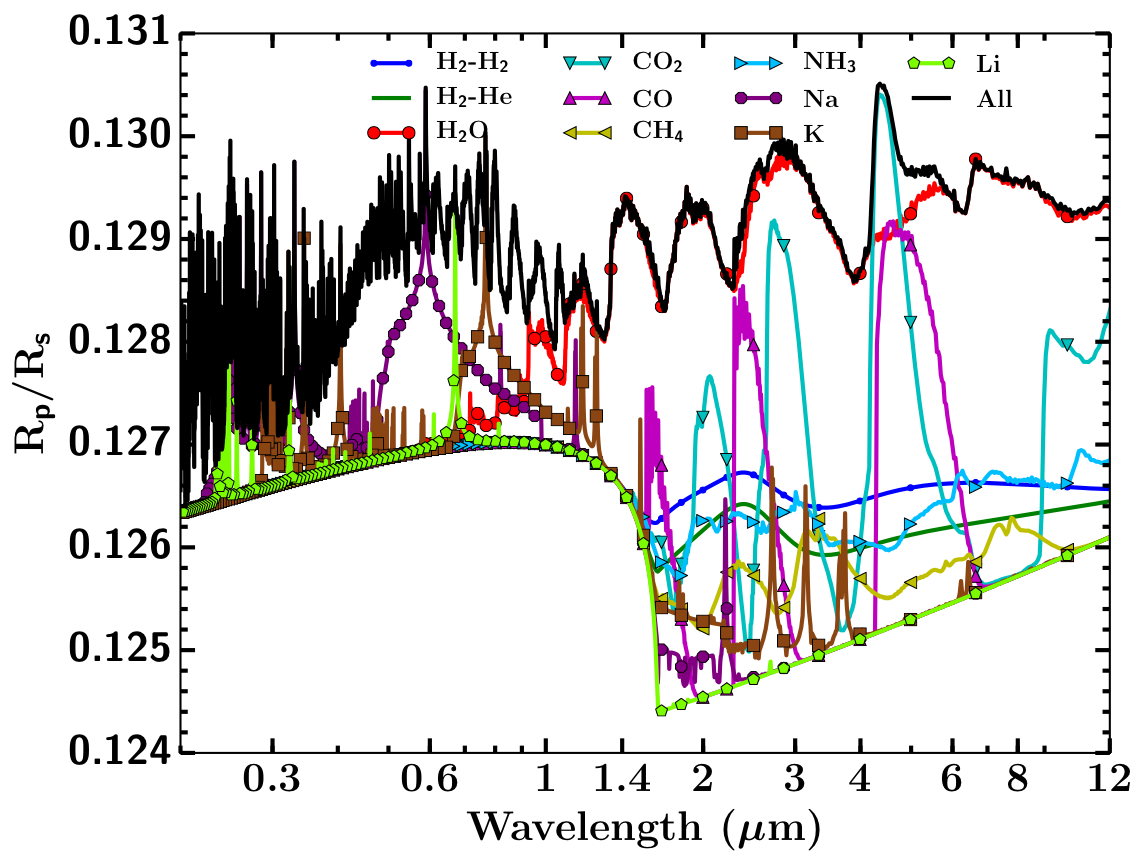}\label{ch5:best_fit3}}
 \subfloat[]{\includegraphics[width=\columnwidth]{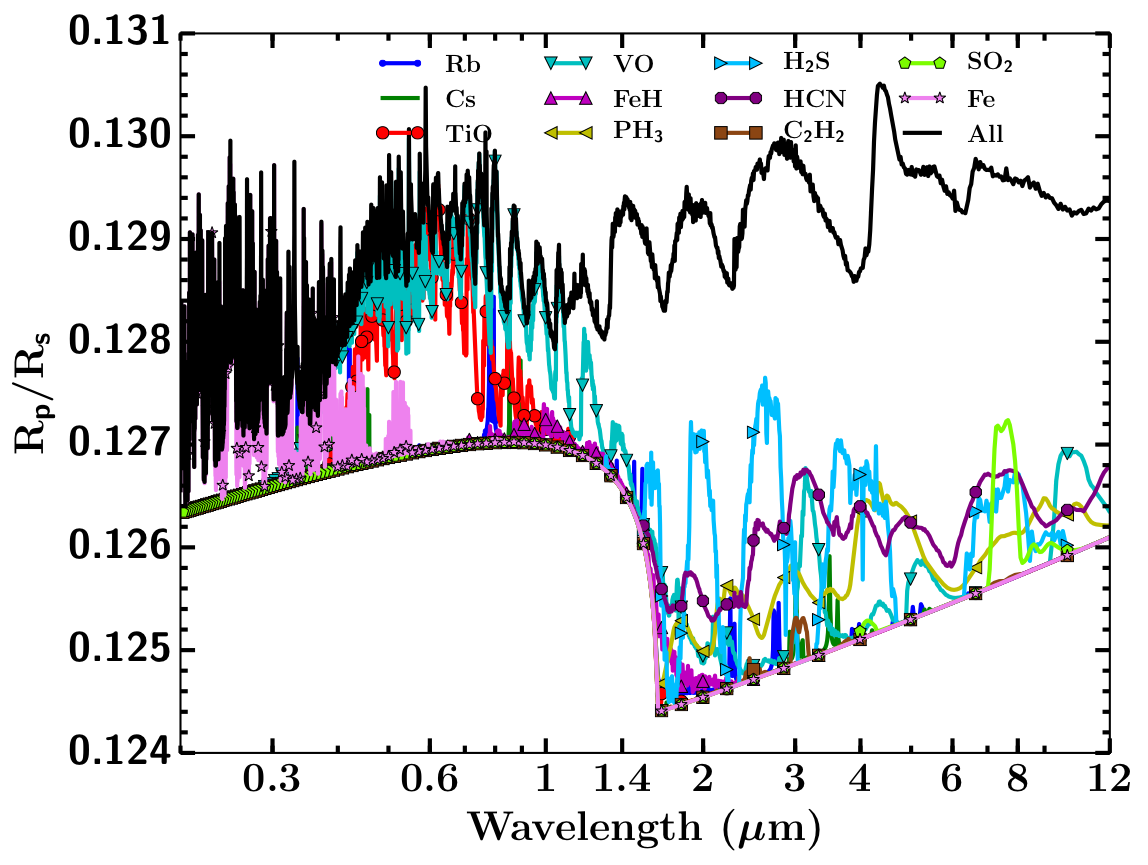}\label{ch5:best_fit4}}
 \caption[Figure showing transmission and emission spectra observations of WASP-121b along with the best fit models from the consistent grid. Decoupled best fit transmission spectra is also shown]{\textbf{(a)} Figure showing best fit model transmission spectra using the grid of model transmission spectra for WASP-121b presented in this paper and observations from \citet{Evans2018} giving $\chi^2$ value of 101.69  with f$_{\textup{c}}$ = 0.25, 100 times solar metallicity and C/O ratio of 0.75. Best-Fit $P$-$T$ profile is shown in the subplot. \textbf{(b)} Figure showing best fit model emission spectra using the grid of model emission spectra for WASP-121b presented in this paper and observations from Evans et al. (2020) with $\chi^2$ value of 88.63 and f$_{\textup{c}}$ = 1.0, solar metallicity and C/O ratio of 0.75. Best-Fit $P$-$T$ profile is shown in the subplot. \textbf{(c)} Best-fit decoupled transmission spectra similar to Figure \ref{ch5:fig_co_wasp17_5}. The lower continuum boundary in the spectrum sharply dropping around 1.6$\mic$ is due to H$^-$ opacity. \textbf{(d)} Best-fit decoupled transmission spectra similar to Figure \ref{ch5:fig_co_wasp17_6}.} 
 \end{center}
 \label{ch5:fig_rc}
\end{figure*}

\section{Interpreting the observations of WASP-121b}
\label{ch5:obs_int}
WASP-121b is not expected to be cloudy due to its extremely high predicted dayside temperature (T$_\textup{eq}$ = 2400\,K). Moreover, \citet{Evans2018} showed that the extremely steep slope in the optical is not due to enhanced Rayleigh scattering from a  haze. In \citet{Evans2018} observations were interpreted using isothermal $P$-$T$ profiles \citep[see Fig. 11 and 12 in][]{Evans2018} resulting in detection of H$_2$O, evidence of VO and the possibility of SH causing significant absorption in the UV and optical. Therefore, here we use our grid of model transmission spectra with RCE $P$-$T$ profiles and additional H$^-$ and Fe opacity, without grey clouds and enhanced scattering, to fit the WASP-121b data from \citet{Evans2018}. We have excluded the data short-ward of 0.47$\mic$ while performing the fitting, since the steep slope in this region is not explained by Rayleigh scattering and is likely in part due to absorption of atomic species like Fe in the thermospheric and higher atmospheric layers \citep{Sing2019, Gibson2020} which are not modeled here.. Fig. \ref{ch5:best_fit1} shows the best fit model transmission spectra for observations from \citet{Evans2018}. It shows a H$_2$O feature at 1.4$\mic$ and features in the optical resembling TiO/VO features. The decoupled spectra in Figure \ref{ch5:best_fit3} and \ref{ch5:best_fit4} shows that the optical spectra is dominated by VO features. This fitting result corroborates the result using isothermal $P$-$T$ profiles discussed in \citet{Evans2018} and consistent with VO being present in the atmosphere of WASP-121b, since the limb $P$-$T$ structure of this planet is in that narrow temperature regime where VO abundance dominates the TiO abundance, as discussed in \citet{Goyal2019}.

The best fit transmission spectra model with RCE $P$-$T$ profiles gives a reduced $\chi^2$ value of 1.37 with 74 DOF (75 data points minus 1). It gives super-solar metallicity and super-solar C/O ratio as shown in  Fig. \ref{ch5:best_fit1} (both when varying O/H and C/H). The \enquote{best-fit} $P$-$T$  profile shown in the sub-plot of  Fig. \ref{ch5:best_fit1} does not show a temperature inversion, since the limb $P$-$T$  profile (in the observed region) is expected to be cooler than the dayside $P$-$T$  profile shown in sub-plot of  Fig. \ref{ch5:best_fit2}. Although the 100 times solar metallicity case is the \enquote{best-fit} (lowest chi-squared), the 50 times solar metallicity case also fits within 3$\sigma$ and a solar metallicity case is consistent with the data at the 6$\sigma$ level, as shown in Figure \ref{fig:chimap_transwasp121b} of Appendix \ref{app:chi-sq-map}, which shows $\chi^{2}$ maps for all the model simulations in the grid for WASP-121b with confidence intervals, similar to those described in details in \citet{Goyal2018}. There are two reasons why high metallicities are preferred, one is the increase in the VO abundances as metallicity increases (more than that of TiO) and second the decrease in the scale height compared to solar metallicity case, both required to better fit the observations, driven by the low R$_{\textup{p}}$/R$_{\textup{s}}$ points in the optical. We also performed tests where we excluded the optical data (< 1$\mic$) and just used the near-infrared data while fitting and in this case the sub-solar metallicity model simulation (0.1 times solar) was the \enquote{best-fit}, thus again concluding that the optical data leads to a higher metallicity being preferred. We note that we have not included relative levels between STIS and WFC3 as a free parameter while fitting, which can potentially alter the results. 

These results demonstrate that a reasonably good match to the observations can be achieved with self-consistent 1D models assuming chemical equilibrium for a range of plausible metallicities and C/O values. The reality is undoubtedly significantly more complex. In particular, we have assumed a single RCE $P$-$T$ profile for the entire limb of the planet, but there should be a strong temperature contrast between the east and west limbs, in turn affecting which species are in the gas phase or condensed. However, limb to limb variations can be incorporated while fitting, simply by the linear combination of transmission spectra from our library for a specific planet \citep{Macdonald2020}. The \enquote{best-fit} limb $P$-$T$  profile of WASP-121b is cold enough for Fe to rainout from the atmosphere, at least in the region of the atmosphere where Local Thermodynamic Equilibrium (LTE) can be assumed. However, \citet{Sing2019, Gibson2020} detected the presence of Fe II and Fe I in UV and optical high resolution observations of WASP-121b, which probes much higher altitudes than the optical transmission spectrum and may well be in the non-LTE region of the atmosphere. The deep interior temperature may also be hotter \citep{Thorngren2019, Sing2019} than assumed in our work (T$_{\textup{int}}$=100\,K), which can prevent Fe from raining out of the atmosphere. The other possibility is that the true planet limb $P$-$T$  profile is much hotter than our \enquote{best-fit} limb $P$-$T$ profile, since we have not modeled the transmission spectrum of WASP-121b short-ward of 0.47$\mic$ for this fitting and not considered inter-terminator variations, as detailed earlier. 

Fig. \ref{ch5:best_fit2} shows the best fit model emission spectra for observations from Evans et al. (2020), with a $\chi^2$ value of 88.63 and reduced $\chi^2$ value of 1.81 with 49 DOF. The best fit model has solar metallically and C/O ratio of 0.75, with f$_{\textup{c}}$ value of 1.0, indicative of very low energy redistribution in the atmosphere.  The dayside best fit $P$-$T$ profile in the sub-plot of  Fig. \ref{ch5:best_fit2} shows a temperature inversion, while the limb $P$-$T$ profile obtained using the transmission spectra fit (see Fig. \ref{ch5:best_fit1} ) is cooler without a temperature inversion, again supporting very low energy redistribution in the atmosphere of WASP-121b. It shows a 1.4$\mic$ H$_2$O emission feature, indicative of inversion layer, as explained in previous sections.

\section{Conclusions}
\label{ch5:summary}

The approximation of an isothermal $P$-$T$ profile has many limitations for a real planetary atmosphere and is a reasonable approximation only to generate model transmission spectra, which probes a very small part of the overall atmosphere. Therefore, in this work we create a library of radiative-convective equilibrium $P$-$T$ profiles and corresponding equilibrium chemical abundances, transmission spectra, emission spectra and contribution functions, for a range of observationally significant exoplanets. Unlike a static table of chemical abundances used in various models, these $P$-$T$ profiles (and all models in this grid) are consistent with equilibrium chemical abundances, meaning  $P$-$T$ profile and chemical abundances are  tied together during each iteration, making it a state of the art model grid. The library of models presented in this work also included H$^-$ and Fe opacity, in addition to opacities included in previous library of models presented in \citet{Goyal2018} and \citet{Goyal2019}. The major conclusions from this work are divided into three parts, first the sensitivity to model choices, second the sensitivity to grid parameters and third the general conclusions. 

\subsection{Sensitivity to Model Choices}
We have to make different model choices while interpreting the observations of exoplanet atmospheres, such as the type of condensation, opacities to be included, source of line-list, line-broadening profile etc. Here we conclude on the effect of these different model choices. 
\begin{itemize}
\item \corr{The investigation of differences in RCE model simulations with local and rainout condensation revealed that adopting different condensation approaches can result in substantial differences in the resulting $P$-$T$ profiles and thereby the model spectra, for a typical hot Jupiter planet like WASP-17b.}

\item \corr{Inclusion of thermal ionic species reduces the abundance of Na and K at atmospheric temperatures greater than $\sim$2000\,K and low pressures (< $\sim$10$^{-1}$ bar), as they are ionized to form Na$^+$ and K$^+$. Therefore, Na and K features are absent in the transmission spectrum at these temperatures. It also allows formation of other ions with significant opacities such as H$^-$. Thermal decomposition leads to reduction in the H$_2$O abundance for planets such as WASP-121b, but only in the region of the atmosphere probed by transmission spectrum and not the emission spectrum.} 

\item \corr{Iron (Fe) opacity leads to very sharp inversion at the top of the model atmosphere ($\sim$10$^{-5}$ to 10$^{-6}$ bar). However, it is unable to produce a temperature inversion in the region of the atmosphere being probed by emission spectrum observations even at 200$\times$ solar metallicity, as produced by TiO/VO or H$^-$ opacities for WASP-121b. Moreover,  Fe opacity tends to cool the $P$-$T$ profile for P > 0.1 bar, due to absorption of radiation in the upper atmosphere. It also imparts narrow spectral features in the UV-optical part of the transmission spectrum short-ward of $\sim$0.4$\mic$. However, Fe opacity is sufficient to create a strong thermal inversion in KELT-9b, the planet with the highest equilibrium temperature in this library of models, even at solar metallicity and without TiO/VO/H$^-$ opacities. Thus, the capability of Fe opacity to form a temperature inversion is system specific showing a strong dependance on the host star effective temperature.}

\item \corr{H$^-$ opacity plays an important role in shaping the  $P$-$T$ profile, transmission and emission spectrum, at temperatures greater then $\sim$2200\,K. In the presence of TiO/VO in the atmosphere H$^-$ opacity leads to cooling in the deeper atmosphere (P > 0.1 bar). At C/O ratios $\geq$ 1, H$^-$ opacity leads to formation of a strong thermal inversion layer, but not at solar C/O ratio even if TiO/VO is absent. Thermal inversion due to H$^-$ opacity lies deeper (higher pressure) in the atmosphere ($\sim$1 to 10$^{-2}$ bar), in comparison to the inversion due to TiO/VO opacities ($\sim$10$^{-1}$ to 10$^{-3}$ bar). In the transmission spectrum, H$^-$ opacity tends to mute the line wings of Na, K and all features up-to 1.6$\mic$ (limit of H$^-$ bound free absorption) mimicking the effect of grey cloud deck opacity. The effect of H$^-$ opacity on $P$-$T$ profiles also substantially alters the emission spectrum, with increase in peak planet flux due to thermal inversion.}

Adopting different line wing profiles of Na and K, led to differences of $\sim$50-100\,K in the deeper atmosphere temperature as shown in the online supplementary material, thus leading to some differences in the emission spectra, which may be detectable with future observations. The differences in the derived $P$-$T$ profiles and the spectra between the simulations, varying C/O ratio by varying O/H and those by varying C/H, shown in the online supplementary material, revealed some differences within the range of C/O ratio adopted in the library, but to a lesser extent compared to other parameters in the library and other model choices.

\end{itemize}

\subsection{Sensitivity to Grid Parameters}

The library of simulations presented in this work was developed for four different recirculation factors, six metallicities and six C/O ratios for 89 hot Jupiter and warm Neptune exoplanets with equilibrium temperatures greater than 1200\,K. The effect of varying the recirculation factor, metallicity and C/O ratio were investigated and shown for WASP-17b and WASP-121b in Sections \ref{ch5:grid_param_sensitivity}, showing the variation of the $P$-$T$ profile, equilibrium chemical abundances and thereby the transmission and the emission spectrum. The major conclusion from these sensitivity tests for different grid parameters are:

\begin{itemize}

\item \corr{The increase in recirculation factor (lower redistribution of energy) leads to transitioning of the $P$-$T$ profile from without a temperature inversion to isothermal to a profile with temperature inversion for WASP-121b, which is also apparent in the emission and transmission spectrum. Therefore, emission spectrum and to a  lesser extent transmission spectrum (due to other degeneracies) can be used to constrain the energy redistribution for all the planets presented in this grid.} 

\item \corr{The increase in metallically leads to the $P$-$T$ profile becoming hotter for both WASP-17b and WASP-121, due to the increase in the total opacity, driven by the increase in the mean molecular weight of the atmosphere.  The increase in metallicity leads to an increase in the size of the transmission spectral features at first (up to 10$\times$ solar metallicity) due to higher abundances and higher temperatures, but then they decrease again as the scale height of the atmosphere decreases (due to increased mean molecular weight) with increased metallicity. However, this effect is not seen in the emission spectrum where absorption/emission features increase with increasing metallicity, since emission spectral features do not have scale height dependance as transmission spectral features. The shift in the atmospheric level probed by emission spectrum to lower pressure levels with increasing metallicity, as shown by contribution functions in the H$_2$O  and CO$_2$ bands, as well as the difference in the level probed by the wings and cores of absorption/emission bands for a given metallicity, both can be used to constrain metallicity and the $P$-$T$ profile of the atmosphere.}

\item \corr{Increasing the C/O ratio leads to a cooling of $P$-$T$ profile for WASP-17b, with a transition seen from a H$_2$O dominated spectrum to a spectrum dominated by carbon bearing species (primarily CH$_4$) between C/O ratios of 0.75 and 1.  For WASP-121b, the strength of the temperature inversion decreases as the abundance of TiO/VO decreases with increasing C/O ratio. At a C/O ratio greater than 1 even though TiO/VO abundances are low, a weak temperature inversion is maintained in the atmosphere due to Na and K at f$_{\textup{c}}$ value of 0.5 as shown by previous works. Whereas, for a f$_{\textup{c}}$ value of 1 (hotter $P$-$T$ profile) H$^-$ opacity contributes substantially to the development of an inversion. Interestingly, extremely irradiated exoplanet atmospheres like WASP-121b show a spectrum dominated by HCN features at infrared wavelengths for high C/O ratios (> 1) in comparison to comparatively cooler planets like WASP-17b which are dominated by CH$_4$ features. HCN and H$^-$ features can be seen as an emission as well as absorption feature at different wavelengths in their emission spectrum. Therefore, HCN and H$^-$ features for these planets can be used to constrain their C/O ratios as well as their thermal structure.} 

\end{itemize}

\subsection{General Conclusions}
The interpretation of observations using the library of models with RCE $P$-$T$ profiles presented in this work was demonstrated using WASP-121b observations. The transmission spectra probing the planetary atmospheric limb reveal H$_2$O features in the infrared with evidence of VO in the optical. The best fit model indicates super-solar metallicity driven by optical observations and VO opacity, with non-inverted limb $P$-$T$ profile. The emission spectra probing the dayside of the planetary atmosphere reveals a H$_2$O feature in emission giving evidence of a potential temperature inversion and very low energy redistribution in the atmosphere. 

One of the major limitations of the model simulations presented in this paper is that they are 1-dimensional. Ideally 2D or 3D approaches would more realistically represent a planetary atmosphere. However, 1D models allow us to explore a very large parameter space in chemistry and radiative transfer for a large number of planetary systems, not possible currently using 2D or 3D approaches due to computational limitations.  Moreover, the combination of different $P$-$T$ profiles, transmission and emission spectra for a given planet can be used to represent different parts of the atmosphere \citep[see for e.g][]{Macdonald2020}, thus paving the way to interpret 2D or 3D observations, like for example using spectral phase curves. We have also not modeled dis-equilibrium processes such as photochemistry and vertical mixing in this library of simulated atmospheres, which is planned for the future development. The current model spectra in the library has a resolution of R$\sim$5000 at 0.2 \textmu m decreasing to R$\sim$35 at 30 \textmu m, which cannot be used to interpret high resolution observations. However, the model atmospheric RCE $P$-$T$ profiles and the corresponding chemical abundances from our library can be used to generate synthetic high resolution spectra (transmission or emission), which can then be used to interpret high resolution observations. We also plan to update the spectra in our library online, to achieve the highest possible JWST resolution over some specific wavelength ranges.

The library of models presented in this work can be used to plan and interpret a broad range of transiting exoplanet observations with current and future facilities such as HST and JWST, since it covers the wavelength range from 0.2 to 30$\mic$.  It represents a valuable resource to guide choices of observational targets and instrument modes. The library of models can also be used to initialize retrieval models as well as to constrain priors in them. Our development of an expansive grid of thermochemically consistent 1D models for exoplanet atmospheres provides the community with another tool with which to tackle our theoretical understanding of hot Jupiter atmospheres and interpret current and future observations of these planets.

\section*{Acknowledgements}
We would like to thank the anonymous referee for their detailed and constructive comments, that improved the clarity and quality of this manuscript. J.M.G and N.M acknowledges funding by a Leverhulme Trust Research Project Grant and University of Exeter College of  Engineering, Mathematics and Physical Sciences PhD studentship. M.W. P. acknowledges support through a UKRI-STFC studentship. This  work  used  the  DiRAC  Complexity  system, operated by the University of Leicester IT Services, which forms part of the  STFC  DiRAC  HPC  Facility. This work also used the University of Exeter Supercomputer, a DiRAC Facility jointly funded by STFC, the Large Facilities Capital Fund of BIS and the University of Exeter.

\section*{Data Availability}
The data underlying this article are currently available publicly here \url{https://drive.google.com/drive/folders/1zCCe6HICuK2nLgnYJFal7W4lyunjU4JE} and here \url{https://noctis.erc-atmo.eu:5001/fsdownload/hq0z4udQJ/goyal2020}
%
%
\bibliographystyle{mnras}
\bibliography{consistent_grid_review}

\begin{thebibliography}{}
\makeatletter
\relax
\def\mn@urlcharsother{\let\do\@makeother \do\$\do\&\do\#\do\^\do\_\do\%\do\~}
\def\mn@doi{\begingroup\mn@urlcharsother \@ifnextchar [ {\mn@doi@}
  {\mn@doi@[]}}
\def\mn@doi@[#1]#2{\def\@tempa{#1}\ifx\@tempa\@empty \href
  {http://dx.doi.org/#2} {doi:#2}\else \href {http://dx.doi.org/#2} {#1}\fi
  \endgroup}
\def\mn@eprint#1#2{\mn@eprint@#1:#2::\@nil}
\def\mn@eprint@arXiv#1{\href {http://arxiv.org/abs/#1} {{\tt arXiv:#1}}}
\def\mn@eprint@dblp#1{\href {http://dblp.uni-trier.de/rec/bibtex/#1.xml}
  {dblp:#1}}
\def\mn@eprint@#1:#2:#3:#4\@nil{\def\@tempa {#1}\def\@tempb {#2}\def\@tempc
  {#3}\ifx \@tempc \@empty \let \@tempc \@tempb \let \@tempb \@tempa \fi \ifx
  \@tempb \@empty \def\@tempb {arXiv}\fi \@ifundefined
  {mn@eprint@\@tempb}{\@tempb:\@tempc}{\expandafter \expandafter \csname
  mn@eprint@\@tempb\endcsname \expandafter{\@tempc}}}

\bibitem[\protect\citeauthoryear{{Alam} et~al.,}{{Alam}
  et~al.}{2018}]{Alam2018}
{Alam} M.~K.,  et~al., 2018, \mn@doi [\aj] {10.3847/1538-3881/aaee89}, \href
  {https://ui.adsabs.harvard.edu/\#abs/2018AJ....156..298A} {156, 298}

\bibitem[\protect\citeauthoryear{{Allard}, {Allard}, {Hauschildt}, {Kielkopf}
  \& {Machin}}{{Allard} et~al.}{2003}]{Allard2003}
{Allard} N.~F.,  {Allard} F.,  {Hauschildt} P.~H.,  {Kielkopf} J.~F.,
  {Machin} L.,  2003, \mn@doi [\aap] {10.1051/0004-6361:20031299}, \href
  {http://adsabs.harvard.edu/abs/2003A%26A...411L.473A} {411, L473}

\bibitem[\protect\citeauthoryear{{Allard}, {Homeier}  \& {Freytag}}{{Allard}
  et~al.}{2012}]{Allard2012}
{Allard} F.,  {Homeier} D.,   {Freytag} B.,  2012, \mn@doi [Philosophical
  Transactions of the Royal Society of London Series A]
  {10.1098/rsta.2011.0269}, \href
  {http://adsabs.harvard.edu/abs/2012RSPTA.370.2765A} {370, 2765}

\bibitem[\protect\citeauthoryear{{Amundsen}}{{Amundsen}}{2015}]{Amundsenthesis}
{Amundsen} D.~S.,  2015, PhD Thesis

\bibitem[\protect\citeauthoryear{{Amundsen}, {Baraffe}, {Tremblin}, {Manners},
  {Hayek}, {Mayne}  \& {Acreman}}{{Amundsen} et~al.}{2014}]{Amundsen2014}
{Amundsen} D.~S.,  {Baraffe} I.,  {Tremblin} P.,  {Manners} J.,  {Hayek} W.,
  {Mayne} N.~J.,   {Acreman} D.~M.,  2014, \mn@doi [\aap]
  {10.1051/0004-6361/201323169}, \href
  {http://adsabs.harvard.edu/abs/2014A%26A...564A..59A} {564, A59}

\bibitem[\protect\citeauthoryear{{Arcangeli} et~al.,}{{Arcangeli}
  et~al.}{2018}]{Arcangeli2018}
{Arcangeli} J.,  et~al., 2018, \mn@doi [\apj] {10.3847/2041-8213/aab272}, \href
  {https://ui.adsabs.harvard.edu/abs/2018ApJ...855L..30A} {855, L30}

\bibitem[\protect\citeauthoryear{{Baraffe}, {Homeier}, {Allard}  \&
  {Chabrier}}{{Baraffe} et~al.}{2015}]{Baraffe2015}
{Baraffe} I.,  {Homeier} D.,  {Allard} F.,   {Chabrier} G.,  2015, \mn@doi
  [\aap] {10.1051/0004-6361/201425481}, \href
  {http://adsabs.harvard.edu/abs/2015A%26A...577A..42B} {577, A42}

\bibitem[\protect\citeauthoryear{{Baudino}, {Molli{\`e}re}, {Venot},
  {Tremblin}, {B{\'e}zard}  \& {Lagage}}{{Baudino} et~al.}{2017}]{Baudino2017}
{Baudino} J.-L.,  {Molli{\`e}re} P.,  {Venot} O.,  {Tremblin} P.,  {B{\'e}zard}
  B.,   {Lagage} P.-O.,  2017, \mn@doi [\apj] {10.3847/1538-4357/aa95be}, \href
  {http://adsabs.harvard.edu/abs/2017ApJ...850..150B} {850, 150}

\bibitem[\protect\citeauthoryear{{Bell} \& {Berrington}}{{Bell} \&
  {Berrington}}{1987}]{Bell1987}
{Bell} K.~L.,  {Berrington} K.~A.,  1987, \mn@doi [Journal of Physics B Atomic
  Molecular Physics] {10.1088/0022-3700/20/4/019}, \href
  {https://ui.adsabs.harvard.edu/#abs/1987JPhB...20..801B} {20, 801}

\bibitem[\protect\citeauthoryear{{Burgasser}, {Kirkpatrick}, {Liebert}  \&
  {Burrows}}{{Burgasser} et~al.}{2003}]{Burgasser2003}
{Burgasser} A.~J.,  {Kirkpatrick} J.~D.,  {Liebert} J.,   {Burrows} A.,  2003,
  \mn@doi [\apj] {10.1086/376756}, \href
  {https://ui.adsabs.harvard.edu/abs/2003ApJ...594..510B} {594, 510}

\bibitem[\protect\citeauthoryear{{Burrows} \& {Sharp}}{{Burrows} \&
  {Sharp}}{1999}]{Burrows1999}
{Burrows} A.,  {Sharp} C.~M.,  1999, \mn@doi [\apj] {10.1086/306811}, \href
  {http://adsabs.harvard.edu/abs/1999ApJ...512..843B} {512, 843}

\bibitem[\protect\citeauthoryear{{Burrows} et~al.,}{{Burrows}
  et~al.}{1997}]{Burrows1997}
{Burrows} A.,  et~al., 1997, \mn@doi [\apj] {10.1086/305002}, \href
  {http://adsabs.harvard.edu/abs/1997ApJ...491..856B} {491, 856}

\bibitem[\protect\citeauthoryear{{Burrows}, {Marley}  \& {Sharp}}{{Burrows}
  et~al.}{2000}]{Burrows2000}
{Burrows} A.,  {Marley} M.~S.,   {Sharp} C.~M.,  2000, \mn@doi [\apj]
  {10.1086/308462}, \href
  {https://ui.adsabs.harvard.edu/#abs/2000ApJ...531..438B} {531, 438}

\bibitem[\protect\citeauthoryear{{Burrows}, {Ram}, {Bernath}, {Sharp}  \&
  {Milsom}}{{Burrows} et~al.}{2002}]{Burrows2002}
{Burrows} A.,  {Ram} R.~S.,  {Bernath} P.,  {Sharp} C.~M.,   {Milsom} J.~A.,
  2002, \mn@doi [\apj] {10.1086/342242}, \href
  {http://adsabs.harvard.edu/abs/2002ApJ...577..986B} {577, 986}

\bibitem[\protect\citeauthoryear{{Burrows}, {Budaj}  \& {Hubeny}}{{Burrows}
  et~al.}{2008}]{Burrows2008}
{Burrows} A.,  {Budaj} J.,   {Hubeny} I.,  2008, \mn@doi [\apj]
  {10.1086/533518}, \href
  {https://ui.adsabs.harvard.edu/abs/2008ApJ...678.1436B} {678, 1436}

\bibitem[\protect\citeauthoryear{{Carter} et~al.,}{{Carter}
  et~al.}{2020}]{Carter2020}
{Carter} A.~L.,  et~al., 2020, \mn@doi [\mnras] {10.1093/mnras/staa1078}, \href
  {https://ui.adsabs.harvard.edu/abs/2020MNRAS.494.5449C} {494, 5449}

\bibitem[\protect\citeauthoryear{{Chamberlain} \& {Hunten}}{{Chamberlain} \&
  {Hunten}}{1987}]{Chamberlain1987}
{Chamberlain} J.~W.,  {Hunten} D.~M.,  1987, {Theory of planetary atmospheres.
  An introduction to their physics andchemistry.}.
 Vol. 36

\bibitem[\protect\citeauthoryear{{Chubb}, {Min}, {Kawashima}, {Helling}  \&
  {Waldmann}}{{Chubb} et~al.}{2020}]{Chubb2020}
{Chubb} K.~L.,  {Min} M.,  {Kawashima} Y.,  {Helling} C.,   {Waldmann} I.,
  2020, arXiv e-prints, \href
  {https://ui.adsabs.harvard.edu/abs/2020arXiv200413679C} {p. arXiv:2004.13679}

\bibitem[\protect\citeauthoryear{{Drummond}, {Tremblin}, {Baraffe}, {Amundsen},
  {Mayne}, {Venot}  \& {Goyal}}{{Drummond} et~al.}{2016}]{Drummond2016}
{Drummond} B.,  {Tremblin} P.,  {Baraffe} I.,  {Amundsen} D.~S.,  {Mayne}
  N.~J.,  {Venot} O.,   {Goyal} J.,  2016, \mn@doi [\aap]
  {10.1051/0004-6361/201628799}, \href
  {http://adsabs.harvard.edu/abs/2016A%26A...594A..69D} {594, A69}

\bibitem[\protect\citeauthoryear{{Drummond}, {Mayne}, {Manners}, {Baraffe},
  {Goyal}, {Tremblin}, {Sing}  \& {Kohary}}{{Drummond}
  et~al.}{2018}]{Drummond2018b}
{Drummond} B.,  {Mayne} N.~J.,  {Manners} J.,  {Baraffe} I.,  {Goyal} J.,
  {Tremblin} P.,  {Sing} D.~K.,   {Kohary} K.,  2018, \mn@doi [\apj]
  {10.3847/1538-4357/aaeb28}, \href
  {https://ui.adsabs.harvard.edu/abs/2018ApJ...869...28D} {869, 28}

\bibitem[\protect\citeauthoryear{{Drummond}, {Carter}, {H{\'e}brard}, {Mayne},
  {Sing}, {Evans}  \& {Goyal}}{{Drummond} et~al.}{2019}]{Drummond2019}
{Drummond} B.,  {Carter} A.~L.,  {H{\'e}brard} E.,  {Mayne} N.~J.,  {Sing}
  D.~K.,  {Evans} T.~M.,   {Goyal} J.,  2019, \mn@doi [\mnras]
  {10.1093/mnras/stz909}, \href
  {https://ui.adsabs.harvard.edu/abs/2019MNRAS.486.1123D} {486, 1123}

\bibitem[\protect\citeauthoryear{{Evans} et~al.,}{{Evans}
  et~al.}{2017}]{Evans2017}
{Evans} T.~M.,  et~al., 2017, \mn@doi [\nat] {10.1038/nature23266}, \href
  {http://adsabs.harvard.edu/abs/2017Natur.548...58E} {548, 58}

\bibitem[\protect\citeauthoryear{{Evans} et~al.,}{{Evans}
  et~al.}{2018}]{Evans2018}
{Evans} T.~M.,  et~al., 2018, \mn@doi [\aj] {10.3847/1538-3881/aaebff}, \href
  {https://ui.adsabs.harvard.edu/\#abs/2018AJ....156..283E} {156, 283}

\bibitem[\protect\citeauthoryear{{Fortney}}{{Fortney}}{2005}]{Fortney2005}
{Fortney} J.~J.,  2005, \mn@doi [\mnras] {10.1111/j.1365-2966.2005.09587.x},
  \href {http://adsabs.harvard.edu/abs/2005MNRAS.364..649F} {364, 649}

\bibitem[\protect\citeauthoryear{{Fortney} \& {Marley}}{{Fortney} \&
  {Marley}}{2007}]{Fortney2007}
{Fortney} J.~J.,  {Marley} M.~S.,  2007, \mn@doi [\apjl] {10.1086/521603},
  \href {http://adsabs.harvard.edu/abs/2007ApJ...666L..45F} {666, L45}

\bibitem[\protect\citeauthoryear{{Fortney}, {Marley}  \& {Barnes}}{{Fortney}
  et~al.}{2007}]{Fortney2007b}
{Fortney} J.~J.,  {Marley} M.~S.,   {Barnes} J.~W.,  2007, \mn@doi [\apj]
  {10.1086/512120}, \href
  {https://ui.adsabs.harvard.edu/abs/2007ApJ...659.1661F} {659, 1661}

\bibitem[\protect\citeauthoryear{{Fortney}, {Lodders}, {Marley}  \&
  {Freedman}}{{Fortney} et~al.}{2008}]{Fortney2008}
{Fortney} J.~J.,  {Lodders} K.,  {Marley} M.~S.,   {Freedman} R.~S.,  2008,
  \mn@doi [\apj] {10.1086/528370}, \href
  {http://adsabs.harvard.edu/abs/2008ApJ...678.1419F} {678, 1419}

\bibitem[\protect\citeauthoryear{{Fortney} et~al.,}{{Fortney}
  et~al.}{2016}]{Fortney2016}
{Fortney} J.~J.,  et~al., 2016, preprint, \href
  {http://adsabs.harvard.edu/abs/2016arXiv160206305F} {} (\mn@eprint {arXiv}
  {1602.06305})

\bibitem[\protect\citeauthoryear{{Gandhi} \& {Madhusudhan}}{{Gandhi} \&
  {Madhusudhan}}{2017}]{Gandhi2017}
{Gandhi} S.,  {Madhusudhan} N.,  2017, \mn@doi [\mnras]
  {10.1093/mnras/stx1601}, \href
  {https://ui.adsabs.harvard.edu/\#abs/2017MNRAS.472.2334G} {472, 2334}

\bibitem[\protect\citeauthoryear{{Gandhi} \& {Madhusudhan}}{{Gandhi} \&
  {Madhusudhan}}{2019}]{Gandhi2019}
{Gandhi} S.,  {Madhusudhan} N.,  2019, \mn@doi [\mnras] {10.1093/mnras/stz751},
  \href {https://ui.adsabs.harvard.edu/abs/2019MNRAS.485.5817G} {485, 5817}

\bibitem[\protect\citeauthoryear{{Gibson} et~al.,}{{Gibson}
  et~al.}{2020}]{Gibson2020}
{Gibson} N.~P.,  et~al., 2020, \mn@doi [\mnras] {10.1093/mnras/staa228}, \href
  {https://ui.adsabs.harvard.edu/abs/2020MNRAS.493.2215G} {493, 2215}

\bibitem[\protect\citeauthoryear{{Goyal}}{{Goyal}}{2019}]{Goyalthesis2019}
{Goyal} J.~M.,  2019, University of Exeter

\bibitem[\protect\citeauthoryear{{Goyal} et~al.,}{{Goyal}
  et~al.}{2018}]{Goyal2018}
{Goyal} J.~M.,  et~al., 2018, \mn@doi [\mnras] {10.1093/mnras/stx3015}, \href
  {https://ui.adsabs.harvard.edu/#abs/2018MNRAS.474.5158G} {474, 5158}

\bibitem[\protect\citeauthoryear{{Goyal} et~al.,}{{Goyal}
  et~al.}{2019a}]{Errgoyal2019}
{Goyal} J.~M.,  et~al., 2019a, \mn@doi [\mnras] {10.1093/mnras/stz755}, \href
  {https://ui.adsabs.harvard.edu/abs/2019MNRAS.tmp..722G} {p.~722}

\bibitem[\protect\citeauthoryear{{Goyal}, {Wakeford}, {Mayne}, {Lewis},
  {Drummond}  \& {Sing}}{{Goyal} et~al.}{2019b}]{Goyal2019}
{Goyal} J.~M.,  {Wakeford} H.~R.,  {Mayne} N.~J.,  {Lewis} N.~K.,  {Drummond}
  B.,   {Sing} D.~K.,  2019b, \mn@doi [\mnras] {10.1093/mnras/sty3001}, \href
  {https://ui.adsabs.harvard.edu/abs/2019MNRAS.482.4503G} {482, 4503}

\bibitem[\protect\citeauthoryear{{Guillot} \& {Showman}}{{Guillot} \&
  {Showman}}{2002}]{Guillot2002}
{Guillot} T.,  {Showman} A.~P.,  2002, \mn@doi [\aap]
  {10.1051/0004-6361:20011624}, \href
  {https://ui.adsabs.harvard.edu/abs/2002A&A...385..156G} {385, 156}

\bibitem[\protect\citeauthoryear{{Heiter} et~al.,}{{Heiter}
  et~al.}{2008}]{Heiter2008}
{Heiter} U.,  et~al., 2008, in Journal of Physics Conference Series. p. 012011,
  \mn@doi{10.1088/1742-6596/130/1/012011}

\bibitem[\protect\citeauthoryear{{Heiter} et~al.,}{{Heiter}
  et~al.}{2015}]{Heiter2015}
{Heiter} U.,  et~al., 2015, \mn@doi [\physscr] {10.1088/0031-8949/90/5/054010},
  \href {https://ui.adsabs.harvard.edu/abs/2015PhyS...90e4010H} {90, 054010}

\bibitem[\protect\citeauthoryear{{Heng} \& {Kitzmann}}{{Heng} \&
  {Kitzmann}}{2017}]{Heng2017}
{Heng} K.,  {Kitzmann} D.,  2017, preprint, \href
  {http://adsabs.harvard.edu/abs/2017arXiv170202051H} {} (\mn@eprint {arXiv}
  {1702.02051})

\bibitem[\protect\citeauthoryear{{Hobbs}, {Shorttle}, {Madhusudhan}  \&
  {Rimmer}}{{Hobbs} et~al.}{2019}]{Hobbs2019}
{Hobbs} R.,  {Shorttle} O.,  {Madhusudhan} N.,   {Rimmer} P.,  2019, \mn@doi
  [\mnras] {10.1093/mnras/stz1333}, \href
  {https://ui.adsabs.harvard.edu/abs/2019MNRAS.487.2242H} {487, 2242}

\bibitem[\protect\citeauthoryear{{Hubeny}, {Burrows}  \& {Sudarsky}}{{Hubeny}
  et~al.}{2003}]{Hubeny2003}
{Hubeny} I.,  {Burrows} A.,   {Sudarsky} D.,  2003, \mn@doi [\apj]
  {10.1086/377080}, \href
  {https://ui.adsabs.harvard.edu/abs/2003ApJ...594.1011H} {594, 1011}

\bibitem[\protect\citeauthoryear{{John}}{{John}}{1988}]{John1988}
{John} T.~L.,  1988, \aap, \href
  {http://adsabs.harvard.edu/abs/1988A%26A...193..189J} {193, 189}

\bibitem[\protect\citeauthoryear{{Knutson} et~al.,}{{Knutson}
  et~al.}{2009}]{Knutson2009}
{Knutson} H.~A.,  et~al., 2009, \mn@doi [\apj] {10.1088/0004-637X/690/1/822},
  \href {http://adsabs.harvard.edu/abs/2009ApJ...690..822K} {690, 822}

\bibitem[\protect\citeauthoryear{{Kreidberg} et~al.,}{{Kreidberg}
  et~al.}{2018}]{Kriedberg2018}
{Kreidberg} L.,  et~al., 2018, \mn@doi [\aj] {10.3847/1538-3881/aac3df}, \href
  {https://ui.adsabs.harvard.edu/\#abs/2018AJ....156...17K} {156, 17}

\bibitem[\protect\citeauthoryear{{Lecavelier Des Etangs}, {Pont},
  {Vidal-Madjar}  \& {Sing}}{{Lecavelier Des Etangs}
  et~al.}{2008}]{Lecavelier2008}
{Lecavelier Des Etangs} A.,  {Pont} F.,  {Vidal-Madjar} A.,   {Sing} D.,  2008,
  \mn@doi [\aap] {10.1051/0004-6361:200809388}, \href
  {http://adsabs.harvard.edu/abs/2008A%26A...481L..83L} {481, L83}

\bibitem[\protect\citeauthoryear{{Lodders} \& {Fegley}}{{Lodders} \&
  {Fegley}}{2002}]{Lodders2002}
{Lodders} K.,  {Fegley} B.,  2002, \mn@doi [\icarus] {10.1006/icar.2001.6740},
  \href {https://ui.adsabs.harvard.edu/\#abs/2002Icar..155..393L} {155, 393}

\bibitem[\protect\citeauthoryear{{Lothringer}, {Barman}  \&
  {Koskinen}}{{Lothringer} et~al.}{2018}]{Lothringer2018}
{Lothringer} J.~D.,  {Barman} T.,   {Koskinen} T.,  2018, \mn@doi [\apj]
  {10.3847/1538-4357/aadd9e}, \href
  {https://ui.adsabs.harvard.edu/abs/2018ApJ...866...27L} {866, 27}

\bibitem[\protect\citeauthoryear{{MacDonald}, {Goyal}  \& {Lewis}}{{MacDonald}
  et~al.}{2020}]{Macdonald2020}
{MacDonald} R.~J.,  {Goyal} J.~M.,   {Lewis} N.~K.,  2020, arXiv e-prints,
  \href {https://ui.adsabs.harvard.edu/abs/2020arXiv200311548M} {p.
  arXiv:2003.11548}

\bibitem[\protect\citeauthoryear{{Madhusudhan}}{{Madhusudhan}}{2012}]{Madhusudhan2012}
{Madhusudhan} N.,  2012, \mn@doi [\apj] {10.1088/0004-637X/758/1/36}, \href
  {http://adsabs.harvard.edu/abs/2012ApJ...758...36M} {758, 36}

\bibitem[\protect\citeauthoryear{{Madhusudhan}, {Ag{\'u}ndez}, {Moses}  \&
  {Hu}}{{Madhusudhan} et~al.}{2016}]{Madhureview2016}
{Madhusudhan} N.,  {Ag{\'u}ndez} M.,  {Moses} J.~I.,   {Hu} Y.,  2016, \mn@doi
  [\ssr] {10.1007/s11214-016-0254-3}, \href
  {https://ui.adsabs.harvard.edu/#abs/2016SSRv..205..285M} {205, 285}

\bibitem[\protect\citeauthoryear{{Malik}, {Kitzmann}, {Mendon{\c{c}}a},
  {Grimm}, {Marleau}, {Linder}, {Tsai}  \& {Heng}}{{Malik}
  et~al.}{2019}]{Malik2018}
{Malik} M.,  {Kitzmann} D.,  {Mendon{\c{c}}a} J.~M.,  {Grimm} S.~L.,  {Marleau}
  G.-D.,  {Linder} E.~F.,  {Tsai} S.-M.,   {Heng} K.,  2019, \mn@doi [\aj]
  {10.3847/1538-3881/ab1084}, \href
  {https://ui.adsabs.harvard.edu/abs/2019AJ....157..170M} {157, 170}

\bibitem[\protect\citeauthoryear{{Mansfield} et~al.,}{{Mansfield}
  et~al.}{2018}]{Mansfield2018}
{Mansfield} M.,  et~al., 2018, \mn@doi [\aj] {10.3847/1538-3881/aac497}, \href
  {https://ui.adsabs.harvard.edu/\#abs/2018AJ....156...10M} {156, 10}

\bibitem[\protect\citeauthoryear{{Marley} \& {Robinson}}{{Marley} \&
  {Robinson}}{2015}]{Marley2014}
{Marley} M.~S.,  {Robinson} T.~D.,  2015, \mn@doi [Annual Review of Astronomy
  and Astrophysics] {10.1146/annurev-astro-082214-122522}, \href
  {https://ui.adsabs.harvard.edu/\#abs/2015ARA&A..53..279M} {53, 279}

\bibitem[\protect\citeauthoryear{{Marley}, {Saumon}, {Guillot}, {Freedman},
  {Hubbard}, {Burrows}  \& {Lunine}}{{Marley} et~al.}{1996}]{Marley1996}
{Marley} M.~S.,  {Saumon} D.,  {Guillot} T.,  {Freedman} R.~S.,  {Hubbard}
  W.~B.,  {Burrows} A.,   {Lunine} J.~I.,  1996, \mn@doi [Science]
  {10.1126/science.272.5270.1919}, \href
  {https://ui.adsabs.harvard.edu/abs/1996Sci...272.1919M} {272, 1919}

\bibitem[\protect\citeauthoryear{{Mikal-Evans} et~al.,}{{Mikal-Evans}
  et~al.}{2019}]{Evans2019}
{Mikal-Evans} T.,  et~al., 2019, \mn@doi [\mnras] {10.1093/mnras/stz1753},
  \href {https://ui.adsabs.harvard.edu/abs/2019MNRAS.488.2222M} {488, 2222}

\bibitem[\protect\citeauthoryear{{Molaverdikhani}, {Henning}  \&
  {Molli{\`e}re}}{{Molaverdikhani} et~al.}{2019}]{Molaverdikhani2019}
{Molaverdikhani} K.,  {Henning} T.,   {Molli{\`e}re} P.,  2019, \mn@doi [\apj]
  {10.3847/1538-4357/aafda8}, \href
  {https://ui.adsabs.harvard.edu/abs/2019ApJ...873...32M} {873, 32}

\bibitem[\protect\citeauthoryear{{Molli{\`e}re}, {van Boekel}, {Dullemond},
  {Henning}  \& {Mordasini}}{{Molli{\`e}re} et~al.}{2015}]{Molliere2015}
{Molli{\`e}re} P.,  {van Boekel} R.,  {Dullemond} C.,  {Henning} T.,
  {Mordasini} C.,  2015, \mn@doi [\apj] {10.1088/0004-637X/813/1/47}, \href
  {http://adsabs.harvard.edu/abs/2015ApJ...813...47M} {813, 47}

\bibitem[\protect\citeauthoryear{{Molli{\`e}re}, {van Boekel}, {Bouwman},
  {Henning}, {Lagage}  \& {Min}}{{Molli{\`e}re} et~al.}{2016}]{Molliere2016}
{Molli{\`e}re} P.,  {van Boekel} R.,  {Bouwman} J.,  {Henning} T.,  {Lagage}
  P.-O.,   {Min} M.,  2016, preprint, \href
  {http://adsabs.harvard.edu/abs/2016arXiv161108608M} {} (\mn@eprint {arXiv}
  {1611.08608})

\bibitem[\protect\citeauthoryear{{Moses}, {Madhusudhan}, {Visscher}  \&
  {Freedman}}{{Moses} et~al.}{2013}]{Moses2013}
{Moses} J.~I.,  {Madhusudhan} N.,  {Visscher} C.,   {Freedman} R.~S.,  2013,
  \mn@doi [\apj] {10.1088/0004-637X/763/1/25}, \href
  {http://adsabs.harvard.edu/abs/2013ApJ...763...25M} {763, 25}

\bibitem[\protect\citeauthoryear{{Nikolov} et~al.,}{{Nikolov}
  et~al.}{2018}]{Nikolov2018}
{Nikolov} N.,  et~al., 2018, \mn@doi [\nat] {10.1038/s41586-018-0101-7}, \href
  {https://ui.adsabs.harvard.edu/#abs/2018Natur.557..526N} {557, 526}

\bibitem[\protect\citeauthoryear{{Parmentier} et~al.,}{{Parmentier}
  et~al.}{2018}]{Parmentier2018}
{Parmentier} V.,  et~al., 2018, \mn@doi [\aap] {10.1051/0004-6361/201833059},
  \href {https://ui.adsabs.harvard.edu/\#abs/2018A&A...617A.110P} {617, A110}

\bibitem[\protect\citeauthoryear{{Rajpurohit}, {Reyl{\'e}}, {Allard},
  {Homeier}, {Schultheis}, {Bessell}  \& {Robin}}{{Rajpurohit}
  et~al.}{2013}]{Rajpurohit2013}
{Rajpurohit} A.~S.,  {Reyl{\'e}} C.,  {Allard} F.,  {Homeier} D.,  {Schultheis}
  M.,  {Bessell} M.~S.,   {Robin} A.~C.,  2013, \mn@doi [\aap]
  {10.1051/0004-6361/201321346}, \href
  {http://adsabs.harvard.edu/abs/2013A%26A...556A..15R} {556, A15}

\bibitem[\protect\citeauthoryear{{Rau}}{{Rau}}{1996}]{Rau1996}
{Rau} A.~R.~P.,  1996, \mn@doi [Journal of Astrophysics and Astronomy]
  {10.1007/BF02702300}, \href
  {https://ui.adsabs.harvard.edu/abs/1996JApA...17..113R} {17, 113}

\bibitem[\protect\citeauthoryear{{Rybicki} \& {Lightman}}{{Rybicki} \&
  {Lightman}}{1986}]{Rybicki1986}
{Rybicki} G.~B.,  {Lightman} A.~P.,  1986, {Radiative Processes in
  Astrophysics}

\bibitem[\protect\citeauthoryear{{Sauval} \& {Tatum}}{{Sauval} \&
  {Tatum}}{1984}]{Sauval1984}
{Sauval} A.~J.,  {Tatum} J.~B.,  1984, \mn@doi [\apjs] {10.1086/190980}, \href
  {http://adsabs.harvard.edu/abs/1984ApJS...56..193S} {56, 193}

\bibitem[\protect\citeauthoryear{{Seager} \& {Sasselov}}{{Seager} \&
  {Sasselov}}{1998}]{Seager1998}
{Seager} S.,  {Sasselov} D.~D.,  1998, \mn@doi [\apjl] {10.1086/311498}, \href
  {http://adsabs.harvard.edu/abs/1998ApJ...502L.157S} {502, L157}

\bibitem[\protect\citeauthoryear{{Sharp} \& {Burrows}}{{Sharp} \&
  {Burrows}}{2007}]{Sharp2007}
{Sharp} C.~M.,  {Burrows} A.,  2007, \mn@doi [\apjs] {10.1086/508708}, \href
  {http://adsabs.harvard.edu/abs/2007ApJS..168..140S} {168, 140}

\bibitem[\protect\citeauthoryear{{Sing} et~al.,}{{Sing}
  et~al.}{2016}]{Sing2016}
{Sing} D.~K.,  et~al., 2016, \mn@doi [\nat] {10.1038/nature16068}, \href
  {http://adsabs.harvard.edu/abs/2016Natur.529...59S} {529, 59}

\bibitem[\protect\citeauthoryear{{Sing} et~al.,}{{Sing}
  et~al.}{2019}]{Sing2019}
{Sing} D.~K.,  et~al., 2019, \mn@doi [\aj] {10.3847/1538-3881/ab2986}, \href
  {https://ui.adsabs.harvard.edu/abs/2019AJ....158...91S} {158, 91}

\bibitem[\protect\citeauthoryear{{Southworth}}{{Southworth}}{2011}]{Tepcat2011}
{Southworth} J.,  2011, \mn@doi [\mnras] {10.1111/j.1365-2966.2011.19399.x},
  \href {http://adsabs.harvard.edu/abs/2011MNRAS.417.2166S} {417, 2166}

\bibitem[\protect\citeauthoryear{{Spiegel}, {Silverio}  \& {Burrows}}{{Spiegel}
  et~al.}{2009}]{Spiegel2009}
{Spiegel} D.~S.,  {Silverio} K.,   {Burrows} A.,  2009, \mn@doi [\apj]
  {10.1088/0004-637X/699/2/1487}, \href
  {http://adsabs.harvard.edu/abs/2009ApJ...699.1487S} {699, 1487}

\bibitem[\protect\citeauthoryear{{Sudarsky}, {Burrows}  \& {Hubeny}}{{Sudarsky}
  et~al.}{2003}]{Sudarsky2003}
{Sudarsky} D.,  {Burrows} A.,   {Hubeny} I.,  2003, \mn@doi [\apj]
  {10.1086/374331}, \href {http://adsabs.harvard.edu/abs/2003ApJ...588.1121S}
  {588, 1121}

\bibitem[\protect\citeauthoryear{{Thorngren}, {Gao}  \& {Fortney}}{{Thorngren}
  et~al.}{2019}]{Thorngren2019}
{Thorngren} D.,  {Gao} P.,   {Fortney} J.~J.,  2019, \mn@doi [\apjl]
  {10.3847/2041-8213/ab43d0}, \href
  {https://ui.adsabs.harvard.edu/abs/2019ApJ...884L...6T} {884, L6}

\bibitem[\protect\citeauthoryear{{Tremblin}, {Amundsen}, {Mourier}, {Baraffe},
  {Chabrier}, {Drummond}, {Homeier}  \& {Venot}}{{Tremblin}
  et~al.}{2015}]{Tremblin2015}
{Tremblin} P.,  {Amundsen} D.~S.,  {Mourier} P.,  {Baraffe} I.,  {Chabrier} G.,
   {Drummond} B.,  {Homeier} D.,   {Venot} O.,  2015, \mn@doi [\apjl]
  {10.1088/2041-8205/804/1/L17}, \href
  {http://adsabs.harvard.edu/abs/2015ApJ...804L..17T} {804, L17}

\bibitem[\protect\citeauthoryear{{Tremblin}, {Amundsen}, {Chabrier}, {Baraffe},
  {Drummond}, {Hinkley}, {Mourier}  \& {Venot}}{{Tremblin}
  et~al.}{2016}]{Tremblin2016}
{Tremblin} P.,  {Amundsen} D.~S.,  {Chabrier} G.,  {Baraffe} I.,  {Drummond}
  B.,  {Hinkley} S.,  {Mourier} P.,   {Venot} O.,  2016, \mn@doi [\apjl]
  {10.3847/2041-8205/817/2/L19}, \href
  {http://adsabs.harvard.edu/abs/2016ApJ...817L..19T} {817, L19}

\bibitem[\protect\citeauthoryear{{Tremblin} et~al.,}{{Tremblin}
  et~al.}{2017}]{Tremblin2017}
{Tremblin} P.,  et~al., 2017, \mn@doi [\apj] {10.3847/1538-4357/aa6e57}, \href
  {https://ui.adsabs.harvard.edu/#abs/2017ApJ...841...30T} {841, 30}

\bibitem[\protect\citeauthoryear{{Visscher}, {Lodders}  \& {Fegley}}{{Visscher}
  et~al.}{2006}]{Visscher2006}
{Visscher} C.,  {Lodders} K.,   {Fegley} Jr. B.,  2006, \mn@doi [\apj]
  {10.1086/506245}, \href {http://adsabs.harvard.edu/abs/2006ApJ...648.1181V}
  {648, 1181}

\bibitem[\protect\citeauthoryear{{Wakeford} et~al.,}{{Wakeford}
  et~al.}{2018}]{Wakeford2018}
{Wakeford} H.~R.,  et~al., 2018, \mn@doi [\aj] {10.3847/1538-3881/aa9e4e},
  \href {https://ui.adsabs.harvard.edu/#abs/2018AJ....155...29W} {155}

\bibitem[\protect\citeauthoryear{{Woitke}, {Helling}, {Hunter}, {Millard},
  {Turner}, {Worters}, {Blecic}  \& {Stock}}{{Woitke}
  et~al.}{2018}]{Woitke2018}
{Woitke} P.,  {Helling} C.,  {Hunter} G.~H.,  {Millard} J.~D.,  {Turner} G.~E.,
   {Worters} M.,  {Blecic} J.,   {Stock} J.~W.,  2018, \mn@doi [\aap]
  {10.1051/0004-6361/201732193}, \href
  {https://ui.adsabs.harvard.edu/\#abs/2018A&A...614A...1W} {614, A1}

\bibitem[\protect\citeauthoryear{Zahnle, Schaefer  \& Fegley}{Zahnle
  et~al.}{2010}]{Zahnle2010}
Zahnle K.,  Schaefer L.,   Fegley B.,  2010, \mn@doi [Cold Spring Harb Perspect
  Biol] {10.1101/cshperspect.a004895}, 2, a004895

\bibitem[\protect\citeauthoryear{{Zhang}, {Chachan}, {Kempton}  \&
  {Knutson}}{{Zhang} et~al.}{2019}]{Zhang2019}
{Zhang} M.,  {Chachan} Y.,  {Kempton} E. M.~R.,   {Knutson} H.~A.,  2019,
  \mn@doi [\pasp] {10.1088/1538-3873/aaf5ad}, \href
  {https://ui.adsabs.harvard.edu/abs/2019PASP..131c4501Z} {131, 034501}

\bibitem[\protect\citeauthoryear{{von Essen}, {Mallonn}, {Welbanks},
  {Madhusudhan}, {Pinhas}, {Bouy}  \& {Weis Hansen}}{{von Essen}
  et~al.}{2019}]{vonEssen2019}
{von Essen} C.,  {Mallonn} M.,  {Welbanks} L.,  {Madhusudhan} N.,  {Pinhas} A.,
   {Bouy} H.,   {Weis Hansen} P.,  2019, \mn@doi [\aap]
  {10.1051/0004-6361/201833837}, \href
  {https://ui.adsabs.harvard.edu/abs/2019A&A...622A..71V} {622, A71}

\makeatother
\end{thebibliography}


\begin{thebibliography}{}
\makeatletter
\relax
\def\mn@urlcharsother{\let\do\@makeother \do\$\do\&\do\#\do\^\do\_\do\%\do\~}
\def\mn@doi{\begingroup\mn@urlcharsother \@ifnextchar [ {\mn@doi@}
  {\mn@doi@[]}}
\def\mn@doi@[#1]#2{\def\@tempa{#1}\ifx\@tempa\@empty \href
  {http://dx.doi.org/#2} {doi:#2}\else \href {http://dx.doi.org/#2} {#1}\fi
  \endgroup}
\def\mn@eprint#1#2{\mn@eprint@#1:#2::\@nil}
\def\mn@eprint@arXiv#1{\href {http://arxiv.org/abs/#1} {{\tt arXiv:#1}}}
\def\mn@eprint@dblp#1{\href {http://dblp.uni-trier.de/rec/bibtex/#1.xml}
  {dblp:#1}}
\def\mn@eprint@#1:#2:#3:#4\@nil{\def\@tempa {#1}\def\@tempb {#2}\def\@tempc
  {#3}\ifx \@tempc \@empty \let \@tempc \@tempb \let \@tempb \@tempa \fi \ifx
  \@tempb \@empty \def\@tempb {arXiv}\fi \@ifundefined
  {mn@eprint@\@tempb}{\@tempb:\@tempc}{\expandafter \expandafter \csname
  mn@eprint@\@tempb\endcsname \expandafter{\@tempc}}}

\bibitem[\protect\citeauthoryear{{Allard}, {Royer}, {Kielkopf}  \&
  {Feautrier}}{{Allard} et~al.}{1999}]{Allard1999}
{Allard} N.~F.,  {Royer} A.,  {Kielkopf} J.~F.,   {Feautrier} N.,  1999,
  \mn@doi [\pra] {10.1103/PhysRevA.60.1021}, \href
  {http://adsabs.harvard.edu/abs/1999PhRvA..60.1021A} {60, 1021}

\bibitem[\protect\citeauthoryear{{Allard}, {Allard}, {Hauschildt}, {Kielkopf}
  \& {Machin}}{{Allard} et~al.}{2003}]{Allard2003}
{Allard} N.~F.,  {Allard} F.,  {Hauschildt} P.~H.,  {Kielkopf} J.~F.,
  {Machin} L.,  2003, \mn@doi [\aap] {10.1051/0004-6361:20031299}, \href
  {http://adsabs.harvard.edu/abs/2003A%26A...411L.473A} {411, L473}

\bibitem[\protect\citeauthoryear{{Allard}, {Spiegelman}  \&
  {Kielkopf}}{{Allard} et~al.}{2007}]{Allard2007}
{Allard} N.~F.,  {Spiegelman} F.,   {Kielkopf} J.~F.,  2007, \mn@doi [\aap]
  {10.1051/0004-6361:20066616}, \href
  {http://adsabs.harvard.edu/abs/2007A%26A...465.1085A} {465, 1085}

\bibitem[\protect\citeauthoryear{{Anderson} et~al.,}{{Anderson}
  et~al.}{2010}]{Anderson:2010aa}
{Anderson} D.~R.,  et~al., 2010, \mn@doi [\apj] {10.1088/0004-637X/709/1/159},
  \href {http://adsabs.harvard.edu/abs/2010ApJ...709..159A} {709, 159}

\bibitem[\protect\citeauthoryear{{Anderson} et~al.,}{{Anderson}
  et~al.}{2011}]{Anderson:2011aa}
{Anderson} D.~R.,  et~al., 2011, \mn@doi [\aap] {10.1051/0004-6361/201016208},
  \href {http://adsabs.harvard.edu/abs/2011A%26A...531A..60A} {531, A60}

\bibitem[\protect\citeauthoryear{{Anderson} et~al.,}{{Anderson}
  et~al.}{2014a}]{Anderson:2014ab}
{Anderson} D.~R.,  et~al., 2014a, preprint, \href
  {http://adsabs.harvard.edu/abs/2014arXiv1410.3449A} {} (\mn@eprint {arXiv}
  {1410.3449})

\bibitem[\protect\citeauthoryear{{Anderson} et~al.,}{{Anderson}
  et~al.}{2014b}]{Anderson:2014ac}
{Anderson} D.~R.,  et~al., 2014b, \mn@doi [\mnras] {10.1093/mnras/stu1737},
  \href {http://adsabs.harvard.edu/abs/2014MNRAS.445.1114A} {445, 1114}

\bibitem[\protect\citeauthoryear{{Anderson} et~al.,}{{Anderson}
  et~al.}{2015}]{Anderson:2015aa}
{Anderson} D.~R.,  et~al., 2015, \mn@doi [\aap] {10.1051/0004-6361/201423591},
  \href {http://adsabs.harvard.edu/abs/2015A%26A...575A..61A} {575, A61}

\bibitem[\protect\citeauthoryear{{Bakos} et~al.,}{{Bakos}
  et~al.}{2007}]{Bakos:2007aa}
{Bakos} G.~{\'A}.,  et~al., 2007, \mn@doi [\apj] {10.1086/509874}, \href
  {http://adsabs.harvard.edu/abs/2007ApJ...656..552B} {656, 552}

\bibitem[\protect\citeauthoryear{{Bakos} et~al.,}{{Bakos}
  et~al.}{2009}]{Bakos:2009aa}
{Bakos} G.~{\'A}.,  et~al., 2009, \mn@doi [\apj] {10.1088/0004-637X/707/1/446},
  \href {http://adsabs.harvard.edu/abs/2009ApJ...707..446B} {707, 446}

\bibitem[\protect\citeauthoryear{{Bakos} et~al.,}{{Bakos}
  et~al.}{2016}]{Bakos:2016aa}
{Bakos} G.~{\'A}.,  et~al., 2016, preprint, \href
  {http://adsabs.harvard.edu/abs/2016arXiv160604556B} {} (\mn@eprint {arXiv}
  {1606.04556})

\bibitem[\protect\citeauthoryear{{Barber}, {Tennyson}, {Harris}  \&
  {Tolchenov}}{{Barber} et~al.}{2006}]{Barber2006}
{Barber} R.~J.,  {Tennyson} J.,  {Harris} G.~J.,   {Tolchenov} R.~N.,  2006,
  \mn@doi [\mnras] {10.1111/j.1365-2966.2006.10184.x}, \href
  {http://adsabs.harvard.edu/abs/2006MNRAS.368.1087B} {368, 1087}

\bibitem[\protect\citeauthoryear{{Barber}, {Strange}, {Hill}, {Polyansky},
  {Mellau}, {Yurchenko}  \& {Tennyson}}{{Barber} et~al.}{2014}]{Barber2014}
{Barber} R.~J.,  {Strange} J.~K.,  {Hill} C.,  {Polyansky} O.~L.,  {Mellau}
  G.~C.,  {Yurchenko} S.~N.,   {Tennyson} J.,  2014, \mn@doi [\mnras]
  {10.1093/mnras/stt2011}, \href
  {http://adsabs.harvard.edu/abs/2014MNRAS.437.1828B} {437, 1828}

\bibitem[\protect\citeauthoryear{{Barros} et~al.,}{{Barros}
  et~al.}{2016}]{Barros:2016aa}
{Barros} S.~C.~C.,  et~al., 2016, \mn@doi [\aap] {10.1051/0004-6361/201526517},
  \href {http://adsabs.harvard.edu/abs/2016A%26A...593A.113B} {593, A113}

\bibitem[\protect\citeauthoryear{{Beatty} et~al.,}{{Beatty}
  et~al.}{2017}]{Beatty2017}
{Beatty} T.~G.,  et~al., 2017, \mn@doi [\aj] {10.3847/1538-3881/aa7511}, \href
  {https://ui.adsabs.harvard.edu/abs/2017AJ....154...25B} {154, 25}

\bibitem[\protect\citeauthoryear{{BelBruno}, {Gelfand}, {Radigan}  \&
  {Verges}}{{BelBruno} et~al.}{1982}]{Belbruno1982}
{BelBruno} J.~J.,  {Gelfand} J.,  {Radigan} W.,   {Verges} K.,  1982, \mn@doi
  [Journal of Molecular Spectroscopy] {10.1016/0022-2852(82)90009-1}, \href
  {http://adsabs.harvard.edu/abs/1982JMoSp..94..336B} {94, 336}

\bibitem[\protect\citeauthoryear{{Bell} \& {Berrington}}{{Bell} \&
  {Berrington}}{1987}]{Bell1987}
{Bell} K.~L.,  {Berrington} K.~A.,  1987, \mn@doi [Journal of Physics B Atomic
  Molecular Physics] {10.1088/0022-3700/20/4/019}, \href
  {https://ui.adsabs.harvard.edu/#abs/1987JPhB...20..801B} {20, 801}

\bibitem[\protect\citeauthoryear{{Bhatti} et~al.,}{{Bhatti}
  et~al.}{2016}]{Bhatti:2016aa}
{Bhatti} W.,  et~al., 2016, preprint, \href
  {http://adsabs.harvard.edu/abs/2016arXiv160700322B} {} (\mn@eprint {arXiv}
  {1607.00322})

\bibitem[\protect\citeauthoryear{{Bieryla} et~al.,}{{Bieryla}
  et~al.}{2015}]{Bieryla:2015aa}
{Bieryla} A.,  et~al., 2015, \mn@doi [\aj] {10.1088/0004-6256/150/1/12}, \href
  {https://ui.adsabs.harvard.edu/abs/2015AJ....150...12B} {150, 12}

\bibitem[\protect\citeauthoryear{{Bouanich}, {Salem}, {Aroui}, {Walrand}  \&
  {Blanquet}}{{Bouanich} et~al.}{2004}]{Bouanich2004}
{Bouanich} J.-P.,  {Salem} J.,  {Aroui} H.,  {Walrand} J.,   {Blanquet} G.,
  2004, \mn@doi [\jqsrt] {10.1016/S0022-4073(03)00143-2}, \href
  {http://adsabs.harvard.edu/abs/2004JQSRT..84..195B} {84, 195}

\bibitem[\protect\citeauthoryear{{Bouchy} et~al.,}{{Bouchy}
  et~al.}{2010}]{Bouchy:2010aa}
{Bouchy} F.,  et~al., 2010, \mn@doi [\aap] {10.1051/0004-6361/201014817}, \href
  {http://adsabs.harvard.edu/abs/2010A%26A...519A..98B} {519, A98}

\bibitem[\protect\citeauthoryear{{Brown} et~al.,}{{Brown}
  et~al.}{2017}]{Brown:2017}
{Brown} D.~J.~A.,  et~al., 2017, \mn@doi [\mnras] {10.1093/mnras/stw2316},
  \href {https://ui.adsabs.harvard.edu/abs/2017MNRAS.464..810B} {464, 810}

\bibitem[\protect\citeauthoryear{{Burke} et~al.,}{{Burke}
  et~al.}{2007}]{Burke:2007aa}
{Burke} C.~J.,  et~al., 2007, \mn@doi [\apj] {10.1086/523087}, \href
  {http://adsabs.harvard.edu/abs/2007ApJ...671.2115B} {671, 2115}

\bibitem[\protect\citeauthoryear{{Burrows}, {Marley}  \& {Sharp}}{{Burrows}
  et~al.}{2000}]{Burrows2000}
{Burrows} A.,  {Marley} M.~S.,   {Sharp} C.~M.,  2000, \mn@doi [\apj]
  {10.1086/308462}, \href
  {https://ui.adsabs.harvard.edu/#abs/2000ApJ...531..438B} {531, 438}

\bibitem[\protect\citeauthoryear{{Carter}, {Winn}, {Gilliland}  \&
  {Holman}}{{Carter} et~al.}{2009}]{Carter:2009aa}
{Carter} J.~A.,  {Winn} J.~N.,  {Gilliland} R.,   {Holman} M.~J.,  2009,
  \mn@doi [\apj] {10.1088/0004-637X/696/1/241}, \href
  {http://adsabs.harvard.edu/abs/2009ApJ...696..241C} {696, 241}

\bibitem[\protect\citeauthoryear{{Ciceri} et~al.,}{{Ciceri}
  et~al.}{2013}]{Ciceri:2013aa}
{Ciceri} S.,  et~al., 2013, \mn@doi [\aap] {10.1051/0004-6361/201321669}, \href
  {http://adsabs.harvard.edu/abs/2013A%26A...557A..30C} {557, A30}

\bibitem[\protect\citeauthoryear{{Collier Cameron} et~al.,}{{Collier Cameron}
  et~al.}{2007}]{Collier-Cameron:2007aa}
{Collier Cameron} A.,  et~al., 2007, \mn@doi [\mnras]
  {10.1111/j.1365-2966.2006.11350.x}, \href
  {http://adsabs.harvard.edu/abs/2007MNRAS.375..951C} {375, 951}

\bibitem[\protect\citeauthoryear{{Collier Cameron} et~al.,}{{Collier Cameron}
  et~al.}{2010}]{Collier-Cameron:2010aa}
{Collier Cameron} A.,  et~al., 2010, \mn@doi [\mnras]
  {10.1111/j.1365-2966.2010.16922.x}, \href
  {http://adsabs.harvard.edu/abs/2010MNRAS.407..507C} {407, 507}

\bibitem[\protect\citeauthoryear{{Collins} et~al.,}{{Collins}
  et~al.}{2014}]{Collins:2014}
{Collins} K.~A.,  et~al., 2014, \mn@doi [\aj] {10.1088/0004-6256/147/2/39},
  \href {https://ui.adsabs.harvard.edu/abs/2014AJ....147...39C} {147, 39}

\bibitem[\protect\citeauthoryear{{Collins}, {Kielkopf}  \& {Stassun}}{{Collins}
  et~al.}{2017}]{Collins:2015aa}
{Collins} K.~A.,  {Kielkopf} J.~F.,   {Stassun} K.~G.,  2017, \mn@doi [\aj]
  {10.3847/1538-3881/153/2/78}, \href
  {https://ui.adsabs.harvard.edu/abs/2017AJ....153...78C} {153, 78}

\bibitem[\protect\citeauthoryear{{Damasso} et~al.,}{{Damasso}
  et~al.}{2015}]{Damasso:2015aa}
{Damasso} M.,  et~al., 2015, \mn@doi [\aap] {10.1051/0004-6361/201425332},
  \href {http://adsabs.harvard.edu/abs/2015A%26A...575A.111D} {575, A111}

\bibitem[\protect\citeauthoryear{{Delrez} et~al.,}{{Delrez}
  et~al.}{2014}]{Delrez:2014aa}
{Delrez} L.,  et~al., 2014, \mn@doi [\aap] {10.1051/0004-6361/201323204}, \href
  {http://adsabs.harvard.edu/abs/2014A%26A...563A.143D} {563, A143}

\bibitem[\protect\citeauthoryear{{Delrez} et~al.,}{{Delrez}
  et~al.}{2016}]{Delrez:2016aa}
{Delrez} L.,  et~al., 2016, \mn@doi [\mnras] {10.1093/mnras/stw522}, \href
  {http://adsabs.harvard.edu/abs/2016MNRAS.458.4025D} {458, 4025}

\bibitem[\protect\citeauthoryear{{Demory} et~al.,}{{Demory}
  et~al.}{2016}]{Demory:2016aa}
{Demory} B.-O.,  et~al., 2016, \mn@doi [\nat] {10.1038/nature17169}, \href
  {http://adsabs.harvard.edu/abs/2016Natur.532..207D} {532, 207}

\bibitem[\protect\citeauthoryear{{Eastman} et~al.,}{{Eastman}
  et~al.}{2016}]{Eastman:2016aa}
{Eastman} J.~D.,  et~al., 2016, \mn@doi [\aj] {10.3847/0004-6256/151/2/45},
  \href {http://adsabs.harvard.edu/abs/2016AJ....151...45E} {151, 45}

\bibitem[\protect\citeauthoryear{{Enoch} et~al.,}{{Enoch}
  et~al.}{2011a}]{Enoch:2011ab}
{Enoch} B.,  et~al., 2011a, \mn@doi [\aj] {10.1088/0004-6256/142/3/86}, \href
  {http://adsabs.harvard.edu/abs/2011AJ....142...86E} {142, 86}

\bibitem[\protect\citeauthoryear{{Enoch} et~al.,}{{Enoch}
  et~al.}{2011b}]{Enoch:2011aa}
{Enoch} B.,  et~al., 2011b, \mn@doi [\mnras]
  {10.1111/j.1365-2966.2010.17550.x}, \href
  {http://adsabs.harvard.edu/abs/2011MNRAS.410.1631E} {410, 1631}

\bibitem[\protect\citeauthoryear{{Evans}, {Southworth}  \& {Smalley}}{{Evans}
  et~al.}{2016}]{Evans:2016aa}
{Evans} D.~F.,  {Southworth} J.,   {Smalley} B.,  2016, \mn@doi [\apjl]
  {10.3847/2041-8213/833/2/L19}, \href
  {http://adsabs.harvard.edu/abs/2016ApJ...833L..19E} {833, L19}

\bibitem[\protect\citeauthoryear{{Evans} et~al.,}{{Evans}
  et~al.}{2017}]{Evans2017}
{Evans} T.~M.,  et~al., 2017, \mn@doi [\nat] {10.1038/nature23266}, \href
  {http://adsabs.harvard.edu/abs/2017Natur.548...58E} {548, 58}

\bibitem[\protect\citeauthoryear{{Evans} et~al.,}{{Evans}
  et~al.}{2018}]{Evans2018}
{Evans} T.~M.,  et~al., 2018, \mn@doi [\aj] {10.3847/1538-3881/aaebff}, \href
  {https://ui.adsabs.harvard.edu/\#abs/2018AJ....156..283E} {156, 283}

\bibitem[\protect\citeauthoryear{{Faedi} et~al.,}{{Faedi}
  et~al.}{2013}]{Faedi:2013aa}
{Faedi} F.,  et~al., 2013, \mn@doi [\aap] {10.1051/0004-6361/201220520}, \href
  {http://adsabs.harvard.edu/abs/2013A%26A...551A..73F} {551, A73}

\bibitem[\protect\citeauthoryear{{Fortney} et~al.,}{{Fortney}
  et~al.}{2011}]{Fortney:2011aa}
{Fortney} J.~J.,  et~al., 2011, \mn@doi [\apjs] {10.1088/0067-0049/197/1/9},
  \href {http://adsabs.harvard.edu/abs/2011ApJS..197....9F} {197, 9}

\bibitem[\protect\citeauthoryear{{Fulton} et~al.,}{{Fulton}
  et~al.}{2015}]{Fulton:2015aa}
{Fulton} B.~J.,  et~al., 2015, \mn@doi [\apj] {10.1088/0004-637X/810/1/30},
  \href {http://adsabs.harvard.edu/abs/2015ApJ...810...30F} {810, 30}

\bibitem[\protect\citeauthoryear{{Gamache}, {Lynch}  \& {Brown}}{{Gamache}
  et~al.}{1996}]{Gamache1996}
{Gamache} R.~R.,  {Lynch} R.,   {Brown} L.~R.,  1996, \mn@doi [\jqsrt]
  {10.1016/0022-4073(96)00098-2}, \href
  {http://adsabs.harvard.edu/abs/1996JQSRT..56..471G} {56, 471}

\bibitem[\protect\citeauthoryear{{Gaudi} et~al.,}{{Gaudi}
  et~al.}{2017}]{Gaudi:2017}
{Gaudi} B.~S.,  et~al., 2017, \mn@doi [\nat] {10.1038/nature22392}, \href
  {https://ui.adsabs.harvard.edu/abs/2017Natur.546..514G} {546, 514}

\bibitem[\protect\citeauthoryear{{Gillon} et~al.,}{{Gillon}
  et~al.}{2012}]{Gillon:2012aa}
{Gillon} M.,  et~al., 2012, \mn@doi [\aap] {10.1051/0004-6361/201218817}, \href
  {http://adsabs.harvard.edu/abs/2012A%26A...542A...4G} {542, A4}

\bibitem[\protect\citeauthoryear{{Gillon} et~al.,}{{Gillon}
  et~al.}{2014}]{Gillon:2014aa}
{Gillon} M.,  et~al., 2014, \mn@doi [\aap] {10.1051/0004-6361/201323014}, \href
  {http://adsabs.harvard.edu/abs/2014A%26A...562L...3G} {562, L3}

\bibitem[\protect\citeauthoryear{{Goyal} et~al.,}{{Goyal}
  et~al.}{2019}]{Errgoyal2019}
{Goyal} J.~M.,  et~al., 2019, \mn@doi [\mnras] {10.1093/mnras/stz755}, \href
  {https://ui.adsabs.harvard.edu/abs/2019MNRAS.tmp..722G} {p.~722}

\bibitem[\protect\citeauthoryear{{Hadded}, {Aroui}, {Orphal}, {Bouanich}  \&
  {Hartmann}}{{Hadded} et~al.}{2001}]{Hadded2001}
{Hadded} S.,  {Aroui} H.,  {Orphal} J.,  {Bouanich} J.-P.,   {Hartmann} J.-M.,
  2001, \mn@doi [Journal of Molecular Spectroscopy] {10.1006/jmsp.2001.8452},
  \href {http://adsabs.harvard.edu/abs/2001JMoSp.210..275H} {210, 275}

\bibitem[\protect\citeauthoryear{{Harris}, {Tennyson}, {Kaminsky}, {Pavlenko}
  \& {Jones}}{{Harris} et~al.}{2006}]{Harris2006}
{Harris} G.~J.,  {Tennyson} J.,  {Kaminsky} B.~M.,  {Pavlenko} Y.~V.,   {Jones}
  H.~R.~A.,  2006, \mn@doi [\mnras] {10.1111/j.1365-2966.2005.09960.x}, \href
  {http://adsabs.harvard.edu/abs/2006MNRAS.367..400H} {367, 400}

\bibitem[\protect\citeauthoryear{{Hartman} et~al.,}{{Hartman}
  et~al.}{2011}]{Hartman:2011ac}
{Hartman} J.~D.,  et~al., 2011, \mn@doi [\apj] {10.1088/0004-637X/742/1/59},
  \href {http://adsabs.harvard.edu/abs/2011ApJ...742...59H} {742, 59}

\bibitem[\protect\citeauthoryear{{Hartman} et~al.,}{{Hartman}
  et~al.}{2012}]{Hartman:2012aa}
{Hartman} J.~D.,  et~al., 2012, \mn@doi [\aj] {10.1088/0004-6256/144/5/139},
  \href {http://adsabs.harvard.edu/abs/2012AJ....144..139H} {144, 139}

\bibitem[\protect\citeauthoryear{{Hartman} et~al.,}{{Hartman}
  et~al.}{2014}]{Hartman:2014aa}
{Hartman} J.~D.,  et~al., 2014, \mn@doi [\aj] {10.1088/0004-6256/147/6/128},
  \href {http://adsabs.harvard.edu/abs/2014AJ....147..128H} {147, 128}

\bibitem[\protect\citeauthoryear{{Hartman} et~al.,}{{Hartman}
  et~al.}{2016}]{Hartman:2016aa}
{Hartman} J.~D.,  et~al., 2016, \mn@doi [\aj] {10.3847/0004-6256/152/6/182},
  \href {http://adsabs.harvard.edu/abs/2016AJ....152..182H} {152, 182}

\bibitem[\protect\citeauthoryear{{Hay} et~al.,}{{Hay}
  et~al.}{2016}]{Hay:2016aa}
{Hay} K.~L.,  et~al., 2016, \mn@doi [\mnras] {10.1093/mnras/stw2090}, \href
  {http://adsabs.harvard.edu/abs/2016MNRAS.463.3276H} {463, 3276}

\bibitem[\protect\citeauthoryear{{Hebb} et~al.,}{{Hebb}
  et~al.}{2009}]{Hebb:2009aa}
{Hebb} L.,  et~al., 2009, \mn@doi [\apj] {10.1088/0004-637X/693/2/1920}, \href
  {http://adsabs.harvard.edu/abs/2009ApJ...693.1920H} {693, 1920}

\bibitem[\protect\citeauthoryear{{Hebb} et~al.,}{{Hebb}
  et~al.}{2010}]{Hebb:2010aa}
{Hebb} L.,  et~al., 2010, \mn@doi [\apj] {10.1088/0004-637X/708/1/224}, \href
  {http://adsabs.harvard.edu/abs/2010ApJ...708..224H} {708, 224}

\bibitem[\protect\citeauthoryear{{H{\'e}brard} et~al.,}{{H{\'e}brard}
  et~al.}{2013}]{Hebrard:2013aa}
{H{\'e}brard} G.,  et~al., 2013, \mn@doi [\aap] {10.1051/0004-6361/201220363},
  \href {http://adsabs.harvard.edu/abs/2013A%26A...549A.134H} {549, A134}

\bibitem[\protect\citeauthoryear{{Heiter} et~al.,}{{Heiter}
  et~al.}{2008}]{Heiter2008}
{Heiter} U.,  et~al., 2008, in Journal of Physics Conference Series. p. 012011,
  \mn@doi{10.1088/1742-6596/130/1/012011}

\bibitem[\protect\citeauthoryear{{Heiter} et~al.,}{{Heiter}
  et~al.}{2015}]{Heiter2015}
{Heiter} U.,  et~al., 2015, \mn@doi [\physscr] {10.1088/0031-8949/90/5/054010},
  \href {https://ui.adsabs.harvard.edu/abs/2015PhyS...90e4010H} {90, 054010}

\bibitem[\protect\citeauthoryear{{Hellier} et~al.,}{{Hellier}
  et~al.}{2009}]{Hellier:2009aa}
{Hellier} C.,  et~al., 2009, \mn@doi [\apjl] {10.1088/0004-637X/690/1/L89},
  \href {http://adsabs.harvard.edu/abs/2009ApJ...690L..89H} {690, L89}

\bibitem[\protect\citeauthoryear{{Hellier} et~al.,}{{Hellier}
  et~al.}{2011}]{Hellier:2011aa}
{Hellier} C.,  et~al., 2011, \mn@doi [\aap] {10.1051/0004-6361/201117081},
  \href {http://adsabs.harvard.edu/abs/2011A%26A...535L...7H} {535, L7}

\bibitem[\protect\citeauthoryear{{Hellier} et~al.,}{{Hellier}
  et~al.}{2012}]{Hellier:2012aa}
{Hellier} C.,  et~al., 2012, \mn@doi [\mnras]
  {10.1111/j.1365-2966.2012.21780.x}, \href
  {http://adsabs.harvard.edu/abs/2012MNRAS.426..739H} {426, 739}

\bibitem[\protect\citeauthoryear{{Hellier} et~al.,}{{Hellier}
  et~al.}{2014}]{Hellier:2014aa}
{Hellier} C.,  et~al., 2014, \mn@doi [\mnras] {10.1093/mnras/stu410}, \href
  {http://adsabs.harvard.edu/abs/2014MNRAS.440.1982H} {440, 1982}

\bibitem[\protect\citeauthoryear{{Hellier} et~al.,}{{Hellier}
  et~al.}{2015}]{Hellier:2015aa}
{Hellier} C.,  et~al., 2015, \mn@doi [\aj] {10.1088/0004-6256/150/1/18}, \href
  {http://adsabs.harvard.edu/abs/2015AJ....150...18H} {150, 18}

\bibitem[\protect\citeauthoryear{{Hellier} et~al.,}{{Hellier}
  et~al.}{2017}]{Hellier:2017aa}
{Hellier} C.,  et~al., 2017, \mn@doi [\mnras] {10.1093/mnras/stw3005}, \href
  {http://adsabs.harvard.edu/abs/2017MNRAS.465.3693H} {465, 3693}

\bibitem[\protect\citeauthoryear{{Henry}, {Marcy}, {Butler}  \& {Vogt}}{{Henry}
  et~al.}{2000}]{Henry:2000aa}
{Henry} G.~W.,  {Marcy} G.~W.,  {Butler} R.~P.,   {Vogt} S.~S.,  2000, \mn@doi
  [\apjl] {10.1086/312458}, \href
  {http://adsabs.harvard.edu/abs/2000ApJ...529L..41H} {529, L41}

\bibitem[\protect\citeauthoryear{{John}}{{John}}{1988}]{John1988}
{John} T.~L.,  1988, \aap, \href
  {http://adsabs.harvard.edu/abs/1988A%26A...193..189J} {193, 189}

\bibitem[\protect\citeauthoryear{{Johnson} et~al.,}{{Johnson}
  et~al.}{2011}]{Johnson:2011aa}
{Johnson} J.~A.,  et~al., 2011, \mn@doi [\apj] {10.1088/0004-637X/735/1/24},
  \href {http://adsabs.harvard.edu/abs/2011ApJ...735...24J} {735, 24}

\bibitem[\protect\citeauthoryear{{Kov{\'a}cs} et~al.,}{{Kov{\'a}cs}
  et~al.}{2007}]{Kovacs:2007aa}
{Kov{\'a}cs} G.,  et~al., 2007, \mn@doi [\apjl] {10.1086/524058}, \href
  {http://adsabs.harvard.edu/abs/2007ApJ...670L..41K} {670, L41}

\bibitem[\protect\citeauthoryear{{Kuhn} et~al.,}{{Kuhn}
  et~al.}{2016}]{Kuhn:2016aa}
{Kuhn} R.~B.,  et~al., 2016, \mn@doi [\mnras] {10.1093/mnras/stw880}, \href
  {http://adsabs.harvard.edu/abs/2016MNRAS.459.4281K} {459, 4281}

\bibitem[\protect\citeauthoryear{{Lam} et~al.,}{{Lam}
  et~al.}{2017}]{Lam:2017aa}
{Lam} K.~W.~F.,  et~al., 2017, \mn@doi [\aap] {10.1051/0004-6361/201629403},
  \href {http://adsabs.harvard.edu/abs/2017A%26A...599A...3L} {599, A3}

\bibitem[\protect\citeauthoryear{{Landrain}, {Blanquet}, {Lep{\`e}re},
  {Walrand}  \& {Bouanich}}{{Landrain} et~al.}{1997}]{Landrain1997}
{Landrain} V.,  {Blanquet} G.,  {Lep{\`e}re} M.,  {Walrand} J.,   {Bouanich}
  J.-P.,  1997, \mn@doi [Journal of Molecular Spectroscopy]
  {10.1006/jmsp.1996.7223}, \href
  {http://adsabs.harvard.edu/abs/1997JMoSp.182..184L} {182, 184}

\bibitem[\protect\citeauthoryear{{Le Moal} \& {Severin}}{{Le Moal} \&
  {Severin}}{1986}]{Lemoal1986}
{Le Moal} M.~F.,  {Severin} F.,  1986, \jqsrt, \href
  {http://adsabs.harvard.edu/abs/1986JQSRT..35..145L} {35, 145}

\bibitem[\protect\citeauthoryear{{Lehmann}, {Guenther}, {Sebastian},
  {D{\"o}llinger}, {Hartmann}  \& {Mkrtichian}}{{Lehmann}
  et~al.}{2015}]{Lehmann:2015aa}
{Lehmann} H.,  {Guenther} E.,  {Sebastian} D.,  {D{\"o}llinger} M.,  {Hartmann}
  M.,   {Mkrtichian} D.~E.,  2015, \mn@doi [\aap]
  {10.1051/0004-6361/201526176}, \href
  {http://adsabs.harvard.edu/abs/2015A%26A...578L...4L} {578, L4}

\bibitem[\protect\citeauthoryear{{Lendl} et~al.,}{{Lendl}
  et~al.}{2012}]{Lendl:2012aa}
{Lendl} M.,  et~al., 2012, \mn@doi [\aap] {10.1051/0004-6361/201219585}, \href
  {http://adsabs.harvard.edu/abs/2012A%26A...544A..72L} {544, A72}

\bibitem[\protect\citeauthoryear{{Lendl} et~al.,}{{Lendl}
  et~al.}{2016}]{Lendl:2016aa}
{Lendl} M.,  et~al., 2016, \mn@doi [\aap] {10.1051/0004-6361/201527594}, \href
  {http://adsabs.harvard.edu/abs/2016A%26A...587A..67L} {587, A67}

\bibitem[\protect\citeauthoryear{{Levy}, {Lacome}  \& {Tarrago}}{{Levy}
  et~al.}{1994}]{Levy1994}
{Levy} A.,  {Lacome} N.,   {Tarrago} G.,  1994, \mn@doi [Journal of Molecular
  Spectroscopy] {10.1006/jmsp.1994.1168}, \href
  {http://adsabs.harvard.edu/abs/1994JMoSp.166...20L} {166, 20}

\bibitem[\protect\citeauthoryear{{Lister} et~al.,}{{Lister}
  et~al.}{2009}]{Lister:2009aa}
{Lister} T.~A.,  et~al., 2009, \mn@doi [\apj] {10.1088/0004-637X/703/1/752},
  \href {http://adsabs.harvard.edu/abs/2009ApJ...703..752L} {703, 752}

\bibitem[\protect\citeauthoryear{{Maciejewski} et~al.,}{{Maciejewski}
  et~al.}{2014}]{Maciejewski:2014aa}
{Maciejewski} G.,  et~al., 2014, \actaa, \href
  {http://adsabs.harvard.edu/abs/2014AcA....64...27M} {64, 11}

\bibitem[\protect\citeauthoryear{{Mancini} et~al.,}{{Mancini}
  et~al.}{2013}]{Mancini:2013aa}
{Mancini} L.,  et~al., 2013, \mn@doi [\mnras] {10.1093/mnras/stt1394}, \href
  {http://adsabs.harvard.edu/abs/2013MNRAS.436....2M} {436, 2}

\bibitem[\protect\citeauthoryear{{Mancini} et~al.,}{{Mancini}
  et~al.}{2017}]{Mancini:2017aa}
{Mancini} L.,  et~al., 2017, \mn@doi [\mnras] {10.1093/mnras/stw1987}, \href
  {http://adsabs.harvard.edu/abs/2017MNRAS.465..843M} {465, 843}

\bibitem[\protect\citeauthoryear{{Mandushev} et~al.,}{{Mandushev}
  et~al.}{2007}]{Mandushev:2007aa}
{Mandushev} G.,  et~al., 2007, \mn@doi [\apjl] {10.1086/522115}, \href
  {http://adsabs.harvard.edu/abs/2007ApJ...667L.195M} {667, L195}

\bibitem[\protect\citeauthoryear{{Mantz}, {Malathy Devi}, {Chris Benner},
  {Smith}, {Predoi-Cross}  \& {Dulick}}{{Mantz} et~al.}{2005}]{Mantz2005}
{Mantz} A.~W.,  {Malathy Devi} V.,  {Chris Benner} D.,  {Smith} M.~A.~H.,
  {Predoi-Cross} A.,   {Dulick} M.,  2005, \mn@doi [Journal of Molecular
  Structure] {10.1016/j.molstruc.2004.11.094}, \href
  {http://adsabs.harvard.edu/abs/2005JMoSt.742...99M} {742, 99}

\bibitem[\protect\citeauthoryear{{Margolis}}{{Margolis}}{1993}]{Margolis1993}
{Margolis} J.~S.,  1993, \mn@doi [\jqsrt] {10.1016/0022-4073(93)90073-Q}, \href
  {http://adsabs.harvard.edu/abs/1993JQSRT..50..431M} {50, 431}

\bibitem[\protect\citeauthoryear{{Maxted} et~al.,}{{Maxted}
  et~al.}{2011}]{Maxted:2011aa}
{Maxted} P.~F.~L.,  et~al., 2011, \mn@doi [\pasp] {10.1086/660007}, \href
  {http://adsabs.harvard.edu/abs/2011PASP..123..547M} {123, 547}

\bibitem[\protect\citeauthoryear{{Maxted} et~al.,}{{Maxted}
  et~al.}{2016}]{Maxted:2016aa}
{Maxted} P.~F.~L.,  et~al., 2016, \mn@doi [\aap] {10.1051/0004-6361/201628250},
  \href {http://adsabs.harvard.edu/abs/2016A%26A...591A..55M} {591, A55}

\bibitem[\protect\citeauthoryear{{McCullough} et~al.,}{{McCullough}
  et~al.}{2006}]{McCullough:2006aa}
{McCullough} P.~R.,  et~al., 2006, \mn@doi [\apj] {10.1086/505651}, \href
  {http://adsabs.harvard.edu/abs/2006ApJ...648.1228M} {648, 1228}

\bibitem[\protect\citeauthoryear{{McKemmish}, {Yurchenko}  \&
  {Tennyson}}{{McKemmish} et~al.}{2016}]{McKemmish2016}
{McKemmish} L.~K.,  {Yurchenko} S.~N.,   {Tennyson} J.,  2016, \mn@doi [\mnras]
  {10.1093/mnras/stw1969}, \href
  {http://adsabs.harvard.edu/abs/2016MNRAS.463..771M} {463, 771}

\bibitem[\protect\citeauthoryear{{McLeod} et~al.,}{{McLeod}
  et~al.}{2017}]{McLeod2017}
{McLeod} K.~K.,  et~al., 2017, \mn@doi [\aj] {10.3847/1538-3881/aa6d5d}, \href
  {https://ui.adsabs.harvard.edu/abs/2017AJ....153..263M} {153, 263}

\bibitem[\protect\citeauthoryear{{Mo{\v{c}}nik}, {Hellier}, {Anderson}, {Clark}
   \& {Southworth}}{{Mo{\v{c}}nik} et~al.}{2017}]{wasp118}
{Mo{\v{c}}nik} T.,  {Hellier} C.,  {Anderson} D.~R.,  {Clark} B.~J.~M.,
  {Southworth} J.,  2017, \mn@doi [\mnras] {10.1093/mnras/stx972}, \href
  {https://ui.adsabs.harvard.edu/abs/2017MNRAS.469.1622M} {469, 1622}

\bibitem[\protect\citeauthoryear{{Neveu-VanMalle} et~al.,}{{Neveu-VanMalle}
  et~al.}{2014}]{Neveu-VanMalle:2014aa}
{Neveu-VanMalle} M.,  et~al., 2014, \mn@doi [\aap]
  {10.1051/0004-6361/201424744}, \href
  {http://adsabs.harvard.edu/abs/2014A%26A...572A..49N} {572, A49}

\bibitem[\protect\citeauthoryear{{Nikolov} et~al.,}{{Nikolov}
  et~al.}{2014}]{Nikolov:2014aa}
{Nikolov} N.,  et~al., 2014, \mn@doi [\mnras] {10.1093/mnras/stt1859}, \href
  {http://adsabs.harvard.edu/abs/2014MNRAS.437...46N} {437, 46}

\bibitem[\protect\citeauthoryear{{Nouri}, {Orphal}, {Aroui}  \&
  {Hartmann}}{{Nouri} et~al.}{2004}]{Nouri2004}
{Nouri} S.,  {Orphal} J.,  {Aroui} H.,   {Hartmann} J.-M.,  2004, \mn@doi
  [Journal of Molecular Spectroscopy] {10.1016/j.jms.2004.05.009}, \href
  {http://adsabs.harvard.edu/abs/2004JMoSp.227...60N} {227, 60}

\bibitem[\protect\citeauthoryear{{Noyes} et~al.,}{{Noyes}
  et~al.}{2008}]{Noyes:2008aa}
{Noyes} R.~W.,  et~al., 2008, \mn@doi [\apjl] {10.1086/527358}, \href
  {http://adsabs.harvard.edu/abs/2008ApJ...673L..79N} {673, L79}

\bibitem[\protect\citeauthoryear{{Padmanabhan}, {Tzanetakis}, {Chanda}  \&
  {Thomson}}{{Padmanabhan} et~al.}{2014}]{Padmanabhan2014}
{Padmanabhan} A.,  {Tzanetakis} T.,  {Chanda} A.,   {Thomson} M.~J.,  2014,
  \mn@doi [\jqsrt] {10.1016/j.jqsrt.2013.07.016}, \href
  {http://adsabs.harvard.edu/abs/2014JQSRT.133...81P} {133, 81}

\bibitem[\protect\citeauthoryear{{Pepper} et~al.,}{{Pepper}
  et~al.}{2017}]{Pepper:2017aa}
{Pepper} J.,  et~al., 2017, \mn@doi [\aj] {10.3847/1538-3881/aa6572}, \href
  {http://adsabs.harvard.edu/abs/2017AJ....153..215P} {153, 215}

\bibitem[\protect\citeauthoryear{{Pine}}{{Pine}}{1992}]{Pine1992}
{Pine} A.~S.,  1992, \mn@doi [\jcp] {10.1063/1.463943}, \href
  {http://adsabs.harvard.edu/abs/1992JChPh..97..773P} {97, 773}

\bibitem[\protect\citeauthoryear{{Pine}, {Markov}, {Buffa}  \&
  {Tarrini}}{{Pine} et~al.}{1993}]{Pine1993}
{Pine} A.~S.,  {Markov} V.~N.,  {Buffa} G.,   {Tarrini} O.,  1993, \mn@doi
  [\jqsrt] {10.1016/0022-4073(93)90069-T}, \href
  {http://adsabs.harvard.edu/abs/1993JQSRT..50..337P} {50, 337}

\bibitem[\protect\citeauthoryear{{Plez}}{{Plez}}{1998}]{Plez1998}
{Plez} B.,  1998, \aap, \href
  {http://adsabs.harvard.edu/abs/1998A%26A...337..495P} {337, 495}

\bibitem[\protect\citeauthoryear{{Plez}}{{Plez}}{1999}]{Plez1999}
{Plez} B.,  1999, in {Le Bertre} T.,  {Lebre} A.,   {Waelkens} C.,  eds,  IAU
  Symposium Vol. 191, Asymptotic Giant Branch Stars. p.~75

\bibitem[\protect\citeauthoryear{{Quinn} et~al.,}{{Quinn}
  et~al.}{2012}]{Quinn:2012aa}
{Quinn} S.~N.,  et~al., 2012, \mn@doi [\apj] {10.1088/0004-637X/745/1/80},
  \href {http://adsabs.harvard.edu/abs/2012ApJ...745...80Q} {745, 80}

\bibitem[\protect\citeauthoryear{{R{\'e}galia-Jarlot}, {Thomas}, {von der
  Heyden}  \& {Barbe}}{{R{\'e}galia-Jarlot} et~al.}{2005}]{Regalia2005}
{R{\'e}galia-Jarlot} L.,  {Thomas} X.,  {von der Heyden} P.,   {Barbe} A.,
  2005, \mn@doi [\jqsrt] {10.1016/j.jqsrt.2004.05.042}, \href
  {http://adsabs.harvard.edu/abs/2005JQSRT..91..121R} {91, 121}

\bibitem[\protect\citeauthoryear{{Richard} et~al.,}{{Richard}
  et~al.}{2012}]{Ciahitranpaper2012}
{Richard} C.,  et~al., 2012, \mn@doi [\jqsrt] {10.1016/j.jqsrt.2011.11.004},
  \href {http://adsabs.harvard.edu/abs/2012JQSRT.113.1276R} {113, 1276}

\bibitem[\protect\citeauthoryear{{Rodriguez} et~al.,}{{Rodriguez}
  et~al.}{2016}]{Rodriguez:2016aa}
{Rodriguez} J.~E.,  et~al., 2016, \mn@doi [\aj] {10.3847/0004-6256/151/6/138},
  \href {http://adsabs.harvard.edu/abs/2016AJ....151..138R} {151, 138}

\bibitem[\protect\citeauthoryear{{Rothman} et~al.,}{{Rothman}
  et~al.}{2009}]{Rothman2009}
{Rothman} L.~S.,  et~al., 2009, \mn@doi [\jqsrt] {10.1016/j.jqsrt.2009.02.013},
  \href {http://adsabs.harvard.edu/abs/2009JQSRT.110..533R} {110, 533}

\bibitem[\protect\citeauthoryear{{Rothman} et~al.,}{{Rothman}
  et~al.}{2010}]{Rothman2010}
{Rothman} L.~S.,  et~al., 2010, \mn@doi [\jqsrt] {10.1016/j.jqsrt.2010.05.001},
  \href {http://adsabs.harvard.edu/abs/2010JQSRT.111.2139R} {111, 2139}

\bibitem[\protect\citeauthoryear{{Rothman} et~al.,}{{Rothman}
  et~al.}{2013}]{Rothman2013}
{Rothman} L.~S.,  et~al., 2013, \mn@doi [\jqsrt] {10.1016/j.jqsrt.2013.07.002},
  \href {http://adsabs.harvard.edu/abs/2013JQSRT.130....4R} {130, 4}

\bibitem[\protect\citeauthoryear{{Ryabchikova}, {Piskunov}, {Kurucz},
  {Stempels}, {Heiter}, {Pakhomov}  \& {Barklem}}{{Ryabchikova}
  et~al.}{2015}]{Ryabchikova2015}
{Ryabchikova} T.,  {Piskunov} N.,  {Kurucz} R.~L.,  {Stempels} H.~C.,  {Heiter}
  U.,  {Pakhomov} Y.,   {Barklem} P.~S.,  2015, \mn@doi [\physscr]
  {10.1088/0031-8949/90/5/054005}, \href
  {https://ui.adsabs.harvard.edu/abs/2015PhyS...90e4005R} {90, 054005}

\bibitem[\protect\citeauthoryear{{Salem}, {Bouanich}, {Walrand}, {Aroui}  \&
  {Blanquet}}{{Salem} et~al.}{2005}]{Salem2005}
{Salem} J.,  {Bouanich} J.-P.,  {Walrand} J.,  {Aroui} H.,   {Blanquet} G.,
  2005, \mn@doi [Journal of Molecular Spectroscopy]
  {10.1016/j.jms.2005.04.014}, \href
  {http://adsabs.harvard.edu/abs/2005JMoSp.232..247S} {232, 247}

\bibitem[\protect\citeauthoryear{{Sato} et~al.,}{{Sato}
  et~al.}{2005}]{Sato:2005aa}
{Sato} B.,  et~al., 2005, \mn@doi [\apj] {10.1086/449306}, \href
  {http://adsabs.harvard.edu/abs/2005ApJ...633..465S} {633, 465}

\bibitem[\protect\citeauthoryear{{Sauval} \& {Tatum}}{{Sauval} \&
  {Tatum}}{1984}]{Sauval1984}
{Sauval} A.~J.,  {Tatum} J.~B.,  1984, \mn@doi [\apjs] {10.1086/190980}, \href
  {http://adsabs.harvard.edu/abs/1984ApJS...56..193S} {56, 193}

\bibitem[\protect\citeauthoryear{{Sharp} \& {Burrows}}{{Sharp} \&
  {Burrows}}{2007}]{Sharp2007}
{Sharp} C.~M.,  {Burrows} A.,  2007, \mn@doi [\apjs] {10.1086/508708}, \href
  {http://adsabs.harvard.edu/abs/2007ApJS..168..140S} {168, 140}

\bibitem[\protect\citeauthoryear{{Sing} et~al.,}{{Sing}
  et~al.}{2016}]{Sing2016}
{Sing} D.~K.,  et~al., 2016, \mn@doi [\nat] {10.1038/nature16068}, \href
  {http://adsabs.harvard.edu/abs/2016Natur.529...59S} {529, 59}

\bibitem[\protect\citeauthoryear{{Skillen} et~al.,}{{Skillen}
  et~al.}{2009}]{Skillen:2009aa}
{Skillen} I.,  et~al., 2009, \mn@doi [\aap] {10.1051/0004-6361/200912018},
  \href {http://adsabs.harvard.edu/abs/2009A%26A...502..391S} {502, 391}

\bibitem[\protect\citeauthoryear{{Smalley} et~al.,}{{Smalley}
  et~al.}{2011}]{Smalley:2011aa}
{Smalley} B.,  et~al., 2011, \mn@doi [\aap] {10.1051/0004-6361/201015992},
  \href {http://adsabs.harvard.edu/abs/2011A%26A...526A.130S} {526, A130}

\bibitem[\protect\citeauthoryear{{Smalley} et~al.,}{{Smalley}
  et~al.}{2012}]{Smalley:2012aa}
{Smalley} B.,  et~al., 2012, \mn@doi [\aap] {10.1051/0004-6361/201219731},
  \href {http://adsabs.harvard.edu/abs/2012A%26A...547A..61S} {547, A61}

\bibitem[\protect\citeauthoryear{{Smith}}{{Smith}}{2015}]{Smith:2015aa}
{Smith} A.~M.~S.,  2015, \actaa, \href
  {http://adsabs.harvard.edu/abs/2015AcA....65..117S} {65}

\bibitem[\protect\citeauthoryear{{Solodov} \& {Starikov}}{{Solodov} \&
  {Starikov}}{2009}]{Solodov2009}
{Solodov} A.~M.,  {Starikov} V.~I.,  2009, \mn@doi [Molecular Physics]
  {10.1080/00268970802698655}, \href
  {http://adsabs.harvard.edu/abs/2009MolPh.107...43S} {107, 43}

\bibitem[\protect\citeauthoryear{{Sousa-Silva}, {Al-Refaie}, {Tennyson}  \&
  {Yurchenko}}{{Sousa-Silva} et~al.}{2014}]{Ph3linelist2014}
{Sousa-Silva} C.,  {Al-Refaie} A.~F.,  {Tennyson} J.,   {Yurchenko} S.~N.,
  2014, VizieR Online Data Catalog, \href
  {http://adsabs.harvard.edu/abs/2014yCat..74462337S} {744}

\bibitem[\protect\citeauthoryear{{Southworth}}{{Southworth}}{2010}]{Southworth:2010aa}
{Southworth} J.,  2010, \mn@doi [\mnras] {10.1111/j.1365-2966.2010.17231.x},
  \href {http://adsabs.harvard.edu/abs/2010MNRAS.408.1689S} {408, 1689}

\bibitem[\protect\citeauthoryear{{Southworth}}{{Southworth}}{2011a}]{Tepcat2011}
{Southworth} J.,  2011a, \mn@doi [\mnras] {10.1111/j.1365-2966.2011.19399.x},
  \href {http://adsabs.harvard.edu/abs/2011MNRAS.417.2166S} {417, 2166}

\bibitem[\protect\citeauthoryear{{Southworth}}{{Southworth}}{2011b}]{Southworth:2011aa}
{Southworth} J.,  2011b, \mn@doi [\mnras] {10.1111/j.1365-2966.2011.19399.x},
  \href {http://adsabs.harvard.edu/abs/2011MNRAS.417.2166S} {417, 2166}

\bibitem[\protect\citeauthoryear{{Southworth}}{{Southworth}}{2012}]{Southworth:2012aa}
{Southworth} J.,  2012, \mn@doi [\mnras] {10.1111/j.1365-2966.2012.21756.x},
  \href {http://adsabs.harvard.edu/abs/2012MNRAS.426.1291S} {426, 1291}

\bibitem[\protect\citeauthoryear{{Southworth} \& {Evans}}{{Southworth} \&
  {Evans}}{2016}]{Southworth:2016ab}
{Southworth} J.,  {Evans} D.~F.,  2016, \mn@doi [\mnras]
  {10.1093/mnras/stw1943}, \href
  {http://adsabs.harvard.edu/abs/2016MNRAS.463...37S} {463, 37}

\bibitem[\protect\citeauthoryear{{Southworth}, {Bruni}, {Mancini}  \&
  {Gregorio}}{{Southworth} et~al.}{2012a}]{Southworth:2012ab}
{Southworth} J.,  {Bruni} I.,  {Mancini} L.,   {Gregorio} J.,  2012a, \mn@doi
  [\mnras] {10.1111/j.1365-2966.2011.20230.x}, \href
  {http://adsabs.harvard.edu/abs/2012MNRAS.420.2580S} {420, 2580}

\bibitem[\protect\citeauthoryear{{Southworth} et~al.,}{{Southworth}
  et~al.}{2012b}]{Southworth:2012ac}
{Southworth} J.,  et~al., 2012b, \mn@doi [\mnras]
  {10.1111/j.1365-2966.2012.21781.x}, \href
  {http://adsabs.harvard.edu/abs/2012MNRAS.426.1338S} {426, 1338}

\bibitem[\protect\citeauthoryear{{Southworth} et~al.,}{{Southworth}
  et~al.}{2013}]{Southworth:2013aa}
{Southworth} J.,  et~al., 2013, \mn@doi [\mnras] {10.1093/mnras/stt1089}, \href
  {http://adsabs.harvard.edu/abs/2013MNRAS.434.1300S} {434, 1300}

\bibitem[\protect\citeauthoryear{{Southworth} et~al.,}{{Southworth}
  et~al.}{2014}]{Southworth:2014aa}
{Southworth} J.,  et~al., 2014, \mn@doi [\mnras] {10.1093/mnras/stu1492}, \href
  {http://adsabs.harvard.edu/abs/2014MNRAS.444..776S} {444, 776}

\bibitem[\protect\citeauthoryear{{Southworth} et~al.,}{{Southworth}
  et~al.}{2016}]{Southworth:2016aa}
{Southworth} J.,  et~al., 2016, \mn@doi [\mnras] {10.1093/mnras/stw279}, \href
  {http://adsabs.harvard.edu/abs/2016MNRAS.457.4205S} {457, 4205}

\bibitem[\protect\citeauthoryear{{Sozzetti} et~al.,}{{Sozzetti}
  et~al.}{2015}]{Sozzetti:2015aa}
{Sozzetti} A.,  et~al., 2015, \mn@doi [\aap] {10.1051/0004-6361/201425570},
  \href {http://adsabs.harvard.edu/abs/2015A%26A...575L..15S} {575, L15}

\bibitem[\protect\citeauthoryear{{Steyert}, {Wang}, {Sirota}, {Donahue}  \&
  {Reuter}}{{Steyert} et~al.}{2004}]{Steyert2004}
{Steyert} D.~W.,  {Wang} W.~F.,  {Sirota} J.~M.,  {Donahue} N.~M.,   {Reuter}
  D.~C.,  2004, \mn@doi [\jqsrt] {10.1016/S0022-4073(02)00300-X}, \href
  {http://adsabs.harvard.edu/abs/2004JQSRT..83..183S} {83, 183}

\bibitem[\protect\citeauthoryear{{Tashkun} \& {Perevalov}}{{Tashkun} \&
  {Perevalov}}{2011}]{Tashkun2011}
{Tashkun} S.~A.,  {Perevalov} V.~I.,  2011, \mn@doi [\jqsrt]
  {10.1016/j.jqsrt.2011.03.005}, \href
  {http://adsabs.harvard.edu/abs/2011JQSRT.112.1403T} {112, 1403}

\bibitem[\protect\citeauthoryear{{Thibault}, {Boissoles}, {Le Doucen},
  {Bouanich}, {Arcas}  \& {Boulet}}{{Thibault} et~al.}{1992}]{Thibault1992}
{Thibault} F.,  {Boissoles} J.,  {Le Doucen} R.,  {Bouanich} J.~P.,  {Arcas}
  P.,   {Boulet} C.,  1992, \mn@doi [\jcp] {10.1063/1.462737}, \href
  {http://adsabs.harvard.edu/abs/1992JChPh..96.4945T} {96, 4945}

\bibitem[\protect\citeauthoryear{{Thibault}, {Calil}, {Boissoles}  \&
  {Launay}}{{Thibault} et~al.}{2000}]{Thibault2000}
{Thibault} F.,  {Calil} B.,  {Boissoles} J.,   {Launay} J.~M.,  2000, \mn@doi
  [Physical Chemistry Chemical Physics (Incorporating Faraday Transactions)]
  {10.1039/B006224N}, \href {http://adsabs.harvard.edu/abs/2000PCCP....2.5404T}
  {2, 5404}

\bibitem[\protect\citeauthoryear{{Tregloan-Reed} et~al.,}{{Tregloan-Reed}
  et~al.}{2018}]{Tregloan-Reed2018}
{Tregloan-Reed} J.,  et~al., 2018, \mn@doi [\mnras] {10.1093/mnras/stx3147},
  \href {https://ui.adsabs.harvard.edu/abs/2018MNRAS.474.5485T} {474, 5485}

\bibitem[\protect\citeauthoryear{{Triaud} et~al.,}{{Triaud}
  et~al.}{2017}]{Triaud:2017}
{Triaud} A. H.~M.~J.,  et~al., 2017, \mn@doi [\mnras] {10.1093/mnras/stx154},
  \href {https://ui.adsabs.harvard.edu/abs/2017MNRAS.467.1714T} {467, 1714}

\bibitem[\protect\citeauthoryear{{Turner} et~al.,}{{Turner}
  et~al.}{2016}]{Turner:2016aa}
{Turner} O.~D.,  et~al., 2016, \mn@doi [\pasp]
  {10.1088/1538-3873/128/964/064401}, \href
  {http://adsabs.harvard.edu/abs/2016PASP..128f4401T} {128, 064401}

\bibitem[\protect\citeauthoryear{{Underwood}, {Tennyson}, {Yurchenko}, {Huang},
  {Schwenke}, {Lee}, {Clausen}  \& {Fateev}}{{Underwood}
  et~al.}{2016}]{Underwood2016}
{Underwood} D.~S.,  {Tennyson} J.,  {Yurchenko} S.~N.,  {Huang} X.,  {Schwenke}
  D.~W.,  {Lee} T.~J.,  {Clausen} S.,   {Fateev} A.,  2016, \mn@doi [\mnras]
  {10.1093/mnras/stw849}, \href
  {http://adsabs.harvard.edu/abs/2016MNRAS.459.3890U} {459, 3890}

\bibitem[\protect\citeauthoryear{{Varanasi} \& {Chudamani}}{{Varanasi} \&
  {Chudamani}}{1990}]{Varanasi1990}
{Varanasi} P.,  {Chudamani} S.,  1990, \mn@doi [\jqsrt]
  {10.1016/0022-4073(90)90060-J}, \href
  {http://adsabs.harvard.edu/abs/1990JQSRT..43....1V} {43, 1}

\bibitem[\protect\citeauthoryear{{Wende}, {Reiners}, {Seifahrt}  \&
  {Bernath}}{{Wende} et~al.}{2010}]{Fehlinelist2010}
{Wende} S.,  {Reiners} A.,  {Seifahrt} A.,   {Bernath} P.~F.,  2010, \mn@doi
  [\aap] {10.1051/0004-6361/201015220}, \href
  {http://adsabs.harvard.edu/abs/2010A%26A...523A..58W} {523, A58}

\bibitem[\protect\citeauthoryear{{West} et~al.,}{{West}
  et~al.}{2009}]{West:2009aa}
{West} R.~G.,  et~al., 2009, \mn@doi [\aj] {10.1088/0004-6256/137/6/4834},
  \href {http://adsabs.harvard.edu/abs/2009AJ....137.4834W} {137, 4834}

\bibitem[\protect\citeauthoryear{{West} et~al.,}{{West}
  et~al.}{2016}]{West:2016aa}
{West} R.~G.,  et~al., 2016, \mn@doi [\aap] {10.1051/0004-6361/201527276},
  \href {http://adsabs.harvard.edu/abs/2016A%26A...585A.126W} {585, A126}

\bibitem[\protect\citeauthoryear{{Wilson} et~al.,}{{Wilson}
  et~al.}{2008}]{Wilson:2008aa}
{Wilson} D.~M.,  et~al., 2008, \mn@doi [\apjl] {10.1086/586735}, \href
  {http://adsabs.harvard.edu/abs/2008ApJ...675L.113W} {675, L113}

\bibitem[\protect\citeauthoryear{{Winn} et~al.,}{{Winn}
  et~al.}{2011}]{Winn:2011aa}
{Winn} J.~N.,  et~al., 2011, \mn@doi [\apjl] {10.1088/2041-8205/737/1/L18},
  \href {http://adsabs.harvard.edu/abs/2011ApJ...737L..18W} {737, L18}

\bibitem[\protect\citeauthoryear{{Yurchenko} \& {Tennyson}}{{Yurchenko} \&
  {Tennyson}}{2014}]{Yurchenko2014}
{Yurchenko} S.~N.,  {Tennyson} J.,  2014, \mn@doi [\mnras]
  {10.1093/mnras/stu326}, \href
  {http://adsabs.harvard.edu/abs/2014MNRAS.440.1649Y} {440, 1649}

\bibitem[\protect\citeauthoryear{{Yurchenko}, {Barber}  \&
  {Tennyson}}{{Yurchenko} et~al.}{2011}]{Yurchenko2011}
{Yurchenko} S.~N.,  {Barber} R.~J.,   {Tennyson} J.,  2011, \mn@doi [\mnras]
  {10.1111/j.1365-2966.2011.18261.x}, \href
  {http://adsabs.harvard.edu/abs/2011MNRAS.413.1828Y} {413, 1828}

\bibitem[\protect\citeauthoryear{{Zhou} et~al.,}{{Zhou}
  et~al.}{2016}]{Zhou:2016aa}
{Zhou} G.,  et~al., 2016, \mn@doi [\aj] {10.3847/0004-6256/152/5/136}, \href
  {http://adsabs.harvard.edu/abs/2016AJ....152..136Z} {152, 136}

\makeatother
\end{thebibliography}


\appendix

\section{High Temperature Additions}
\label{app:high_temp}

\subsection{Thermal Ionization}
\label{ch5:thermal_ions}
\begin{figure}
\centering
\includegraphics[scale=0.42]{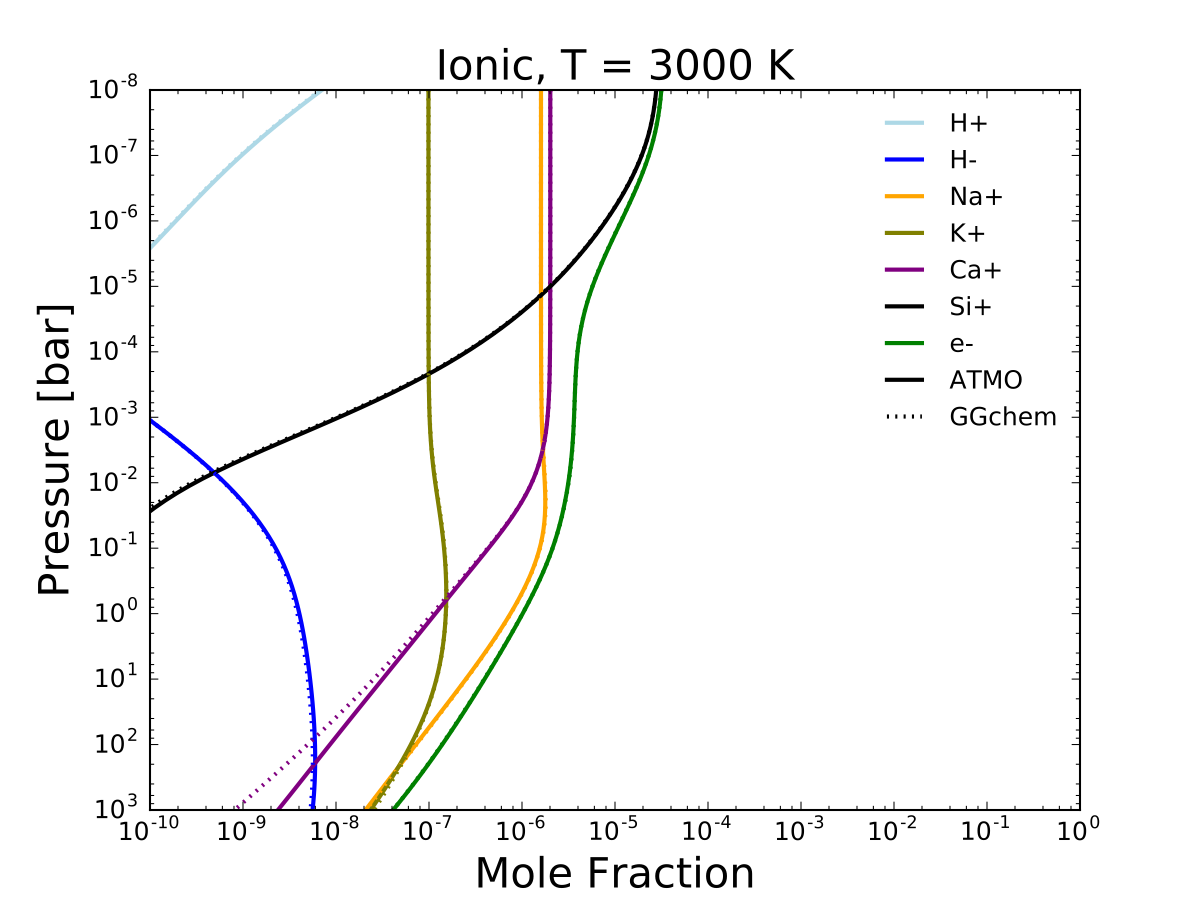}
 \caption{Figure showing validation of ATMO chemical equilibrium abundances of ionic species with GGChem \citep{Woitke2018} at 3000\,K.}
    \label{fig:val_thermal_ion}
\end{figure}

Thermal ionic species were recently included in ATMO to compute the chemical equilibrium abundance of H$^+$, H$^-$, Na$^+$, K$^+$, Ca$^+$, Si$^+$ and e$^-$. This was simply done by including thermodynamic data of these species in our database and using Gibbs energy minimization to compute the equilibrium abundances in each layer of the atmosphere, including the ionic species. The equilibrium chemical abundances of ionic species from ATMO were validated by comparing with those from the GGchem model \citep{Woitke2018} at 2000, 3000 and 4000\,K as shown in Figure \ref{fig:val_thermal_ion} for 3000\,K. We used the same input $P$-$T$ profile for both ATMO and GGchem for this validation. It can be seen that there is a very good agreement between the abundances from both the models. Only small differences exist in the deeper parts of the atmosphere, currently not probed by either emission or transmission spectra. 

\subsection{Implementation and Validation of H$^-$ opacity}
\label{ch5:h-_imp}
Many publications highlight the importance of H$^-$ opacity in the atmosphere of extremely irradiated hot Jupiters \citep[e.g][]{Arcangeli2018,Mansfield2018,Kriedberg2018,Parmentier2018}. We also included H$^-$ opacities in \texttt{ATMO} and investigated its effects on the $P$-$T$ profiles and thereby the spectra. The formation of species H$^-$ (hydrogen anion) is basically a result of electron attachment,  driven by the presence of abundant hydrogen (H) and low energy electrons in the ionised atmospheres of stars or hot Jupiters \citep{Rau1996}. The absorption of electromagnetic radiation by H$^-$ is driven by photo-detachment (bound-free) and free-free transition processes. These are computed using the analytical equations of \citet{John1988} derived from the original derivation of free-free transition in \citet{Bell1987}.

The photo-detachment process of H$^-$ absorption is given by 

\begin{equation} 
\ce{\it{h}{\nu} + H^- -> H + e^-},
\label{ch5:chem1}
\end{equation}

where $h$ is the Planck's constant and $\nu$ is the frequency of radiation. In this process H$^-$ ions absorbs radiation of frequency $\nu$, to form atomic hydrogen and an electron. As shown in \citet{John1988} for wavelengths ($\lambda$) less than the ionisation threshold $\lambda$ of H$^-$  ($\lambda$ < 1.6419 \textmu m) this is computed using,

\begin{align}
k_{bf}(\lambda) = 10^{-18} \lambda^3 \bigg(\frac{1}{\lambda} - \frac{1}{\lambda_0}\bigg)^{3/2} f(\lambda), \\
\textup{where}, \nonumber \\ 
f(\lambda) = \sum_{n=1}^6 C_n \bigg[\frac{1}{\lambda} - \frac{1}{\lambda_0}\bigg]^{(n-1)/2} \nonumber,
\end{align}

where $k_{bf}(\lambda)$ is the bound-free cross-section of H$^-$ in units of $\textup{cm}^2$, $\lambda$ is the wavelength, $\lambda_0$ = 1.6419 \textmu m is the threshold wavelength, $C_n$ are the coefficients for $n$ different values given in the Table 2 of \citet{John1988}. The bound-free absorption cross-section above the threshold $\lambda$ of 1.6419 \textmu m is zero. The total opacity due to bound-free absorption is then computed using,

\begin{align}
\kappa_{\textup{bf}} = k_{\textup{bf}}(\lambda,T) A[H^-] \frac{n_d}{\rho},
\end{align}

where $\kappa_\textup{{bf}}$ is the total bound-free opacity, $A[H^-]$ is the abundance of H$^-$ (mixing ratio), $n_d$ is the atmospheric number density ($\textup{cm}^{-3}$) and $\rho$ is the atmospheric mass density ($g/\textup{cm}^3$). 

The free-free transition process of H$^-$ is given by 

\begin{equation} 
\ce{\it{h}{\nu} + e^- + H  -> H + e^-}.
\label{ch5:chem1}
\end{equation}

In this reaction photons can be absorbed by electrons interacting with neutral hydrogen atom across the whole spectral range (0 < $\lambda$ < $\infty$). This process is solely responsible for H$^-$ opacity beyond the ionisation threshold wavelength (1.6419 \textmu m). This is computed using 

\begin{align}
\label{eq:kff}
\resizebox{0.45\textwidth}{!}{$k_{ff}(\lambda,T) = 10^{-29} \sum_{n=1}^6 \bigg(\frac{5040}{T}\bigg)^{(n+1)/2} (\lambda^2 A_n + B_n + C_n/\lambda + D_n/\lambda^2 + E_n/\lambda^3 + F_n/\lambda^4)$},
\end{align}

where $k_{ff}(\lambda,T)$ is the free-free cross-section of H$^-$ in units of $\textup{cm}^{4}/dyne$ and $T$ is the temperature. A$_n$, B$_n$, C$_n$, D$_n$, E$_n$ and F$_n$ are coefficients as given in table 3a and 3b of \citet{John1988} for $\lambda$ > 0.3645 \textmu m and 0.1823 < $\lambda$ < 0.3645 \textmu m, respectively. By multiplying equation \ref{eq:kff} by the Boltzmann's constant (1.38$\times$10$^{-16}$ erg/s) and the temperature, $k_{ff}(\lambda,T)$ is obtained in the units of $\textup{cm}^5$. The total opacity due to free-free absorption is then computed using,

\begin{align}
\kappa_{ff} = k_{ff}(\lambda,T) A[H] \frac{n_d}{\rho} A[e^-] n_d,
\end{align}

where $\kappa_{ff}$ is the total free-free opacity, $A[H]$ is the abundance of neutral hydrogen (mixing ratio) and $A[e^-]$ is the abundance of electron (mixing ratio). The total opacity of the H$^-$ ion ($\kappa_{tot}$) due to bound-free and free-free transitions is then given by $\kappa_{tot} = \kappa_{bf} + \kappa_{ff}$.

\begin{figure*}
\begin{center}
 \subfloat[]{\includegraphics[width=\columnwidth]{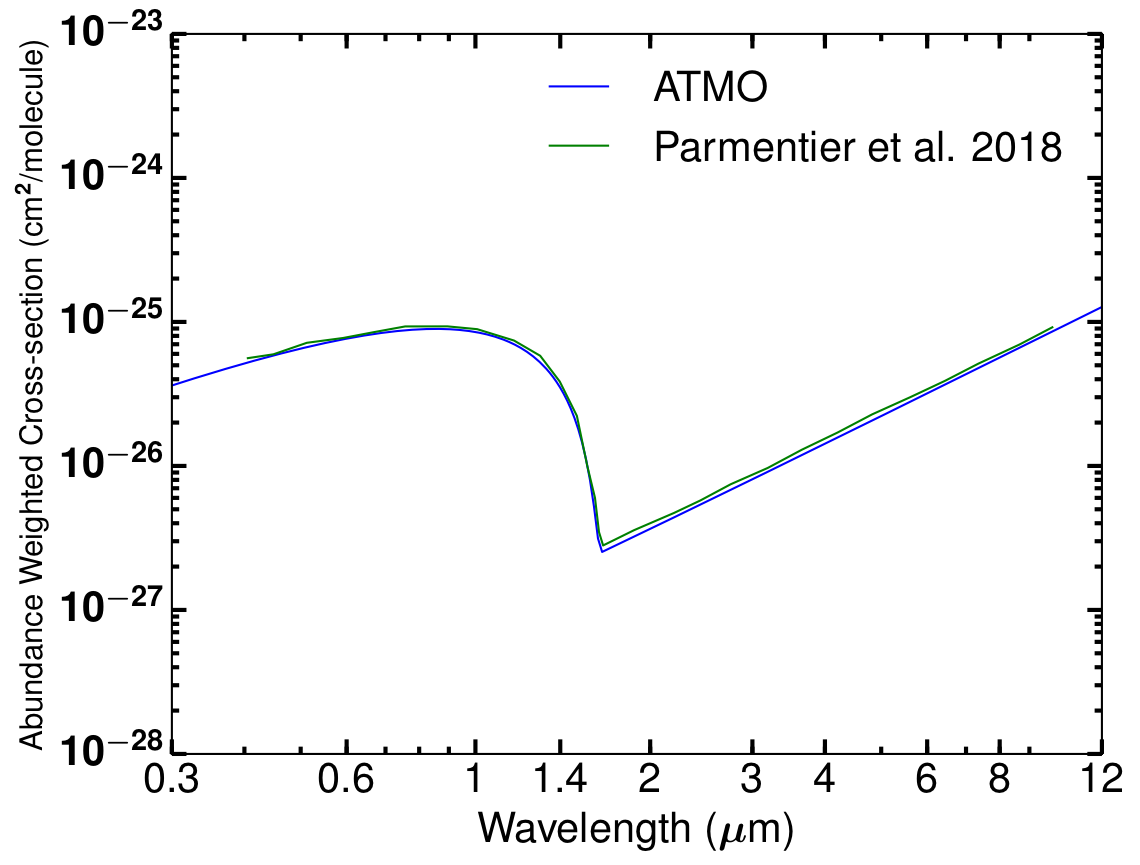}\label{ch5:fig_h-_op1}}
 \subfloat[]{\includegraphics[width=\columnwidth]{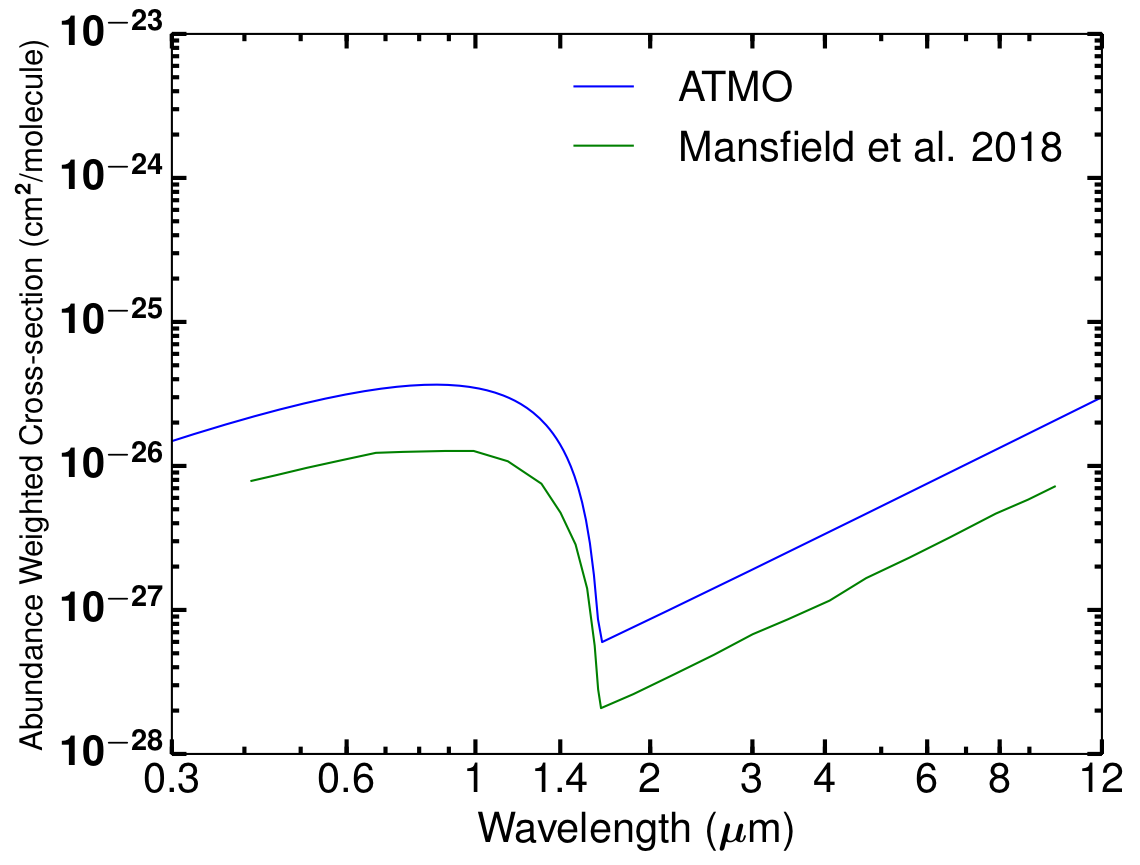}\label{ch5:fig_h-_op2}}
\end{center}
 \caption[Figure comparing H$^-$ opacity from ATMO and that from \citet{Parmentier2018} and \citet{Mansfield2018}]{\textbf{(a)} Figure showing  abundance weighted cross-section of H$^-$ at 0.042 bar and 3100\,K from Figure 4 of \citet{Parmentier2018} (green) and from ATMO (blue). \textbf{(b)} Figure showing abundance weighted cross-section of H$^-$ at 0.084 bar and 2756\,K from Figure 7 of \citet{Mansfield2018} (green) and from ATMO (blue).}
\end{figure*}

We validate the abundance weighted H$^-$ opacity by comparing with the results from \citet{Parmentier2018} and \citet{Mansfield2018}. Figure 4 of \citet{Parmentier2018} shows the abundance weighted absorption cross-section of H$^-$ (cm$^2$/molecule) as a function of wavelength at the $P$-$T$ point of 0.042 bar and 3100\,K. We also compute the abundance weighted absorption cross-section of H$^-$ at this $P$-$T$ point within ATMO using $\kappa_{\textup{tot}}\times\frac{\mu_{\textup{mean}}}{N_A}$, where $\kappa_{\textup{tot}}$ is the total opacity of the H$^-$ ion computed as shown in the previous section, $\mu_{\textup{mean}}$ is the mean molecular weight of the atmosphere and N$_A$ is the Avogadro's constant. The comparison is shown in Figure \ref{ch5:fig_h-_op1}, demonstrating that the agreement is good and even the equilibrium chemical abundances at this $P$-$T$ point are similar in ATMO and \citet{Parmentier2018} (from Figure 3 in their paper), thus validating the implementation of H$^-$ opacity in ATMO. 

Figure 7 of \citet{Mansfield2018} shows the abundance weighted absorption cross-section of H$^-$ at a $P$-$T$ point of 0.084 bar and 2756\,K. When compared with this, there is a substantial difference between the abundance weighted absorption cross-section of H$^-$ as shown in Figure \ref{ch5:fig_h-_op1}. The primary reason being the difference in equilibrium chemical abundances at this $P$-$T$ point, which is $\sim$3.4 times larger in ATMO as compared to \citet{Mansfield2018}. However, when this factor  of 3.4 is taken into consideration while comparing the abundance weighted absorption cross-section of H$^-$ from both models, there is a good agreement (not shown here). The reason for differences in equilibrium chemical abundances is still unclear and can be due to many factors, such as the differences in input elemental abundances, polynomial coefficients etc. as shown in \citet{Errgoyal2019}. We note that the equilibrium chemistry scheme used  in ATMO has been validated by comparing to various numerical and analytical equilibrium chemistry models, with local and rainout condensation \citep{Drummond2016, Errgoyal2019}.

\subsection{Fe Absorption cross-sections}
\label{ch5:fe_imp}
Absorption cross-sections of gaseous iron (Fe) have been included in ATMO using the Fe line list from the VALD database \citep{Heiter2008,Heiter2015} and the partition function from \citet{Sauval1984}.  The H$_2$ and He pressure broadening line-widths for Fe have been computed using the van der Waals coefficient contained within the VALD database and using Equation 23 in \citet{Sharp2007}. The absorption cross-sections of Fe included in ATMO have also been validated by comparing with the absorption cross-sections of \citet{Sharp2007}.

\section{WASP-121b $\chi^{2}$ map}
\label{app:chi-sq-map}
\begin{figure}
\includegraphics[width=\columnwidth]{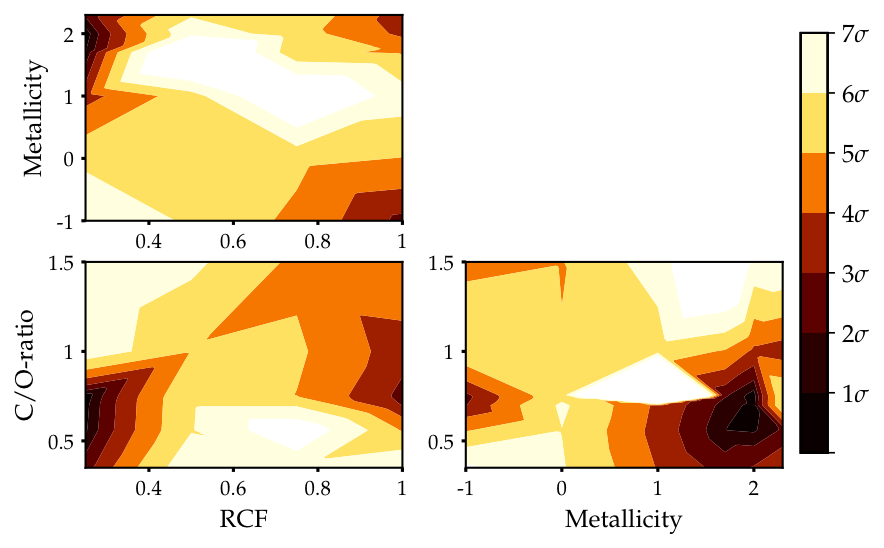}
 \caption{Figure showing $\chi^{2}$ map for WASP-121b for the transmission spectra fitting described in Section \ref{ch5:obs_int}. Contours of $\chi^{2}$ are shown for all the combinations of grid parameters. Metallicity is also log-scaled, 0 being solar metallicity and 2 being 100 times solar metallicity. Colours indicate confidence intervals as shown in colormap to the right. See \citet{Goyal2018} for more details.}
    \label{fig:chimap_transwasp121b}
\end{figure}

\bsp	
\label{lastpage}
\end{document}


\label{firstpage}
\pagerange{\pageref{firstpage}--\pageref{lastpage}}
\maketitle


\section{Decoupled emission spectrum}
\label{ch5:decon_emiss_spec}

\begin{figure*}
\begin{center}
 \subfloat[]{\includegraphics[width=\columnwidth]{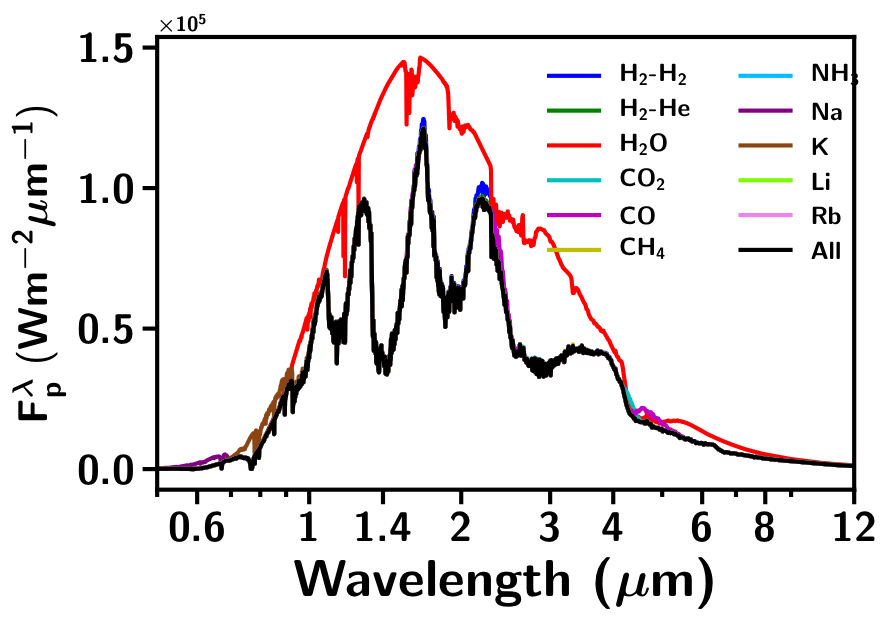}\label{ch5:fig_nkap_1}}
 \subfloat[]{\includegraphics[width=\columnwidth]{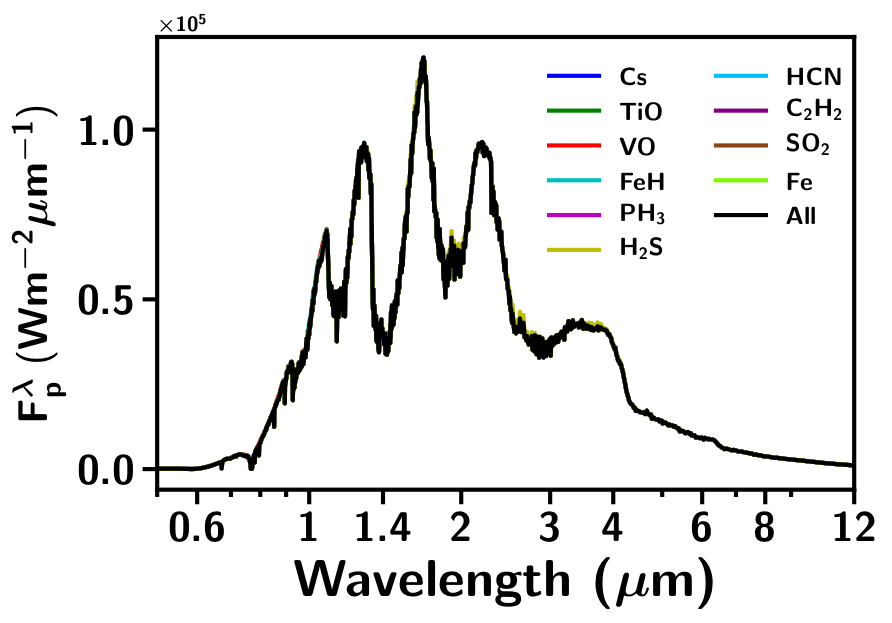}\label{ch5:fig_nkap_2}}
\end{center}
 \caption[Figure showing decoupled emission spectra for all the opacity species used in ATMO, using WASP-17b as the test case]{Figure showing WASP-17b emission spectrum with 0.25 f$_{\textup{c}}$, solar metallicity and solar C/O ratio, when removing opacity contributions of single species at a time from the calculation, shown by their respective colours.  Emission spectrum including all 21 opacities is shown in black. \textbf{(a)} Emission Spectrum when removing opacity due to species H$_2$-H$_2$ (blue), H$_2$-He (green), H$_2$O (red), CO$_2$ (cyan), CO (magenta), CH$_4$ (yellow), NH$_3$ (lightblue), Na (purple), K (brown), Li (lightgreen) and Rb (violet). \textbf{(b)} Same as \ref{ch5:fig_nkap_1} but when removing opacity due to species Cs (blue), TiO (green), VO (red), FeH (cyan), PH$_3$ (magenta), H$_2$S (yellow), HCN (lightblue), C$_2$H$_2$ (purple), SO$_2$ (brown) and Fe (lightgreen)}
 \label{ch5_decoup_emiss_wasp017b}
\end{figure*}

\begin{figure*}
\begin{center}
 \subfloat[]{\includegraphics[width=\columnwidth]{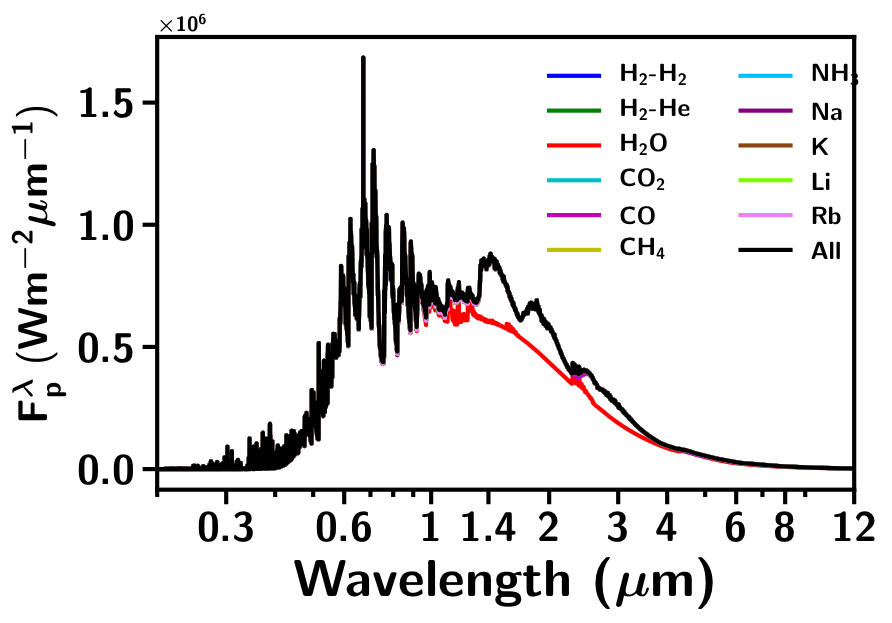}\label{ch5:fig_nkap_3}}
 \subfloat[]{\includegraphics[width=\columnwidth]{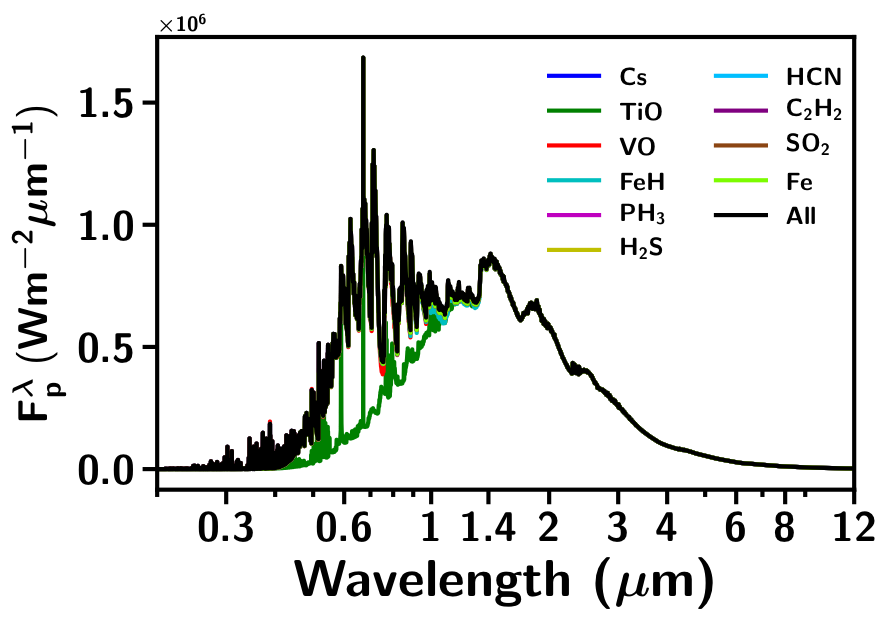}\label{ch5:fig_nkap_4}}
\end{center}
 \caption[Same as \ref{ch5_decoup_emiss_wasp017b}, but using WASP-121b as the test case]{\textbf{(a)} Same as Figure \ref{ch5:fig_nkap_1} but for WASP-121b, at solar metallicity and solar C/O ratio with 0.5 f$_{\textup{c}}$. \textbf{(b)} Same as \ref{ch5:fig_nkap_2} but for WASP-121b, at solar metallicity and solar C/O ratio with 0.5 f$_{\textup{c}}$.}
\end{figure*}

To identify the features of various absorbing/emitting species in the emission spectrum, we compute the spectrum by removing the opacity contribution due to one species at a time, which we term the decoupled emission spectrum. In the emission spectrum as the radiation travelling radially outward from the planetary atmosphere is measured, spectral features are indicated by a dip due to absorption as opposed to in a transmission spectrum, unless there is an inversion, in which case it shows an emission feature (bump). We show decoupled spectra for two planets, first is WASP-17b, a hot Jupiter with T$_\textup{eq}$ = $\sim$ 1750\,K in Fig. \ref{ch5:fig_nkap_1} and \ref{ch5:fig_nkap_2}. Second an extremely irradiated hot Jupiter WASP-121b, with T$_\textup{eq}$ = $\sim$ 2400\,K and a strong evidence of an inversion layer in its atmosphere \citep{Evans2017} in Fig. \ref{ch5:fig_nkap_3} and \ref{ch5:fig_nkap_4}. For WASP-17b it can be seen that most of the contribution to the emission spectrum is from H$_2$O, since when \water is removed while computing the emission spectrum as shown in Figure \ref{ch5:fig_nkap_1} most of the absorption features vanish as compared to where all opacities sources are included (black). The absence of \water also allows probing the hotter and deeper regions of the atmospheres, as can be seen from higher flux for models without \water opacity. There is also a strong CO absorption feature at 4.5$\mic$. The $P$-$T$ profile used to generate the spectrum for WASP-17b is shown in Figure 3a (Rainout Condensation) in the main paper. 

Figure \ref{ch5:fig_nkap_3} and \ref{ch5:fig_nkap_4} show the decoupled spectra for WASP-121b with the corresponding $P$-$T$ profile shown in Figure 9a (0.5 f$_{\textup{c}}$) in the main paper.  For WASP-121b also it can be seen that most of the contribution to the emission spectrum is from H$_2$O\, since when the opacity of \water is removed, shown in Figure \ref{ch5:fig_nkap_3} most of the emission features vanish when compared to the spectrum including all opacity sources (black). However, it is interesting to note that due to the temperature inversion in the simulated atmosphere of WASP-121b the H$_2$O features are seen in emission, as opposed to absorption as in the case of WASP-17b. There is also a decrease in the overall flux after removal of \water opacity as emission is now from the deeper regions of the atmosphere which are cooler. The emission from CO can also be noticed near $\sim$4.5$\mic$. The optical spectrum is dominated by emission features of TiO as seen in Figure \ref{ch5:fig_nkap_4}. It must be noted here that we remove TiO only while computing emission spectra for this test shown in Figure \ref{ch5:fig_nkap_4}, not while computing the RCE $P$-$T$ profile used to generate the emission spectrum. 

\begin{figure*}
\begin{center}
 \subfloat[]{\includegraphics[width=\columnwidth]{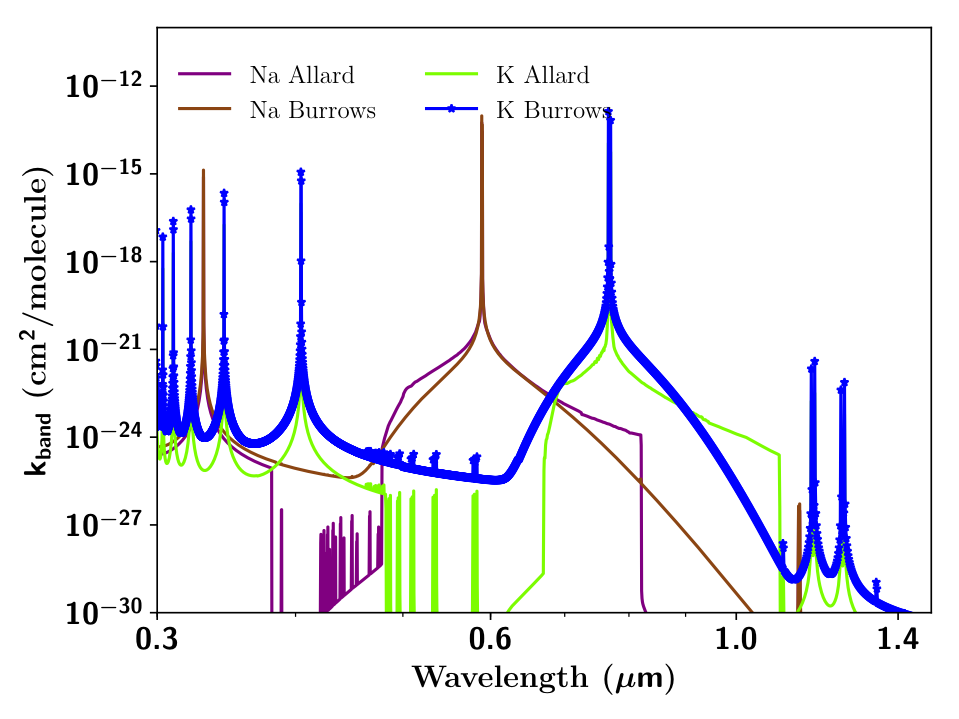}\label{ch5:fig_na_k_1}}
 \subfloat[]{\includegraphics[width=\columnwidth]{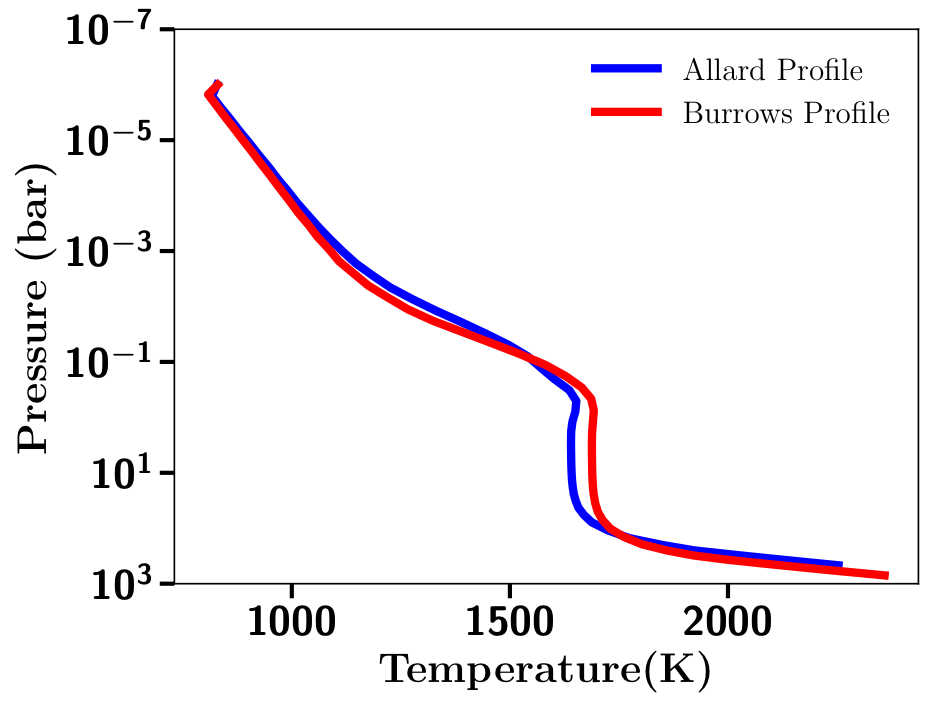}\label{ch5:fig_na_k_2}}
 \newline
 \subfloat[]{\includegraphics[width=\columnwidth]{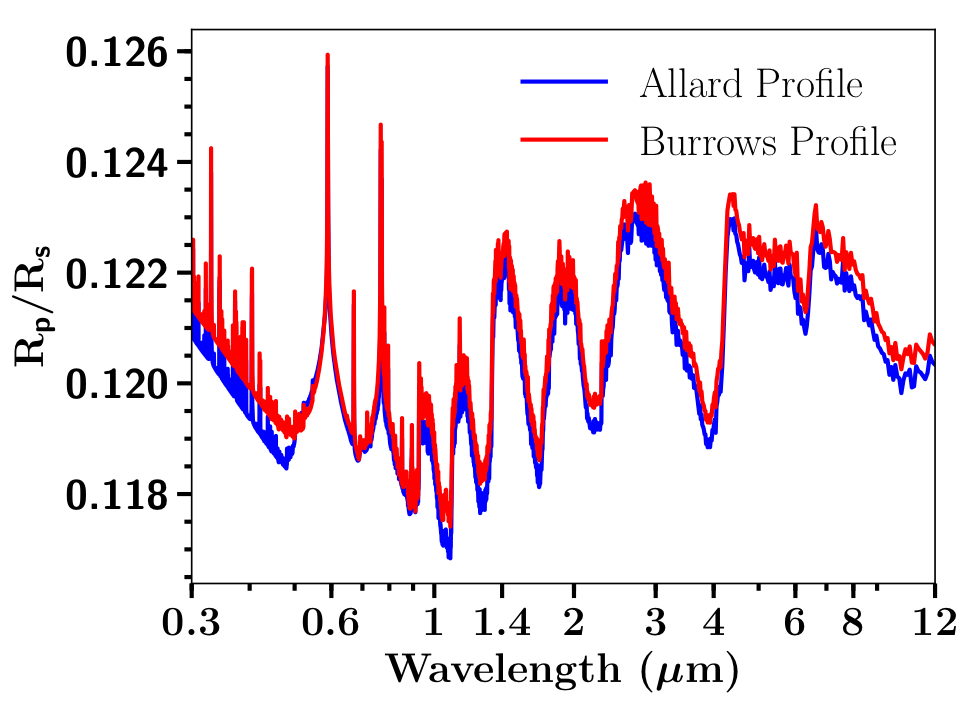}\label{ch5:fig_na_k_3}}
 \subfloat[]{\includegraphics[width=\columnwidth]{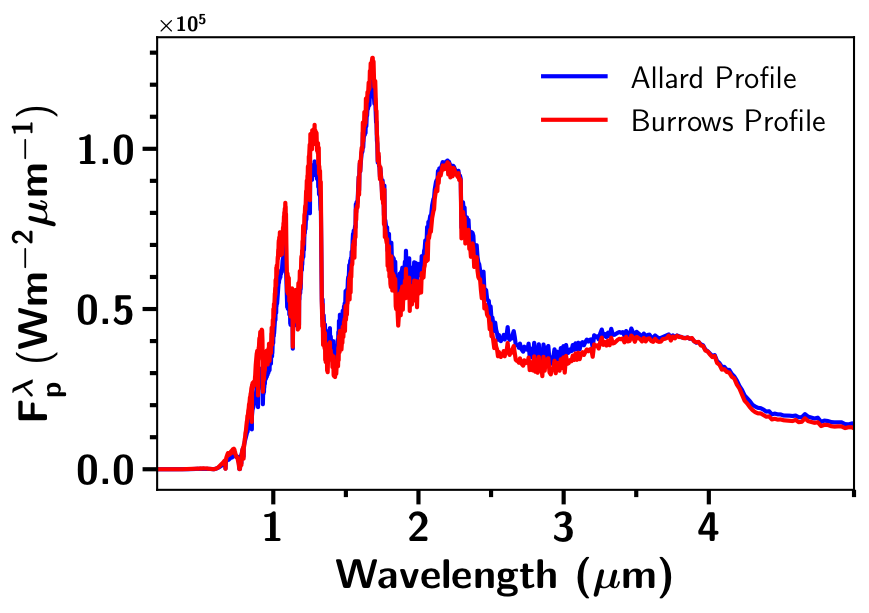}\label{ch5:fig_na_k_4}}
\end{center}
 \caption[Figure comparing absorption cross-sections of Na and K, RCE $P$-$T$ profiles and the corresponding transmission and emission spectra, with two different Na and K pressure broadening line wing profiles]{\textbf{(a)} Figure showing absorption cross-sections (cm$^2$/molecule) of Na and K using two different pressure broadening treatments from \citet{Allard2007} and \citet{Burrows2000} \textbf{(b)} Figure showing RCE $P$-$T$ profiles using two different pressure broadening profiles (Allard and Burrows) for Na and K, shown in Figure \ref{ch5:fig_na_k_1}, for WASP-17b at solar metallicity and solar C/O ratio with 0.25 f$_{\textup{c}}$. \textbf{(c)} Transmission spectra for WASP-17b using  $P$-$T$ profiles shown in Figure \ref{ch5:fig_na_k_2}, and two different pressure broadening profiles, namely, Allard and Burrows. \textbf{(d)} Emission spectra for WASP-17b using $P$-$T$ profiles shown in Figure \ref{ch5:fig_na_k_2}, and two different pressure broadening profiles, namely, Allard and Burrows.} 
\end{figure*}

\section{Sensitivity to Model Choices}
\label{app:sensitivity_model_choice}

\subsection{Comparing model simulations with different Na and K line wing profiles}
\label{ch5:eff_Na_K}

Figure \ref{ch5:fig_na_k_1} shows the Na and K cross-sections at 1000\,K and 1 mbar with line profiles from \citet{Burrows2000} which are primarily Lorentzian and from \citet{Allard2007} which uses more detailed quantum mechanical calculations. It can be noticed from this figure that the differences are primarily in the pressure broadened wings. 

The effect of these line profiles on the RCE $P$-$T$ profiles, transmission and emission spectra are shown in Figure \ref{ch5:fig_na_k_2}, \ref{ch5:fig_na_k_3} and \ref{ch5:fig_na_k_4}, respectively, for WASP-17b at 0.25 f$_{\textup{c}}$, solar metallicity and solar C/O ratio. The differences are negligible and likely not to be detectable with current observations, in all cases. In the emission spectrum, adopting a profile from \citet{Allard2007} henceforth \enquote{Allard Profile} leads to a slightly higher flux than when adopting a  profile from \citet{Burrows2000} henceforth \enquote{Burrows profile}, due to a slight increase in temperature for the case using the Allard profile between 100 to 0.1 mbar, as shown in Figure \ref{ch5:fig_na_k_2}. 

\begin{figure*}
\begin{center}
 \subfloat[]{\includegraphics[width=\columnwidth]{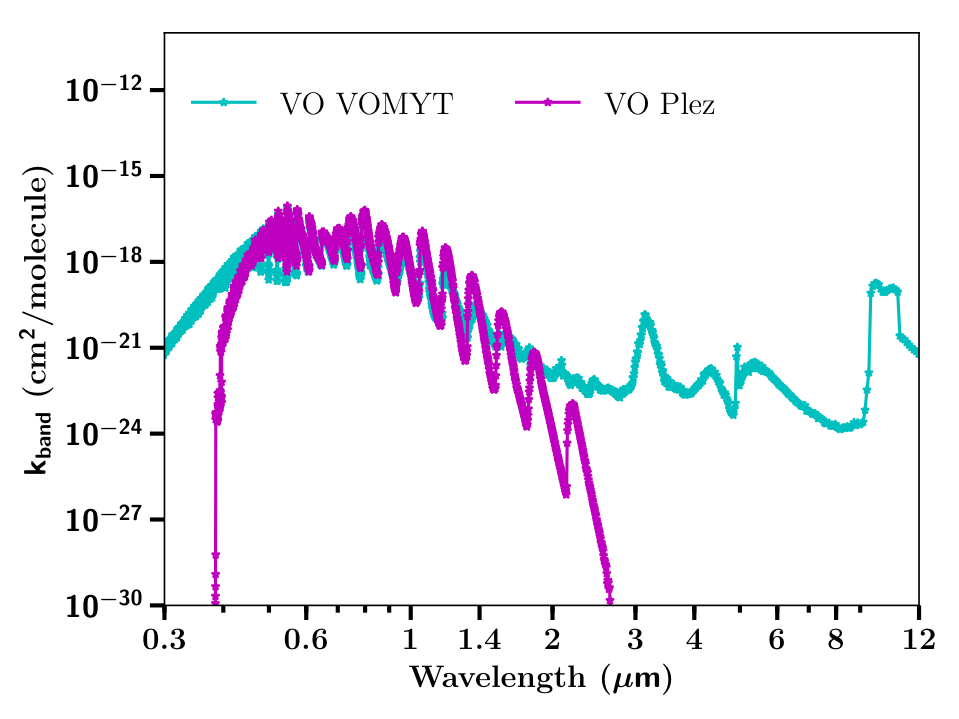}\label{ch5:fig_vo1}}
 \subfloat[]{\includegraphics[width=\columnwidth]{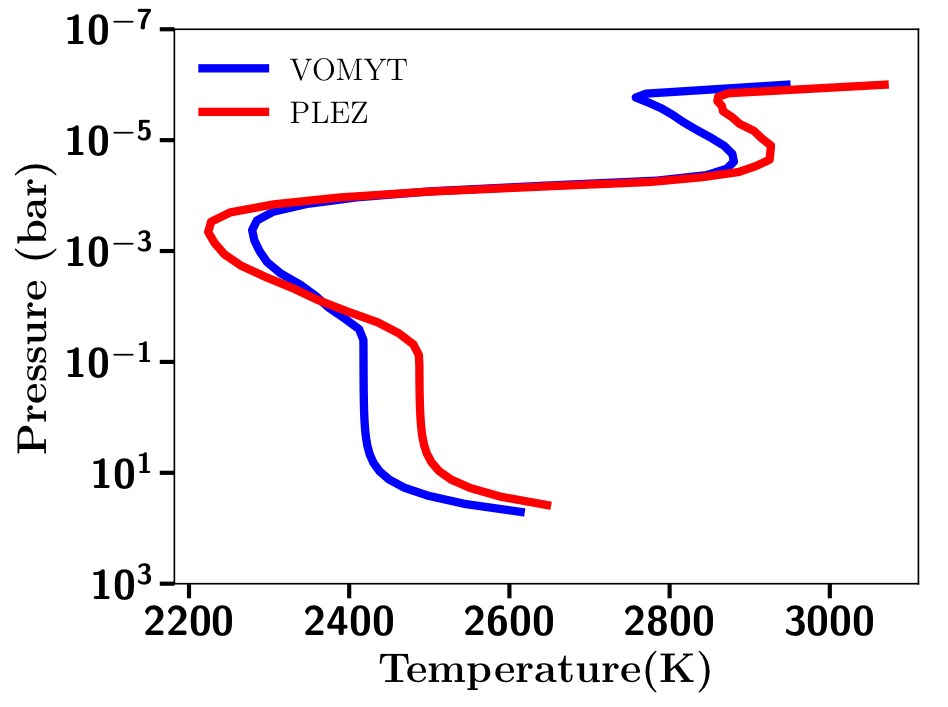}\label{ch5:fig_vo2}}
  \newline
  \subfloat[]{\includegraphics[width=\columnwidth]{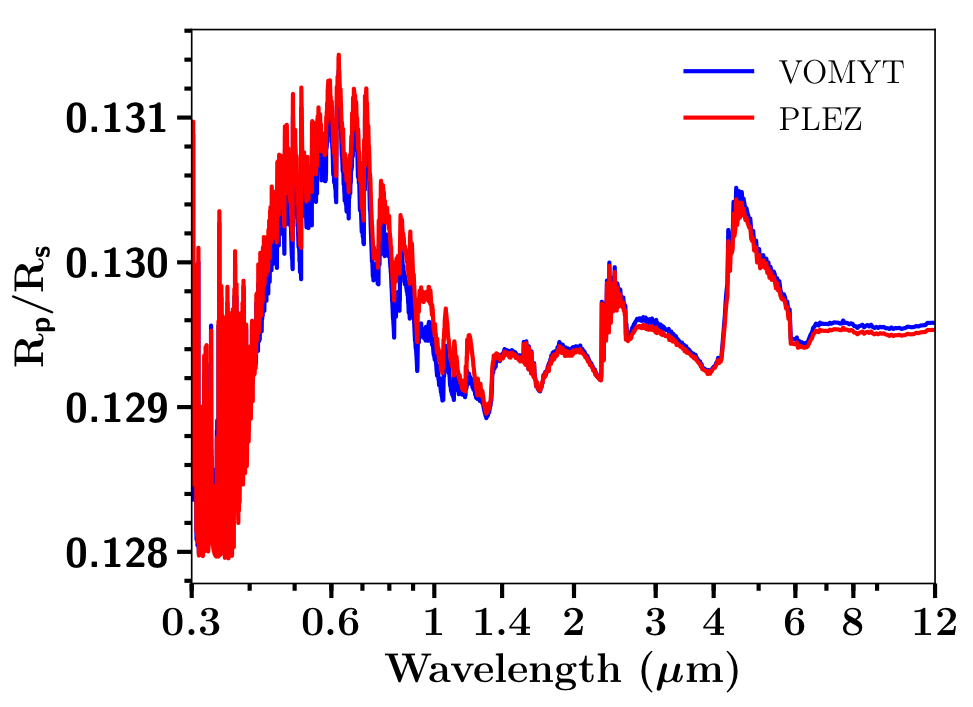}\label{ch5:fig_vo3}}
 \subfloat[]{\includegraphics[width=\columnwidth]{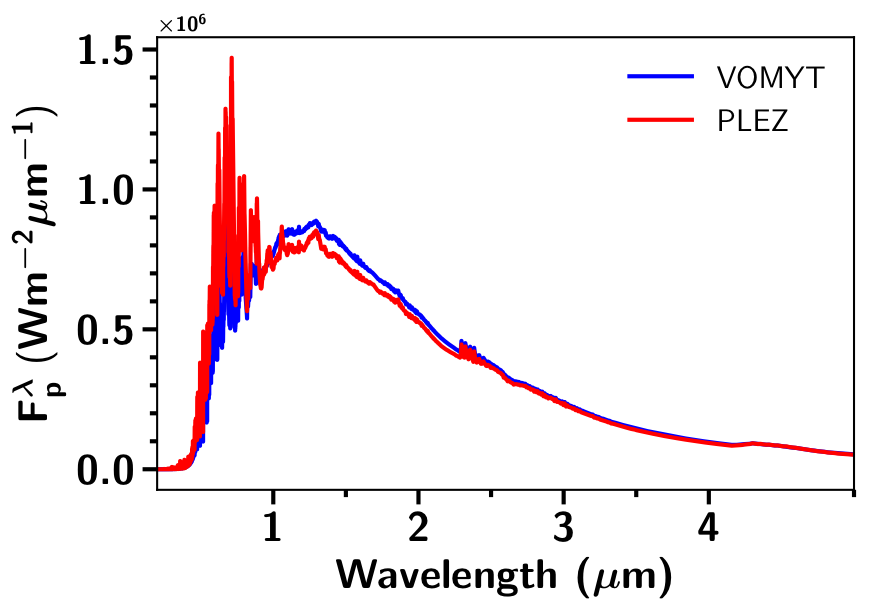}\label{ch5:fig_vo4}}
\end{center}
 \caption[Figure comparing absorption cross-sections of VO, RCE $P$-$T$ profiles and the corresponding transmission and emission spectra with two different sources of line-lists for VO]{\textbf{(a)} Figure showing absorption cross-sections (cm$^2$/molecule) of VO using different two different line-lists VOMYT \citep{McKemmish2016} and PLEZ \citep{Plez1999} \textbf{(b)} Figure showing RCE $P$-$T$ profiles using different VO line-list named PLEZ and VOMYT for WASP-121b at 200x solar metallicity and solar C/O ratio with 0.5 f$_{\textup{c}}$. \textbf{(c)} Transmission spectra for WASP-121b using  $P$-$T$ profiles shown in Figure \ref{ch5:fig_vo2}, and two different VO line-lists. \textbf{(d)} Emission spectra for WASP-121b using $P$-$T$ profiles shown in Figure \ref{ch5:fig_vo2}, and two different VO line-lists.} 
\end{figure*}

\begin{figure*}
\begin{center}
 \subfloat[]{\includegraphics[width=\columnwidth]{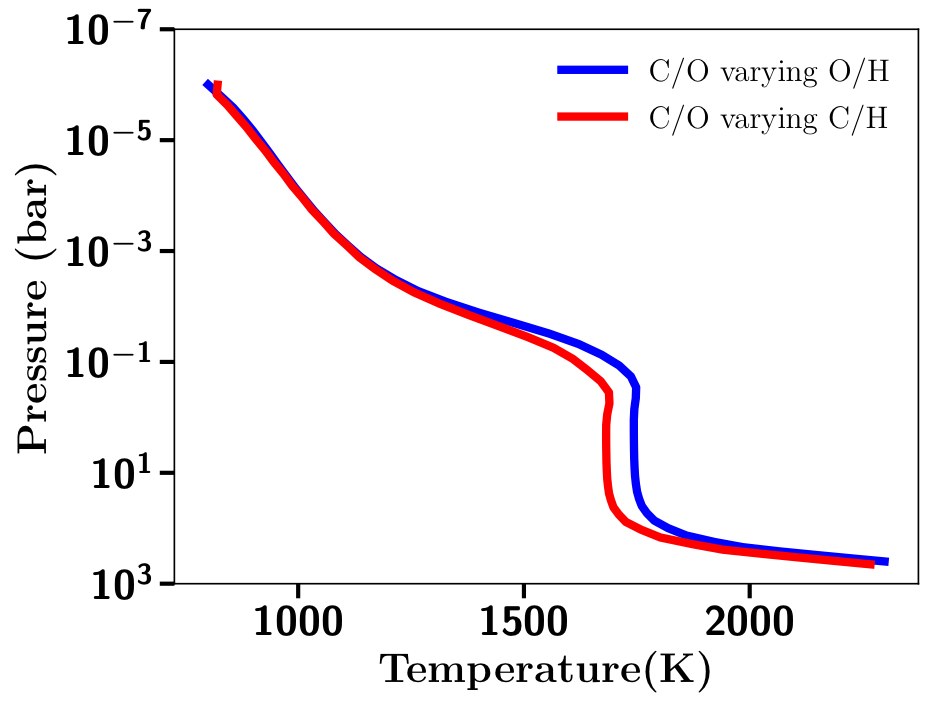}\label{ch5:fig_reverse_CO_1}}
 \subfloat[]{\includegraphics[width=\columnwidth]{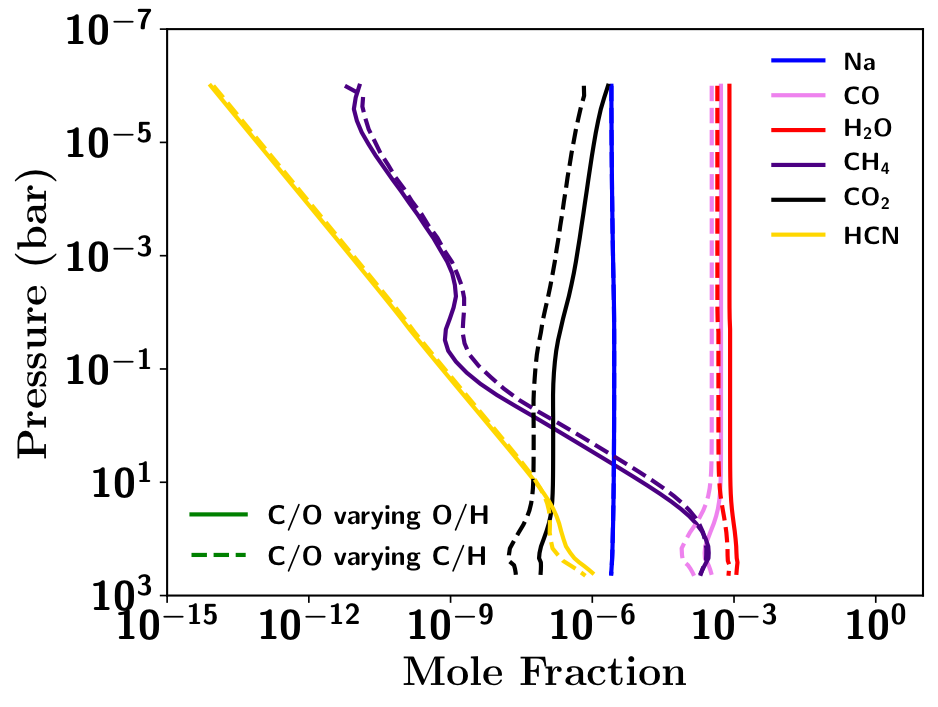}\label{ch5:fig_reverse_CO_2}}
 \newline
  \subfloat[]{\includegraphics[width=\columnwidth]{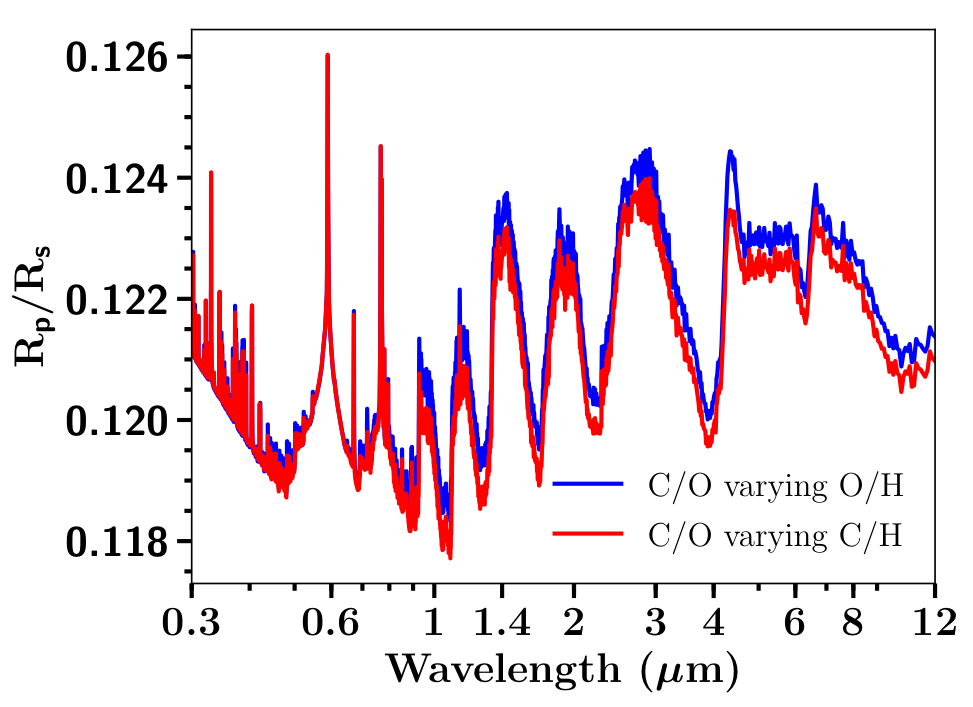}\label{ch5:fig_reverse_CO_3}}
 \subfloat[]{\includegraphics[width=\columnwidth]{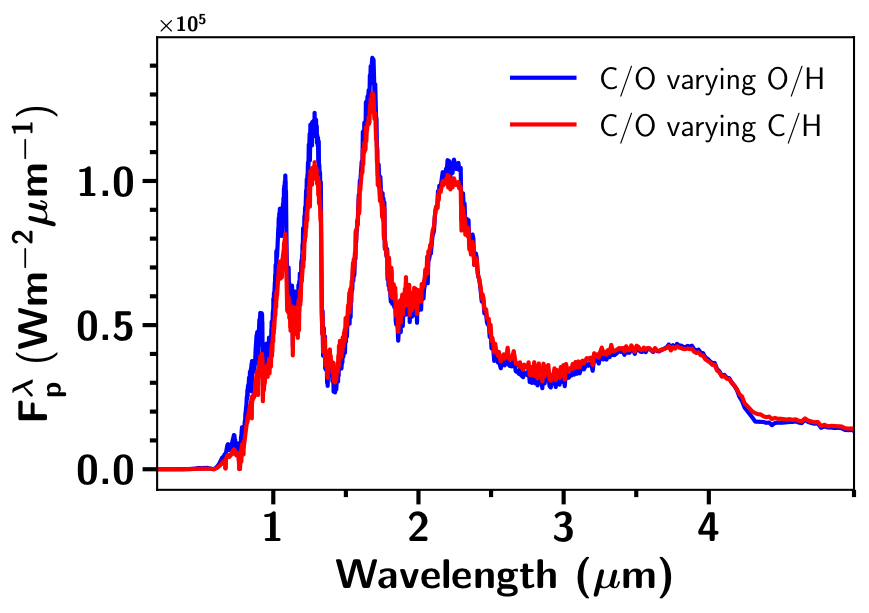}\label{ch5:fig_reverse_CO_4}}
  \newline
  \subfloat[]{\includegraphics[width=\columnwidth]{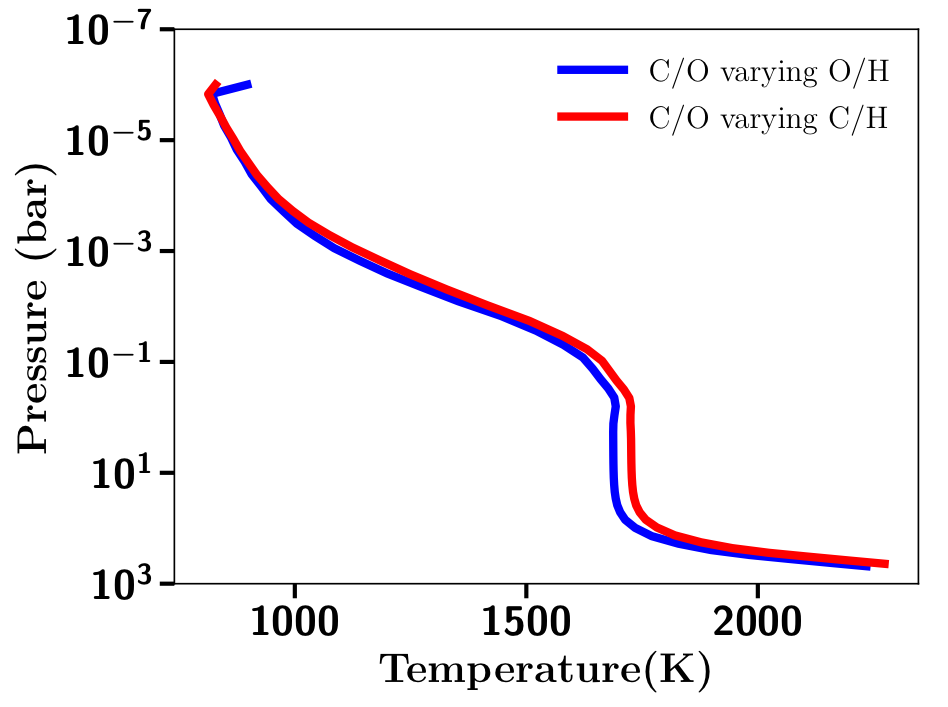}\label{ch5:fig_reverse_CO_5}}
 \subfloat[]{\includegraphics[width=\columnwidth]{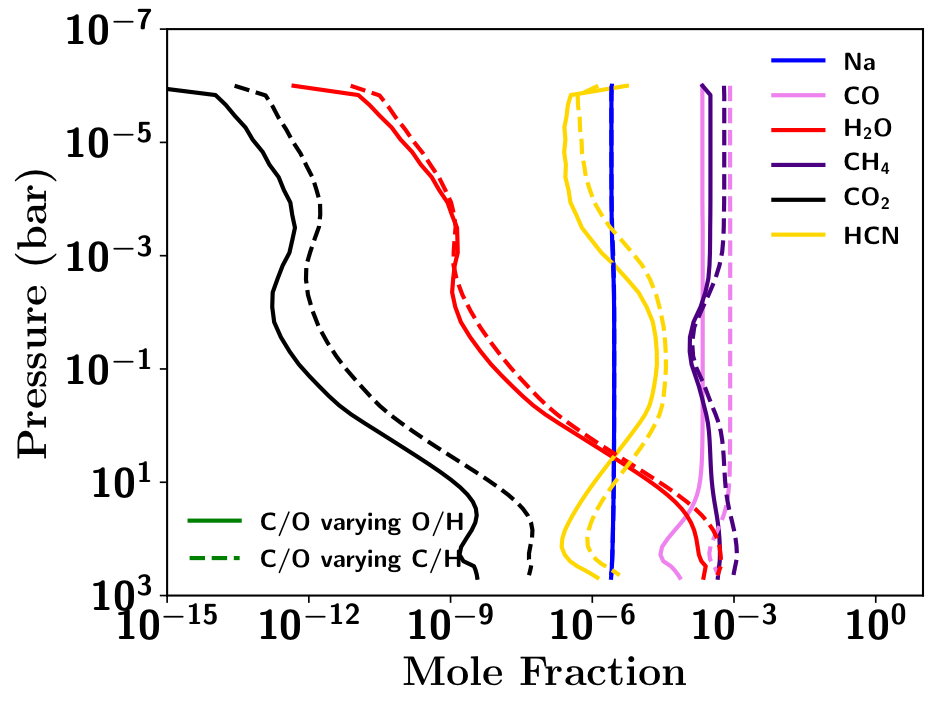}\label{ch5:fig_reverse_CO_6}}
\end{center}
 \caption[Figure comparing RCE $P$-$T$ profiles and the corresponding equilibrium chemical abundances, transmission and emission spectra,  varying C/O ratio by varying O/H and those by varying C/H using WASP-17b as the text case]{\textbf{(a)} Figure showing $P$-$T$ profiles with C/O ratio of 0.35 by varying C/H (red) and O/H (blue) relative to solar C/O ratio (0.55) at 0.25 f$_{\textup{c}}$ and solar metallicity for WASP-17b. \textbf{(b)} Figure showing equilibrium chemical abundances for  $P$-$T$ profiles shown in Fig. \ref{ch5:fig_reverse_CO_1} by varying O/H (solid) and C/H (dashed)\textbf{(c)} Figure showing transmission spectra using $P$-$T$ profiles shown in  Fig. \ref{ch5:fig_reverse_CO_1} and chemical abundances shown in Fig. \ref{ch5:fig_reverse_CO_2} \textbf{(d)} Figure showing emission spectra using $P$-$T$ profiles shown in  Fig. \ref{ch5:fig_reverse_CO_1} and chemical abundances shown in Fig. \ref{ch5:fig_reverse_CO_2} \textbf{(e)} Same as Fig. \ref{ch5:fig_reverse_CO_1} but for C/O ratio of 1.5. \textbf{(f)} Same as Fig. \ref{ch5:fig_reverse_CO_2}} but for C/O ratio of 1.5.
 \label{ch5:fig_co_comp_oxy_car_wasp17b}
\end{figure*}

\subsection{Comparing model simulations with different VO line-list sources}
\label{ch5:vo}

Vanadium Oxide (VO) has been predicted to be the major absorber in the atmosphere of WASP-121b leading to an inversion, with the hints of VO features detected in the transmission and emission spectrum of WASP-121b \citep{Evans2017, Evans2018}. In ATMO we initially included the VO line-list from \citet{Plez1999} (PLEZ), this was updated to new high temperature VO line-list from Exomol named as VOMYT \citep{McKemmish2016}. Figure \ref{ch5:fig_vo1} show the differences in the cross-sections computed using both line-lists. It can be seen that the new line-list (VOMYT) is more complete, especially in the infrared. Surprisingly, there is also a strong VO absorption band in the infrared between 10-12$\mic$ in the new VO line list \citep{McKemmish2016} in comparison to the old line-list. There are two important peaks one near 3$\mic$ and other a broad band peak between 10-12$\mic$, offering a potential wavelength region to detect VO using JWST.  The effect of using different line-lists on the RCE $P$-$T$ profiles is shown in Figures \ref{ch5:fig_vo2} at 200 times solar metallicity for WASP-121b. At solar metallicity and solar C/O ratio with 0.5 f$_{\textup{c}}$, there is no difference in the RCE $P$-$T$ profiles generated using the two different line-lists (not shown here), due to the comparatively low abundance of VO. However, when the metallicity is increased to 200x solar, the abundance of VO increases, \citet[see Figure 13b in][]{Errgoyal2019} rising to $\sim$1000 times than that at solar metallicity. These increased abundances are then sufficient to create differences between the $P$-$T$ profiles obtained using the two different line-lists. Moreover, retrieval models have also predicted the presence of substantially large abundances of VO ($\sim$ 1000x solar abundance), required to produce an inversion in the atmosphere of WASP-121b \citep{Evans2017}. At 200$\times$ solar metallicity,  the $P$-$T$ profile obtained using the VOMYT line-list is hotter by $\sim$ 50\,K around 1 millibar region, but this absorption of radiation in the upper atmosphere leads to cooling in the deeper atmosphere from 0.1 to 10 bar by about 100\,K. 

The effect of different VO line-lists on the transmission and emission spectra of WASP-121b are shown in Figure \ref{ch5:fig_vo3} and \ref{ch5:fig_vo4}, respectively. In the transmission spectra the differences are only in the size of the features which is a combination of differences in $P$-$T$ profiles, abundances and small differences in line-lists. In the emission spectra, the difference is a result of differences in the $P$-$T$ profile. As the temperature in the region of the inversion which is probed by emission spectra is higher for VOMYT than PLEZ, it leads to higher value of F$_p$ (planetary flux), thus producing differences. 

\subsection{Comparing model simulations varying O/H with those varying C/H to vary C/O ratio.}
\label{ch5:eff_co_car}
 The differences in the $P$-$T$ profiles and the spectra between the simulations, varying C/O ratio by varying O/H and those by varying C/H, are shown here for WASP-17b. We note that when we vary C/H or O/H we keep the abundances of other species to their solar value or to a metallicity scaled value. 

Figures \ref{ch5:fig_reverse_CO_1} shows the $P$-$T$ profiles at a C/O ratio of 0.35 resulting from varying O/H and C/H for WASP-17b. We can see there are differences of the order of 50\,K between the two profiles. The differences can also be seen in the equilibrium chemical abundances shown in \ref{ch5:fig_reverse_CO_2}.  The abundances of H$_2$O, CO and CO$_2$ are slightly larger when O/H is varied, compared to the same C/O ratio model where C/H is varied. This is because for C/O ratios less than 1, abundances of these species are limited by the carbon abundance (more so for CO as it is the dominant molecule), therefore when O/H is varied to reach a C/O ratio of 0.35, O/H is increased thus favouring an increase in oxygen bearing species. On the contrary if C/H is varied, it has to be decreased (keeping O/H constant) to reach a C/O ratio of 0.35 thus limiting (decreasing) the formation of carbon bearing species. There is also an effect caused by the changing of $P$-$T$ profile which can be seen for CH$_4$ as its abundances are mainly different in the region where there are larger differences in $P$-$T$ profiles, between simulations varying either O/H or C/H, at the same C/O ratio. The  changing $P$-$T$ profile will also have an effect on other species. These differences can also be seen in the transmission and emission spectra shown in Fig. \ref{ch5:fig_reverse_CO_3} and \ref{ch5:fig_reverse_CO_4}, respectively, primarily at 4.5$\mic$ due to CO. 

Figures \ref{ch5:fig_reverse_CO_5} and Figures \ref{ch5:fig_reverse_CO_6} show the $P$-$T$ profiles and equilibrium chemical abundances similar to Fig. \ref{ch5:fig_reverse_CO_1} and \ref{ch5:fig_reverse_CO_2}, respectively, but at a C/O ratio of 1.5. At a C/O ratio of 1.5 differences can again be seen in the $P$-$T$ profiles and abundances. However, in this case since the C/O ratio is greater than 1, the abundances are limited by the O/H abundance.  Therefore, the abundance of carbon bearing species is larger when C/H is varied as compared to when O/H is varied at C/O ratios > 1. However, it must be noted that the effect of changing $P$-$T$ profiles is also embedded in this difference. 

In this section we presented the differences in the $P$-$T$ profiles and the spectra, at extreme values of C/O ratio parameter space in the grid (0.35 and 1.5) by adopting two different approaches, one by varying O/H and other by varying C/H. Although there are some differences in the results obtained using these two different methodologies, in the parameter space we consider, they are smaller compared to the effects of other parameters in the grid and other model choices (for e.g. local or rainout condensation). The effects of varying C/O ratio by varying O/H across the parameter space is shown in Section 5.3.3 of the main paper.

\begin{figure}
 \includegraphics[width=\columnwidth]{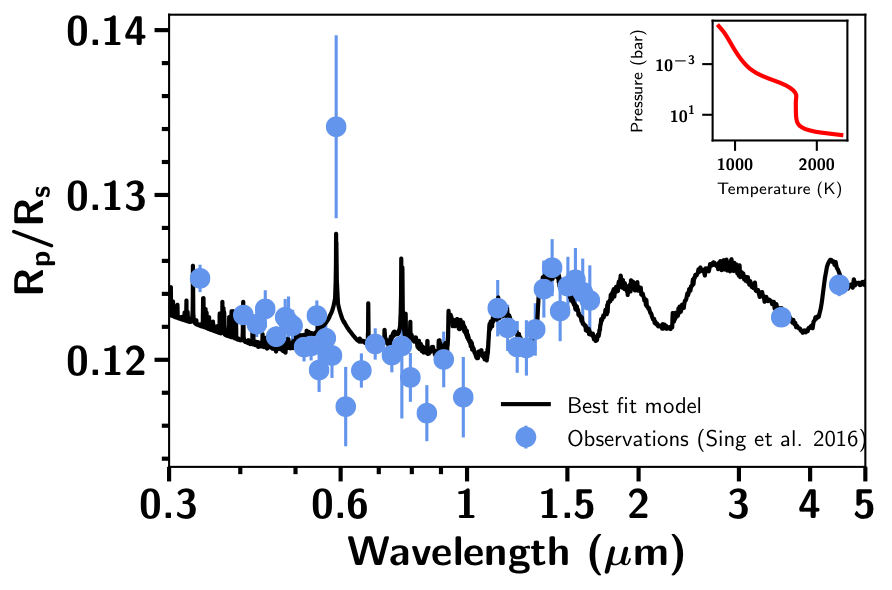}
 \caption{Figure showing best fit model transmission spectra using the grid of model transmission spectra for WASP-17b presented in this paper and observations from \citet{Sing2016} giving $\chi^2$ value of 64.64 (reduced $\chi^2$ = 1.7)  with f$_{\textup{c}}$ = 0.25, solar metallicity and C/O ratio of 0.35 (sub-solar). Best-Fit $P$-$T$ profile is shown in the subplot.} 
 \label{ch5:best_fit_wasp17}
\end{figure}

\begin{table*}
	\centering	
	\begin{tabular}{|p{2cm}|p{4cm}|p{4cm}l}    
		\hline
		Species & Line list & Partition Function \\
		\hline
		H$_2$O & \citet{Barber2006} & \citet{Barber2006} \\
		CO$_2$ & \citet{Tashkun2011} & \citet{Rothman2009} \\
		CO & \citet{Rothman2010} & \citet{Rothman2009}\\
		CH$_4$ & \citet{Yurchenko2014} & \citet{Yurchenko2014} \\
		NH$_3$ & \citet{Yurchenko2011} & \citet{Sauval1984} \\
		Na & VALD3$^1$ & \citet{Sauval1984} \\ 
 		K & VALD3$^1$ & \citet{Sauval1984} \\
		Li & VALD3$^1$ & \citet{Sauval1984} \\
		Rb & VALD3$^1$ & \citet{Sauval1984} \\
		Cs & VALD3$^1$ & \citet{Sauval1984} \\
		Fe & VALD3$^1$ & \citet{Sauval1984} \\
		TiO & \citet{Plez1998} & \citet{Sauval1984} \\
		VO & \citet{McKemmish2016} & \citet{Sauval1984} \\
		FeH & \citet{Fehlinelist2010} & \citet{Fehlinelist2010} \\
	        PH$_3$ & \citet{Ph3linelist2014} & \citet{Ph3linelist2014} \\
	        HCN & \citet{Harris2006} & \citet{Harris2006} \\
		        & \citet{Barber2014} & \citet{Barber2014} \\
		C$_2$H$_2$ & \citet{Rothman2013} & \citet{Rothman2013} \\
		H$_2$S & \citet{Rothman2013} & \citet{Rothman2013} \\
		SO$_2$ & \citet{Underwood2016} & \citet{Underwood2016} \\
		H$_2$-H$_2$ CIA & \citet{Ciahitranpaper2012} & N/A \\
		H$_2$-He CIA & \citet{Ciahitranpaper2012} & N/A \\
		H$^-$ & \citet{John1988, Bell1987} (Analytical) & N/A \\
		\hline
	\end{tabular}
	\centering
	\caption{Line list (opacity) sources of different species used in this library. Pressure broadening sources are shown in Table  \ref{table:Broadening Source}. \newline $^1$\citet{Heiter2008, Heiter2015, Ryabchikova2015} (\url{http://vald.astro.uu.se/~vald/php/vald.php}).} 
	\label{table:Linelist}
\end{table*}

\label{app:Pressure Broadening Sources}
\begin{table*}
	\centering	
	\begin{tabular}{|p{2cm}|p{2cm}|p{8cm}lp{4cm}l}        
		\hline
		Molecule & Broadener & Line Width Source & Exponent Source \\
	        \hline
		& H$_2$ & \citet{Gamache1996} & \citet{Gamache1996}  \\ [-1ex]\raisebox{1.5ex}{H$_2$O} & He & \citet{Solodov2009,Steyert2004} & \citet{Gamache1996}  \\
		\hline
	       & H$_2$ & \citet{Padmanabhan2014} & \citet{Sharp2007}  \\ [-1ex]\raisebox{1.5ex}{CO$_2$} & He & \citet{Thibault1992} & \citet{Thibault2000}  \\       
	       \hline
	       & H$_2$ & \citet{Regalia2005} & \citet{Lemoal1986}  \\ [-1ex]\raisebox{1.5ex}{CO} & He & \citet{Belbruno1982, Mantz2005} & \citet{Mantz2005}  \\
	       \hline
	       & H$_2$ & \citet{Pine1992, Margolis1993} & \citet{Margolis1993}  \\ [-1ex]\raisebox{1.5ex}{CH$_4$} & He & \citet{Pine1992} & \citet{Varanasi1990}  \\
	       \hline
	       & H$_2$ & \citet{Hadded2001, Pine1993} & \citet{Nouri2004}  \\ [-1ex]\raisebox{1.5ex}{NH$_3$} & He & \citet{Hadded2001, Pine1993} & \citet{Sharp2007}  \\
	       \hline
	       & H$_2$ & \citet{Allard1999, Allard2003, Allard2007} & \citet{Sharp2007}  \\ [-1ex]\raisebox{1.5ex}{Na} & He & \citet{Allard1999, Allard2003, Allard2007} & \citet{Sharp2007}  \\
	       \hline
	       & H$_2$ & \citet{Allard1999, Allard2003, Allard2007} & \citet{Sharp2007}  \\ [-1ex]\raisebox{1.5ex}{K} & He & \citet{Allard1999, Allard2003, Allard2007} & \citet{Sharp2007}  \\
	       \hline
	       & H$_2$ & \citet{Allard1999} & \citet{Sharp2007}  \\ [-1ex]\raisebox{1.5ex}{Li, Rb, Cs} & He & \citet{Allard1999} & \citet{Sharp2007}  \\
	       \hline
	       & H$_2$ & \citet{Sharp2007} & \citet{Sharp2007}  \\ [-1ex]\raisebox{1.5ex}{TiO, VO} & He & \citet{Sharp2007} & \citet{Sharp2007}  \\
	       \hline
	       & H$_2$ & \citet{Sharp2007} & \citet{Sharp2007}  \\ [-1ex]\raisebox{1.5ex}{FeH, Fe} & He & \citet{Sharp2007} & \citet{Sharp2007}  \\       
	       \hline
	       & H$_2$ & \citet{Bouanich2004} & \citet{Levy1994}  \\ [-1ex]\raisebox{1.5ex}{PH$_3$} & He & \citet{Salem2005} & \citet{Levy1994}  \\
	       \hline
	       & H$_2$ & \citet{Landrain1997} & \citet{Sharp2007}  \\ [-1ex]\raisebox{1.5ex}{HCN} & He & \citet{Landrain1997} & \citet{Sharp2007}  \\
	       \hline
	       {C$_2$H$_2$,H$_2$S, SO$_2$}  & Air & \citet{Rothman2009} & \citet{Rothman2009}\\
		\hline
	\end{tabular}
	\centering
	\caption{Type and source of pressure broadening for all opacities used in this library.}
	\label{table:Broadening Source}
\end{table*}

\begin{landscape}
\label{tab:tepcat_table}
\begin{table}
\begin{tabular}{|p{1.5cm}|p{1cm}|p{1.25cm}|p{0.75cm}|p{0.75cm}|p{0.75cm}|p{0.75cm}|p{0.75cm}|p{0.75cm}|p{0.75cm}|p{0.75cm}|p{0.75cm}|p{4cm}|p{4cm}|}
\hline
System&T$_{\textup{star}}$ &[Fe/H]$_{\textup{star}}$ &M$_{\textup{star}}$ &R$_{\textup{star}}$ &logg$_{\textup{star}}$ & a &M$_\textup{p}$&R$_\textup{p}$&g$_\textup{p}$&Teq$_\textup{p}$&V$_{\textup{mag}}$&Discovery Paper&Updated Reference\\
\hline
&(K)&&(M$_{\textup{sun}}$)&(R$_{\textup{sun}}$)&($m/s^2$)&(AU)&(M$_{\textup{jup}}$)&(R$_{\textup{jup}}$)&($m/s^2$)&(K)&&&\\
\hline
55-Cnc-e	&	5196.00	&	0.31	&	0.91	&	0.94	&	4.43	&	0.02	&	0.03	&	0.17	&	21.40	&	2349.00	&	5.95	&	\citet{Winn:2011aa}	&	\citet{Demory:2016aa}	\\
HAT-P-01	&	5975.00	&	0.13	&	1.15	&	1.17	&	4.36	&	0.06	&	0.53	&	1.32	&	7.46	&	1322.00	&	10.40	&	\citet{Bakos:2007aa}	&	\citet{Nikolov:2014aa}	\\
HAT-P-04	&	6036.00	&	0.28	&	1.27	&	1.60	&	4.13	&	0.04	&	0.68	&	1.34	&	9.42	&	1691.00	&	11.00	&	\citet{Kovacs:2007aa}	&	\citet{Southworth:2011aa}	\\
HAT-P-06	&	6570.00	&	-0.13	&	1.29	&	1.52	&	4.19	&	0.05	&	1.06	&	1.40	&	13.50	&	1704.00	&	10.54	&	\citet{Noyes:2008aa}	&	\citet{Southworth:2012aa}	\\
HAT-P-13	&	5653.00	&	0.41	&	1.32	&	1.76	&	4.07	&	0.04	&	0.91	&	1.49	&	10.15	&	1725.00	&	10.62	&	\citet{Bakos:2009aa}	&	\citet{Southworth:2012ab}	\\
HAT-P-25	&	5500.00	&	0.31	&	1.01	&	0.96	&	4.48	&	0.05	&	0.57	&	1.19	&	10.00	&	1202.00	&	13.19	&	\citet{Quinn:2012aa}	&	\citet{Quinn:2012aa}	\\
HAT-P-30	&	6338.00	&	0.12	&	1.24	&	1.22	&	4.36	&	0.04	&	0.71	&	1.34	&	9.80	&	1630.00	&	10.36	&	\citet{Johnson:2011aa}	&	\citet{Johnson:2011aa}	\\
HAT-P-32	&	6269.00	&	-0.04	&	1.18	&	1.23	&	4.33	&	0.03	&	0.80	&	1.81	&	6.00	&	1801.00	&	11.29	&	\citet{Hartman:2011ac}	&	\citet{Tregloan-Reed2018}	\\
HAT-P-33	&	6446.00	&	0.07	&	1.38	&	1.64	&	4.15	&	0.05	&	0.76	&	1.69	&	6.60	&	1782.00	&	11.19	&	\citet{Hartman:2011ac}	&	\citet{Hartman:2011ac}	\\
HAT-P-39	&	6340.00	&	0.19	&	1.40	&	1.63	&	4.16	&	0.05	&	0.60	&	1.57	&	5.90	&	1752.00	&	12.42	&	\citet{Hartman:2012aa}	&	\citet{Hartman:2012aa}	\\
HAT-P-40	&	6080.00	&	0.22	&	1.51	&	2.21	&	3.93	&	0.06	&	0.61	&	1.73	&	5.13	&	1770.00	&	11.34	&	\citet{Hartman:2012aa}	&	\citet{Hartman:2012aa}	\\
HAT-P-41	&	6390.00	&	0.21	&	1.42	&	1.68	&	4.14	&	0.04	&	0.80	&	1.69	&	6.90	&	1941.00	&	11.36	&	\citet{Hartman:2012aa}	&	\citet{Hartman:2012aa}	\\
HAT-P-45	&	6330.00	&	0.07	&	1.26	&	1.32	&	4.30	&	0.05	&	0.89	&	1.43	&	10.70	&	1652.00	&	12.79	&	\citet{Hartman:2014aa}	&	\citet{Hartman:2014aa}	\\
HAT-P-46	&	6120.00	&	0.30	&	1.28	&	1.40	&	4.25	&	0.06	&	0.49	&	1.28	&	7.30	&	1458.00	&	11.94	&	\citet{Hartman:2014aa}	&	\citet{Hartman:2014aa}	\\
HAT-P-47	&	6703.00	&	0.00	&	1.39	&	1.51	&	4.22	&	0.06	&	0.21	&	1.31	&	2.95	&	1605.00	&	10.69	&	\citet{Bakos:2016aa}	&	\citet{Bakos:2016aa}	\\
HAT-P-48	&	5946.00	&	0.02	&	1.10	&	1.22	&	4.30	&	0.05	&	0.17	&	1.13	&	3.24	&	1361.00	&	12.16	&	\citet{Bakos:2016aa}	&	\citet{Bakos:2016aa}	\\
HAT-P-65	&	5835.00	&	0.10	&	1.21	&	1.86	&	3.98	&	0.04	&	0.53	&	1.89	&	3.63	&	1930.00	&	13.15	&	\citet{Hartman:2016aa}	&	\citet{Hartman:2016aa}	\\
HATS-19	&	5896.00	&	0.24	&	1.30	&	1.75	&	4.07	&	0.06	&	0.43	&	1.66	&	3.80	&	1570.00	&	13.03	&	\citet{Bhatti:2016aa}	&	\citet{Bhatti:2016aa}	\\
HATS-21	&	5695.00	&	0.30	&	1.08	&	1.02	&	4.45	&	0.05	&	0.33	&	1.12	&	6.50	&	1284.00	&	12.19	&	\citet{Bhatti:2016aa}	&	\citet{Bhatti:2016aa}	\\
HD-149026	&	6147.00	&	0.36	&	1.34	&	1.54	&	4.19	&	0.04	&	0.37	&	0.81	&	13.55	&	1634.00	&	8.16	&	\citet{Sato:2005aa}	&	\citet{Carter:2009aa}	\\
HD-209458	&	6117.00	&	0.02	&	1.15	&	1.16	&	4.37	&	0.05	&	0.71	&	1.38	&	9.30	&	1459.00	&	7.65	&	\citet{Henry:2000aa}	&	\citet{Southworth:2010aa}	\\
KELT-04	&	6206.00	&	-0.12	&	1.20	&	1.60	&	4.11	&	0.04	&	0.90	&	1.70	&	7.74	&	1823.00	&	10.47	&	\citet{Eastman:2016aa}	&	\citet{Eastman:2016aa}	\\
KELT-06	&	6102.00	&	-0.28	&	1.08	&	1.58	&	4.07	&	0.08	&	0.43	&	1.19	&	7.40	&	1313.00	&	10.42	&	\citet{Collins:2014}	&	\citet{Collins:2014}	\\
KELT-07	&	6789.00	&	0.14	&	1.53	&	1.73	&	4.15	&	0.04	&	1.28	&	1.53	&	13.50	&	2048.00	&	8.54	&	\citet{Bieryla:2015aa}	&	\citet{Bieryla:2015aa}	\\
KELT-08	&	5754.00	&	0.27	&	1.21	&	1.67	&	4.08	&	0.05	&	0.87	&	1.86	&	6.20	&	1675.00	&	10.83	&	\citet{Fulton:2015aa}	&	\citet{Fulton:2015aa}	\\
KELT-09	&	10170.00	&	-0.03	&	2.52	&	2.36	&	4.09	&	0.03	&	2.88	&	1.89	&	20.00	&	4050.00	&	7.56	&	\citet{Gaudi:2017}	&	\citet{Gaudi:2017}	\\
KELT-10	&	5948.00	&	0.09	&	1.11	&	1.21	&	4.32	&	0.05	&	0.68	&	1.40	&	8.57	&	1377.00	&	10.70	&	\citet{Kuhn:2016aa}	&	\citet{Kuhn:2016aa}	\\
KELT-11	&	5375.00	&	0.17	&	1.44	&	2.69	&	3.70	&	0.06	&	0.17	&	1.35	&	2.55	&	1712.00	&	8.03	&	\citet{Pepper:2017aa}	&	\citet{Beatty2017}	\\
KELT-15	&	6003.00	&	0.05	&	1.18	&	1.48	&	4.23	&	0.05	&	1.20	&	1.52	&	12.80	&	1904.00	&	11.44	&	\citet{Rodriguez:2016aa}	&	\citet{Rodriguez:2016aa}	\\
KELT-17	&	7454.00	&	-0.02	&	1.64	&	1.65	&	4.22	&	0.05	&	1.31	&	1.52	&	13.90	&	2087.00	&	9.29	&	\citet{Zhou:2016aa}	&	\citet{Zhou:2016aa}	\\
KELT-18	&	6670.00	&	0.09	&	1.52	&	1.91	&	4.06	&	0.05	&	1.18	&	1.57	&	11.80	&	2085.00	&	10.21	&	\citet{McLeod2017}	&	\citet{McLeod2017}	\\
Kepler-012	&	5947.00	&	0.07	&	1.16	&	1.49	&	4.16	&	0.06	&	0.43	&	1.71	&	3.66	&	1485.00	&	13.53	&	\citet{Fortney:2011aa}	&	\citet{Southworth:2012aa}	\\
TrES-4	&	6295.00	&	0.28	&	1.45	&	1.83	&	4.09	&	0.05	&	0.49	&	1.84	&	2.82	&	1795.00	&	11.59	&	\citet{Mandushev:2007aa}	&	\citet{Sozzetti:2015aa}	\\
WASP-001	&	6160.00	&	0.14	&	1.24	&	1.47	&	4.20	&	0.04	&	0.85	&	1.48	&	9.80	&	1830.00	&	11.31	&	\citet{Collier-Cameron:2007aa}	&	\citet{Maciejewski:2014aa}	\\
WASP-002	&	5170.00	&	0.04	&	0.85	&	0.82	&	4.54	&	0.03	&	0.88	&	1.06	&	19.31	&	1286.00	&	11.98	&	\citet{Collier-Cameron:2007aa}	&	\citet{Southworth:2012aa}	\\
WASP-004	&	5540.00	&	-0.03	&	0.93	&	0.91	&	4.49	&	0.02	&	1.25	&	1.36	&	16.64	&	1673.00	&	12.46	&	\citet{Wilson:2008aa}	&	\citet{Southworth:2012aa}	\\
WASP-007	&	6520.00	&	0.00	&	1.32	&	1.48	&	4.22	&	0.06	&	0.98	&	1.37	&	12.90	&	1530.00	&	9.48	&	\citet{Hellier:2009aa}	&	\citet{Southworth:2012aa}	\\
WASP-012	&	6313.00	&	0.21	&	1.43	&	1.66	&	4.16	&	0.02	&	1.47	&	1.90	&	10.09	&	2580.00	&	11.69	&	\citet{Hebb:2009aa}	&	\citet{Collins:2015aa}	\\
WASP-013	&	6025.00	&	0.11	&	1.22	&	1.66	&	4.09	&	0.06	&	0.51	&	1.53	&	5.44	&	1531.00	&	10.51	&	\citet{Skillen:2009aa}	&	\citet{Southworth:2012aa}	\\
\hline
\end{tabular}
\end{table}
\end{landscape}

\begin{landscape}
\begin{table}
\begin{tabular}{|p{1.5cm}|p{1cm}|p{1.25cm}|p{0.75cm}|p{0.75cm}|p{0.75cm}|p{0.75cm}|p{0.75cm}|p{0.75cm}|p{0.75cm}|p{0.75cm}|p{0.75cm}|p{4cm}|p{4cm}|}
\hline
System&T$_{\textup{star}}$ &[Fe/H]$_{\textup{star}}$ &M$_{\textup{star}}$ &R$_{\textup{star}}$ &logg$_{\textup{star}}$ & a &M$_\textup{p}$&R$_\textup{p}$&g$_\textup{p}$&Teq$_\textup{p}$&V$_{\textup{mag}}$&Discovery Paper&Updated Reference\\
\hline
&(K)&&(M$_{\textup{sun}}$)&(R$_{\textup{sun}}$)&($m/s^2$)&(AU)&(M$_{\textup{jup}}$)&(R$_{\textup{jup}}$)&($m/s^2$)&(K)&&&\\
\hline
WASP-015	&	6573.00	&	0.09	&	1.30	&	1.52	&	4.19	&	0.05	&	0.59	&	1.41	&	7.39	&	1676.00	&	10.92	&	\citet{West:2009aa}	&	\citet{Southworth:2013aa}	\\
WASP-016	&	5630.00	&	0.07	&	0.98	&	1.09	&	4.36	&	0.04	&	0.83	&	1.22	&	13.92	&	1389.00	&	11.31	&	\citet{Lister:2009aa}	&	\citet{Southworth:2013aa}	\\
WASP-017	&	6550.00	&	-0.25	&	1.29	&	1.58	&	4.15	&	0.05	&	0.48	&	1.93	&	3.16	&	1755.00	&	11.50	&	\citet{Anderson:2010aa}	&	\citet{Southworth:2012ac}	\\
WASP-019	&	5460.00	&	0.14	&	0.94	&	1.02	&	4.39	&	0.02	&	1.14	&	1.41	&	14.21	&	2077.00	&	12.31	&	\citet{Hebb:2010aa}	&	\citet{Mancini:2013aa}	\\
WASP-020	&	6000.00	&	-0.01	&	1.09	&	1.14	&	4.36	&	0.06	&	0.38	&	1.28	&	5.80	&	1282.00	&	10.68	&	\citet{Anderson:2015aa}	&	\citet{Evans:2016aa}	\\
WASP-021	&	5924.00	&	-0.22	&	0.89	&	1.14	&	4.28	&	0.05	&	0.28	&	1.16	&	5.07	&	1333.00	&	11.59	&	\citet{Bouchy:2010aa}	&	\citet{Ciceri:2013aa}	\\
WASP-025	&	5736.00	&	0.06	&	1.05	&	0.92	&	4.53	&	0.05	&	0.60	&	1.25	&	9.54	&	1210.00	&	11.85	&	\citet{Enoch:2011aa}	&	\citet{Southworth:2014aa}	\\
WASP-031	&	6175.00	&	-0.20	&	1.16	&	1.25	&	4.31	&	0.05	&	0.48	&	1.55	&	4.56	&	1575.00	&	11.94	&	\citet{Anderson:2011aa}	&	\citet{Anderson:2011aa}	\\
WASP-033	&	7430.00	&	0.10	&	1.56	&	1.51	&	4.27	&	0.03	&	2.16	&	1.68	&	19.00	&	2710.00	&	8.18	&	\citet{Collier-Cameron:2010aa}	&	\citet{Lehmann:2015aa}	\\
WASP-034	&	5704.00	&	0.08	&	1.01	&	0.93	&	4.50	&	0.05	&	0.59	&	1.22	&	9.10	&	1250.00	&	10.37	&	\citet{Smalley:2011aa}	&	\citet{Smalley:2011aa}	\\
WASP-035	&	6072.00	&	-0.05	&	1.07	&	1.09	&	4.40	&	0.04	&	0.72	&	1.32	&	9.50	&	1450.00	&	10.95	&	\citet{Enoch:2011ab}	&	\citet{Enoch:2011ab}	\\
WASP-041	&	5546.00	&	0.06	&	0.99	&	0.89	&	4.54	&	0.04	&	0.98	&	1.18	&	17.45	&	1242.00	&	11.64	&	\citet{Maxted:2011aa}	&	\citet{Southworth:2016aa}	\\
WASP-043	&	4520.00	&	-0.01	&	0.72	&	0.67	&	4.64	&	0.02	&	2.03	&	1.04	&	47.00	&	1440.00	&	12.37	&	\citet{Hellier:2011aa}	&	\citet{Gillon:2012aa}	\\
WASP-049	&	5600.00	&	-0.23	&	1.00	&	1.04	&	4.41	&	0.04	&	0.40	&	1.20	&	7.13	&	1399.00	&	11.36	&	\citet{Lendl:2012aa}	&	\citet{Lendl:2016aa}	\\
WASP-052	&	5000.00	&	0.03	&	0.80	&	0.79	&	4.55	&	0.03	&	0.43	&	1.25	&	6.85	&	1315.00	&	12.20	&	\citet{Hebrard:2013aa}	&	\citet{Mancini:2017aa}	\\
WASP-054	&	6296.00	&	0.00	&	1.21	&	1.83	&	4.00	&	0.05	&	0.64	&	1.65	&	5.32	&	1759.00	&	10.42	&	\citet{Faedi:2013aa}	&	\citet{Faedi:2013aa}	\\
WASP-055	&	6070.00	&	0.09	&	1.16	&	1.10	&	4.42	&	0.06	&	0.63	&	1.33	&	8.73	&	1300.00	&	11.76	&	\citet{Hellier:2012aa}	&	\citet{Southworth:2016aa}	\\
WASP-058	&	5800.00	&	-0.45	&	0.94	&	1.17	&	4.27	&	0.06	&	0.89	&	1.37	&	10.70	&	1270.00	&	11.66	&	\citet{Hebrard:2013aa}	&	\citet{Hebrard:2013aa}	\\
WASP-062	&	6230.00	&	0.04	&	1.25	&	1.28	&	4.32	&	0.06	&	0.57	&	1.39	&	6.76	&	1440.00	&	10.22	&	\citet{Hellier:2012aa}	&	\citet{Hellier:2012aa}	\\
WASP-063	&	5715.00	&	0.28	&	1.32	&	1.88	&	4.01	&	0.06	&	0.38	&	1.43	&	4.17	&	1540.00	&	11.16	&	\citet{Hellier:2012aa}	&	\citet{Hellier:2012aa}	\\
WASP-070	&	5700.00	&	-0.01	&	1.11	&	1.22	&	4.31	&	0.05	&	0.59	&	1.16	&	10.00	&	1387.00	&	10.79	&	\citet{Anderson:2014ac}	&	\citet{Anderson:2014ac}	\\
WASP-074	&	5990.00	&	0.39	&	1.48	&	1.64	&	4.18	&	0.04	&	0.95	&	1.56	&	8.91	&	1910.00	&	9.76	&	\citet{Hellier:2015aa}	&	\citet{Hellier:2015aa}	\\
WASP-076	&	6250.00	&	0.23	&	1.46	&	1.73	&	4.13	&	0.03	&	0.92	&	1.83	&	6.31	&	2160.00	&	9.53	&	\citet{West:2016aa}	&	\citet{West:2016aa}	\\
WASP-078	&	6100.00	&	-0.35	&	1.39	&	2.35	&	3.84	&	0.04	&	0.86	&	2.06	&	4.69	&	2470.00	&	11.96	&	\citet{Smalley:2012aa}	&	\citet{Brown:2017}	\\
WASP-079	&	6600.00	&	0.03	&	1.39	&	1.51	&	4.23	&	0.05	&	0.86	&	1.53	&	8.39	&	1716.00	&	10.04	&	\citet{Smalley:2012aa}	&	\citet{Brown:2017}	\\
WASP-081	&	5890.00	&	-0.36	&	1.08	&	1.28	&	4.26	&	0.04	&	0.73	&	1.43	&	9.27	&	1623.00	&	12.29	&	\citet{Triaud:2017}	&	\citet{Triaud:2017}	\\
WASP-082	&	6500.00	&	0.12	&	1.64	&	2.22	&	3.96	&	0.04	&	1.25	&	1.71	&	9.75	&	2202.00	&	10.08	&	\citet{West:2016aa}	&	\citet{Smith:2015aa}	\\
WASP-088	&	6430.00	&	-0.08	&	1.45	&	2.08	&	3.96	&	0.06	&	0.56	&	1.70	&	4.68	&	1772.00	&	11.39	&	\citet{Delrez:2014aa}	&	\citet{Delrez:2014aa}	\\
WASP-090	&	6440.00	&	0.11	&	1.55	&	1.98	&	4.03	&	0.06	&	0.63	&	1.63	&	5.37	&	1840.00	&	11.69	&	\citet{West:2016aa}	&	\citet{West:2016aa}	\\
WASP-093	&	6700.00	&	0.07	&	1.33	&	1.52	&	4.20	&	0.04	&	1.47	&	1.60	&	13.20	&	1942.00	&	10.97	&	\citet{Hay:2016aa}	&	\citet{Hay:2016aa}	\\
WASP-094	&	6170.00	&	0.26	&	1.45	&	1.62	&	4.18	&	0.06	&	0.45	&	1.72	&	3.48	&	1604.00	&	10.06	&	\citet{Neveu-VanMalle:2014aa}	&	\citet{Neveu-VanMalle:2014aa}	\\
WASP-095	&	5830.00	&	0.14	&	1.11	&	1.13	&	4.38	&	0.03	&	1.13	&	1.21	&	21.80	&	1570.00	&	10.09	&	\citet{Hellier:2014aa}	&	\citet{Hellier:2014aa}	\\
WASP-096	&	5500.00	&	0.14	&	1.06	&	1.05	&	4.42	&	0.05	&	0.48	&	1.20	&	7.59	&	1285.00	&	12.19	&	\citet{Hellier:2014aa}	&	\citet{Hellier:2014aa}	\\
WASP-097	&	5670.00	&	0.23	&	1.12	&	1.06	&	4.43	&	0.03	&	1.32	&	1.13	&	23.40	&	1555.00	&	10.58	&	\citet{Hellier:2014aa}	&	\citet{Hellier:2014aa}	\\
WASP-101	&	6380.00	&	0.20	&	1.34	&	1.29	&	4.34	&	0.05	&	0.50	&	1.41	&	5.75	&	1560.00	&	10.34	&	\citet{Hellier:2014aa}	&	\citet{Hellier:2014aa}	\\
WASP-103	&	6110.00	&	0.06	&	1.21	&	1.41	&	4.22	&	0.02	&	1.47	&	1.65	&	14.34	&	2489.00	&	12.39	&	\citet{Gillon:2014aa}	&	\citet{Southworth:2016ab}	\\
WASP-108	&	6000.00	&	0.05	&	1.17	&	1.22	&	4.34	&	0.04	&	0.89	&	1.28	&	12.39	&	1590.00	&	11.22	&	\citet{Anderson:2014ab}	&	\citet{Anderson:2014ab}	\\
WASP-109	&	6520.00	&	-0.22	&	1.20	&	1.35	&	4.26	&	0.05	&	0.91	&	1.44	&	10.00	&	1685.00	&	11.44	&	\citet{Anderson:2014ab}	&	\citet{Anderson:2014ab}	\\
\hline
\end{tabular}
\end{table}
\end{landscape}

\begin{landscape}
\begin{table}
\begin{tabular}{|p{1.5cm}|p{1cm}|p{1.25cm}|p{0.75cm}|p{0.75cm}|p{0.75cm}|p{0.75cm}|p{0.75cm}|p{0.75cm}|p{0.75cm}|p{0.75cm}|p{0.75cm}|p{4cm}|p{4cm}|}
\hline
System&T$_{\textup{star}}$ &[Fe/H]$_{\textup{star}}$ &M$_{\textup{star}}$ &R$_{\textup{star}}$ &logg$_{\textup{star}}$ & a &M$_\textup{p}$&R$_\textup{p}$&g$_\textup{p}$&Teq$_\textup{p}$&V$_{\textup{mag}}$&Discovery Paper&Updated Reference\\
\hline
&(K)&&(M$_{\textup{sun}}$)&(R$_{\textup{sun}}$)&($m/s^2$)&(AU)&(M$_{\textup{jup}}$)&(R$_{\textup{jup}}$)&($m/s^2$)&(K)&&&\\
\hline
WASP-113	&	5890.00	&	0.10	&	1.32	&	1.61	&	4.20	&	0.06	&	0.47	&	1.41	&	5.50	&	1496.00	&	11.77	&	\citet{Barros:2016aa}	&	\citet{Barros:2016aa}	\\
WASP-118	&	6410.00	&	0.16	&	1.32	&	1.75	&	4.10	&	0.05	&	0.52	&	1.39	&	5.71	&	1753.00	&	11.02	&	\citet{Hay:2016aa}	&	\citet{wasp118}	\\
WASP-121	&	6460.00	&	0.13	&	1.35	&	1.46	&	4.24	&	0.03	&	1.18	&	1.86	&	9.40	&	2358.00	&	10.52	&	\citet{Delrez:2016aa}	&	\citet{Delrez:2016aa}	\\
WASP-122	&	5720.00	&	0.32	&	1.24	&	1.52	&	4.17	&	0.03	&	1.28	&	1.74	&	9.66	&	1970.00	&	11.00	&	\citet{Turner:2016aa}	&	\citet{Turner:2016aa}	\\
WASP-123	&	5740.00	&	0.18	&	1.17	&	1.28	&	4.29	&	0.04	&	0.90	&	1.32	&	11.70	&	1520.00	&	11.03	&	\citet{Turner:2016aa}	&	\citet{Turner:2016aa}	\\
WASP-124	&	6050.00	&	-0.02	&	1.07	&	1.02	&	4.44	&	0.04	&	0.60	&	1.24	&	8.90	&	1400.00	&	12.70	&	\citet{Maxted:2016aa}	&	\citet{Maxted:2016aa}	\\
WASP-126	&	5800.00	&	0.17	&	1.12	&	1.27	&	4.28	&	0.04	&	0.28	&	0.96	&	6.80	&	1480.00	&	11.20	&	\citet{Maxted:2016aa}	&	\citet{Maxted:2016aa}	\\
WASP-127	&	5750.00	&	-0.18	&	1.08	&	1.39	&	4.18	&	0.05	&	0.18	&	1.37	&	2.14	&	1400.00	&	10.16	&	\citet{Lam:2017aa}	&	\citet{Lam:2017aa}	\\
WASP-131	&	5950.00	&	-0.18	&	1.06	&	1.53	&	4.09	&	0.06	&	0.27	&	1.22	&	4.17	&	1460.00	&	10.08	&	\citet{Hellier:2017aa}	&	\citet{Hellier:2017aa}	\\
WASP-140	&	5300.00	&	0.12	&	0.90	&	0.87	&	4.51	&	0.03	&	2.44	&	1.44	&	25.00	&	1320.00	&	11.13	&	\citet{Hellier:2017aa}	&	\citet{Hellier:2017aa}	\\
XO-1	&	5750.00	&	0.02	&	1.04	&	0.94	&	4.51	&	0.05	&	0.92	&	1.21	&	15.80	&	1210.00	&	11.14	&	\citet{McCullough:2006aa}	&	\citet{Southworth:2010aa}	\\
XO-2	&	5332.00	&	0.43	&	0.96	&	1.00	&	4.44	&	0.04	&	0.60	&	1.02	&	14.13	&	1328.00	&	11.25	&	\citet{Burke:2007aa}	&	\citet{Damasso:2015aa}	\\
\hline
\end{tabular}
\caption{All the stellar and planetary parameters adopted from TEPCat \citep{Tepcat2011} database, for the model simulations of 89 exoplanets in the grid are listed here. First column shows planet names with 'b' omitted indicating first planet of the stellar system as in TEPCat database. Subsequent columns show, stellar temperature (T$_{\textup{star}}$) in Kelvin, stellar metallicity ([Fe/H]$_{\textup{star}}$ ), stellar mass (M$_{\textup{star}}$ ) in units of solar mass,  stellar radius (R$_{\textup{star}}$ ) in units of solar radius, logarithmic (base 10) stellar gravity (logg$_{\textup{star}}$ ) in $m/s^2$, semi-major axis (a) in AU, planetary mass (M$_\textup{p}$) in units of Jupiter mass,  planetary radius (R$_\textup{p}$) in units of Jupiter radius, planetary surface gravity (g$_\textup{p}$) in $m/s^2$, planetary equilibrium temperature (Teq$_\textup{p}$) in Kelvin assuming 0 albedo and efficient redistribution, V magnitude (V$_{\textup{mag}}$) of the host star, discovery paper reference (Discovery Paper) and finally the most updated reference.}
\end{table}
\end{landscape}


%
%
\bibliographystyle{mnras}
\bibliography{consistent_grid_review}


\bsp	
\label{lastpage}